\documentclass[a4 paper,11pt]{article}

\usepackage{fancyhdr}
\usepackage{latexsym}
\usepackage{mathrsfs}
\usepackage{cancel}
\usepackage{color}
\usepackage{multirow}
\usepackage{eso-pic}
\usepackage{dsfont}
\usepackage{amssymb}
\usepackage{graphicx}
\usepackage{setspace}
\usepackage{geometry}
\usepackage{enumitem}
\usepackage{amsthm}
\usepackage{hyperref}
\usepackage[intlimits]{amsmath}
\usepackage{hyperref}
\usepackage{tikz-cd}
\usepackage{amssymb,amsfonts}
\usepackage[all,arc]{xy}
\usepackage{mathrsfs}
\usepackage{tikz}
\usepackage{bm}
\usepackage{bbm}
\usepackage{graphicx}
\usetikzlibrary{braids}

\graphicspath{{./figures/}}

\newcommand{\CDCNd}{\raisebox{-.15cm}{\includegraphics[width=.5cm]{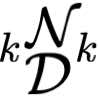}}}
\newcommand{\CDCN}{\raisebox{-.15cm}{\includegraphics[width=.5cm]{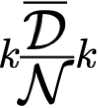}}}
\newcommand{\kDk}{{}_k\overline{\mathcal D}_k}
\newcommand{\kNk}{{}_k\overline{\mathcal N}_k}
\newcommand{\kNkd}{{}_k{\mathcal N}_k}

\newcommand{\cf}{\emph{cf.}}
\newcommand{\ie}{\emph{i.e.}}
\newcommand{\eg}{\emph{e.g.}}
\newcommand{\ol}{\overline}
\newcommand{\wt}{\widetilde}
\newcommand{\wh}{\widehat}

\newcommand{\End}{\mathrm{End}}

\newcommand{\Hom}{\mathrm{Hom}}

\newcommand{\Mod}{\text{-mod}}

\newcommand{\lp}{\left(}
\newcommand{\rp}{\right)}

\newcommand{\Ind}{\mathrm{Ind}}

\newcommand{\Coh}{\mathrm{Coh}}
\newcommand{\QCoh}{\mathrm{QCoh}}

\newcommand{\MF}{\mathrm{MF}}

\newcommand{\mb}{\mathbf}

\newcommand{\pd}{{\partial}}
\newcommand{\id}{{\mathds{1}}}

\newcommand{\AT}{{\textbf T}}
\newcommand{\AD}{{\textbf{D}}}
\newcommand{\ADline}{{\textbf{D}$_{\rm line}$}}
\newcommand{\AC}{{\textbf C}}

\newcommand{\C}{\mathbb C}
\newcommand{\R}{\mathbb R}
\newcommand{\Z}{\mathbb Z}

\newcommand{\E}{\mathbb E}

\newcommand{\fg}{\mathfrak{g}}
\newcommand{\fh}{\mathfrak{h}}

\newcommand{\CA}{{\mathcal A}}
\newcommand{\CB}{{\mathcal B}}
\newcommand{\CC}{{\mathcal C}}
\newcommand{\CD}{{\mathcal D}}
\newcommand{\CE}{{\mathcal E}}
\newcommand{\CF}{{\mathcal F}}
\newcommand{\CG}{{\mathcal G}}
\newcommand{\CH}{{\mathcal H}}
\newcommand{\CI}{{\mathcal I}}

\newcommand{\CK}{{\mathcal K}}
\newcommand{\CL}{{\mathcal L}}

\newcommand{\CN}{{\mathcal N}}
\newcommand{\CO}{{\mathcal O}}
\newcommand{\CP}{{\mathcal P}}

\newcommand{\CR}{{\mathcal R}}
\newcommand{\CS}{{\mathcal S}}
\newcommand{\CT}{{\mathcal T}}

\newcommand{\CV}{{\mathcal V}}
  
\newcommand{\CX}{{\mathcal X}}

\newcommand{\CZ}{{\mathcal Z}}

\newcommand{\be}{\begin{equation}}
\newcommand{\ee}{\end{equation}}

\newcommand{\btik}{\begin{tikzcd}}
\newcommand{\etik}{\end{tikzcd}}

\begin{document}

\title{Tannakian QFT: from spark algebras to quantum groups}

\author{Tudor Dimofte and Wenjun Niu}

\newtheorem{Def}{Definition}[section]
\newtheorem{Thm}[Def]{Theorem}
\newtheorem{Prop}[Def]{Proposition}
\newtheorem{Cor}[Def]{Corollary}
\newtheorem{Lem}[Def]{Lemma}
\newtheorem{Rem}[Def]{Remark}
\newtheorem{Asp}[Def]{Assumption}

\setcounter{tocdepth}{2}

\numberwithin{equation}{section}

\maketitle

{\abstract{We propose a nonperturbative construction of Hopf algebras that represent categories of line operators in topological quantum field theory, in terms of semi-extended operators (spark algebras) on pairs of transverse topological boundary conditions. The construction is a direct implementation of Tannakian formalism in QFT. Focusing on $d=3$ dimensional theories, we find topological definitions of R-matrices, ribbon twists, and the Drinfeld double construction for generalized quantum groups. We illustrate our construction in finite-group gauge theory, and apply it to obtain new results for B-twisted 3d $\mathcal N=4$ gauge theories, \emph{a.k.a.} equivariant Rozansky-Witten theory, or supergroup BF theory (including ordinary BF theory with compact gauge group). We reformulate our construction mathematically in terms of abelian and dg tensor categories, and discuss connections with Koszul duality.
}}

\tableofcontents

\newpage

\section{Introduction}\label{sec:Intro}

In a topological quantum field theory, local operators are well known to have the structure of an algebra $A$ over the complex numbers. The underlying vector space of $A$ is in principle straightforward to construct, from a physical perspective. For example, in $d$-dimensional QFT, $A$ may be defined (via state-operator correspondence) as the space of states on a sphere $S^{d-1}$, and there are multiple explicit ways to compute this. Similarly, the algebra multiplication $A\otimes A\to A$ is defined by the amplitude, or path integral, on a $d$-dimensional pair-of-pants. Perturbatively, it bears the same information as inserting a pair of local operators at two nearby points and quantum-correcting the product with Feynman diagrams.%
\footnote{There may be higher infinity-like products on $A$ as well, \emph{cf.} \cite{Lurie}, understood decades ago in $d=2$ \cite{WittenZwiebach,LianZuckerman,Getzler,PenkavaSchwarz}, and more recently given a physical construction for $d\geq 3$ \cite{descent}.}

Somewhat analogously, $k$-dimensional extended operators in topological QFT are expected to carry the structure of $k$-categories. The presence of extra dimensions transverse to the operators endows the categories with a tensor product and other higher operations. For example, line operators ($k=1$) in $d=2$ should form a monoidal category, and line operators ($k=1$) in $d=3$ should form a braided monoidal category. In contrast to the case of local operators, it is far more challenging to explicitly ``compute,'' or systematically describe, the categories of extended operators and their products or higher structure in a given physical QFT.

A number of methods have been developed to tackle this problem. A standard ``bottom-up'' approach is to construct and classify large collections of physical extended operators, then assemble them into a category analyzing the structure of their junctions, and their collisions. For example, in a topological gauge theory, Wilson lines are natural physical operators to consider; and in Chern-Simons theory they turn out to saturate the set of objects in the category of line operators \cite{Witten-Jones}. In topological twists of supersymmetric theories, the extended operators are much richer, and have led to intricate classification problems --- a small sample of classic work on boundaries in 2d, lines/boundaries in 3d, lines/surfaces/boundaries in 4d includes \cite{HoriIqbalVafa,Douglas-Dbranes}; \cite{RW,KRS,KapustinRozansky};  \cite{KapustinWitten,GukovWitten,GW-bc}.

Alternatively, a ``top-down'' approach --- attempting mathematically to capture an entire category of extended operators at once --- may involve higher-dimensional analogues of the state-operator correspondence and powerful techniques from representation theory. In 2d, Kontsevich's proposal \cite{Kontsevich-MS} for categories of boundary conditions in A and B models was of this flavor.
More recent 2d examples include Kapustin-Witten's description of A-branes as D-modules \cite{KapustinWitten} (related to \cite{KapustinOrlov,NadlerZaslow}) as well as Gaiotto-Moore-Witten's construction of boundary conditions in massive theories \cite{GMW,GMW-intro}, which categorified \cite{HoriIqbalVafa}.
In 3d Chern-Simons theory, introducing holomorphic boundary conditions allowed line operators to be represented as modules for the WZW vertex operator algebra \cite{Witten-Jones,MooreSeiberg-taming,EMSS}; and introducing an infinite-dimensional Wilson line \cite{Witten-vertex} gave a partial physical explanation for Reshetikhin-Turaev's proposal \cite{RT} to represent line operators as modules for the quantum group $U_q(\mathfrak g)$. More recently, Costello \cite{Costello-4dCS} proposed to use Koszul duality to represent line operators in perturbative 4d Chern-Simons theory as modules for Yangians (see also \cite{CWY-I,CWY-II}); this perspective was extended by Costello-Paquette  \cite{CostelloPaquette}, who outlined a general approach to represent line operators in perturbative, partially topological QFT's via the Koszul-dual $A^!$ of the algebra $A$ of local operators,
\be \text{perturbative line operators} \; \simeq \; A^!\text{-mod}\,. \label{CP} \ee

In this paper, we introduce a different, though closely related, method for representing line operators as modules for an algebra. Its main advantages are that the method is non-perturbative, and the algebra in question is formulated explicitly in terms of physical operators. A disadvantage is that the method requires a QFT to possess some extra structures, whose existence is far from  guaranteed. In the simplest (but most restrictive) version of our proposal, we require a $d$-dimensional topological QFT $\CT$ to admit a \emph{pair} of topological boundary conditions, denoted $\CD$ and $\CN$, such that
\begin{itemize}
\item $\CD$ and $\CN$ are \emph{transverse}, roughly meaning that $\CT$ compactified on an interval with $\CD$ on one side and $\CN$ on the other is a trivial $(d-1)$-dimensional QFT; and
\item $\CD$ and $\CN$ are \emph{complete}, roughly meaning that if $\CT$ is placed on a space with a $\CD$ boundary of topology $\Sigma$ and an $\CN$ boundary of topology $\overline\Sigma$, the two boundaries can be glued together by inserting appropriate operators in the product $\CN\circ \CD$.
\end{itemize}
\be \raisebox{-.4in}{\includegraphics[width=5.5in]{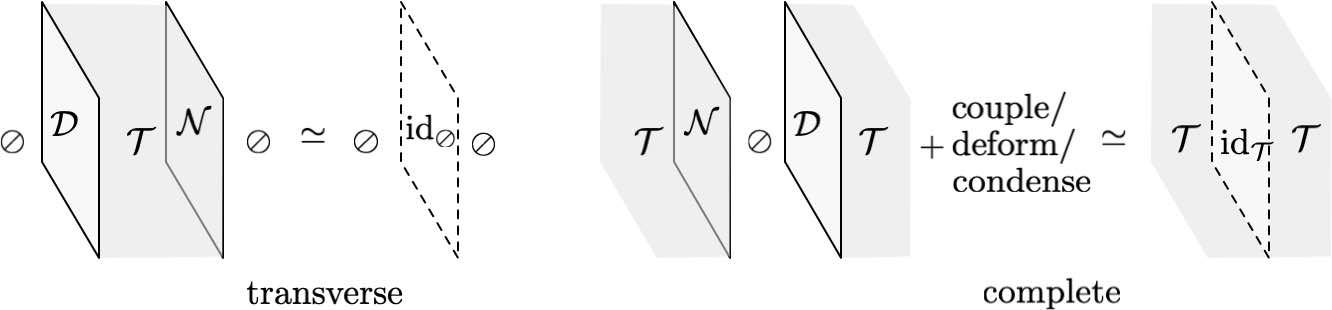}} \label{TC-intro} \ee
The assumptions are described in much greater detail in Section \ref{sec:setup}.

A heuristic example that illustrates this setup, and motivates the notation, is a topological gauge theory.
In gauge theory, one can take $\CD$ to be a Dirichlet boundary condition, trivializing gauge bundles and connections along the boundary. A transverse boundary condition $\CN$ is typically given by Neumann, which lets gauge bundles and connections along the boundary fluctuate, so long as boundary gauge anomalies vanish. Completeness --- gluing $\CN$ back to $\CD$ --- is achieved by coupling the gauge symmetry on $\CN$ to the global symmetry on~$\CD$. Topological gauge theory with $\CN$ and $\CD$ boundary conditions controls gauging and ungauging of symmetries in one dimension lower (\cf\ \cite{Teleman-ICM}), and has been central to much recent work on generalized symmetries, following general structure laid out in \cite{GKSW,FMT}.

Given a setup as in \eqref{TC-intro}, we construct three algebras as follows. We consider the bulk theory on $S^{d-2}\times [0,\epsilon)\times \R_t$, \ie\ a spacetime whose spatial slices are one-sided neighborhoods of a sphere $S^{d-2}$. The significance of this geometry is that it's the punctured neighborhood of a line extended along $\R^t$. We split the boundary of $S^{d-2}\times [0,\epsilon)\times \R_t$ into two discs,
\be \text{boundary} = S^{d-2}\times \R_t \simeq \underbrace{\big(D_{\rm left}^{d-2} \times \R_t\big)}_\CD\cup \underbrace{\big(D_{\rm right}^{d-2} \times \R_t\big)}_\CN\,, \ee
placing $\CD$ along one half and $\CN$ along the other, with a junction between them that's canonically defined due to transversality. Then we define
\be \label{HU-intro}
\begin{array}{r@{\;:=\;}l}
\CH_\CD & \{\text{operators supported in an infinitesimal nhd. of $D_{\rm left}^{d-2}$, at fixed time}\} \\
\CH_\CN & \{\text{operators supported in an infinitesimal nhd. of $D_{\rm right}^{d-2}$, at fixed time}\} \\
U &  \{\text{operators supported in an infinitesimal nhd. of $S^{d-2}$, at fixed time}\}\,. \end{array}
\ee
These are all algebras, with a product defined by collision/composition of operators in the time direction $\R_t$. We refer to $\CH_\CD,\CH_\CN,U$ as \emph{spark algebras}, since when $d=3$ the operators in question resemble electric sparks moving along a pair of wires:
\be \raisebox{-.4in}{\includegraphics[width=4.2in]{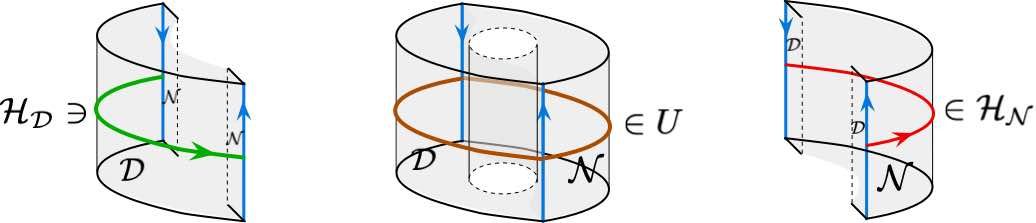}} \label{sparks-intro} \ee

In this paper, we will focus on the case of dimension $d=3$. Our main theoretical results can be summarized as follows. Assuming some precise conditions on topological invariance and finiteness of 3d topological QFT's, outlined in Section \ref{sec:setup} (roughly, that there are no framing anomalies, and state spaces are well defined, though partition functions on closed 3-manifolds need not be), and also assuming transversality of $\CD$ and $\CN$, we will argue in Sections \ref{sec:linefunctor}---\ref{sec:bulk} that
\begin{itemize}
    \item $\CH_\CD,\CH_\CN,U$ are all Hopf algebras, with coproducts, counits, and antipodes all defined topologically, and antipodes satisfying $S^2=\text{id}$ (Prop. \ref{Prop:Hopf}).
    \item $U$ is a ribbon, quasi-triangular Hopf algebra --- \emph{a.k.a.} a generalized quantum group --- with an R-matrix and ribbon element defined topologically, and its ribbon element coincides with the ``Drinfeld element'' (Prop. \ref{Prop:R}). When the algebras are infinite dimensional, the R-matrix and ribbon element live in a completion of $U$.
    \item There's a nondegenerate bilinear pairing $h:\CH_\CD\times \CH_\CN\to \C$, defined by placing sparks on the boundary of a 3-ball \vspace{-.2in}
\be \label{pair-intro} \raisebox{-.3in}{\includegraphics[width=1.5in]{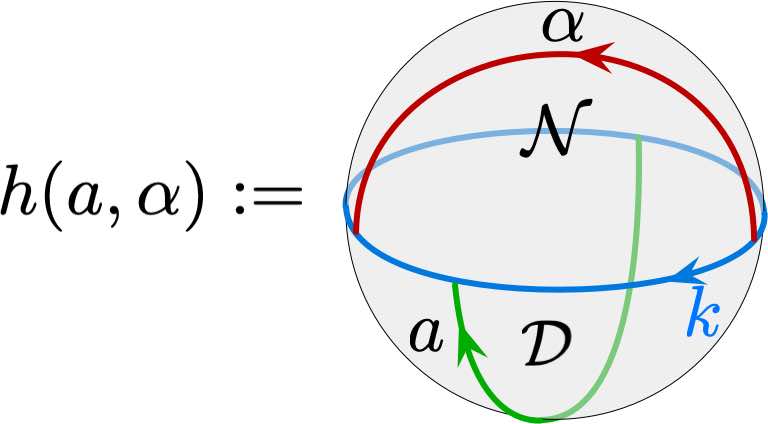}} \ee
    This pairing identifies $\CH_\CD\simeq \CH_\CN^*$ as dual Hopf algebras (interpreted as continuous duals of topological vector spaces when the algebras are infinite dimensional), and identifies $U$ as the Drinfeld double of either $\CH_\CD$ or $\CH_\CN$ (Prop. \ref{Prop:decomp}, \ref{Prop:Hopf}, Thm. \ref{Thm:double}).
\end{itemize}
Finally, adding in completeness, we argue that
\begin{itemize}
    \item There are equivalences of monoidal categories with duals,
\be \label{modules-intro} \begin{array}{r@{}c@{}l}
 \text{lines on an $\CN$ boundary} & \;\;\overset{\sim}{\longrightarrow}\;\; &  \CH_\CD\text{-mod}\,, \\
  \text{lines on a $\CD$ boundary} &  \;\;\overset{\sim}{\longrightarrow}\;\; & \CH_\CN\text{-mod}\,, \\
 \text{lines in the bulk of $\CT$} &  \;\;\overset{\sim}{\longrightarrow}\;\; &  U\text{-mod}\,,
\end{array} \ee
additionally preserving a braided ribbon structure in the bulk (Thms. \ref{Thm:faithful}, \ref{Thm:Rep}). Without completeness, the maps in \eqref{modules-intro} make sense, but need not be equivalences.
\end{itemize}

Our main perspective, and goal throughout, is to introduce a setup that is of practical use to a physicist who may have limited information about a topological QFT, and would like to know how to go about analyzing its line operators. We have thus kept our formalism and assumptions as adaptable as possible. 
For example, the module categories on the RHS of \eqref{modules-intro} can come in many versions. One may choose whether to consider finite-dimensional or infinite-dimensional modules --- this is related to whether one considers dualizable line operators or all line operators on the LHS. Moreover, in a topological QFT of cohomological type, such as the twist of a supersymmetric theory, the spark algebras are naturally dg  (differential graded) or $A_\infty$ algebras, and their module categories are dg or $A_\infty$ categories. Indeed, the main examples that motivated this paper, line operators in twisted 3d $\CN=4$ gauge theories,  are of dg type.

The development of our physical setup has been guided mathematically by a fundamental set of results in representation theory known as \emph{Tannaka duality}, or categorical reconstruction. In the remainder of the introduction, we review the basic principles of Tannaka duality (Section \ref{sec:Tannaka}), and explain how they we implement them (Section \ref{sec:Tannaka-phys}). We'll also describe some other perspectives on spark algebras, and how they connect to other parts of the literature --- including previous physical applications of Tannaka duality, over the past three decades. In Section \ref{sec:examples}, we'll survey our main examples.

This work grew out of trying to better understand the braided monoidal structure of line operators in topological twists of 3d $\CN=4$ gauge theories. In the case of abelian theories, we described this structure in \cite{BCDN} (joint with A. Ballin and T. Creutzig) using boundary VOA's, and fusion/braiding in their module categories. In \cite{creutzig2024kazhdan}, the second author and T. Creutzig proved a Kazhdan-Lusztig-like correspondence relating the VOA categories to modules for certain generalized quantum groups. In the current paper, we \emph{bypass} VOA's, and obtain the quantum groups directly from 3d field theory, for both abelian and nonabelian examples.
The particular approach we take here was inspired by work of  N. Aamand \emph{et al.} \cite{Aamand-CS,Aamand-Wilson,Aamand-BF}, which extracted quantum groups in perturbative Chern-Simons/BF theory by considering correlators of Wilson lines in the presence of a pair of transverse boundary conditions. (This in turn was inspired by similar constructions in 4d Chern-Simons \cite{Costello-4dCS,CWY-I,CWY-II}.)

Two mathematical works currently in preparation, one by Johnson-Freyd and Reutter \cite{JFR-talk, JFR} (extending previous work of Reutter \cite{Reutter-slides} on topological manifestations of Hopf algebras) and another by C. Bae \cite{Bae} (generalizing a construction of Hopf algebras governing the Ising model by Freed-Teleman \cite{FreedTeleman-Ising}, which was also alluded to in \cite[Sec. 5]{FMT}) present theoretical constructions very similar to the one in this paper.  \cite{JFR} uses an extended TQFT framework similar to ours, while \cite{Bae} works in a finite semisimple context. In addition, \cite{JFR} generalizes to theories with framing anomalies.
Neither \cite{JFR} nor \cite{Bae} (to our knowledge) considers completeness, \emph{i.e.} whether maps \eqref{modules-intro} are equivalences, or the quasi-triangular and ribbon structure and on bulk algebras $U$.\label{JFR-page}

On the physics side, the work \cite{CHO-solitons} by C\'ordova-Holfester-Ohmori appeared while our manuscript was in preparation. It defines ``strip algebras'' that are analogous to our ``boundary spark algebras.'' The main differences are that \cite{CHO-solitons} restricts to a finite semisimple context, but does not require transversality; with the consequence that \cite{CHO-solitons}  constructs \emph{weak} Hopf algebras from fusion categories. Quasi-triangular structures are not considered.

\subsection{Tannaka duality}
\label{sec:Tannaka}

The principles of what is now called Tannaka duality, or categorical reconstruction theory, were introduced by Tannaka and Krein in the 1930's \cite{Tannaka,Krein} to analyze categories of representation of a group, then generalized in a number of successive developments. These include a general theory developed for monoidal categories \cite{SaavedraRivano,DeligneMilne}, and later (following the advent of quantum groups \cite{Drinfeld-QG,Jimbo-QG}) braided monoidal categories, \eg\ \cite{Lyubashenko-TD,Ulbrich,Woronowicz-TD,Majid-TD}. Further history and references are reviewed in \cite{JoyalStreet}, as well as in the book \cite{egno}, where reconstruction theory is a central theme.
More recently, Tannakian ideas have been extended to the dg/infinity setting, \emph{cf.} \cite{Wallbridge,Iwanari}; a treatment covering settings we are interested in appears in Chapter~9 of the book \cite{Lurie-SAG} and \cite[Sec. 3]{lurie2007derived}.

To illustrate the basic idea of Tannaka duality, suppose that $\CC$ is a $\C$-linear (abelian or dg) category, and that we'd like to find an (ordinary or dg) algebra $A$ over $\C$ such that $\CC\simeq A\text{-mod}$. One way to approach this is to construct a functor
\be \CF:\CC\to \text{Vect} \label{FF-intro} \ee
to the ``trivial'' category of (ordinary or dg) vector spaces, called a \emph{fiber functor}.
Then we define
\be A := \text{End}(\CF) \label{defA-intro} \ee
to be the algebra of natural transformations from $\CF$ to itself, which one should think of as the symmetries of $\CF$. It follows from the definition of natural transformations that for all objects $\ell\in \CC$ the vector space $\CF(\ell)$ has an action of $A$; and moreover for all morphisms $\mu:\ell\to \ell'$ the $A$-actions commute with $\CF(\mu)$. Therefore, by endowing the vector spaces on the RHS of \eqref{FF-intro} with the structure of $A$-modules, the fiber functor lifts to a functor
\be \wt\CF:\CC \to A\text{-mod}\,. \label{FFA-intro} \ee

Reconstruction theorems give conditions under which the lift \eqref{FFA-intro} is an equivalence of categories. This always requires $\CF$ to be faithful, as well as sufficiently well behaved, depending on the setting one is in. For example, for abelian categories \eqref{FFA-intro} is an equivalence if and only if $\CF$ is faithful and exact, \cf\ \cite[Sec 1.10]{egno}. In the dg case, one further needs to assume that $\CF$ is continuous, and usually that it admits a left adjoint (see \cite[section 3]{lurie2007derived}). In both cases, if the category $\CC$ is infinite (\eg\ an abelian category with infinitely many simple objects) one also needs to consider topology when defining $A$ and its category of modules, \emph{e.g.} taking continuous modules for a topological algebra.

When a category $\CC$ has additional algebraic structures, such as a tensor product or duals, and a fiber functor preserves these structures, the endomorphism algebra $A$ gets correspondingly enhanced --- in a unique, systematic way. Roughly,
\be \label{mon-intro}
\begin{array}{l@{\quad\leadsto\quad}l}
\text{$\CC$ monoidal} & \text{$A$ is a bialgebra (with coproduct $\Delta$)} \\
\text{$\CC$ monoidal w/ unit $\id$} & \text{$A$ is a bialgebra (with coproduct and co-unit $\varepsilon$)} \\
\text{$\CC$ monoidal w/ unit \& duals} & \text{$A$ is Hopf (in addition, an antipode $S$)} \\
\text{$\CC$ braided} & \text{$A$ is quasitriangular Hopf (in addition, an R-matrix $R$)} \\
\text{$\CC$ ribbon} & \text{$A$ is ribbon (in addition, a ribbon element $v$)}
\end{array}
\ee
(See \cite[Sec 5.1-4, 8.3, 8.11]{egno} or \cite[Ch. 9]{Lurie-SAG}, \cite[Sec. 3]{lurie2007derived} or our Appendix \ref{sec:math} for further details.) The additional structures on $A$ are engineered precisely so that the lifted functor \eqref{FFA-intro} becomes a map of the desired type of category. For example, if $\CC$ is braided then $\wt \CF$ becomes a functor of braided categories, with the R-matrix in $A$ controlling the braiding on the RHS.

\subsection{Tannaka duality in $d\geq 3$ physics}
\label{sec:Tannaka-phys}

Let us now explain, in very rough terms, how our physical implementation of Tanaka duality works. 

Let $\CC_\CT$ be the category of line operators in a $d$-dimensional topological QFT $\CT$ (we always assume $d\geq 3$, and in the main body of the paper $d=3$). Given a pair of topological boundary conditions $\CD,\CN$ for $\CT$, we may construct a functor $\CF_\CT:\CC_\CT\to\text{Vect}$ by sending any line operator $\ell\in \CC_\CT$ to the state space of the topological quantum mechanics on a solid cylinder $D^{d-1}\times \R_t$ bounded by $\CD$ and $\CN$, with $\ell$ in the core:
\be \raisebox{-.4in}{\includegraphics[width=2.1in]{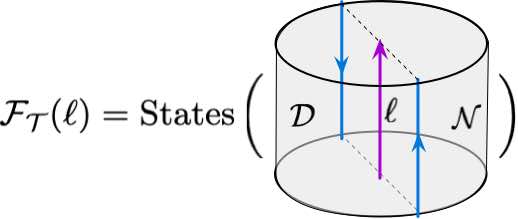}} \label{FT-intro} \ee
When the two boundary conditions are \emph{transverse}, as in \eqref{TC-intro}, there is a canonical choice of ``seam'' to place at the junction of $\CD$ and $\CN$; and the functor $\CF_\CT$ will preserve tensor products and associators. (The functor would not have preserved tensor products had we just used a single boundary condition to `wrap' lines.)
Furthermore, when the boundary conditions are \emph{complete}, the functor $\CF_\CT$ will be faithful.

These are the basic conditions needed for Tannakian formalism to proceed for a monoidal category.
Thus, we will phrase our setup from \eqref{TC-intro} as conditions for a topological QFT to be ``Tannakian.''

Since we introduce boundary conditions, we can also consider monoidal categories $\CC_\CD$, $\CC_\CN$ of line operators on the respective boundaries. We get fiber functors for the boundary categories by modifying \eqref{FT-intro}:
\be \raisebox{-.4in}{\includegraphics[width=4.3in]{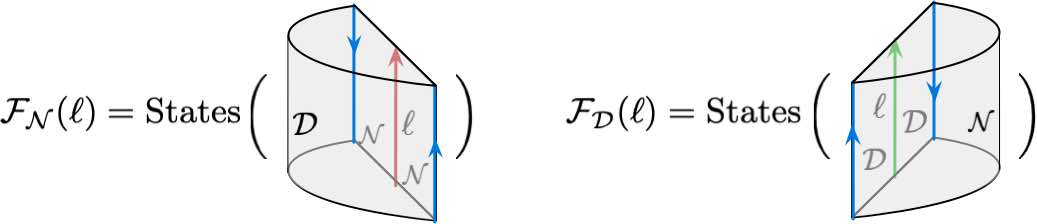}} \label{FD-intro} \ee

The physical setup also gives us an explicit way to describe the symmetry algebras of the fiber functors. They are precisely given by the spark algebras from \eqref{HU-intro}. Namely,
\be \CH_\CD = \text{End}(\CF_\CN)\,,\qquad \CH_\CN = \text{End}(\CF_\CD)\,,\qquad U = \text{End}(\CF_\CT)\,. \ee
For example, symmetries of $\CF_\CT$ come from operators supported anywhere in the solid cylinder on the RHS of \eqref{FT-intro}, but localized in time and disjoint from the line $\ell$. Since the theory is topological, any such operators can be deformed to sparks as in \eqref{sparks-intro}.

The fact that spark algebras are Hopf may now be established, equivalently, via formal algebraic reconstruction as in \eqref{mon-intro}; or by purely topological manipulations, which we'll describe in detail in Sections \ref{sec:Hopf-bdy}--\ref{sec:bulk}. For example, the Hopf operations on $U$ look like \vspace{-.2in}
\be \raisebox{-.9in}{\includegraphics[width=4.3in]{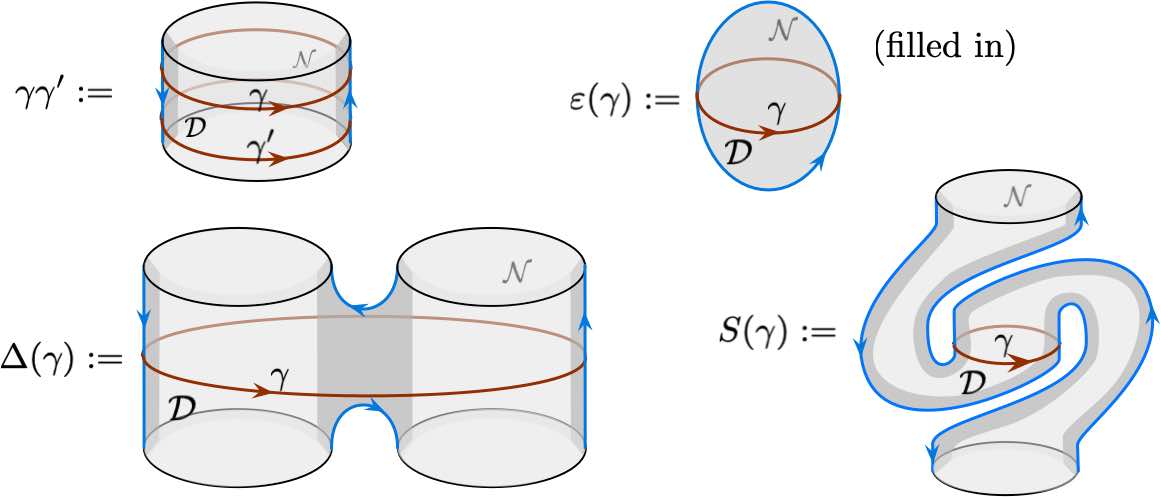}} \label{sparkops-intro} \ee
The topological operations lend themselves well to direct computation in a physical QFT. Similarly, when $d=3$, $U$ acquires an R-matrix and ribbon, either element by reconstruction, or equivalently defined by the topological cobordisms
\be \raisebox{-.4in}{\includegraphics[width=4.5in]{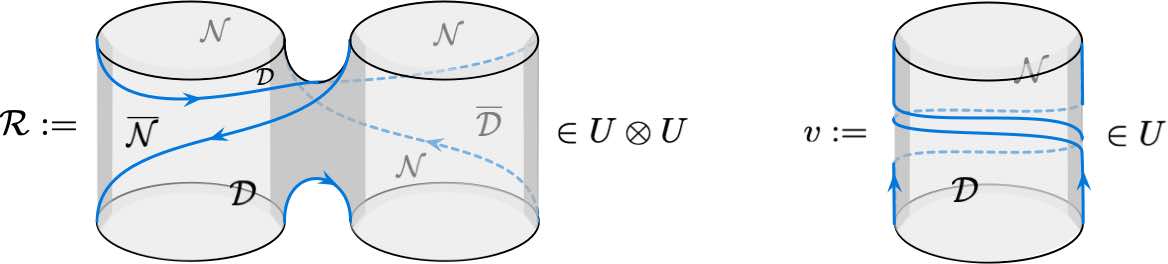}} \label{Rv-intro} \ee
We will argue topologically that inserting $\CH_\CD$ and $\CH_\CN$ sparks into the solid ball \eqref{pair-intro} defines a nondegenerate Hopf pairing $h$, such that $\CR \simeq h^{-1}$. This is a key part of our argument that $U$ is the Drinfeld double of either $\CH_\CD$ or $\CH_\CN$.

Finally, faithfulness of the fiber functors --- due to completeness of the $\CD,\CN$ boundary conditions --- ultimately ensures equivalences of categories $\CC_\CT\simeq U\text{-mod}$, $\CC_\CN\simeq \CH_\CD\text{-mod}$, $\CC_\CD\simeq\CH_\CN\text{-mod}$, as in \eqref{modules-intro}.

Our spark algebras turn out to have several equivalent descriptions, which are useful in their own right and also help connect our construction to other parts of the literature.

\subsubsection{Sparks as state spaces}

A state-operator correspondence relates the underlying vector spaces of spark algebras $\CH_\CD,\CH_\CN,U$ to state spaces of the theory $\CT$ on certain $(d-1)$-manifolds with boundary:
\be \begin{array}{c@{\;\simeq\;}l}
\CH_\CD & \text{States}(D^{d-2}\times I) \quad \text{with $\CD$ on $D^{d-2}\sqcup D^{d-2}$, $\CN$ on $S^{d-3}\times I$} \\
\CH_\CN & \text{States}(D^{d-2}\times I) \quad \text{with $\CN$ on $D^{d-2}\sqcup D^{d-2}$, $\CD$ on $S^{d-3}\times I$}  \\
U & \text{States}(S^{d-2}\times I) \quad \text{with $\CD$ on $D_{\rm left}^{d-2}\sqcup D_{\rm left}^{d-2}$,  $\CN$ on $D_{\rm right}^{d-2}\sqcup D_{\rm right}^{d-2}$}\,. 
\end{array}
\ee
When $d=3$, the first two both look like rectangles (with opposite orientation, if one is careful), and the last looks like an annulus glued from the two rectangles: \vspace{-.1in}
\be \vspace{-.1in} \raisebox{-.4in}{\includegraphics[width=4.7in]{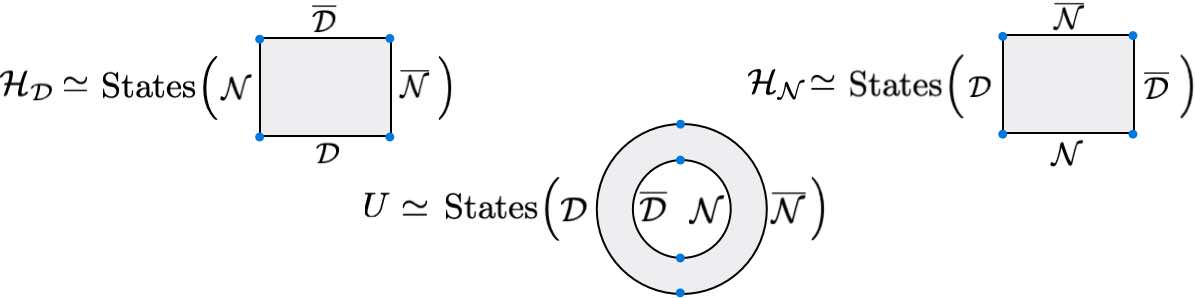}} \label{fig:rectangles}\ee 
This description leads to a quick argument that $\CH_\CD\simeq \CH_\CN^*$ and (using transversality to cut the annulus) that $U\simeq \CH_\CD\otimes \CH_\CN$.

This description connects with work of Reutter, who observed around 2017 \cite{Reutter-slides} that state spaces on a rectangle in $d=3$ TQFT, with two transverse boundary conditions, naturally have the structure of a Hopf algebra. Similar observations were made by Freed and Teleman while studying (and generalizing to a nonabelian setting) dualities of the Ising model in 2+1 and higher dimensions \cite[Sec. 8]{FreedTeleman-Ising}; an instance of our boundary fiber functor $\CF_\CD$ appears in Figure 20 of \cite{FreedTeleman-Ising}. Hopkins \cite{Hopkins-slides} considered state spaces associated to picture similar to \eqref{fig:rectangles}, but using a single boundary condition rather than two transverse ones; the cobordisms drawn for these state spaces are extremely similar to our pictures as well as those found in \cite{Reutter-slides}. The rectangles of \eqref{fig:rectangles} also appear in the very recent physics work \cite{CHO-solitons}, and will play an important role in the upcoming \cite{JFR,Bae}, mentioned on page \pageref{JFR-page}.

Spark algebras also seem closely related to the construction by Schrader and Shapiro \cite{SchraderShapiro} of quantum groups via quantization of framed local systems on a punctured disc. The punctured disc in question corresponds to a spatial slice of the geometry on the RHS of \eqref{sparks-intro} (for $d=3$), and framing is similar --- but not quite identical --- to our specification of transverse boundary conditions. \cite{SchraderShapiro} constructs an algebra of operators acting on their disc, with Hopf operations and R-matrix induced from topology much as in \eqref{sparkops-intro}, \eqref{Rv-intro}. It would be interesting to understanding the connection more precisely.

\subsubsection{Sparks, strips, and tubes}

We'll explain in Section \ref{sec:linefunctor} that spark algebras can also (in principle) be defined using Hochschild homology, or factorization homology \cite{AyalaFrancis-fact} (or blob homology \cite{MorrisonWalker}, etc.). The basic idea for $d=3$ is as follows.

Transversality of $\CD$ and $\CN$ guarantees the existence of a canonical interface `$k$' between these two boundary conditions (which appeared as a blue line in figures above). One can then construct Hochschild/factorization homology of either boundary category $\CC_\CD$ or $\CC_\CN$ on an interval $I$ that's bounded by $k$ on both ends, reproducing the spark algebras
\be \CH_\CD =  \int_{{}_kI_k} \CC_\CD\,,\qquad \CH_\CN = \int_{{}_kI_k} \CC_\CN\,.  \label{Hoch-intro} \ee
These have products induced from the tensor product on $\CC_\CD,\CC_\CN$, and coproducts induced from splitting the interval in half. Our basic claim for boundary categories is that, under our transversality and completeness assumptions, there is a duality $\CH_\CD\text{-mod}\simeq \CC_\CN$ and $\CH_\CN\text{-mod}\simeq \CC_\CD$.

The algebras \eqref{Hoch-intro} are also instances of the `ladder categories' of \cite[Sec. 6]{MorrisonWalker}, \cite{BBJ}. In a finite semisimple context, the ``strip algebras'' of the recent \cite{CHO-solitons} gave an explicit construction of \eqref{Hoch-intro}.

The algebra $U$ has an analogous construction using Hochschild/factorization homology on a circle, cut twice by $k$, with $\CC_\CD$ on one half and $\CC_\CN$ on the other,
\be U = \int_{{}_kS^1\!{}_k} (\CC_\CD,\CC_\CN)\,. \label{US1-intro} \ee
Precisely such a construction was the main focus of \cite{Hoek}, where it was shown, in a finite semisimple context, that modules for $U$ are in fact what we are calling bulk line operators $\CC_\CT$ (described there in terms of a Drinfeld center). This had been an expected result, or `folk theorem,' in the TQFT community. To the best of our knowledge, a quasitriangular ribbon Hopf structure on such a $U$ has not been previously constructed using topology.

The construction \eqref{US1-intro} of $U$ is closely related to the \emph{tube algebras} of \cite{Ocneanu-tube}. 
Tube algebras correspond to the circle factorization homology of a \emph{single} boundary category, \emph{e.g.} $\int_{S^1} \CC_\CD$. It is known that, under suitable conditions, their modules reproduce the Drinfeld center of a boundary category, \emph{e.g.} $\Big(\int_{S^1} \CC_\CD\Big)\text{-mod}\simeq Z_{\text{Drin}}(\CC_\CD)\simeq \CC_\CT$, \emph{cf.} \cite{EK-tube, Izumi-tube, Muger-tube, PSV,KST-tube}. Thus, tube algebras provide an alternative way to represent bulk lines. This has played an important role in recent work on generalized symmetries in QFT \cite{LOST-tube,BBG-tube,BSN-tube}. However, in general, tube algebras at best expected to be \emph{weak} Hopf algebras, and not obviously quasitriangular; from a Tannakian perspective, they are associated to a lax-monoidal rather than a monoidal fiber functor. This makes it harder to use them to recover the braided monoidal structure on $\CC_\CT$.

\subsubsection{Sparks as Koszul duals}

In Appendix \ref{sec:Koszul}, we will explain that for a \emph{perturbative} topological QFT, spark algebras reproduce the Koszul-duality approach of \cite{Costello-4dCS,CostelloPaquette} to representing line operators. This has mathematical precursors in work of Tamarkin \cite{Tamarkin-formality, tamarkin2007quantization} and Lurie \cite{Lurie-DAGVI}, with related observations were developed in the context of twisted holography, \emph{e.g.}  \cite{CostelloLi-twisted, Costello-M2, IMZ-twisted, GaiottoOh}. See also the review \cite{PaquetteWilliams}.

Namely, let $A_\CT$ denote the algebra of local operators in the bulk theory $\CT$ and $A_\CD, A_\CN$ denote the algebras of local operators on the respective boundary conditions $\CD,\,\CN$. Then our transversality and completeness conditions on boundary conditions imply that --- perturbatively --- spark algebras coincide with the Koszul duals
\be \CH_\CD \simeq A_\CN^!\,,\qquad \CH_\CN\simeq A_\CD^!\,,\qquad U \simeq A_\CT^! \,. \label{spark!-intro} \ee
Moreover, in a perturbative TQFT in $d=3$, we find that the Hopf pairing \eqref{pair-intro} is controlled by the higher $E_3$/Poisson bracket on $A_\CT$ (defined using topological descent \cite{descent}), and thus the R-matrix in $U$ inverts the $E_3$ bracket. This recovers mathematical expectations about Koszul duals of $E_3$ algebras \cite{Lurie-DAGVI,CostelloFrancisGwilliam}. We also find that the relations \eqref{spark!-intro} provide a physical manifestation of Tamarkin's formal quantization of Lie bialgebras \cite{tamarkin2007quantization} using the little disc operad.

\subsubsection{Sparks and VOA's}

For a $d=3$ dimensional topological QFT, there is yet another well-trod approach to understanding line operators. If one introduces a \emph{holomorphic} boundary condition, supporting a vertex operator algebra (VOA) $\CV$, then bulk line operators map to VOA modules,
\be (\CC_\CT,\otimes,\text{braiding}) \;\to\; (\CV\text{-mod},\text{fusion},\text{VOA braiding/KZ})\,. \ee
For a sufficiently rich holomorphic boundary condition, this is expected to be an equivalence of braided monoidal categories. A classic example is Chern-Simons theory with its WZW boundary \cite{Witten-Jones, MooreSeiberg-taming}; analogous examples involving topological twists of 3d $\CN=4$ theories and logarithmic VOA's have appeared in \cite{CG,CDGG}.

When a bulk theory $\CT$ admits both a holomorphic boundary condition \emph{and} a topological $(\CD,\CN)$ pair, there is a rich interplay between spark algebras and VOA's, leading in particular to a Kazhdan-Lusztig-like correspondence  $\CV\text{-mod}\simeq U\text{-mod}$ between VOA modules and `quantum group' modules. We find indications that boundary spark algebras $\CH_\CD,\CH_\CN$ generalize the Nichols algebras, or ``algebras of screenings'' that have been used to produce Kazhdan-Lusztig correspondences, given a free-field realization of a VOA \cite{semikhatov2012nichols,semikhatov2013logarithmic,lentner2021quantum,CLR23}. We will explore this further in \cite{sparkVOA}.

\subsubsection{Tannaka duality in 4d}

We note that in $d=4$, Kapustin and Witten used transverse boundary conditions in almost the same way as we propose, to analyze line operators in Langlands-twisted Yang-Mills theory. 
In particular, placing 't Hooft lines of $G$ gauge theory in a geometry $S^2\times I\times \R_t$ with transverse Neumann and Dirichlet b.c. at the two ends of $I$ engineered a fiber functor that identified (part of) the category of 't Hooft lines with representations of the Langlands dual group ${}^LG$; see \cite[Sec. 11]{KapustinWitten} and \cite[Sec. 2]{Witten-Nahm}. This provided a physical interpretation of the geometric Satake correspondence \cite{MV-Satake}.
More recently, Tannaka duality and boundary conditions have played a central role in \cite{BZSV}, relating 4d twisted Yang-Mills, geometric Langlands, and arithmetic Langlands.

\subsection{Our main examples}
\label{sec:examples}

We apply our construction to four increasingly complex examples, in Sections \ref{sec:DW}--\ref{sec:GM}. We briefly review these examples and comment on the connection between our construction and the vast amount of existing literature.

\subsubsection{Dijkgraaf-Witten theory}

In Section \ref{sec:DW}, we consider 3d gauge theory with finite gauge group $G$, \emph{a.k.a.} Dijkgraaf-Witten theory \cite{DW}. This is a theory in which everything one could hope for is already known about line operators and boundary conditions. Tannakian ideas were applied early on to represent line operators \cite{RPD-quasiHopf,Freed-DW}, though with a slightly different setup than ours. Later, the theory was defined mathematically as a fully extended TQFT \cite{FreedQuinn,Freed-DW,FHLT}, allowing full mathematical control over boundary conditions, their junctions, etc. For us, Dijkgraaf-Witten theory serves as a toy model to explicitly illustrate our definitions, assumptions, and results --- in a finite semisimple context. 

In Dijkgraaf-Witten theory, we take $\CD$ to be the canonical Dirichlet boundary condition, and $\CN$ to be Neumann.%
\footnote{This same $(\CD,\CN)$ pair has played a central role in the recent study of generalized global symmetries and their anomalies, \cf\ \cite{GKSW,FMT}.} %
The $\CN$ boundary exists, and the theory is ``Tannakian,'' only when the 3-cocycle defining its action is trivial. We show by direct computation of operators as in \eqref{sparks-intro} that $\CH_\CD$ is the group algebra and $\CH_\CN$ is algebra of functions on $G$,
\be \CH_\CD\simeq \C G = \C\langle g\rangle_{g\in G}\,,\qquad \CH_\CN\simeq \CO(G) = \C\langle \delta_g\rangle_{g\in G}\,, \ee
which are dual Hopf algebras; whereas $U \simeq \CH_\CD\otimes \CH_\CN$ is the double, with
\be \CR = \sum_{g\in G} g\otimes \delta_g\,,\qquad v = \sum_{g\in G} \delta_g g^{-1}\,. \ee
We will also explain how completeness of the $(\CD,\CN)$ pair is implemented by an explicit condensation procedure. Completeness then guarantees equivalences of categories
\be \begin{array}{c} \CC_\CN \simeq \CH_\CD\text{-mod} = \text{Rep}(G)=\text{Shv}^G(pt)\,,\qquad \CC_\CD \simeq \CH_\CN\text{-mod} = \text{Vect}_G = \text{Shv}(G)\,,\\[.1cm]  \CC_\CT \simeq U\text{-mod} = (\text{Vect}_G)^G= \text{Shv}^G(G)\,.\end{array} \vspace{-.1in} \ee

\subsubsection{3d $\CN=4$ in the B twist}

We then look at 3d $\CN=4$ gauge theory in the topological B twist \cite{BlauThompson}. We consider a fairly general situation, with any compact gauge group $G_c$, acting on linear hypermultiplet matter in a symplectic representation $T^*V=V\oplus V^*$. (We write $G$ for the algebraic group complexifying $G_c$.) There are several different ways to think about this bulk theory $\CT$: either as a gauged/equivariant version of Rozansky-Witten TQFT \cite{RW} (which traditionally takes a smooth hyperk\"ahler target); or as a derived enhancement%
\footnote{Here $\Pi V$ denotes the parity-shifted, or fermionic, version of $V$. By ``derived enhancement,'' we mean that we quantize in the BV-BRST formalism, introducing ghosts, and keep states/operators of all ghost numbers in the spectrum.}%
of 3d BF theory with a gauge supergroup $G_c\ltimes \Pi V$; or as (derived) Chern-Simons theory with partially noncompact gauge supergroup $(G_c\ltimes \Pi V)\ltimes(\fg^*\times \Pi V^*)$.

We split this example into three different cases, each of which introduces new features of spark algebras. In Section \ref{sec:matter}, we consider pure matter theory (trivial $G$, any linear $V$). This is the simplest 3d topological QFT of cohomological type; its spark algebras still turn out to be finite-dimensional. In Section \ref{sec:gauge}, we consider pure gauge theory (any $G$, trivial $V$). In addition to being of cohomological type, this example has the added feature of infinite-dimensional spark algebras, which must be carefully handled as topological vector spaces. In Section \ref{sec:GM}, we combine gauge and matter. A particularly interesting feature in this case is the physical interpretation of its spark algebras, via interacting matter, Wilson lines, and symmetry defects.

Much is already known about the B twist of 3d $\CN=4$ gauge theory with matter. Rozansky-Witten theory was described as an extended TQFT in \cite{KRS,KapustinRozansky} (recently revisited in \cite{BCR-extended,BCFR-defect}), in a way that has a fairly straightforward gauge-theory generalization, \cf\ \cite{BDGH,OR-Chern,OR-TQFT,hilburn2021tate,HG-Betti}. Its category of line operators is well understood, as a linear (dg) category. Line operators in our gauge theory include Wilson lines, Gukov-Witten-like monodromy defects, and various couplings to the matter fields \cite{Mikhaylov,AsselGomis,lineops,BCDN}. One standard model for bulk lines is given by equivariant matrix factorizations
\be \CC_\CT \simeq \text{MF}^G\big(G\times T^*V,\,W = \langle g-1,\mu\rangle\big)\,, \label{MF-intro} \ee
with a superpotential $W$ formed from $g\in G$ and the moment map $\mu:T^*V\to \mathfrak g^*$.

There are several ways one might introduce a braided monoidal structure on $\CC_\CT$. In principle, it should be obtained by generalizing to an equivariant setting 
a classic construction of Roberts and Willerton \cite{RobertsWillerton} for sheaves on complex-symplectic spaces, based on relating Rozansky-Witten theory to formal quantization of metric Lie algebras \cite{Kapranov-RW,Kontsevich-RW}. This approach sheafifies Drinfeld's quantization of metric Lie algebras by solving a KZ equation \cite{drinfeld1990quasitriangular}. (For us, the relevant Lie algebra is $\mathfrak g\oplus \Lambda^\bullet(V\oplus V^*)\oplus \mathfrak g^*$.) However, this approach is also famously difficult to implement explicitly (some progress for nonlinear targets appeared just recently in \cite{GHR}).
Even when successful, its output is a braided monoidal category with nontrivial Drinfeld associators --- thus not obviously modules for a Hopf algebra, which would have trivial associators.

A simpler approach is to recognize that the category \eqref{MF-intro} happens to be a derived Drinfeld center, in the sense of \cite{Ben-Zvi:2008vtm}, of either category of sheaves
\be \CC_\CN = \text{Coh}^G(V)\qquad\text{or}\qquad \CC_\CD = \text{Coh}(G\times V^*)\,, \label{Coh-intro} \ee
with standard tensor product of sheaves in $\CC_\CN$, and a convolution product on $G$ in $\CC_\CD$. (Equivalently, one may express $\CC_\CT$ as sheaves on the derived loop space of the stack $V/G$, as in \cite{ben2012loop}.)
$\CC_\CN$ and $\CC_\CD$ do happen to be known categories of lines on boundaries; and taking Drinfeld center is well known to be the correct operation for extracting bulk lines from a boundary condition; and the derived Drinfeld center is indeed braided monoidal, this time with trivial associators. However, it is not obvious that the naive tensor products used in this construction are correct physically --- \emph{i.e.} that they would not acquire any additional quantum corrections.

We clarify the situation using spark algebras.
We take $(\CD,\CN)$ to be boundary conditions that preserve 2d $\CN=(2,2)$ SUSY and are compatible with the B twist, as classified in \cite{KRS,BDGH}. Roughly, \vspace{-.3cm}
\be \notag
\begin{array}{l}
\text{$\CD=$ Dirichlet for gauge fields; Dirichlet (Neumann) for matter bosons in $V$ ($V^*$)} \\[0cm]
\text{$\CN=$ Neumann for gauge fields; Neumann (Dirichlet) for matter bosons in $V$ ($V^*$)}  \vspace{-.2cm}
\end{array} \ee
These are transverse, and completeness is achieved by deforming the superposition $\CN\circ\CD$ with a term in the 2d boundary action, \emph{a.k.a.} a ``2-Maurer-Cartan'' element.

We then compute and find topologically-dual Hopf algebras
\be  \CH_\CD = \text{Dist}(G)\ltimes \Lambda^\bullet V\,,\qquad \CH_\CN =  \CO(G) \otimes \Lambda^\bullet V^*\,. \label{H-N4-intro} \ee
Here $\CO(G)$ denotes algebraic functions on $G$, which arise from Wilson lines on the~$\CN$ boundary; while $\text{Dist}(G)$ denotes the dual algebra of distributions, a topological version of the group algebra of $G$, which arises from global symmetry defects on~$\CD$. The exterior algebras $\Lambda^\bullet V$, $\Lambda^\bullet V^*$ arise from hypermultiplet fermions integrated along sparks. We show that the product, coproduct, etc. on \eqref{H-N4-intro} can all be obtained by exact quantum computations, and are such that their derived module categories are equivalent to \eqref{Coh-intro}. Our general construction then produces a ribbon (quasi-triangular) Hopf algebra
\be U\simeq \CH_\CD\otimes \CH_\CN  \simeq  \text{Dist}(G)\ltimes \big( \Lambda^\bullet(V\oplus V^*) \otimes \CO(G)\big)\,. \label{U-N4-intro} \ee
We argue that its derived module category, with trivial associators, is equivalent to~\eqref{MF-intro}. The algebra \eqref{U-N4-intro}, together with its explicit coproduct, R-matrix, and ribbon element (Section \ref{sec:GM-bulk}), is new; to our knowledge, it provides the most direct available access to the braided ribbon structure of line operators in B-twisted gauge theories.

This construction connects to many other results in the literature, including:

\begin{itemize}[leftmargin=*]

\item In the setting of ordinary metric Lie algebras (as opposed to the sheaves of dg Lie algebras that appear in 3d B twists), there also existed two different methods of quantization \cite{drinfeld1986quantum,drinfeld1990quasitriangular}: either by solving the KZ equation (leading to nontrivial associators, defined as power series in $\hbar$) or by taking the Drinfeld double of a Lagrangian half of the algebra (leading to trivial associators, and exact in $\hbar$). Etingof and Kazhdan ultimately proved that these two quantizations were equivalent, by engineering a fiber functor on modules of the former whose endomorphism algebra reproduced the latter \cite{etingof1996quantization}. Our fiber functor on $\CC_\CT$, built from $(\CD,\CN)$ boundary conditions, generalizes the construction of \cite{etingof1996quantization}.

A mathematically rigorous construction of our fiber functor on \eqref{MF-intro}, relating Roberts-Willerton-style quantization to a Drinfeld double, will appear in \cite{WNTangent}.

\item For abelian gauge group $G$, quite a lot is already known. The bulk categories of line operators in this case may be accessed via boundary VOA's \cite{CG, CCG, BCDN}, and thereafter represented by quantum groups via recently established logarithmic Kazhdan-Lusztig correspondences \cite{creutzig2024kazhdan}. Alternatively, theories of this type may be directly related to supergroup Chern-Simons theories, from which quantum groups may again be extracted \cite{Mikhaylov, CDGG, GeerYoung, GarnerNiu, garner2024btwisted}. For example, $G_c=U(1)$ and $V=\C$ (\emph{a.k.a.} B-twisted SQED) leads to $U(1|1)$ Chern-Simons theory. The abelian version of \eqref{U-N4-intro} agrees with all these results.

The work of \cite{GarnerNiu,garner2024btwisted} extends beyond our examples, to ``Gaiotto-Witten theories'' --- roughly speaking, B twists of 3d $\CN=4$ gauge theories with additional Chern-Simons terms, which can be engineered from interfaces in 4d Langlands-twisted super-Yang-Mills \cite{gaiotto2010janus}. This would be interesting to explore further. In particular, we expect that a Chern-Simons deformation of pure gauge theory ($V=0$) is controlled by the quantum double $U_q(\mathfrak g)\otimes \CO_q(G)$ rather than \eqref{U-N4-intro}.

\item For pure hypermultiplet matter (\ie\ $G=1$ and $V=\C$) one obtains $\mathfrak{psl}(1|1)$ Chern-Simons theory, which has been very well studied in various guises. The theory is mirror to the A twist of SQED, \emph{a.k.a.} the 3d reduction of Seiberg-Witten TQFT.
 Our quantum group in this case is an exterior algebra in two generators
\be \begin{array}{c} U = \C[\psi_+,\psi_-]\,,\qquad \Delta(\psi_\pm) = \psi_\pm\otimes 1+1\otimes \psi_{\pm}\,,\quad S(\psi_\pm) = -\psi_\pm\,,\\[.1cm] \CR = e^{-\psi_+\otimes\psi_-}\,,\quad v= e^{\psi_-\psi_+}\,. \end{array}\ee
Expectation values of certain line operators in this theory compute Alexander polynomials \cite{RozanskySaleur-QFT,RozanskySaleur,Viro-Alex,Mikhaylov}, and suitably regulated partition functions on 3-manifolds compute torsion. Indeed, a version of this theory was used by Donaldson \cite{Donaldson-TQFT} in giving a TQFT-like proof of the Meng-Taubes theorem~\cite{MengTaubes}.

Nonetheless, honest partition functions of this theory on 3-manifolds typically diverge, due to a noncompact moduli space. One  way to deal with this issue --- killing the moduli space --- might be to include small boundary components labelled by our transverse $(\CD,\CN)$ pairs. It seems that exactly our $(\CD,\CN)$ pairs have appeared this way (in the A-twisted mirror) in what's known as \emph{sutured} Floer homology \cite{Juhasz-sutured} and its 3d reduction \cite{FJR-sutured}.%
\footnote{We thank J. Rasmussen for pointing out this potential connection to us.} %
It would be interesting to expand this connection.

\item In the opposite extreme of trivial matter $V=0$, we have the B twist of pure 3d $\CN=4$ $G_c$ gauge theory, which is a derived enhancement of standard 3d BF theory with bosonic gauge group $G_c$. An underived, unitary truncation of BF theory with $(\CD,\CN)$ boundary conditions recently played a role as a `SymTFT' for continuous symmetries \cite{BS-continuous,BZM-continuous}. In this context, the important bulk defects were Wilson lines (elements of $\text{Rep}(G)=\text{Dist}(G)\Mod$) and monodromy or `Gukov-Witten' defects labelled by conjugacy classes, which are 1d representations of $U=\text{Dist}(G)\ltimes\CO(G)$.

Using perturbative BF theory with $(\CD,\CN)$ boundary conditions, \cite{Aamand-BF} produced a perturbative quantum group $U\approx U(\mathfrak g \oplus \mathfrak g^*)$ with $\CR =\exp\Big(\sum_a t^a\otimes t_a\Big)$ (where $t^a$ is a basis of $\mathfrak g$ and $t_a$ a dual basis). This inspired our construction, and indeed agrees with a perturbative limit of \eqref{U-N4-intro}.

Similarly, lifting to 4d, the framed segments of Wilson lines that generate $\CO(G)$ in \eqref{U-N4-intro} are direct analogues of the framed Wilson lines employed in \cite[Sec. 2]{CWY-II} to generate the ``RTT presentation'' of the Yangian in perturbative 4d Chern-Simons theory; for example, the vertical segments in Fig. 1, 2, 6 of \cite{CWY-II} depict sparks, their products, and coproducts.

\end{itemize}

\subsection{Future directions}

Many potential directions for future investigation have arisen from this work. Some were already indicated above. A handful of others that we hope to pursue include:

\begin{enumerate}
    \item Generalizing our setup of ``Tannakian QFT'' (a topological QFT with two topological boundary conditions $\CN$ and $\CD$) to allow topological interfaces to other 3d theories, rather than strict boundary conditions. We expect that such a generalization could be used to analyze perturbative Chern-Simons theory, recovering classic quantum groups $U_\hbar(\fg)$ as relative Drinfeld doubles of their Borels. Algebraically, a setup with topological interfaces would lead to fiber functors valued in nontrivial braided tensor categories, rather than Vect.

    \item Generalizing our setup by relaxing transversality, so that the $(\CD,\CN)$ sandwich is a nontrivial 2d topological QFT. We would expect this to lead to Hopf algebroids, rather than Hopf algebras, with additional idempotents corresponding to vacua in the 2d QFT.

    \item Applying spark-algebra technology to A-twisted 3d $\CN=4$ gauge theories. All the same $\CD,\CN$ boundary conditions that are used here in the B twist should also work in the A twist \cite{BDGH}, and satisfy transversality and completeness. However, the computation of sparks could have instanton corrections, and would generally lead to nontrivial $A_\infty$ algebras. Recent results on 2-categories of A-twisted boundary conditions \cite{gammage2022perverse, HG-Betti} may help. It would also be interesting to relate A- and B-twisted spark algebras of 3d-mirror theories.

    \item Deforming our B-twisted 3d $\CN=4$ gauge-theory examples by a nonzero FI parameter, and relating them to Rozansky-Witten theory on smoothly resolved Higgs branches. (In this paper, the FI parameters are strictly zero.)

    \item Extending to higher dimensions. Working out the additional structures on spark algebras that represent bulk and boundary line operators in $d>3$, as well as generalizing spark algebras to ``spark categories'' that can represent extended operators with higher-dimensional support. (For example: surface operators in $d=4$ form a braided 2-category, which should be represented as modules for a monoidal 1-category with extra structure.)

    \item Extending to higher holomorphic dimensions. What are the analogues of spark algebras in a QFT that is partially holomorphic and partially topological, and allows transverse and complete boundary conditions, such as a twist of 4d $\CN=2$ gauge theory?
    In the case of 4d Chern-Simons theory, with sparks supported on topological interfaces, does this shed additional light on the Yangians of \cite{Costello-4dCS,CWY-I,CWY-II}?
    
\end{enumerate}

\subsection{Organization}

The first half of the paper is general and theoretical. We begin in Section \ref{sec:setup} by setting up the conditions that we will place on a topological QFT for it to be ``Tannakian'': being precise about what topological invariance and finiteness conditions the bulk $\CT$ and boundaries $\CD,\CN$ should obey, and what transversality and completeness mean. Then in Section \ref{sec:linefunctor} we construct fiber functors topologically, prove they are monoidal, and define spark algebras (their endomorphism algebras) in several ways. In Sections \ref{sec:Hopf-bdy}--\ref{sec:bulk} we establish, at a physical level of rigor, our main results: that spark algebras are Hopf algebras; that bulk sparks are ribbon-Hopf and the Drinfeld double of boundary sparks; and that (given completeness) sparks faithfully represent bulk and boundary categories of line operators.

A reader might be mainly interested in these early sections; or might instead wish to review the results summarized at the beginning of each section and then skip directly to the second half of the paper: applications. 

In Sections \ref{sec:DW}--\ref{sec:GM}, we derive spark algebras in a series of related but increasingly complex examples. We begin with finite-group gauge theory (where categories are abelian), and then proceed to B-twisted 3d $\CN=4$ matter, 3d $\CN=4$ pure gauge theory, and general 3d $\CN=4$ gauge theory with matter (all of which have dg categories of lines).

The Appendices contain alternative perspectives on Tannakian QFT.

In Appendix \ref{sec:math}, we translate the setup of Section \ref{sec:setup} to the language of tensor categories (as opposed to fully extended 3d TQFT). This may be useful since, mathematically, many 3d TQFT's are defined purely via tensor categories. Moreover, this perspective allows us to give a fully mathematically rigorous formulation of all the main results of Sections \ref{sec:linefunctor}--\ref{sec:bulk}, in both abelian and dg settings.

In Appendix \ref{sec:bdyE2}, we make some brief remarks on how boundary conditions define $E_2$ algebra objects in the bulk category, and what our setup means from this perspective. This is a particularly useful perspective for connecting with VOA's.

Finally, in Appendix \ref{sec:Koszul}, we revisit and expand on the relation between spark algebras and Koszul duality, in a QFT that is both Tannakian and perturbative.

\subsection{Acknowledgements}

We are grateful to many friends and colleagues for discussions and advice that have helped us develop the ideas in this paper; and in particular to Chris Beem, Mat Bullimore, Kevin Costello, Davide Gaiotto, Eugene Gorsky, Justin Hilburn, Theo Johnson-Freyd, David Jordan, Lukas M\"uller, and Constantin Teleman. We are especially grateful to Thomas Creutzig for collaboration on an earlier state of this project, which began as a companion to joint work on Kazhdan-Lusztig correspondences for 3d $\CN=4$ boundary VOA's. TD’s research is supported by EPSRC Open Fellowship EP/W020939/1.  WN's research is supported by Perimeter Institute for Theoretical Physics. Research at Perimeter Institute is supported in part by the Government of Canada through the Department of Innovation, Science and Economic Development Canada and by the Province of Ontario through the Ministry of Colleges and Universities.

\section{Tannakian QFT}
\label{sec:setup}

In this section, we describe in greater detail the properties that we require of a topological QFT in spacetime dimension $d=3$, and its boundary conditions, in order to allow a simple physical implementation of the Tannakian formalism. 

Let $\CT$ be a 3d topological QFT, and $(\CD,\CN)$ a pair of topological boundary conditions. The theory $\CT$ may be topological on the nose (sometimes called a Schwarz-type TQFT) or topological only in the cohomology of a BRST symmetry $Q$ (sometimes called a Witten-type TQFT, such as a twist of a supersymmetric theory).
In Section \ref{sec:top} we'll explain exactly what we mean by ``topological'' -- roughly that we require the theory to be free of framing anomalies, and require state spaces to be defined; but we do \emph{not} require partition functions on closed 3-manifolds, possibly bounded by $\CN$ and $\CD$, to be finite (assumption \AD).

In Section \ref{sec:lines}, we'll introduce the categories of line operators on the boundaries and in the bulk, and discuss their expected structure as monoidal categories, with braiding and ribbon twists in the bulk. We'll also introduce an extra assumption \ADline\ to control the ``density'' of dualizable objects in the categories. This is particularly important in the dg setting.

In Section \ref{sec:trans}, we break down the assumption of \emph{transversality} (\AT) of the $(\CD,\CN)$ pair --- that the 3d theory on a $(\CD,\CN)$ sandwich is trivial. This is the main assumption that will make sparks have the structure of Hopf algebras, with R-matrices and ribbon elements (in the bulk), in later sections. We also introduce a subtle property \AD$_k$ involving dualizability of the interface that witnesses transversality. Whether or not \AD$_k$ holds will determine whether spark algebras are finite dimensional.

Finally, in Section \ref{sec:comp}, we describe our second main assumption of \emph{completeness} (\AC) -- the ability to glue a theory back together along an $(\CN,\CD)$ pair. A special instance of completeness, which we call strip gluing (\AC$_{\rm strip}$), will allow us to faithfully represent categories of line operators as modules for spark algebras.

\subsection{Topological invariance}
\label{sec:top}

In order for our constructions in this paper to make sense, we will need to assume 
\begin{itemize}
\item[\AD] (Topological Invariance) The bulk theory $\CT$ is defined locally on smooth, oriented 3-manifolds. If a metric is chosen in the construction of $\CT$, the theory is locally invariant under small deformations of the metric --- \emph{i.e.} the stress tensor either vanishes, or more generally, in a cohomological theory, the stress tensor is BRST-exact. State spaces on closed 2-manifolds must be defined but need not be finite-dimensional (in a cohomological setting: need not have finite cohomology). Partition functions on closed 3-manifolds need not be defined.

Implicit here is that $\CT$ has no `framing anomaly' --- no dependence on a trivialization of the 3d tangent bundle.

Similarly, $\CD$ and $\CN$ are required to be topologically invariant boundary conditions, defined on neighborhoods of smooth, oriented 2-manifolds (with no framing anomaly). We'll need certain bordisms involving $\CD$ and $\CN$ to be defined, and say more about them momentarily.
\end{itemize}

These assumptions can be placed within the mathematical formalism of axiomatic TQFT, as laid out \emph{e.g.} in \cite{Lurie}. The bulk 3d theory $\CT$ is an object in the symmetric monoidal 3-category $\mathbb T$ of all oriented 3d  TQFT's. There are some options for $\mathbb T$, \emph{e.g.} depending on whether we work with QFT's that are topological ``on the nose'' (in which case $\mathbb T$ is abelian) or cohomological TQFT's (in which case $\mathbb T$ is a dg or infinity category). 
We will eventually consider examples of both types. 
The sort of conditions we put on the bulk theory in \AD\ amount to it being 2-dualizable, and orientable.

We initially think of $\CD$ as a left boundary condition for $\CT$, and $\CN$ as a right boundary condition, meaning that they are 1-morphisms
\be \CD \in \text{Hom}_{\mathbb T}(\CT,\oslash)\,,\qquad \CN \in \text{Hom}_{\mathbb T}(\oslash,\CT) \ee
between $\CT$ and the empty/trivial 3d TQFT `$\oslash$'. (Each ``space of 1-morphisms'' appearing here is a 2-category.)
However, we will eventually need to bend and rotate $\CD$ and $\CN$ continuously. We assume that such deformations can be performed with no ambiguity or monodromy. Formally, this means that there exist ``dual'' objects
\be \ol \CD\in \text{Hom}_{\mathbb T}(\oslash,\CT)\,,\qquad \ol\CN\in \text{Hom}_{\mathbb T}(\CT,\oslash) \ee
such that the bordisms 
\be  \raisebox{-.5in}{\includegraphics[width=4.7in]{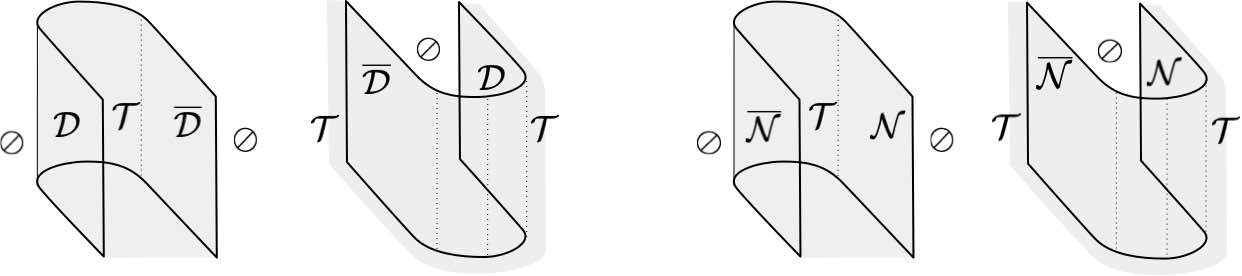}} \label{dualDN} \ee
make sense, independent of bending axis, and there exist pinch-isomorphisms that straighten out an S-bend (\cf\ Eqn. (2.11) of \cite{BCFR-defect}). Mathematically, this amounts to requiring that both $\CD$ and $\CN$ are \emph{one-dualizable} morphisms in $\mathbb T$.

\subsection{The category of line operators}
\label{sec:lines}

At this point, we re-introduce categories of line operators, as these are the main objects that we would like to probe. 

As in the Introduction, let $\CC_\CT$ denote the category of topological line operators in the bulk theory $\CT$, and let $\CC_\CD$ and $\CC_\CN$ denote the categories of line operators on the respective $\CD,\CN$ boundary conditions. Formally,
\be \CC_\CD = \text{End}(\CD)\,,\qquad \CC_\CN = \text{End}(\CN)\,,\qquad \CC_{\CT} = \text{End}(\text{id}_\CT)\,, \label{def-C} \ee
where $\text{id}_\CT \in \text{End}_{\mathbb T}(\CT)$ denotes the identity interface between $\CT$ and itself, and ``op'' denotes a reversed monoidal structure, discussed below. These are each either abelian or dg/infinity categories, depending on the type of QFT we are dealing with. Pictorially:
\be \raisebox{-.5in}{\includegraphics[width=3.4in]{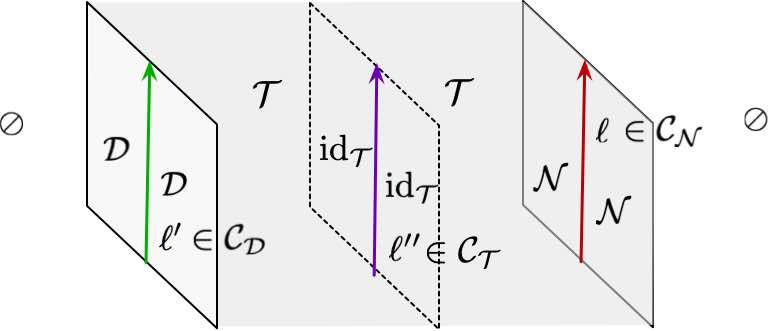}}  \label{HomNN} \ee

The line operators in $\CC_\CD,\CC_\CN,\CC_\CT$ may depend on a choice of orientation along the line, as well as a choice of normal framing: a trivialization of the normal bundle to the line. In the case of $\CC_\CD,\CC_\CN$ there is a canonical choice of normal framing induced by the orientation, so we never speak of it; but for $\CC_\CT$ the normal framing may be relevant --- it appears physically when defining correlation functions of line operators via point-splitting regularization, \cf\ \cite{Witten-Jones}.

The line operators in $\CC_\CD,\CC_\CN,\CC_\CT$ are also invariant under small deformations of their support, and of their normal framing. This endows the categories with the following structures:
\begin{itemize}
\item Each of $\CC_\CD,\CC_\CN,\CC_\CT$ are monoidal categories, with tensor product defined by collision of parallel lines:
\be \hspace{-.3in} \raisebox{-.5in}{\includegraphics[width=5.2in]{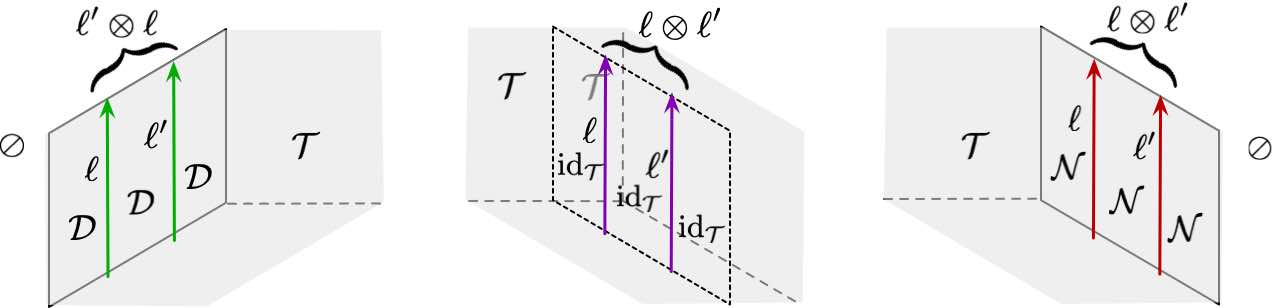}} \label{TensorN} \ee
Note the reversed order of tensor product in $\CC_{\CD}$, as viewed ``from the inside'' of the bulk theory. A monoidal category requires associativity morphisms, which in turn are defined by small topological deformations of triples of lines, of the form
\be \raisebox{-.5in}{\includegraphics[width=4in]{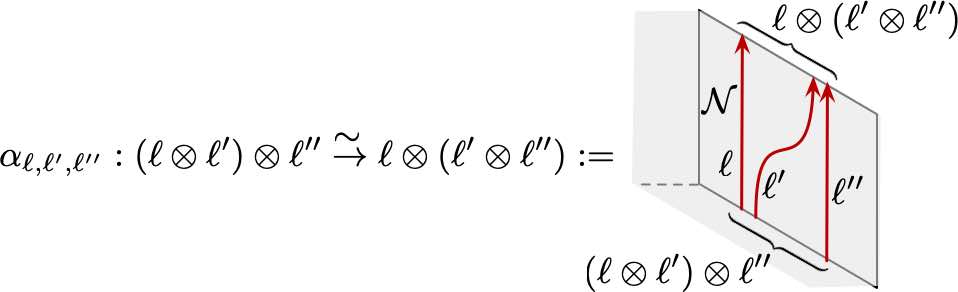}} \label{TensorN-ass} \ee

\item We \emph{assume} that the tensor product for $\CC_\CD,\CC_\CN,\CC_\CT$ is an exact functor, in each of its two slots. This is automatic in the dg/infinity case: the correct notion of tensor product is a derived tensor product, which is exact by construction. For abelian categories, exactness of the tensor product is also automatic when categories are semisimple.

We believe that exactness of the tensor product is a fairly reasonable assumption to make physically.  The one case in which it is not automatic is for categories that are abelian but non semisimple. However, as elaborated on in (\eg) the introduction to \cite{CDGG}, abelian but non-semisimple categories seem to be unnatural in physics --- it's their derived categories that are more naturally embedded in QFT.

Mathematically, monoidal categories with an exact tensor product are usually called tensor categories.

\item The bulk category $\CC_\CT$ is braided, with braiding isomorphisms defined by moving lines out of the plane of the trivial interface $\text{id}_\CT$ (formally using $\text{id}_\CT\simeq  \text{id}_\CT\otimes \text{id}_\CT$ to accomplish this):
\be \raisebox{-.5in}{\includegraphics[width=5in]{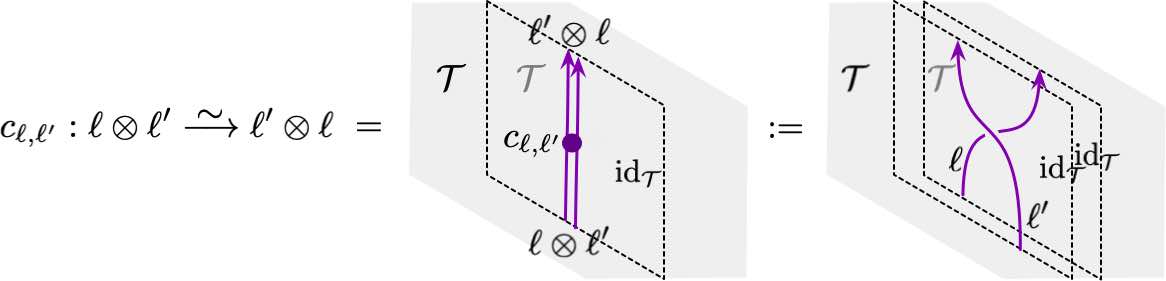}} \label{def-braid} \ee

\item Finally, the bulk category should be equipped with twist isomorphisms, constructed by continuously rotating a line's normal framing by 360$^\circ$:
\be \raisebox{-.3in}{\includegraphics[width=1.1in]{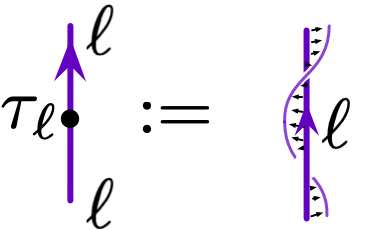}}  \label{def-twist} \ee
\end{itemize}
Altogether, these structures make $\CC_\CD,\CC_\CN$ tensor categories, and $\CC_\CT$ a so-called ribbon category --- a tensor category with braiding and twists isomorphisms, satisfying compatibility relations, \cf\ \cite[Ch. 8]{egno}.

There is one further piece of structure that we might like line operators to have: a notion of duals. Physically, the dual of a line is defined by bending it 180$^\circ$. This is not always possible: standard definitions of topological line operators guarantee that they are invariant under infinitesimal deformations of their support, but the infinitesimal deformations may not always be integrable to such a large, finite displacement. One way to clarify the distinction is to consider a QFT in which line operators are defined by coupling bulk local operators to 1d topological quantum mechanics along the line. When the 1d quantum mechanics has a finite-dimensional Hilbert space, 180$^\circ$ bending can always be done. When the 1d quantum mechanics has an infinite-dimensional Hilbert space, 180$^\circ$ bending may or may not be possible, depending on its particular couplings to the bulk.

When the dual $\ell^*$ of a line operator in any of the categories $\CC_\CD,\CC_\CN,\CC_\CT$ \emph{does} exist, it is canonical, due to our assumptions on the absence of framing anomalies for $\CD,\CN,\CT$: bending $\ell$ to the left or right produces the same $\ell^*$, \eg\ 
\be
\raisebox{-.5in}{\includegraphics[width=5.2in]{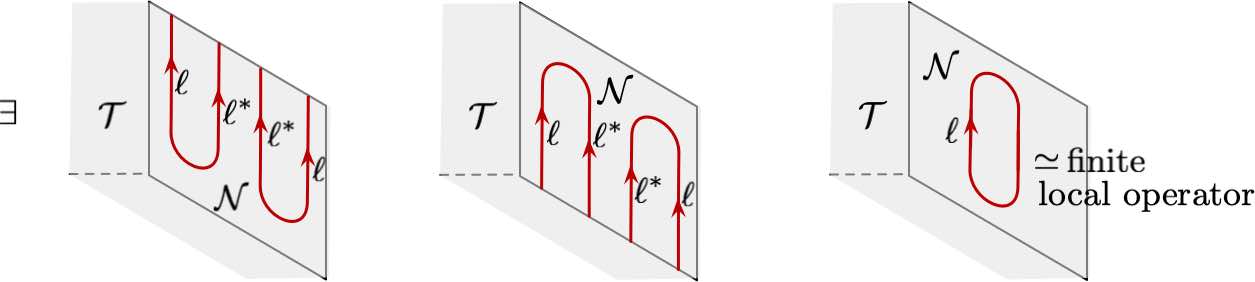}} \label{line-duals}
\ee
and bending $\ell$ 360$^\circ$ produces a line operator that is canonically isomorphic to $\ell$.
Topological invariance further implies that the duals satisfy S-moves such as
\be
\raisebox{-.5in}{\includegraphics[width=3in]{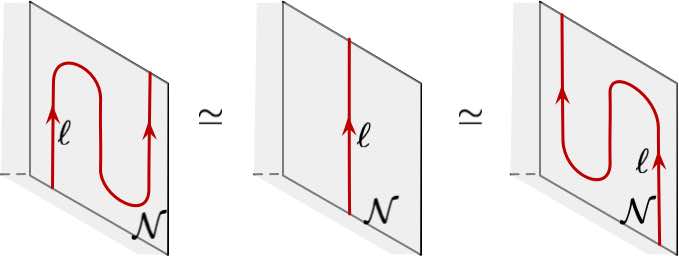}} \label{line-duals2}
\ee

Mathematically, objects that admit duals as in \eqref{line-duals}, satisfying \eqref{line-duals2}, are called rigid, or just dualizable. It's useful for us to introduce full monoidal subcategories 
\be
\CC_\CD^{fd}\subseteq \CC_\CD\,,\qquad \CC_\CN^{fd}\subseteq \CC_\CN\,,\qquad \CC_\CT^{fd}\subseteq \CC_\CT\,.
\ee
of dualizable objects -- where ``fd'' stands for ``fully dualizable,'' or (if one likes the quantum-mechanics analogy) ``finite dimensional.'' The canonical equivalence of left and right duals due to absence of framing anomaly means that $\CC_\CD^{fd}, \CC_\CN^{fd},\CC_\CT^{fd}$ all have canonical pivotal structures, \cf\ \cite[Ch. 4.7]{egno}

Tannaka duality, and other natural TQFT manipulations, are only well behaved when the difference between dualizable objects and the entire category is reasonably small. Aspects of this philosophy were explained in \cite{Ben-Zvi:2008vtm}. The following assumption will make things as nice as possible for us, and covers theories of both abelian and dg/infinity type:
\begin{itemize}
\item[\ADline] For each category $\CC_\CD,\CC_\CN,\CC_\CT$, compact objects are dualizable, and generate the full categories of lines by taking ind-completions. Namely, for $*=\CN,\CD,\CT$: 
\be  \CC_*=\Ind (\CC^c_*)\,, \qquad \CC^c_*\subseteq \CC^{fd}_*\,. \ee
Here $\CC^c$ is the subcategory of compact objects. 
(Technically, we should add the assumption that $\CC_\CD,\CC_\CN,\CC_\CT$ themselves admit arbitrary limits and colimits.)
\end{itemize}
Heuristically, a compact object $\ell$ in a category is an object that is ``small," in the sense that:
\be
\Hom (\ell, \varinjlim_i \ell_i)=\varinjlim_i \Hom (\ell, \ell_i)\,.
\ee
Namely, any map from $\ell$ to a colimit (which is a kind of infinite union) of objects $\ell_i$ has to factor through an element $\ell_i$. In the case when $\CC$ is the category of modules of an algebra, being compact is the same as being a finite chain complex of projective modules (also known as perfect complexes). When we assume $\CC=\Ind (\CC^{c})$, we are assuming that any object in $\CC$ is a colimit of compact objects, which are also dualizable. 

Categories satisfying \ADline\ are called \emph{rigid monoidal categories} in the infinity setting of \cite{gaitsgory2015sheaves}. By default, we will assume \ADline. However, it \emph{does not hold} for some main examples of interest, such as B-twisted 3d $\CN=4$ gauge theories (Sections \ref{sec:gauge}, \ref{sec:GM}). We'll comment in Appendix \ref{app:weakenD} how it can be relaxed slightly, and the consequences thereof.

\subsection{Transversality}
\label{sec:trans}

In order to construct fiber functors, we further assume that the two boundary conditions $\CD$ and $\CN$ are \emph{transverse}. Schematically, this means that sandwiching the bulk theory $\CT$ between a $(\CD,\CN)$ pair produces a trivial 2d TQFT:
\be \hspace{-.3in}   \raisebox{-.5in}{\includegraphics[width=2.7in]{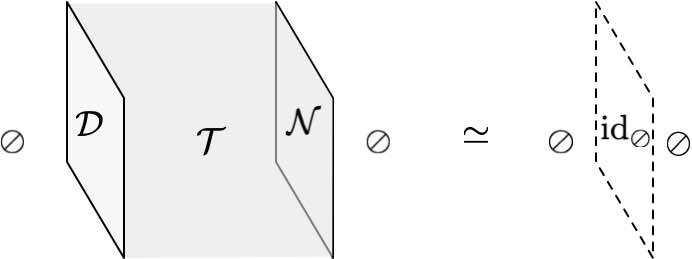}} \label{trans-glob} \ee
We will require that this trivialization holds \emph{locally}, in as strong a way as possible. Formally, we assume:

\begin{itemize}
\item[\AT:] (Transversality) The composition of 1-morphisms $\CD\circ \CN\in \text{Hom}_{\mathbb T}(\oslash,\oslash)$ (representing the sandwich) is isomorphic to the identity 1-morphism $\text{id}_\oslash\in \text{Hom}_{\mathbb T}(\oslash,\oslash)$.
\end{itemize}
Quite a lot is implied in saying that $\CD\circ \CN$ and $\text{id}_\oslash$, which are objects in a 2-category (\emph{a.k.a.} 2d extended operators), are isomorphic. It means there must exist an interface
\be k\in \text{Hom}(\text{id}_\oslash,\CD\circ \CN)\qquad\qquad   \raisebox{-.6in}{\includegraphics[width=1in]{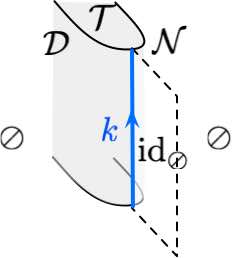}}  \label{trans-loc}  \ee
(defined heuristically by annihilating the $\CD\circ\CN$ sandwich in a half-space) that witnesses the isomorphism.
In turn, the dual $k^*\in \text{Hom}(\CD\circ \CN,\text{id}_\oslash)$ (the orientation-reversal, or 180$^\circ$ bend of $k$) must also be defined, as are cup, cap, and saddle cobordisms
\be \raisebox{-.7in}{\includegraphics[width=5.2in]{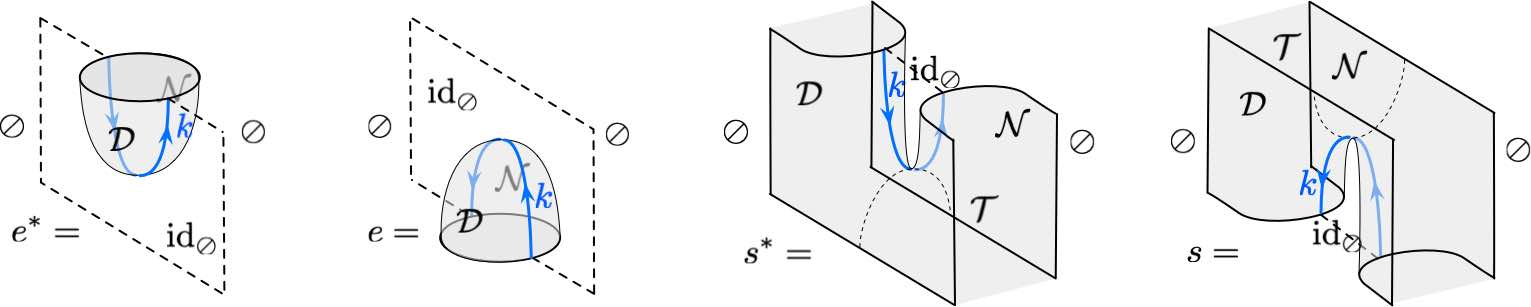}} \label{hole-a} \ee
satisfying standard S-moves analogous to \eqref{line-duals2}, together with identities
\be ee^*=1_{\text{id}_\oslash}\,,\qquad e^*e=\text{id}_{k^*\otimes k}\,,\qquad ss^*=1_{\CD\circ\CN}\,,\qquad s^*s = \text{id}_{k^*\otimes k}\,  \label{k-ids}\ee
manifesting the fact that $k$ and $k^*$ are inverses to each other --- as required if they are to implement the transversality isomorphism $\CD\circ\CN\simeq \text{id}_\oslash$. The first two identities in \eqref{k-ids} allow us to eliminate ``bubbles'' of $\CT$ capped by a $(\CD,\CN)$ pair and to merge bubbles: 
\begin{subequations} 
\be \hspace{0in}  \raisebox{-.8in}{\includegraphics[width=5in]{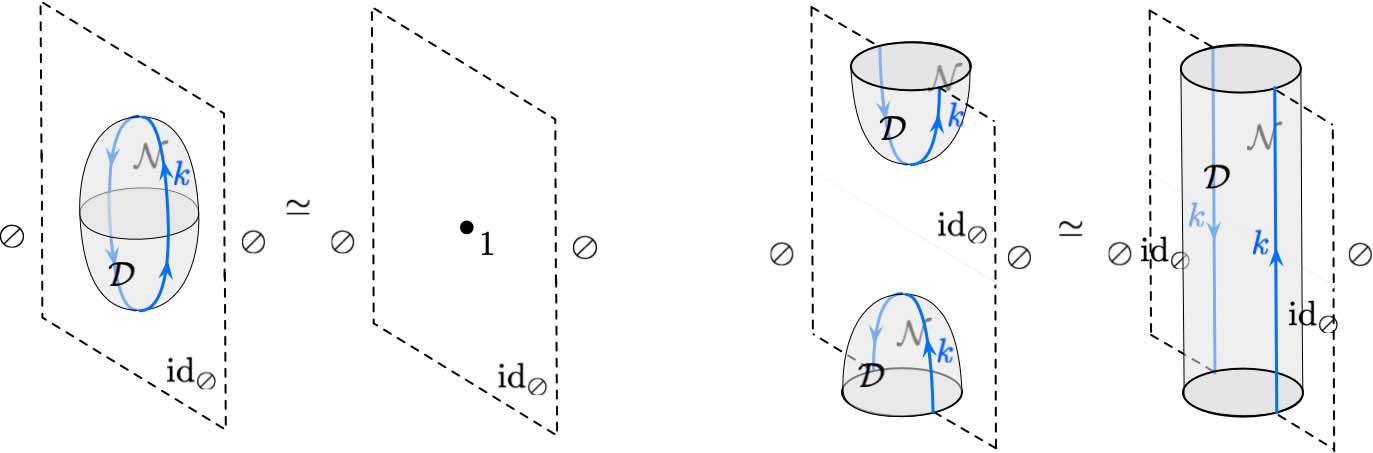}} \label{hole1} \ee
The second two identities allow us to ``punch holes'' in a $(\CD,\CN)$ sandwich, and to merge holes:
\label{hole}
\be \hspace{-.25in}  \raisebox{-.8in}{\includegraphics[width=5.4in]{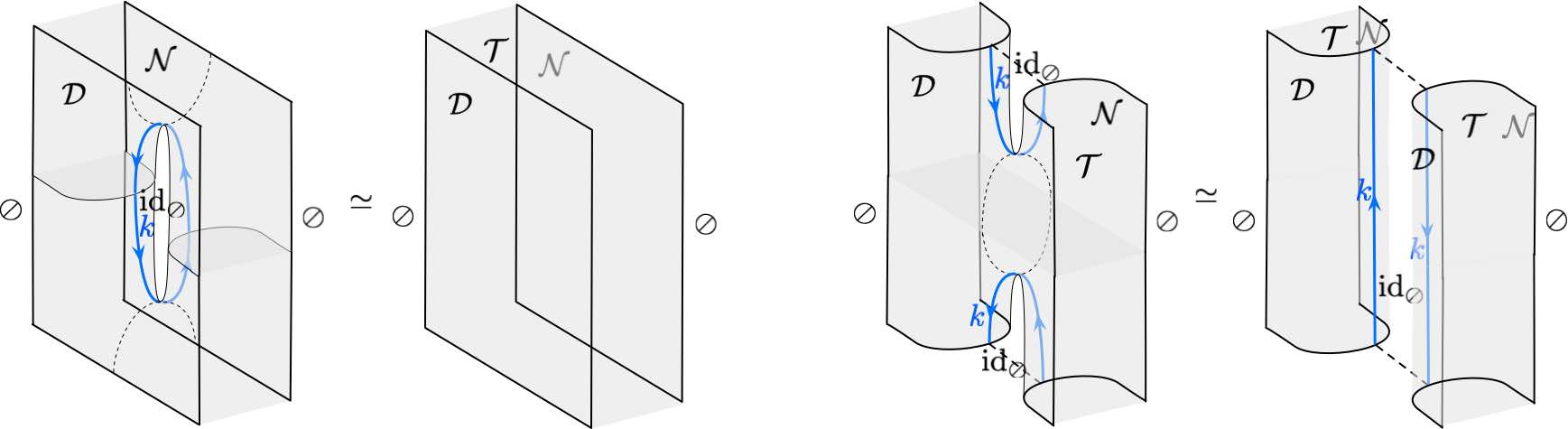}} \label{hole2} \ee
\end{subequations}

There are some useful observations to make about $k$. The category $\text{Hom}(\text{id}_\phi,\CD\circ\CN)$ is the category of boundary conditions for the 2d sandwich theory $\CD\circ\CN$; but since transversality requires $\CD\circ\CN$ to be a trivial 2d TQFT, this category must be Vect --- the category of vector spaces (either ordinary or dg vector spaces, as appropriate). Thus,
\be \AT\quad\Rightarrow\quad \text{Hom}(\text{id}_\phi,\CD\circ\CN) \simeq \text{Vect} \label{DN-Vect} \ee
However, there is a unique invertible object in Vect, namely the one-dimensional vector space `$\C$'. Our interface $k$ must be the image of $\C\in \text{Vect}$ under the equivalence \eqref{DN-Vect}. Similarly, $k^*$ is the image of $\C$ under the dual equivalence $\text{Hom}(\CD\circ\CN,\text{id}_\phi) \simeq \text{Vect}$.

Moreover, note that by starting with the $\CD\circ\CN$ sandwich in \eqref{trans-loc} and rotating $\CD$ to the right (using 1-dualizability of $\CD$) we get an isomorphism of categories $\text{Hom}(\text{id}_\varphi,\CD\circ\CN) \simeq \text{Hom}(\ol \CD,\CN)$. In other words, boundary conditions for $\CD\circ\CN$ are the same as interfaces between $\CN$ and $\ol\CD$. Then we can --- and often will --- interpret $k$ as a special interface in between $\CN$ and $\ol\CD$, isomorphic to $\C\in \text{Vect}$:
\be \raisebox{-.8in}{\includegraphics[width=4in]{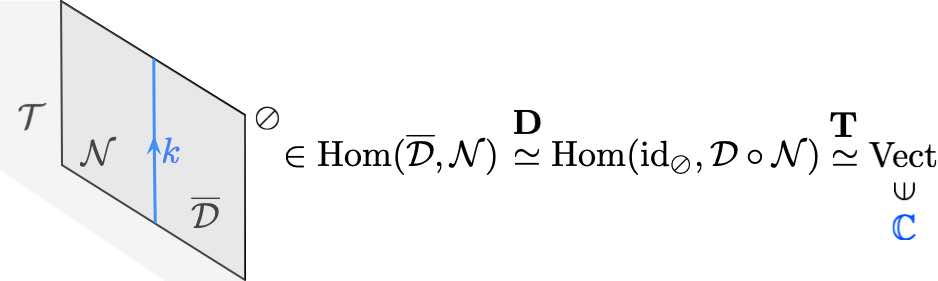}} \label{k-bdy} \ee
Similarly, swinging boundaries around in other ways, we can interpret $k$ as an element of $\text{Hom}(\ol \CN,\CD)$; and $k^*$ as an element of $\text{Hom}(\CN,\ol\CD)$ or $\text{Hom}(\CD,\ol\CN)$, all categories that are isomorphic to Vect.

Physically, $k$ is a locus where the entire QFT is completely trivialized. Any local operators that can be brought there simply evaluate to constants. Any line operator $\ell$ that can be brought to a $k$ line simply ``evaluates'' to a vector space.

Finally, we note that there a stronger dualizability condition we could impose~on~$k$:
\begin{itemize}
\item[\AD$_k$:] $k$ is dualizable as an element of $\text{Hom}(\ol \CN,\CD)$, or equivalently as an element of $\text{Hom}(\ol \CD,\CN)$. In other words, we can bend the interface $k$ between $\CD$ and $\ol\CN$ (and between $\CN$ and $\ol \CD$) by 180$^\circ$, obeying S-moves such as
\be \hspace{-.2in} \raisebox{-.4in}{\includegraphics[width=5in]{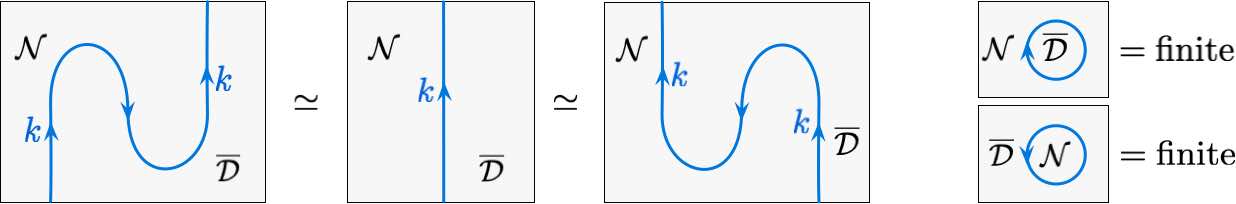}} \label{k-superdual} \ee
\end{itemize}
The condition \AD$_k$ does not follow automatically from dualizability of $k\in\text{Hom}(\text{id}_\oslash,\CD\circ\CN)$ in transversality. In fact, \AD$_k$ turns out to be remarkably strong --- it will lead later in the paper to \emph{finite-dimensional} Hopf algebras. We will only sometimes impose it, as not all our examples satisfy it, and it is interesting to see what happens when it fails.

\subsection{Completeness}
\label{sec:comp}

There is one more assumption that we will use, in order to guarantee faithfulness of various fiber functors, as well as to construct bulk quantum groups as Drinfeld doubles.
It is a sort of adjoint to transversality, and roughly amounts to saying that we can glue a bulk theory back together along pairs of $\CN$ and $\CD$ boundaries:
\begin{itemize}
\item[\AC] (Completeness) Consider the interface $\CN\circ \CD$ between $\CT$ and itself, formed from superposing the left boundary condition $\CN$ with the right boundary condition~$\CD$. We assume that there exists a \emph{deformation} of $\CN\circ \CD$ that is equivalent to the trivial interface:
\be \raisebox{-.5in}{\includegraphics[width=4in]{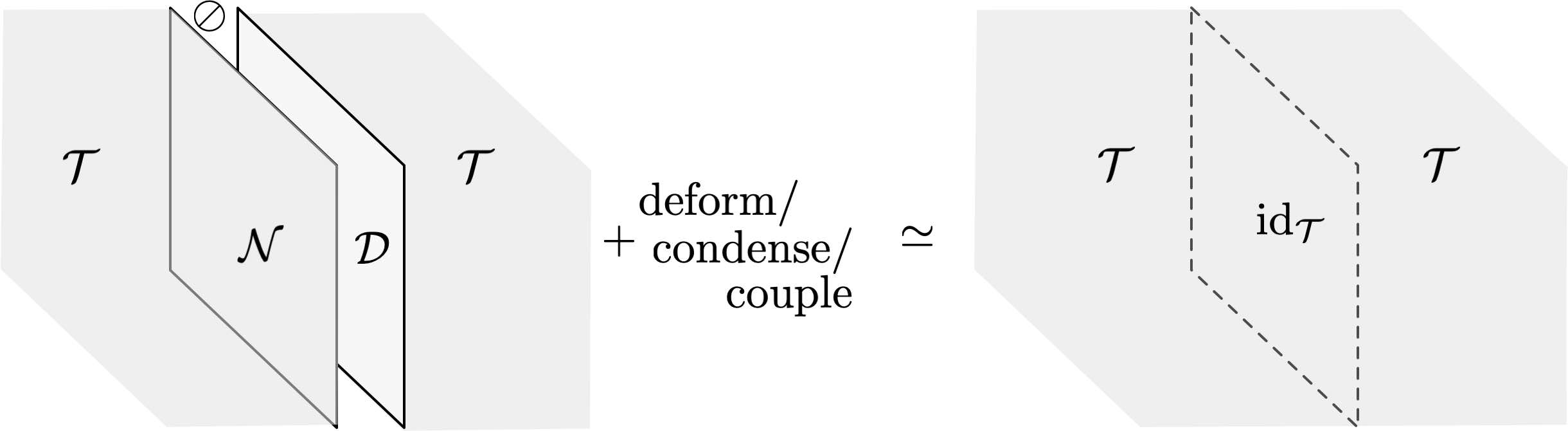}} \label{C-global}\ee
(Equivalently, by dualizability \AD, there is a deformation of $\ol\CD\circ \ol\CN$ to the trivial interface between $\CT$ and itself, in the opposite direction.)

By ``deformation'' of $\CN\circ \CD$ we mean any insertion of operators belonging to the tensor product of 2d boundary theories on $\CN$ and $\CD$.
In the case of a dg TQFT defined perturbatively, this amounts to the choice of a degree-two Maurer-Cartan element in the product of boundary operator algebras, which deforms the boundary action. Non-perturbatively, the deformation may be expressed as an insertion of a codimension-one web of defects along the boundary, or a limit of such, sometimes referred to as a condensation.
\end{itemize}

We should be more precise about how locally we require the equivalence in \eqref{C-global} to hold. The strongest local version, analogous to what we required for transversality, would posit the existence of a fully dualizable morphism 
\be \exists\;?\quad  \varphi\in \text{Hom}(\text{id}_\CT,\CN\circ\CD) \qquad  \raisebox{-.45in}{\includegraphics[width=1in]{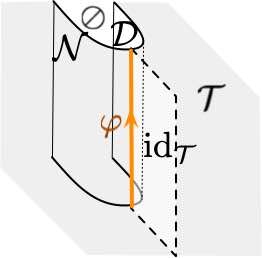}}, \ee
representing the interface between the non-interacting $\CN\circ \CD$ superposition and the invisible interface $\text{id}_\CT$, \cf\ \cite[Def. 1.3.1]{GaiottoJF}. The morphism $\varphi$ would then need to have the property that bubbles of the empty theory can be removed,
\be  \raisebox{-.45in}{\includegraphics[width=2.2in]{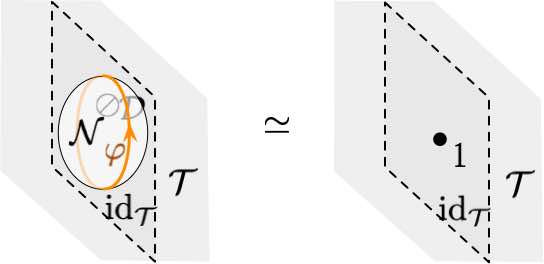}} \label{C-loc} \ee

Unfortunately, this fully local version rarely seems to hold. For example, in gauge theory the putative interface $\varphi$ does not exist. It would be constructed by coupling $\CN$ and $\CD$ together along \emph{half} of the common $\R^2$ that they share, say along $\R_+\times \R$. However, if $G$ is continuous, the perturbative coupling
\be \int_{\R_+\times \R} A\big|_\CN\cdot *J\big|_\CD \label{gauge-glue} \ee
is not gauge invariant due to a boundary term. If $G$ is finite, the condensation webs defined later in Section \ref{sec:DW-C} again do not make sense on a half-space, due to a lack of proper boundary condition.

The most we can ask for is a deformation/condensation of $\CN\circ\CD$ to $\text{id}_\CT$ along 2d regions \emph{with appropriate boundary conditions}. Suppose that our bulk theory $\CT$ is initially placed on a 3-manifold (possibly disconnected) with right boundary $\Sigma_R$ and left boundary $\Sigma_L$. Let $R\subseteq \Sigma_L$ be a region of the left boundary, with $\pd R$ a finite disjoint union of piecewise smooth curves (possibly $\pd R=\oslash$); and let $\ol R\subseteq \Sigma_R$ be a region of the right boundary such that there is an orientation-reversing diffeomorphism
\be h:R\overset\sim\longrightarrow \ol R \ee
that extends to an orientation-preserving diffeomorphism $h:\pd R \overset\sim\longrightarrow \pd\ol R  $. Suppose that $\CR$ is labelled entirely by the boundary condition $\CD$ and that $\ol R$ is labelled entirely by $\CN$.
\begin{Def} \label{def-bounded} We say that the pair $R,\ol R$ as above is \emph{well-bounded} if for every smooth segment $\gamma\subseteq \pd R$ exactly one of the following holds:
\begin{itemize}
\item $\gamma$ lies on a $k$ (or $k^*$) interface between $\CD$ and $\ol \CN$ on the surface $\Sigma_L$, and $h(\gamma)$ lies in the interior of an $\CN$ b.c. on $\Sigma_R$; or
\item $\gamma$ lies in the interior of a $\CD$ b.c. on $\Sigma_L$, and $h(\gamma)$ lies on a $k$ (or $k^*$) interface between $\CN$ and $\ol\CD$ on $\Sigma_R$.
\end{itemize}
\end{Def}

\noindent Note that if $\pd R=\oslash$ then $R,\ol R$ are automatically well bounded. We then require
\begin{itemize}
\item[\AC$_{\rm R}$:] For any well-bounded pair $R,\ol R$ labelled respectively by $\CD,\CN$, there exists a deformation/condensation such that $\CN\circ\CD$ is isomorphic to $\text{id}_\CT$ along $R\simeq \ol R$. In other words, we can glue the bulk theory back together along $R$.
\end{itemize}

\noindent There are two special cases that will play a key role:
\begin{itemize}
\item[\AC$_{\rm strip}$:] (Strip gluing) There exists a deformation $\CN\circ \CD\leadsto \text{id}_\CT$ along an infinite strip
\be \hspace{-.1in} \label{stripglue} \raisebox{-1.3in}{\includegraphics[width=4.8in]{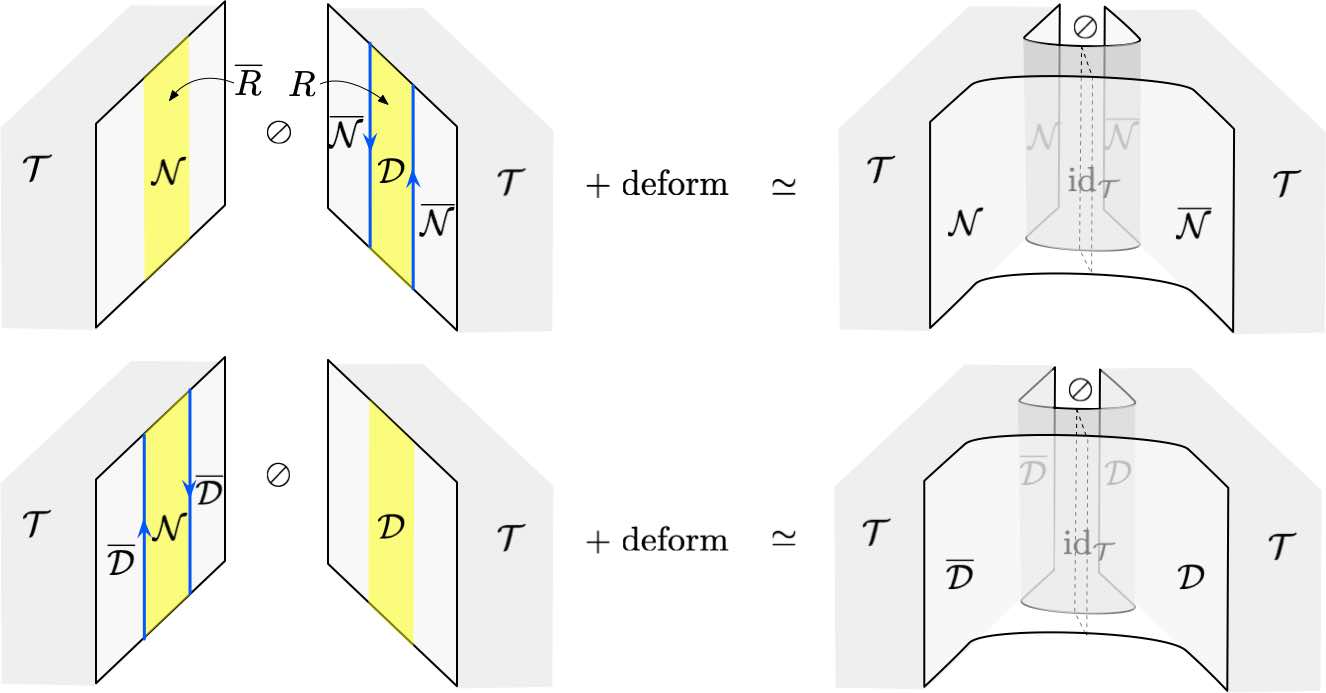}} \ee
where $\CD$ is bounded by $\ol\CN$, or $\CN$ is bounded by $\ol\CD$, as depicted above.
(The strip $R$ along which the deformation is performed is highlighted in yellow.)

\item[\AC$_{\rm box}$:] (Box gluing) There exists a deformation $\CN\circ \CD\leadsto \text{id}_\CT$ along a rectangle $R$ formed from the intersection of transverse bounded strips on $\CN$ and $\CD$, as shown here:
\be \hspace{-.2in}  \label{boxglue} \raisebox{-.7in}{\includegraphics[width=5in]{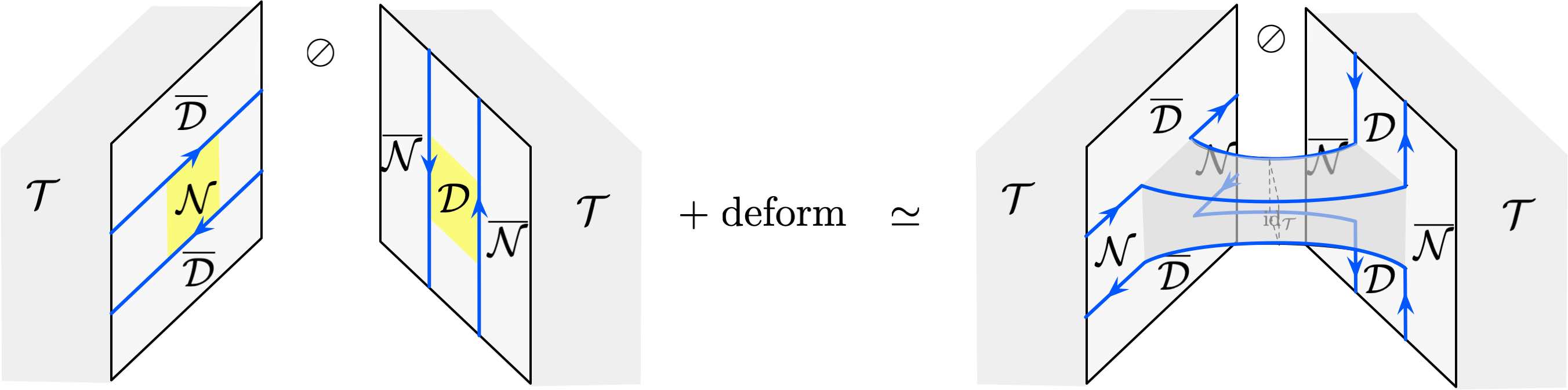}} \ee
(We will argue later in Section \ref{sec:h-box} that box gluing actually does not actually require any special completeness assumptions on $\CD$ and $\CN$, aside from a degree of dualizability, essentially because the region $R$ in this case is compact.)
\end{itemize}

\section{Fiber functors and spark algebras}
\label{sec:linefunctor}

Suppose we have 3d topological QFT $\CT$ and a pair of boundary conditions $\CD,\CN$ satisfying the assumptions of Section \ref{sec:setup}. In this section, we begin implementing Tannaka duality by constructing functors
\be \CF_\CD:\CC_\CD\to\text{Vect}\,,\qquad \CF_\CT:\CC_\CT\to\text{Vect}\,,\qquad \CF_\CN:\CC_\CN\to\text{Vect}\, \label{threeFF} \ee
for each of the boundary and bulk categories of line operators. We define the functors in Section \ref{sec:FF} by ``wrapping'' line operators in boundary conditions, and show

\begin{Prop} \label{Prop:F}
Assuming topological invariance (\AD, Section \ref{sec:top}) and transversality (\AT, Section \ref{sec:trans}), the  functors \eqref{threeFF} are monoidal, send $\id$ to $\C$, and preserve duals.
\end{Prop}

\noindent We'll also give alternative definitions of the fiber functors in Section \ref{sec:FF-Hom}, as taking `Homs' with three special objects $\kDk,\kNk,\CDCN$ in the respective categories $\CC_\CD,\CC_\CT,\CC_\CN$.

We then ask what the symmetry algebras of these functors are. There is a natural physical answer to this question, which we present in Section \ref{sec:spark}. We define three ``spark algebras'' roughly as
\be \label{threeH} \begin{array}{lcl} \CH_\CD  &=& \text{\{operators supported near an interval on $\CD$ with ends on $\CN$\}}\,,  \\[.1cm]
 \CH_\CN  &=& \text{\{operators supported near an interval on $\CN$ with ends on $\CD$\}}\,,  \\[.1cm]
 U &=& \text{\{operators supported near an $S^1$ traversing both $\CN$ and $\CD$\}}\,,
\end{array} \ee
proposing that they give the desired symmetries $\CH_\CD\simeq \text{End}(\CF_\CN)$, $\CH_\CN\simeq \text{End}(\CF_\CD)$, $U\simeq \text{End}(\CF_\CT)$. In particular we explain that the fiber functors \eqref{threeFF} lift naturally to functors of module categories
\be \wt\CF_\CD:\CC_\CD\to\CH_\CN\text{-mod}\,,\qquad \wt\CF_\CT:\CC_\CT\to U\text{-mod}\,,\qquad \wt\CF_\CN:\CC_\CN\to\CH_\CD\text{-mod}\,. \label{threeFFH} \ee
We'll give several other equivalent definitions of the spark algebras as well, including as state spaces on rectangles and annuli, connecting with work of \cite{Reutter-slides, FreedTeleman-Ising}; and as interval factorization homology, connecting with ladder categories \cite{MorrisonWalker,Hoek}, tube algebras \cite{Ocneanu-tube}, and strip algebras \cite{CHO-solitons}.

Along the way, we will reveal several simple structural properties of the spark algebras. We'll argue in Section \ref{sec:spark-alt} that
\begin{Prop} \label{Prop:decomp}
Assuming topological invariance (\AD) and transversality (\AT), bulk sparks decompose as
\be U \simeq \CH_\CD\otimes \CH_\CN \qquad \text{(as vector spaces)}\,. \label{TND-vect}  \ee
and there is a continuous nondegenerate bilinear pairing
\be h:\CH_\CD\otimes \CH_\CN\to \C \ee
defined by inserting sparks into a solid 3-ball with $\CN$ and $\CD$ boundaries.
\end{Prop}

More physically, one may think of $\CH_\CD$ and $\CH_\CN$ as incoming and outgoing states on the same rectangle, and the pairing $h$ is their inner product. In Section \ref{sec:spark-finite}, we show that the strong dualizability condition \AD$_k$ guarantees that spark algebras are in fact finite dimensional. Then
\be \CH_\CN \simeq \CH_\CD^* \qquad \text{(as vector spaces)}\,. \label{ND-dual-vect} \ee
More generally, $\CH_\CD$ and $\CH_\CN$ are continuous duals, as topological vector spaces.

Finally, in Section \ref{sec:faithful}, we delve into faithfulness of the fiber functors. After working out an explicit general form of the strip gluing \AC$_{\rm strip}$ from \eqref{stripglue}, we argue that
\begin{Thm}\label{Thm:faithful}
Assuming \AD, \AT, \AD$_{k}$, completeness \AC$_{\rm strip}$, and a mild technical condition (exactness and continuity of fiber functors), the lifted fiber functors \eqref{threeFFH} become equivalences of categories --- either abelian or dg/infinity as appropriate.
\end{Thm}
\noindent Here the \AD$_{k}$ assumption allows us to represent fiber functors as Homs out of the objects $\kDk,\kNk,\CDCN$, and we show that completeness implies that $\kDk,\kNk,\CDCN$ are generators of their respective categories. We suspect that a more sophisticated argument could circumvent \AD$_{k}$.

\subsection{Fiber functors}
\label{sec:FF}

Consider the following operation. Given any line $\ell$ on $\CN$, we can associate a vector space to it by ``capping'' a strip of the $\CN$ boundary (containing $\ell$) with a half-cylinder of $\CD$ boundary, and completely filling in the resulting half-cylinder with the bulk theory $\CT$. Call the resulting vector space $\CF_\CN(\ell)$. In pictures:
\be \raisebox{-.4in}{\includegraphics[width=2.7in]{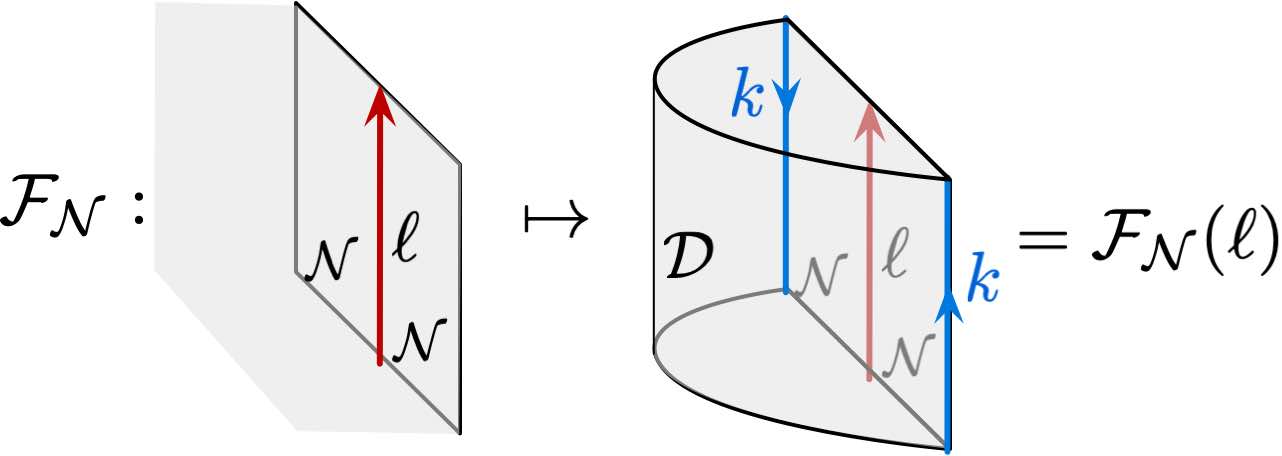}} \label{FiberBdy-N} \ee
or if we draw just a spatial slice
\be \CF_\CN:\ell\mapsto \CF_\CN(\ell)= \text{States}\Big( \raisebox{-.3in}{\includegraphics[width=.75in]{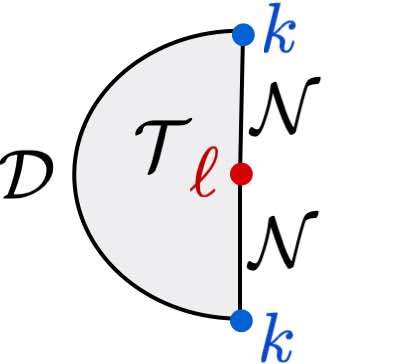}}\!\!\!\Big)\,. \ee
(A similar fiber functor was described in \cite[Figure 20]{FreedTeleman-Ising}.)
Note that at the junctions between $\CN$ and $\CD$ we have placed our transversality interface $k$ (Section \ref{sec:trans}). In theories that are topological on the nose, \eqref{FiberBdy-N} is a standard vector space. In cohomological theories, the RHS is naturally a dg vector space, \ie\ a chain complex.

This ``capping off'' operation is naturally compatible with morphisms, and defines a functor
\be \CF_\CN:\CC_\CN\to \text{Vect}\,. \ee
Namely, given a pair of lines $\ell,\ell'$ and a morphism (local operator) $\mathcal O\in \text{Hom}_{\CC_\CN}(\ell,\ell')$ between them, 
we can cap off $\CO$ to get a map $\CF_\CN(\CO)$ from the state space $\CF_\CN(\ell)$ to the state space $\CF_\CN(\ell')$:
\be \raisebox{-.4in}{\includegraphics[width=4in]{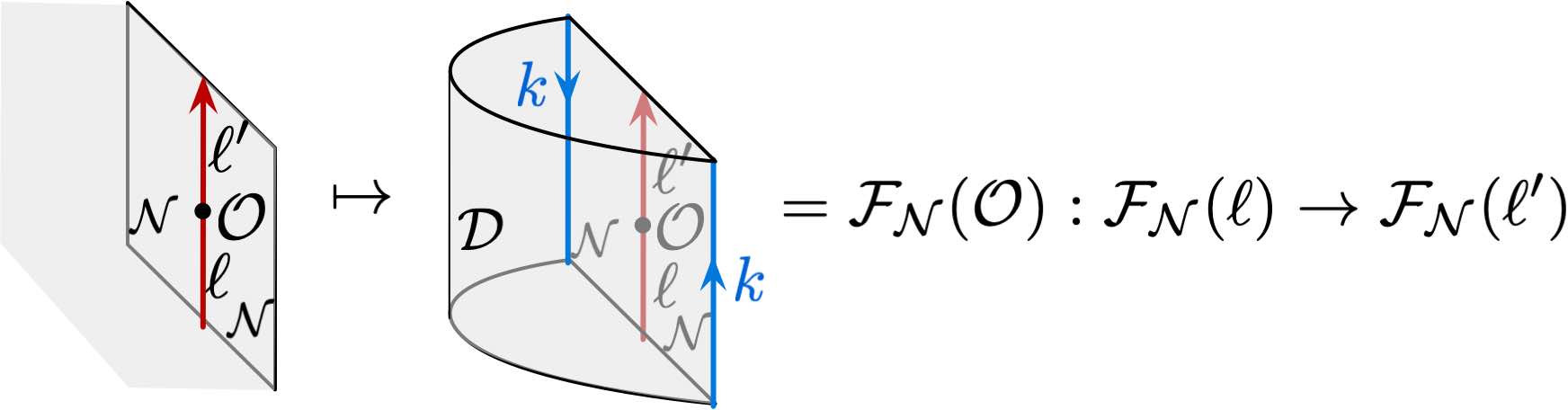}} \label{Falpha} \ee
Inserting multiple local operators $\CO_n...\CO_2\CO_1$ along the line gives $\CF_\CN(\CO_n...\CO_2\CO_1) \simeq \CF_\CN(\CO_n)...\CF_\CN(\CO_2)\CF_\CN(\CO_1)$, preserving products (up to quasi-isomorphism of chain complexes in the dg setting). In the dg/infinity setting, the functor also preserves higher operations, defined (\eg) by integrated descendants of bulk and boundary operators, and defines a dg/infinity functor.

In a similar way, we can define functors
\be \CF_\CD:\CC_\CD\to\text{Vect}\,, \qquad\qquad \CF_\CT:\CC_\CT\to\text{Vect}\,. \ee
The functor $\CF_\CD$ sends a line on the $\CD$ boundary to a vector space by wrapping it in a half-cylinder bounded by $\CN$:
\be \raisebox{-.45in}{\includegraphics[width=2.1in]{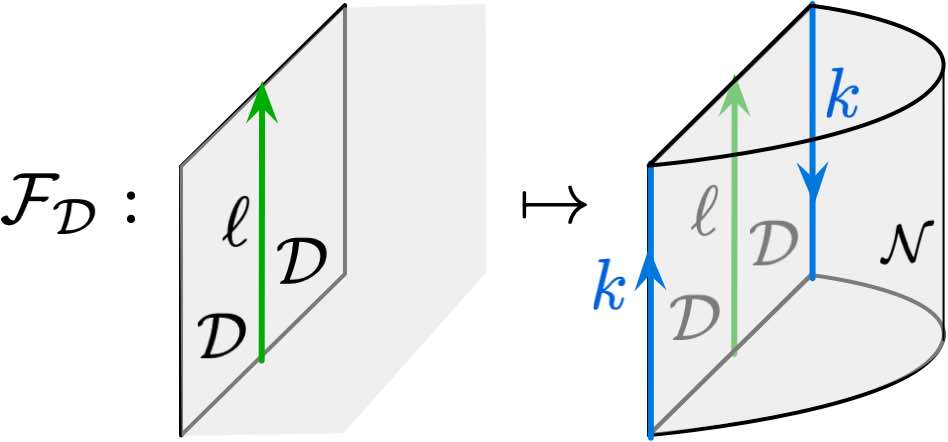}} \quad\text{or}\quad \CF_\CD:\ell\mapsto\text{States}\Big(\;\raisebox{-.32in}{\includegraphics[width=.6in]{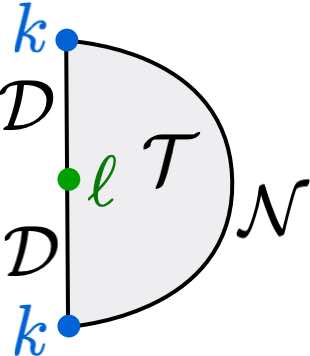}} \Big) =\CF_\CD(\ell) \label{defFD} \ee
The functor $\CF_\CT$ sends a line in the bulk to a vector space by wrapping it in a full cylinder bounded half by $\CD$ and half by $\CN$:
\be \raisebox{-.39in}{\includegraphics[width=5in]{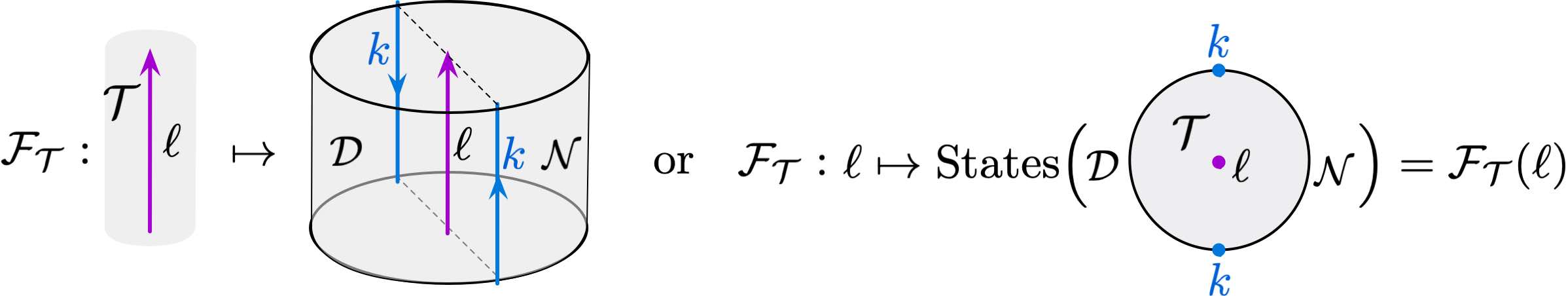}} \label{defFT} \ee
The three functors $\CF_\CN,\CF_\CD,\CF_\CT$ are essentially the same -- they only difference is where the line operator appears on the RHS: on the $\CN$ boundary, the $\CD$ boundary, or in the bulk.

\subsubsection{Preserving tensor products and duals}
\label{sec:mon}

Transversality now implies that the three functors $\CF_\CN,\CF_\CD,\CF_\CT$ have additional properties. They 1) send the trivial line $\id$ in each line-operator category to the trivial line $\id_{\rm Vect}=\C$ in vector spaces; 2) are monoidal, preserving tensor products; and 3) preserve duals, when duals exist. Let's explain how this works topologically, establishing Prop. \ref{Prop:F}. We'll focus on $\CF_\CN$, as the operations for the other two functors work essentially the same way.

First, the isomorphism $e:\CF_\CN(\id)\overset\sim\to \C$ (and its inverse $e^*: \C \overset\sim\to \CF_\CN(\id)$)
are given precisely by the cup and cap diagrams for the transversality interface $k$, from \eqref{hole-a}:
\be \raisebox{-.4in}{\includegraphics[width=4.7in]{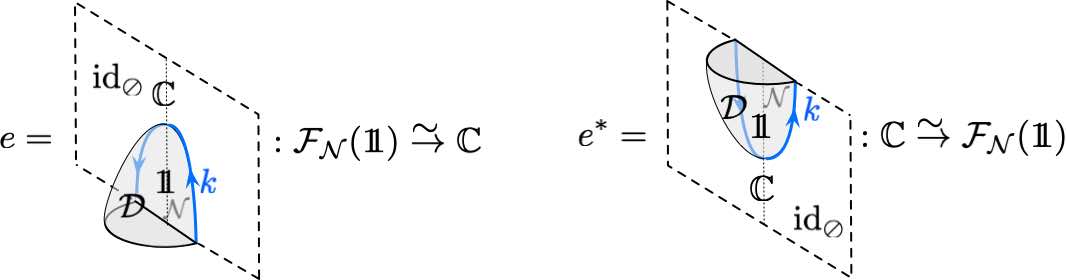}} \label{e-counit} \ee

Second, recall that being monoidal means there exists a natural isomorphism $J:\CF_\CN(-)\otimes\CF_\CN(-)\overset\sim\to \CF_\CN(-\otimes-)$, or explicitly
\be J_{\ell, \ell'}: \CF_\CN(\ell)\otimes \CF_\CN(\ell')   \overset\sim\longrightarrow \CF_\CN(\ell\otimes\ell') \qquad (\forall\,\ell,\ell'\in \CC_\CN) \,, \label{FN-monoidal} \ee
where the LHS is the tensor product of vector spaces, and the RHS is the tensor product in $\CC_\CN$. Moreover, $J$ is required to intertwine the (trivial) associator in vector spaces with the associator in $\CC_\CN$, as discussed in \cite[Sec. 2.4]{egno}.

In our case, the isomorphism $J$ and its inverse are defined by
\be \raisebox{-.4in}{\includegraphics[width=4in]{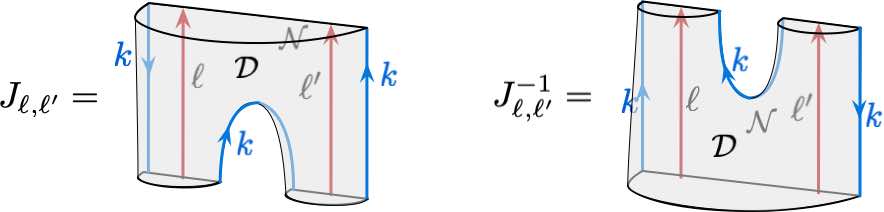}} \label{defJ} \ee
using the saddles from transversality \eqref{hole-a}.
The identities  $J_{\ell,\ell'}\circ J_{\ell,\ell'}^{-1} \simeq \text{id}_{\CF_\CN(\ell\otimes\ell')}$ and  $J_{\ell,\ell'}^{-1}\circ J_{\ell,\ell'} \simeq \text{id}_{\CF_\CN(\ell)\otimes \CF_\CN(\ell')}$ come directly from applying the LHS and RHS of the transversality axioms \eqref{hole2}; for example
\be \raisebox{-.4in}{\includegraphics[width=4.2in]{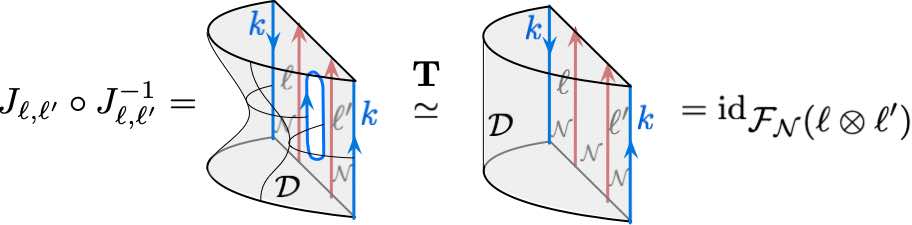}} \label{FN-Mon-3d} \ee
A purely topological manipulation shows that associators are preserved; it looks like:
\be \raisebox{-.4in}{\includegraphics[width=3.6in]{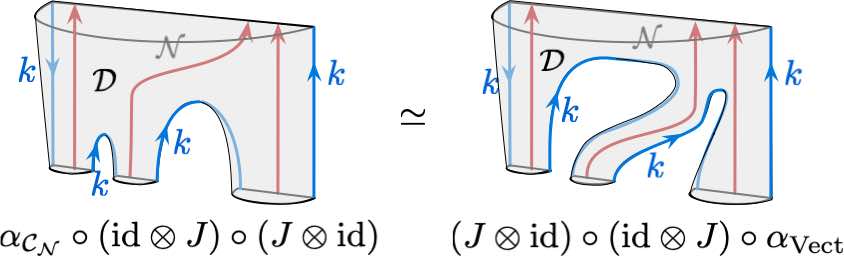}} \label{FN-Mon-ass} \ee

For the functors $\CF_\CD$ and $\CF_\CT$, the $J$ isomorphisms are defined the same way. For example, for $\CF_\CT$ we have
\be J_{\ell,\ell'} =  \raisebox{-.39in}{\includegraphics[width=1.2in]{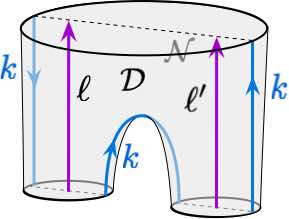}} \label{J-bulk}   \ee

Finally, preserving duals means heuristically that $\CF_\CN(\ell^*)\simeq \CF_\CN(\ell)^*$ for any dualizable object $\ell\in \CC_\CN$. To spell this out, recall that for $\ell$ to be dualizable in $\CC_\CN$ means that there are evaluation and coevaluation maps (caps and cups) $\text{ev}_\ell:\ell^*\otimes\ell\to \id$, $\text{ev}_\ell':\ell\otimes{}^*\ell\to \id$,  $\text{coev}_\ell:\id\to\ell\otimes\ell^*$, $\text{coev}_\ell':\id\to{}^*\ell\otimes\ell$ satisfying S-moves. In addition, due to a lack of framing anomaly, left and right duals are canonically isomorphic: $\ell^*={}^*\ell$, so we only talk about $\ell^*$. For $\CF_\CN$ to preserve duals means that there exist four corresponding natural morphisms
\be \text{Ev}_\ell : \CF_\CN(\ell^*)\otimes\CF_\CN(\ell)\to \C\,,\quad \text{Coev}_\ell:\C\to\CF_\CN(\ell)\otimes\CF_\CN(\ell^*)\,, \ee
and similarly $\text{Ev}_\ell',\text{Coev}_\ell'$, for all \emph{dualizable} $\ell$, obeying S-moves, etc. 

The existence of the four natural morphisms $\text{Ev}_\ell$, etc. follows automatically from the fact that $\CF_\CN$ is monoidal and $\ell$ is dualizable in $\CC_\CN$. For example,
\be \hspace{-.1in} \text{Ev}_\ell := e\circ \CF_\CN(\text{ev}_\ell) \circ J_{\ell^*,\ell} = \raisebox{-.3in}{\includegraphics[width=1in]{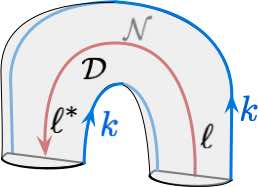}}\,,\quad \text{Coev}_\ell := J^{-1}_{\ell,\ell^*} \circ \CF_\CN(\text{coev}_\ell)\circ e^* = \raisebox{-.3in}{\includegraphics[width=1in]{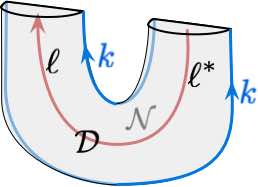}} 
\label{defEv} \ee

\subsubsection{Exactness and continuity}
\label{sec:cts}

We are often going to need to impose some mild technical assumptions on the fiber functors $\CF_\CN,\CF_\CD,\CF_\CT$ in order for later arguments to work mathematically.

First, we assume that the fiber functors are exact. This is automatic in the dg/infinity setting. It is also automatic for semisimple abelian theories. It is not automatic for non-semisimple abelian theories, in which case exactness must be imposed by hand. However, we note again that non-semisimple abelian theories might be somewhat unnatural in physics, as QFT seems to naturally replace them with their derived ``completions.'' (We already encountered this in the context of of tensor products in Section \ref{sec:lines}.)

Second, we will assume that the fiber functors are continuous. This means that they preserve (commute with) infinite limits and colimits of objects. This is a trivial requirement for finite abelian categories, but is important to impose in the infinite and dg setting in order for functors to be sufficiently well behaved. It will hold in the examples of this paper.

\subsection{Fiber functors as Homs}
\label{sec:FF-Hom}

The three fiber functors of the previous section can be given an alternative construction that uses the Hom functors within the categories $\CC_\CN,\CC_\CD,\CC_\CT$ themselves, provided that we assume the extra dualizability condition \AD$_k$ on $k$ from \eqref{k-superdual}. Physically, this alternative construction comes from using state-operator correspondences to relate the state spaces on half-discs or discs from the previous section to spaces of local operators at junctions of certain lines (\emph{a.k.a.} Hom's). The alternative construction is especially useful when analyzing properties of endomorphism algebras of the fiber functors mathematically.

Let ${}_k\ol\CD_k$ denote the line operator on a $\CN$ b.c. created from a strip of orientation-reversed $\CD$ b.c. that has been squeezed to infinitesimal thickness, and similarly for ${}_k\ol\CN_k$:
\be \raisebox{-.4in}{\includegraphics[width=4.4in]{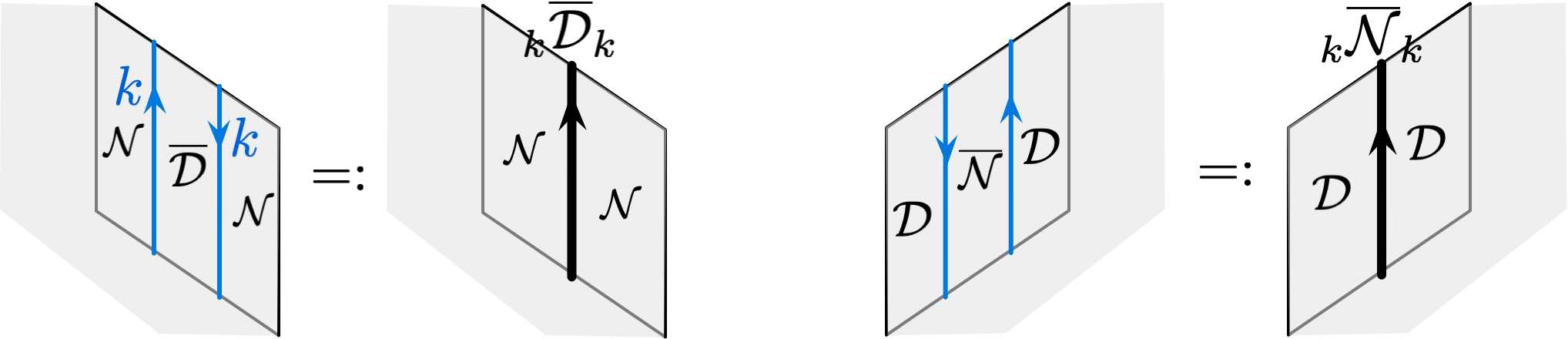}} \label{def-kDk} \ee
Then a state-operator correspondence relates the half-disc state space $\CF_\CN(\ell)$ to the space of local operator at a junction of the lines $\ell$ and ${}_k\ol\CD_k$, on the boundary $\CN$:
\be \raisebox{-.5in}{\includegraphics[width=4.4in]{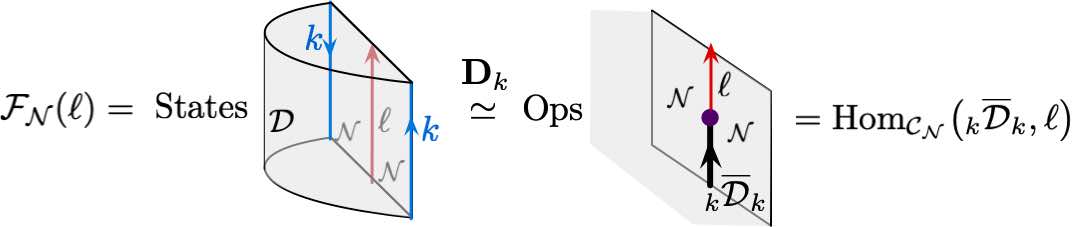}} \label{CDC-stateop} \ee
If we are careful about the topological manipulation that ``inverts'' or ``opens up'' the cylinder into a half-space in \eqref{CDC-stateop}, we find that it requires bending $k$ interfaces around as in \eqref{k-superdual}, and thus requires the strong dualizability condition \AD$_k$ on $k$.

The RHS of \eqref{CDC-stateop} is a morphism space in the category of line operators on $\CN$. Thus the functor $\CF_\CN$ may be represented as
\begin{subequations} \label{FHomCDC}
\be \CF_\CN = \text{Hom}_{\CC_\CN}({}_k\ol\CD_k,-)\,. \ee
Similarly, the functor $\CF_\CD$ may be represented as
\be \CF_\CD = \text{Hom}_{\CC_\CD}({}_k\ol\CN_k,-)\,. \ee
\end{subequations}

For the bulk fiber functor, we create a special line operator $\raisebox{-.15cm}{\includegraphics[width=.5cm]{CDCNline.jpg}} \in \CC_\CT$ by 
first drilling a solid cylinder out of the bulk, then splitting its boundary into two infinite strips labelled by $\ol\CD$ and $\ol\CN$, and finally taking a limit in which this drilled-out cylinder becomes infinitesimally thick:
\be \raisebox{-.35in}{\includegraphics[width=2in]{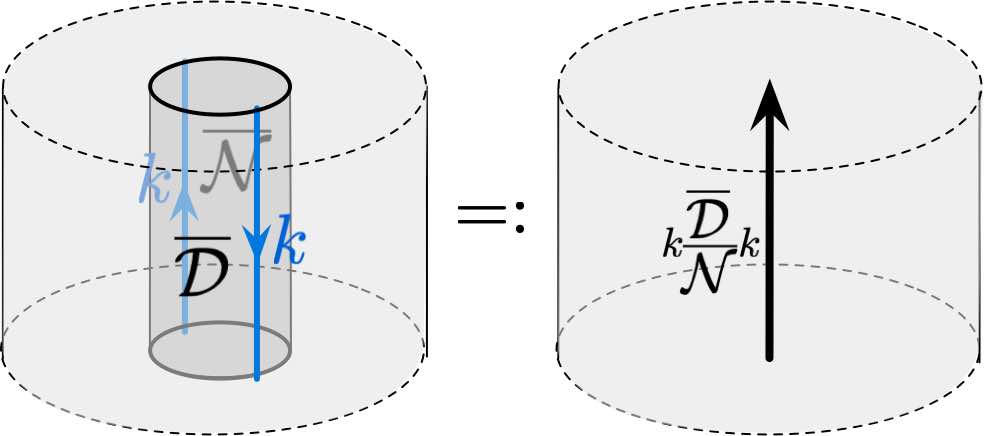}} \label{CDCN} \ee
Now a state-operator correspondence identifies states on a solid disc with operators at a junction
\be \raisebox{-.35in}{\includegraphics[width=4.2in]{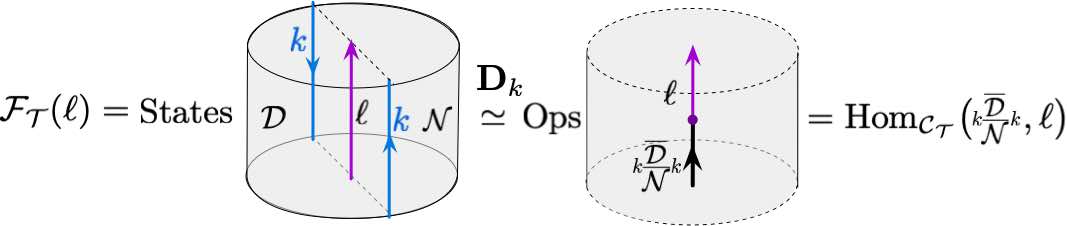}} \label{CDCN-stateop} \ee
so that
\be \CF_\CT = \text{Hom}_{\CC_\CT}\big(\raisebox{-.15cm}{\includegraphics[width=.5cm]{CDCNline.jpg}},-)\,. \label{FHomT} \ee

We remark that, once we assume \AD$_k$, the objects $\kDk,\kNk$, and $\CDCN$ have a lot of extra structure! It makes sense to bend them by 180$^\circ$, so we find that they must be dualizable in their respective categories. Moreover, they are automatically \emph{self-dual}, as their definition is independent of orientation, \eg\
\be \raisebox{-.3in}{\includegraphics[width=2.8in]{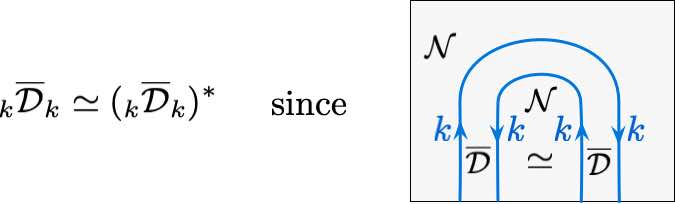}}  \label{kDk-dual} \ee
They are also bi-algebra objects, with multiplication and co-multiplication morphisms defined by pairs of pants, \eg
\be \raisebox{-.3in}{\includegraphics[width=5.5in]{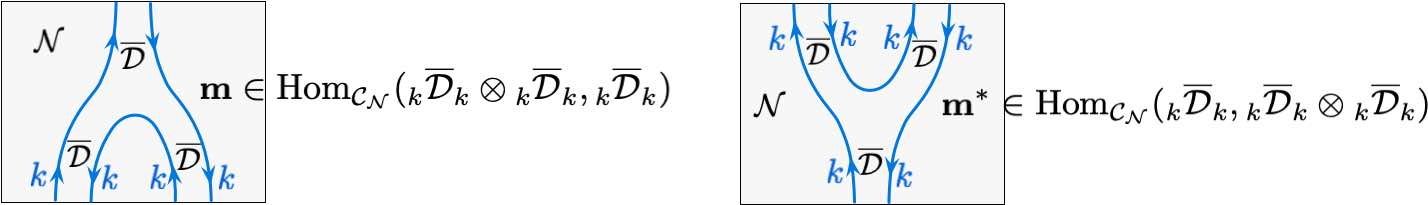}}  \label{kDk-alg} \ee
More so, they are Frobenius algebra objects, with evaluation, coevaluation, and a trace defined (respectively) by caps (as in \eqref{kDk-dual}), cups, and closed loops.

Mathematically, \AD$_k$ implies that the fiber functors $\CF$ have well-defined (and continuous) left and right adjoints, given by bending the interface $k$ to the left or to the right. The three special objects $\kDk, \kNk$ and $\CDCN$ are simply the left adjoint of fiber functors applied to the trivial vector space $\C$. This is explained in Appendix \ref{subsubsec:appendixfiber}. 

The assumption that fiber functors are exact (in an abelian setting) translates to requiring that $\kDk,\kNk,\CDCN$ are projective objects. The assumption that fiber functors are continuous translates to requiring that $\kDk,\kNk,\CDCN$ are compact objects, which is guaranteed by \AD$_k$ since they are the image of a compact object under a left adjoint.

\subsection{Spark algebras}
\label{sec:spark}

Given a fiber functor $\CF:\CC\to \text{Vect}$, Tannaka duality tells us to expect that $\CC$ may be represented as modules for the algebra of symmetries of $\CF$, \emph{a.k.a.} natural transformations from $\CF$ to itself, denoted $\text{End}(\CF)$.
Our next item of business is to describe, in physical terms, such algebras of symmetries for the fiber functors $\CF_\CN,\CF_\CD,\CF_\CT$ of the previous sections.

Let's start with the fiber functor $\CF_\CN:\CC_\CN\to \text{Vect}$, defined by ``capping off with $\CD$'' as in \eqref{FiberBdy-N}. We claim that all symmetries of $\CF_\CN$ correspond to operators supported in a region of the solid half-cylinder on the RHS of \eqref{FiberBdy-N} that 1) has finite extent in time (so that it does not alter the half-disc state space far below and far above it) and 2) is disjoint from the support of the line operator(s) $\ell$ (ensuring that it commutes with morphisms $\CO$ between lines as in \eqref{Falpha}). The region we have in mind thus looks like a punctured half-disc in space, times a finite interval $(-\epsilon,\epsilon)$ in time:
\be \raisebox{-.35in}{\includegraphics[width=3.2in]{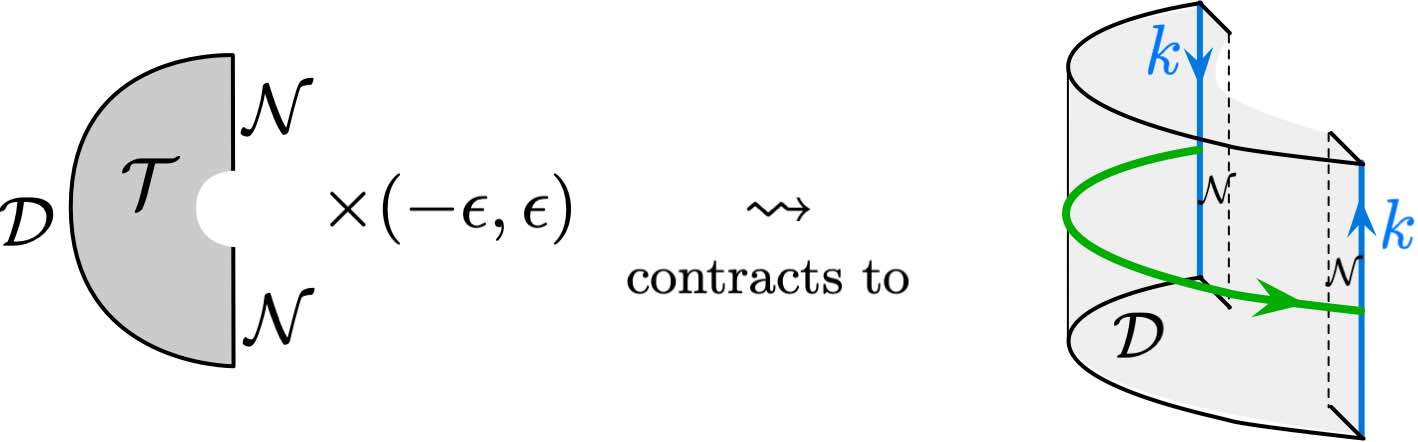}} \label{spark-region} \ee
Such a region is topologically equivalent (by a contraction) to the neighborhood of an arc on the boundary condition $\CD$, stretched between two junctions with $\CN$, as shown (in green) on the RHS of \eqref{spark-region}.

We let $\CH_\CD$ denote the vector space of all operators supported in the neighborhood of such an arc on $\CD$, with endpoints on $\CN$. $\CH_\CD$ could in principle contain local operators supported anywhere in the neighborhood of the arc; however, in practice, we'd expect that any such local operator can be brought to the interface $k$ where $\CD$ and $\CN$ meet, and that (due to transversality) it then becomes equivalent to a constant multiple of the identity operator. Hence we expect local operators to contribute in a trivial way to $\CH_\CD$. More interestingly, $\CH_\CD$ may contain genuine extended operators --- for example, line operators on $\CD$ turned sideways and stretched from one $(\CD,\CN)$ junction to the other.

To see that $\CH_\CD$ is a good candidate for the symmetry algebra of $\CF_\CN$, note first that $\CH_\CD$ \emph{is} an algebra: it is naturally endowed with an associative product%
\footnote{As we will see in the twisted gauge-theory examples at the end of the paper, spark algebras can be very infinite. Mathematically, products of sparks -- and other Hopf, etc. operations we'll construct -- may only be defined in a topological sense, \ie\ after suitable completions. We'll discuss the size of spark algebras further in Section \ref{sec:spark-finite}.} %
from vertical collision of disjoint (neighborhoods of) arcs on $\CD$:
\be \raisebox{-.5in}{\includegraphics[width=1.3in]{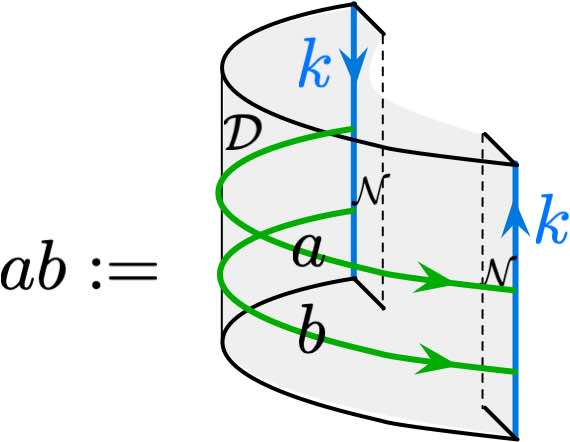}} \label{spark-prod} \ee
We call $\CH_\CD$ the \emph{spark algebra} (on $\CD$), due to the suggestive similarity of its elements with electric sparks/arcs moving along a pair of wires. More so, note that every vector space $\CF_\CN(\ell)$ is naturally endowed with an action of $\CH_\CD$: inserting a spark in the fiber-functor configuration maps the state space $\CF_\CN(\ell)$ below the spark to the state space $\CF_\CN(\ell)$ above the spark:
\be \raisebox{-.75in}{\includegraphics[width=4.5in]{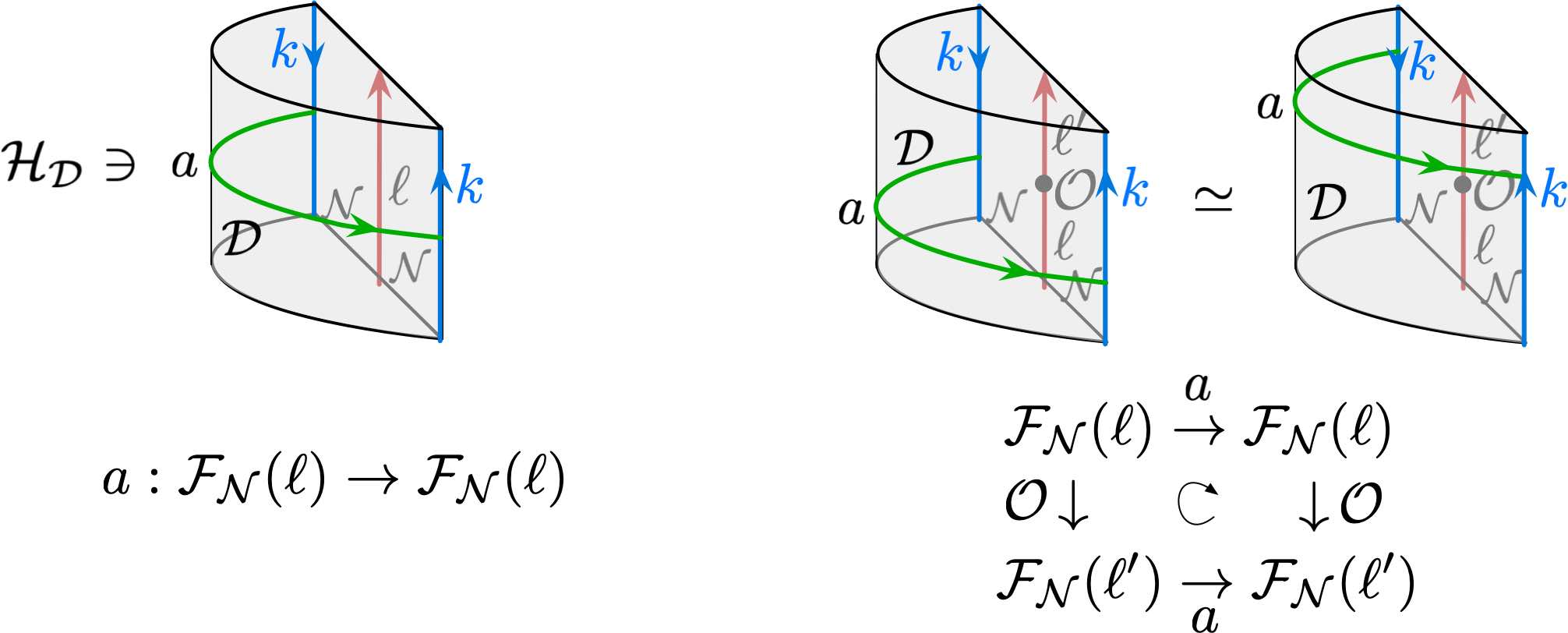}} \label{HD-act} \ee
The action of $\CH_\CD$ on the $\CF_\CN(\ell)$ also commutes with morphisms $\CF_\CN(\CO):\CF_\CN(\ell)\to \CF_\CN(\ell')$, since sparks are disjoint in space from local operators between lines $\ell,\ell'$, as on the the RHS of \eqref{HD-act}. The latter property ensures that each spark $a\in \CH_\CD$ is indeed a natural transformation of $\CF_\CN$. More importantly for us, it also ensures that the functor $\CF_\CN:\CC_\CN\to \text{Vect}$ lifts to a functor
\be \wt \CF_\CN:\CC_\CN\to \CH_\CD\text{-mod}\,, \label{FN-lift} \ee
where we now remember the $\CH_\CD$ action on all the spaces $\CF_\CN(\ell)$.

In a very similar way, we define $\CH_\CN$, the \emph{spark algebra on $\CN$}, to be the vector space of operators supported in the neighborhood of an arc along $\CN$, with endpoints on $\CD$:
\be \raisebox{-.35in}{\includegraphics[width=3.8in]{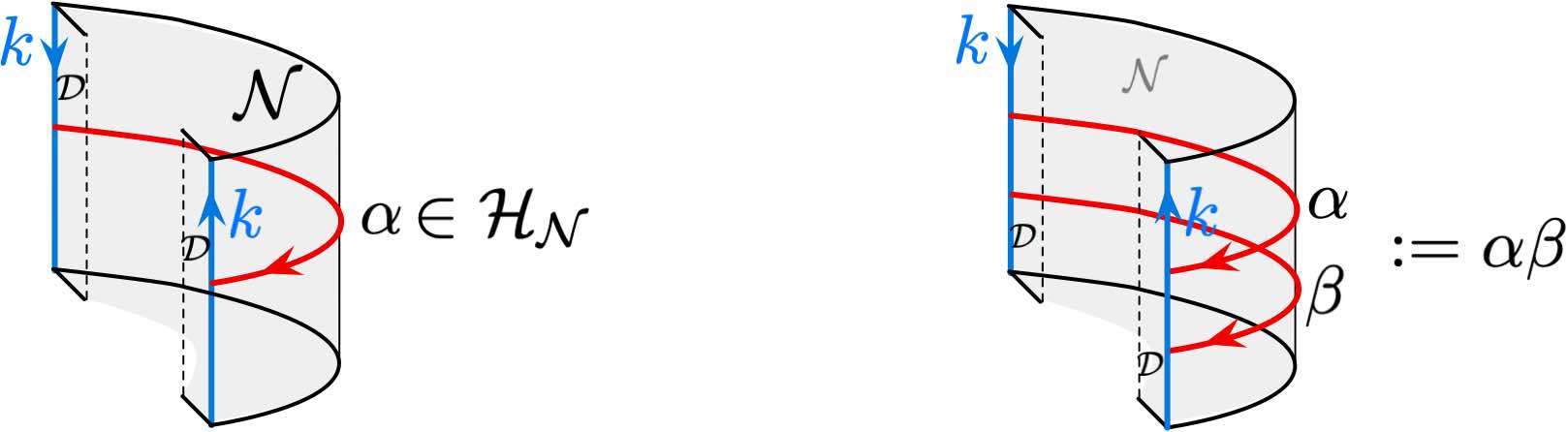}} \label{spark-prodN} \ee
It gains an associative product from vertical collision, and acts on all spaces $\CF_\CD(\ell)$ (in a way that commutes with morphisms $\CO:\CF_\CD(\ell)\to \CF_\CD(\ell')$, giving rise to a lift of the fiber functor
\be \wt \CF_\CD:\CC_\CD\to \CH_\CN\text{-mod}\,. \label{FD-lift}\ee

Finally, we define the \emph{bulk spark algebra} $U$ to be the vector space of operators supported in the neighborhood of a loop that crosses both $\CN$ and $\CD$.
\be \raisebox{-.35in}{\includegraphics[width=3.8in]{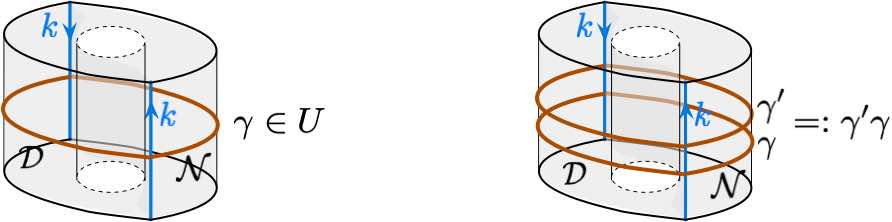}} \label{spark-prodU} \ee
It gains an associative product from vertical collision and acts on spaces $\CF_\CT(\ell)$, lifting the functor
\be \wt \CF_\CT:\CC_\CT\to U\text{-mod}\,. \label{FT-lift}\ee
Since the loop around which bulk sparks are supported crosses the $k$ interfaces between $\CN$ and $\CD$ (where entire theory is trivialized), we expect that any bulk spark can be represented as sum of products of sparks on $\CD$ and sparks on $\CN$. Put differently, we expect that the map $\CH_\CD\otimes \CH_\CN\to U$ defined by including boundary sparks in the neighborhood of a loop traversing both $\CN$ and $\CD$, as in \eqref{DN-T}, is an isomorphism:
\be \raisebox{-.35in}{\includegraphics[width=3.4in]{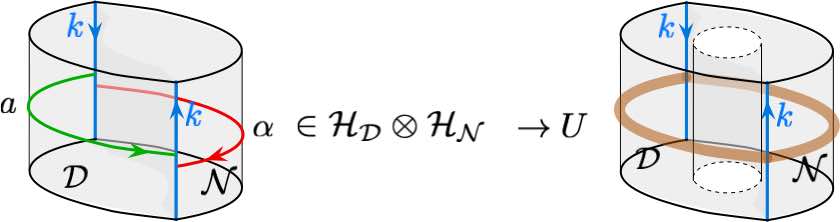}} \label{DN-T} \ee
We will justify this more precisely in the next section, using state-operator correspondences and transversality.

\subsection{Alternative definitions of sparks}
\label{sec:spark-alt}

A few equivalent definitions of spark algebras help reveal some of their properties.

\subsubsection{Sparks as state spaces}
\label{sec:spark-space}

First, state-operator correspondences help us relate the underlying vector spaces of spark algebras to ordinary state spaces. To set up the requisite correspondences we need to identify the topological link --- essentially the boundary --- of the region on which sparks are supported.

The link of a spark on $\CD$ is the boundary of its neighborhood, shown on the LHS of \eqref{SparkLink}. This region is topologically a rectangle, with two `long' edges on $\CD$, two short edges on $\ol \CN$, and corners on the $k$ interfaces between them:
\be \raisebox{-.5in}{\includegraphics[width=2.7in]{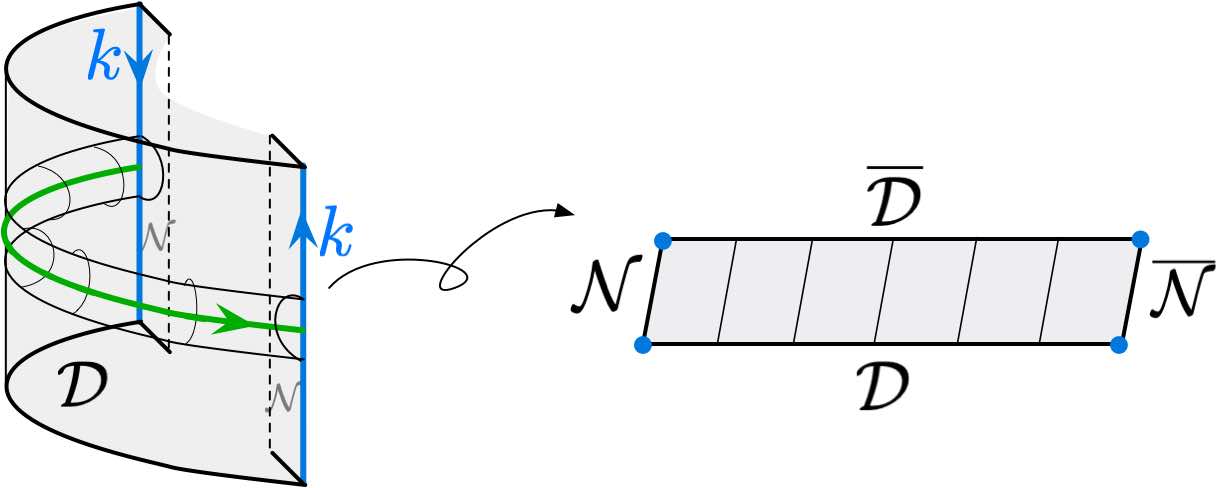}} \label{SparkLink} \ee
Precisely which edges we label $\CD$ (vs. $\ol\CD$) and $\CN$ (vs. $\ol\CN$) is somewhat a matter of convention. We can fix the convention by breaking the symmetry of the rectangle a bit, deforming it slightly into a parallelogram, as on the RHS of \eqref{SparkLink}.

The state-operator correspondence now identifies $\CH_\CD$ (as a vector space) with the space of states on a rectangle. 
Similarly, $\CH_\CN$ is identified with the space of states on a rectangle with oppositely labelled edges:
\be \raisebox{-.34in}{\includegraphics[width=4.8in]{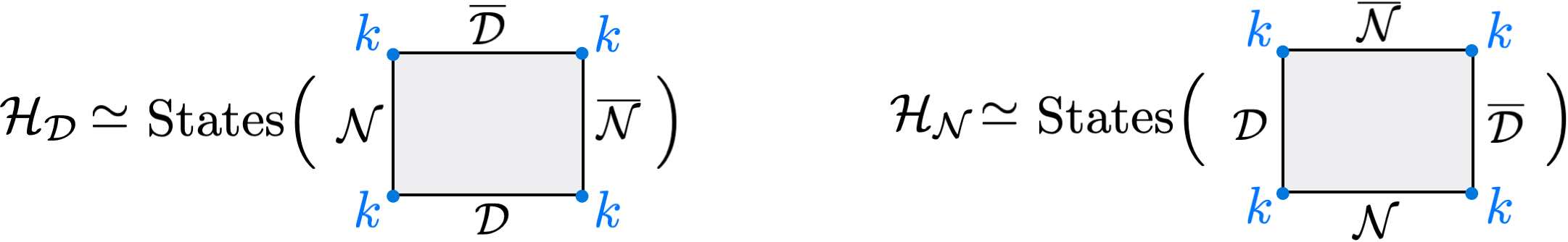}} \label{Rectangles} \ee
The two rectangles appearing here are identical, up to a reversal of orientation, so we find that $\CH_\CD$ and $\CH_\CN$ must be linear-dual vector spaces (at least when they are finite-dimensional),
\be \CH_\CD \simeq \CH_\CN^* \label{DN-dualspaces} \ee

Another way to picture this duality is to observe that the tubular neighborhood of a spark on $\CD$, and the tubular neighborhood of a spark on $\CN$, glue together perfectly (after a 90$^\circ$ twist) into a solid 3-ball, with half its boundary labelled by $\CD$ and half is boundary labelled by $\CN$:
\be \raisebox{-.4in}{\includegraphics[width=4.7in]{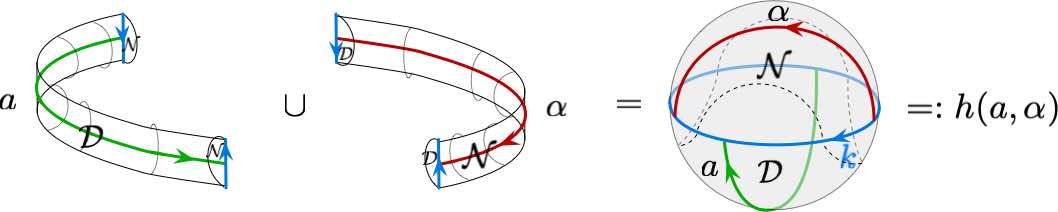}} \label{link-Hopf} \ee
We may think of $\CH_\CD$ as spanning the space of incoming states on the rectangle, and $\CH_\CN$ as spanning the space of outgoing states; and the 3-ball partition function with two sparks inserted as measuring their natural, nondegenerate, bilinear pairing 
\be h:\CH_\CD\times \CH_\CN \to \C\,, \label{h-glued} \ee
which is the second part of Prop. \ref{Prop:decomp}.

When $\CH_\CD$ and $\CH_\CN$ are finite dimensional they are of course isomorphic. When they are infinite dimensional, the pairing \eqref{h-glued} is still well defined. Physically, one expects in this case that $\CH_\CD$ and $\CH_\CN$ are topological vector spaces and that $h$ is a continuous bilinear functional, establishing an isomorphism \eqref{DN-dualspaces} as continuous duals.

We may also use a state-operator correspondence to analyze bulk sparks, and establish the first part of Prop. \ref{Prop:decomp}. The link of a bulk spark is an annulus, with two long boundaries on $\CD$ and two on $\CN$:
\be \raisebox{-.4in}{\includegraphics[width=3.6in]{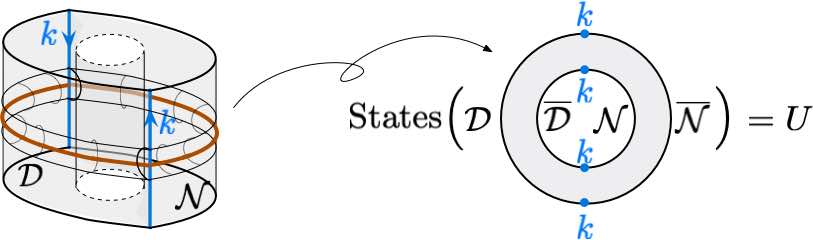}} \label{Annulus} \ee
Therefore, we identify the bulk spark algebra $U$ (as a vector space) with the space of states on this annulus. We can now use tranversality to break the annulus into the two rectangles representing boundary sparks, by first twisting its inner and outer boundaries slightly relative to each other, and then annihilating small segments of $\CD$ against opposing segments of $\CN$:
\be \raisebox{-.8in}{\includegraphics[width=5in]{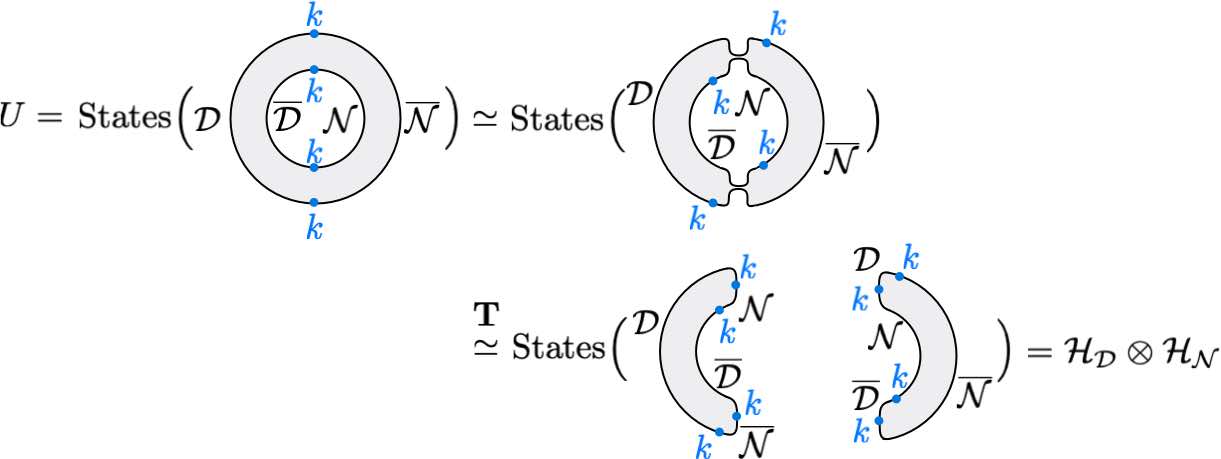}} \label{T-DN-proof} \ee
This proves that $U\simeq \CH_\CD\otimes \CH_\CN$, as vector spaces, just as we had anticipated in \eqref{DN-T}. The relation $U\simeq \CH_\CD\otimes \CH_\CN$ holds for both finite and infinite-dimensional spark algebras.

\subsubsection{Sparks via factorization homology}
\label{sec:spark-HH}

Next, we observe that sparks on $\CD$ (say) look nearly the same as line operators on $\CD$, but turned horizontal and compactified on an interval. Thus, it should be possible to derive the algebra $\CH_\CD$ if one knows sufficient information about the category $\CC_\CD$ of lines on $\CD$.

The operation of ``compactifying'' a category can been formulated (in slightly different settings) in terms of Hochschild homology, or factorization homology \cite{AyalaFrancis-fact}, or blob homology \cite{MorrisonWalker}. Keeping this discussion at a heuristic level, we won't differentiate between these (though we will give an example momentarily). If we let $\int_{{}_kI_k}$ denote the factorization (etc.) homology operation that compactifies a category on an interval $I$ bounded by $k$ interfaces, then we expect
\be  \CH_\CD \simeq \int\!\!\!\!\! \underset{{}_kI_k}{\phantom{\sum}} \CC_\CD\,,\qquad \CH_\CN = \int\!\!\!\!\! \underset{{}_kI_k}{\phantom{\sum}} \CC_\CN\,, \ee
with fiber functors expressing a beautiful symmetry
\be \CF_\CN:\CC_\CN \to \Big(  \int\!\!\!\!\! \underset{{}_kI_k}{\phantom{\sum}} \CC_\CD\Big)\text{-mod}\,,\qquad \CF_\CD:\CC_\CD\to  \Big(  \int\!\!\!\!\! \underset{{}_kI_k}{\phantom{\sum}} \CC_\CN\Big)\text{-mod}\,.\ee

We give a simple example to illustrate the compactification operation. Suppose that the category $\CC_\CD$ (say) admits a  compact generator $L\in \CC_\CD$. That means that every  $\ell\in \CC_\CD$ is a sub-quotient of a direct sum of $L$'s, or in the dg case every $\ell$ is a deformation of a sum of $L$'s by a Maurer-Cartan element. For example, if $\CC_\CD$ is a semisimple abelian category, then $L$ can be chosen to be a direct sum of all the simple objects.

Let $A = \text{End}_{\CC_\CD}(L)$ be the algebra of local operators on $L$. Let $M_l$ and $M_r$ be the vector spaces of local operators at the junctions of $L$ with the ends of a $\CD$ strip:
\be \raisebox{-.3in}{\includegraphics[width=1.15in]{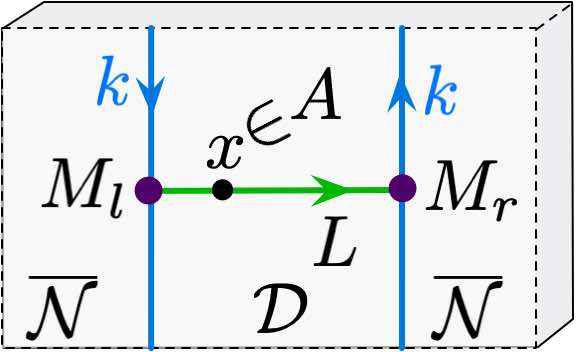}} \label{L-fact} \ee
Then $M_l$ and $M_r$ are left and right modules for the algebra $A$, and the spark algebra (as a vector space) is the relative tensor product
\be \CH_\CD = \int_{{}_kI_k} \CC_\CD \simeq M_l \otimes_A M_r\,. \label{DHH} \ee
In a dg setting, \eqref{DHH} must be interpreted as a derived tensor product, which is the Hochschild homology of $A$ with coefficients in the module $M_l\otimes M_r$. Physically, this means that $\CH_\CD$ may contain descendants of local operators on $L$ integrated along the interval. 

Note that the algebra product in $\CH_\CD$ is not at all manifest in \eqref{DHH}. It is extra structure that is induced from the tensor product in the category  $\CC_\CD$. In a finite semisimple setting, the construction of $\CH_\CD$ as in \eqref{DHH}, as well as its algebra product (and other Hopf operations) was spelled out very explicitly in the recent work \cite{CHO-solitons}, where it was called a ``strip algebra.''

\subsubsection{Sparks from Yoneda functors}
\label{sec:spark-Yo}

Finally, recall that (assuming \AD$_k$) all our fiber functors can be represented as Homs with special objects of each category $\CC_\CN,\CC_\CD,\CC_\CT$, constructed in Section \ref{sec:FF-Hom}. Such functors are sometimes known as Yoneda functors. In general, whenever a functor $\CF:\CC\to \text{Vect}$ is represented as $\CF=\text{Hom}_\CC(X,-)$ for some  object $X$, the symmetry algebra of $\CF$ can be recast (via the Yoneda Lemma) in terms of the endomorphism algebra of $X$ itself:
\be \text{End}(\CF) \simeq \text{End}_\CC(X)^{\rm op}\,, \ee
where `op' reverses the order of the associative algebra product in $\text{End}_\CC(X)$.

Thus, combined with Section \ref{sec:FF-Hom}, we expect that
\be \CH_\CD \simeq \text{End}_{\CC_\CN}(\kDk)^{\rm op}\,,\qquad  \CH_\CN \simeq \text{End}_{\CC_\CD}(\kNk)^{\rm op}\,,  \qquad
   U\simeq \text{End}_{\CC_\CT}\big(\CDCN\big)^{\rm op}\,.
 \label{EndCDC} \ee
This is not actually that surprising or mysterious. For example, the endomorphisms of ${}_k\ol\CD_k$ (say) in \eqref{EndCDC} are just the result of squeezing sparks on $\CD$ into point-like local operators at a junction of ${}_k\ol\CD_k$ and itself. Similarly, endomorphisms of $\raisebox{-.15cm}{\includegraphics[width=.5cm]{CDCNline.jpg}}$ just come from bulk sparks that have been squeezed down into point-like local operators on the line operator $\raisebox{-.15cm}{\includegraphics[width=.5cm]{CDCNline.jpg}}$.

\subsection{Finiteness of spark algebras}
\label{sec:spark-finite}

We've noted above that sometimes spark algebras are finite-dimensional, and sometimes they are not. The finite-dimensional case is especially well behaved, and it's useful to recognize when it occurs. We would like to show here that the dualizability condition \AD$_k$ from \eqref{k-superdual} is enough to guarantee it. This is notable since, as we've just shown, the same condition \AD$_k$ also ensures that fiber functors are represented as Homs out of dualizable objects (Sec. \ref{sec:FF-Hom}).

First, observe that due to the non-degenerate pairing $h:\CH_\CD\times\CH_\CN\to \C$ and the relation $U\simeq \CD_\CD\otimes \CH_\CN$, it follows that if \emph{any} of the three spark algebras $\CH_\CD,\CH_\CN,U$ are finite-dimensional then they all are.

Now let's consider $\CH_\CD$ (say), as the state space on a rectangle. We can interpret this space as an object of Vect, the category of line operators in the trivial 3d TQFT:
\be \label{Rect-dualizable} \hspace{-5.7in}
\makebox[0pt][l]{%
  \raisebox{-1.5in}[0pt][0pt]{%
    \includegraphics[width=5.5in]{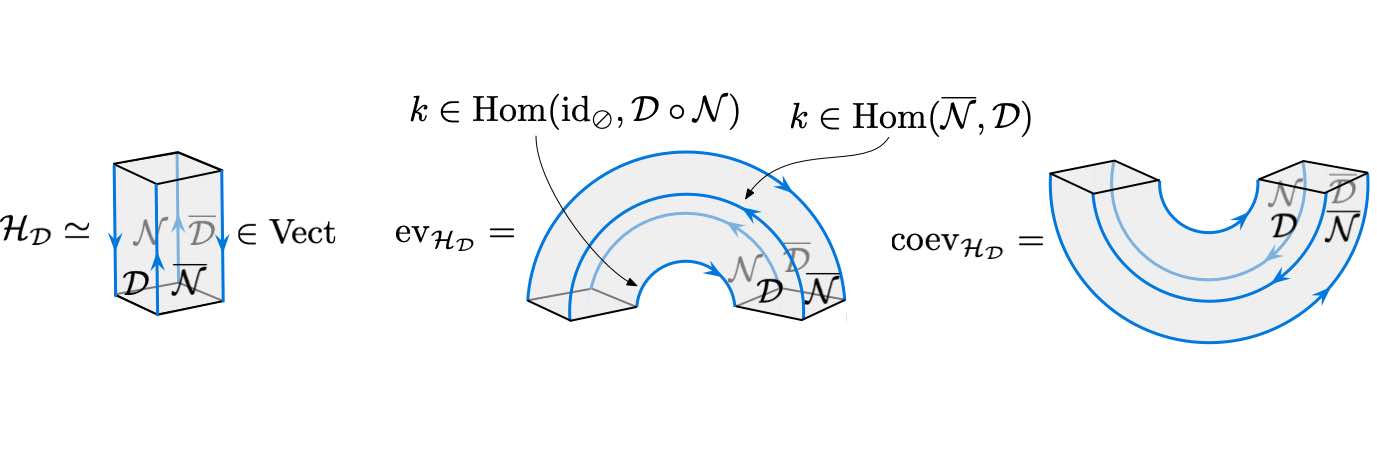}}}
\vspace{.9in}
\ee
Then we can reinterpret finite-dimensionality of $\CH_\CD$ as whether the object $\CH_\CD\in\text{Vect}$ is dualizable. Recall that the original transversality interfaces $k\in \text{Hom}(\text{id}_\oslash,\CD\circ\CN)$ and $k\in \text{Hom}(\ol\CN\circ\ol\CD,\text{id}_\oslash)$ are always dualizable. If we further assume \AD$_k$, then $k\in\text{Hom}(\ol\CN,\CD)$ and $k\in\text{Hom}(\ol\CD,\CN)$ are also dualizable, and we can define the cups and caps needed to make $\CH_\CD\in\text{Vect}$ dualizable, as shown in \eqref{Rect-dualizable}. (In addition to what's shown,  we also have $\text{ev}_{\CH_\CD}^*$ and $\text{coev}_{\CD_\CD}^*$, given by orientation reversals of $\text{ev}_{\CH_\CD}$ and $\text{coev}_{\CH_{\CD}}$.) \AD$_k$ further guarantees that the cups and caps of $\CH_\CD$ obey S-moves, and hence is a dualizable object of Vect.

In the abelian case, $\CH_\CD\in\text{Vect}$ being dualizable means it's finite dimensional. Its dimension is computed as the partition function on a rectangle $\times S^1$, namely
\be \text{dim}\,\CH_\CD = \CZ(\text{Rectangle}\times S^1)= \text{ev}_{\CH_\CD}\circ \text{coev}^*_{\CH_\CD}\,. \ee
In the dg case, $\CH_\CD\in\text{Vect}$ being dualizable means it has finite-dimensional cohomology. Its Euler character is computed by $\chi(\CH_\CD)= \CZ(\text{Rectangle}\times S^1)= \text{ev}_{\CH_\CD}\circ \text{coev}^*_{\CH_\CD}$\,.

\subsection{Faithfulness}
\label{sec:faithful}

Finally, we address the faithfulness of the fiber functors and their lifts to module categories.

We have commented in the introduction that depending on what flavor of TQFT we are dealing with, there are different results that give sufficient conditions for the lifted functors \eqref{FN-lift}, \eqref{FD-lift}, \eqref{FT-lift} to be equivalences (\cite{egno} for abelian categories, and \cite[Ch. 9]{Lurie-SAG}, \cite[Sec. 3]{lurie2007derived} for dg categories). Unfortunately, all of these conditions are actually quite hard to check in practice. At the very least they require fiber functors to be conservative, meaning $\CF(\ell)\neq 0$ for all $\ell$. In this paper, we take the perspective that we we \emph{don't know} very much about the categories of line operators to begin with, so checking things like $\CF(\ell)\neq 0$ for all $\ell$ directly is not possible.

In contrast, the completeness condition from Section \ref{sec:comp} is something that can be checked directly in examples of topological QFT's. We will argue in this section that completeness (specifically, strip-gluing \AC$_{\rm strip}$), together with the mild technical assumptions of Section \ref{sec:cts} on fiber functors, implies that the lifts
\be \wt\CF_\CN:\CC_\CN\overset\sim\to\CH_\CD\text{-mod}\,,\qquad  \wt\CF_\CD:\CC_\CD\overset\sim\to\CN_\CD\text{-mod}\,,\qquad  \wt\CF_\CT:\CC_\CT\overset\sim\to U\text{-mod} \label{comp-equiv} \ee
are all invertible, and thus equivalences of categories (Theorem \ref{Thm:faithful}).

For concreteness, we will work under the assumption that dualizability \AD$_k$ of the $k$ interface holds; so in particular fiber functors are represented by Hom's with the objects $\kDk,\kNk,\CDCN$; and (as a side effect) spark algebras are finite dimensional.%
\footnote{We expect that this assumption can be removed, and replaced by other weaker conditions. We will see in Sections \ref{sec:gauge}, \ref{sec:GM} examples in which \AD$_k$ and finite-dimensionality do not hold, but fiber functors still induce equivalences.} %
Then the statement we land on, as a consequence of completeness, is that $\kDk,\kNk,\CDCN$ are \emph{generators} of their respective categories. Moreover, when (say) $\wt\CF_\CN(\ell)= \text{Hom}_{\CC_\CN}(\kDk,\ell)$ (with the structure of an $\CH_\CD$ module coming from $\CH_\CD = \End_{\CC_\CN}(\kDk)$) then the inverse functor is given by
\be \wt\CF_\CN^\vee:\CH_\CD\text{-mod} \to \CC_\CN\,,\qquad  \wt\CF_\CN^\vee(M) = \kDk\otimes_{\CH_\CD} M\,, \label{hom-tensor-N} \ee
and similarly for $\wt\CF_\CD$, $\wt\CF_\CT$.

\subsubsection{A closer look at strip gluing}
\label{sec:strip}

Our first item of business is to make the strip gluings \AC$_{\rm strip}$ more explicit. In \eqref{stripglue}, we stated that there existed condensations or deformations along the highlighted strips that glued them together. We can simplify this procedure, in several steps.

For concreteness, suppose that the deformation is implemented by a condensation operation: by inserting sums of products of webs of operators on the two sides of the gluing.  Notice that for the strip that is bounded by two $k$ interfaces, any potential web can be simplified so that it is expressed as an element of the spark algebra. This is illustrated in \eqref{strip-web} schematically for a web on a $\CD$ strip:
\be \raisebox{-.7in}{\includegraphics[width=4.5in]{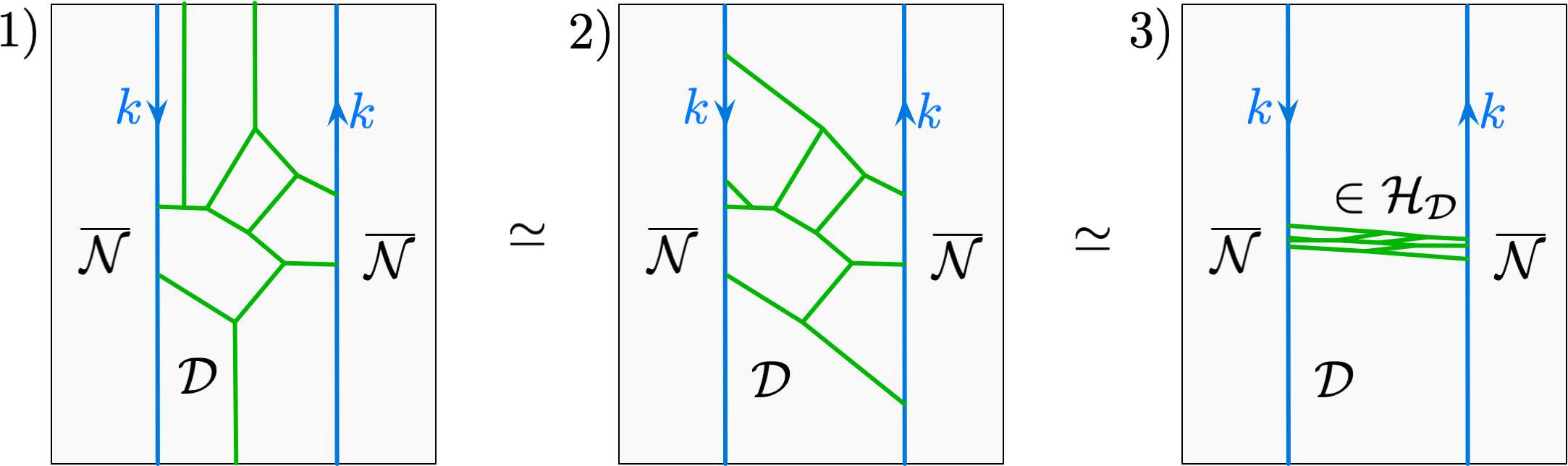}} \label{strip-web} \ee
Namely,
\begin{itemize}
\item[1)] Any web has a finite number of vertices, which are supported in a region of finite vertical extent. Beyond this region, the web can only contain semi-infinite vertical lines.
\item[2)] The semi-infinite vertical lines can be tilted to collide with the $k$ interfaces, where they trivialize. They may leave behind vector spaces, schematically denoted $V_i$ in the figure. This gives the entire web a finite vertical extent.
\item[3)] The web can then be squeezed into a very narrow horizontal band, which is the support of a spark on $\CD$. There may also be nontrivial vector spaces $V,W$ tensored into the top and bottom. We may assume without loss of generality that $V=W$, as $V\neq W$ can be obtained by further inserting a projection operator into a $V=W$ configuration.
\end{itemize}
We learn that any insertion on a bounded $\CD$ strip can be reduced to sparks, and endomorphisms of some additional vector space $V$.

If gluing is implemented by adding an interaction term to the boundary action (\emph{a.k.a.} adding a 2-Maurer-Cartan element), the argument looks a little different, but the conclusion is the same: due to topological invariance and trivializations at the two $k$ interfaces, any insertion on a bounded strip can be expressed entirely in terms of sparks.  In the case of a cohomological TQFT, descendants of sparks integrated along the entire vertical extent of the strip may also appear.

On the other side of the strip gluing in \eqref{stripglue} there is a strip that is \emph{not} bounded by $k$'s. Thus webs have no boundary vertices. Any web supported in this opposing strip can be squeezed into a thin vertical line, representing a collection of line operators and morphisms between them. The morphisms can further be composed into a single morphism $\CO$, separating a single pair of lines $\ell,\ell'$. This is depicted here for the unbounded $\CN$ strip opposite the $\CD$ strip of \eqref{strip-web}:
\be \raisebox{-.7in}{\includegraphics[width=4.8in]{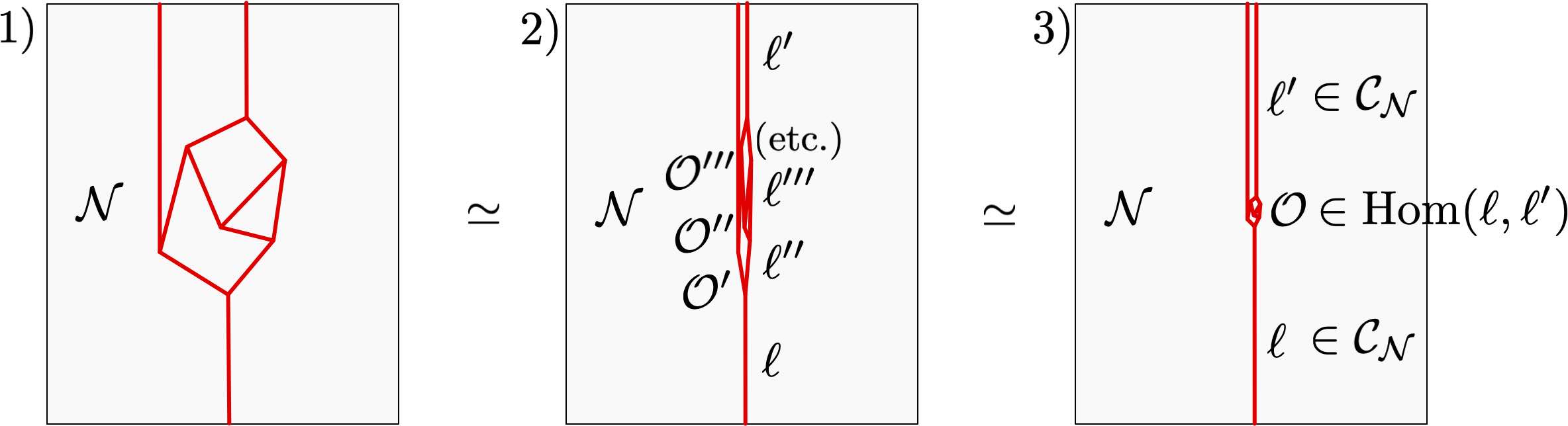}} \label{strip-web-line} \ee

More so, any configuration involving a morphism $\ell\overset{\CO}\longrightarrow  \ell'$ on the RHS of \eqref{strip-web-line}  may be replaced by an equivalent configuration where the line operators at the top and bottom are the same. To achieve this, let $\tilde \ell:=\ell\oplus \ell'$, let $p_\ell:\tilde\ell\to \ell$ and $p_{\ell'}^\vee:\ell'\to \tilde\ell$ be the projection and inclusion of the respective summands, and define $\wt\CO = p_{\ell'}^\vee \CO p_\ell = \left(\begin{smallmatrix} 0 & \CO \\ 0 & 0 \end{smallmatrix}\right)$\,. Then by `sliding' $p_\ell$ to the bottom and $p_{\ell'}^\vee$ to the top we can relate  $\tilde \ell \overset{\widetilde\CO}\longrightarrow \tilde \ell$ to $\ell\overset{\CO}\longrightarrow  \ell'$\,:
\be \raisebox{-.2in}{\includegraphics[width=2.5in]{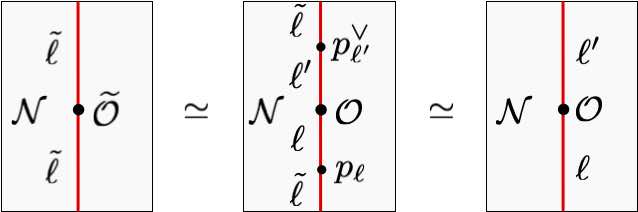}} \label{strip-web-simp} \ee

Now consider the actual gluing of an unbounded strip to a bounded strip, which in general involves a sum of products of insertions on the two sides. We replace insertions on the bounded strip with sparks, and insertions on the unbounded strip line operators and endomorphisms thereof. Any additional vector space $V$ appearing on the bounded strip can be reinterpreted as a trivial line operator on the unbounded strip. Then we find that a general strip gluing can be expressed as
\be \raisebox{-.6in}{\includegraphics[width=5in]{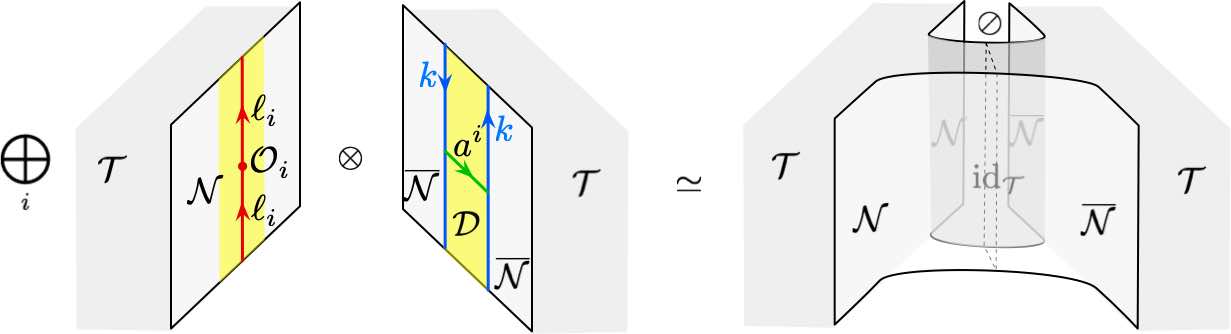}}\,, \label{stripglue-spark} \ee
for \emph{some} sum of operators $\bigoplus_i   \big(\ell_i\overset{\CO_i}\to \ell_i\big)\otimes a^i$, with $\ell_i\in\CC_\CN$ and $a^i\in \CH_\CD$. In cohomological TQFT, there may be integrated descendants of the $\CO_i$ along $\ell_i$, and $a^i$ along the strip.

This can even be made a bit nicer by introducing the total direct sum
\be \mb L_\CN := \bigoplus_i \ell_i \label{def-LN} \ee
of all the line operators appearing on the $\CN$ boundary. (We assume that the sum is either sufficiently finite or that $\CC_\CN$ is appropriately completed in order for $\mb L_\CN$ to be a genuine object.) Then each $\CO_i$ can be promoted to an endomorphism of the single $\mb L_\CN$, acting as zero on all summands except $\ell_i$; and the strip gluing can expressed by inserting
\be \mu_\CN :=  \sum_i \CO_i \otimes a^i \in \text{End}_{\CC_\CN}(\mb L_\CN)\otimes \CH_\CD \label{def-muN} \ee
on $\mb L_\CN$ tensored with the bounded $\CD$ strip.

We note that $\mu_\CN$ is subject to the important constraint
\be \mu_\CN\cdot \mu_\CN = \mu_\CN\,. \label{abelian-MC} \ee
This expresses the fact that if we simplify webs using \eqref{strip-web} and \eqref{strip-web-line} in two distinct `vertically' separated regions we get an equivalent gluing as if we had simplified all at once. (This is a standard constraint on elements used to define condensations). In the case of a cohomological TQFT, \eqref{abelian-MC} is replaced by the condition that $\mu_\CN$ is a Maurer-Cartan element that deforms the tensor product of $\mb L_\CN$ and the $\CD$ strip.

\subsubsection{Generators}
\label{sec:generators}

Next, we argue that the special objects $\kDk,\kNk,\CDCN$ are all generators in their respective categories.

In the abelian case, for $X$ to be a generator means that any object can be written as a subquotient of $X$ tensored with a (possibly infinite) vector space. In the dg/infinity case, being a generator means that any object may be constructed by tensoring $X$ with a (possibly infinite) dg vector space, and deforming the product by a Maurer-Cartan element (\eg\ turning on a new differential in the product).

Consider boundary lines first, say on an $\CN$ boundary. For any $\ell\in \CC_\CN$, let $M_\ell$ denote the vector space of states in the quantum mechanics on a solid cylinder with $\CN$ all around its boundary, and an insertion of $\ell$:
\be M_\ell :=\;\; \raisebox{-.2in}{\includegraphics[width=1.9in]{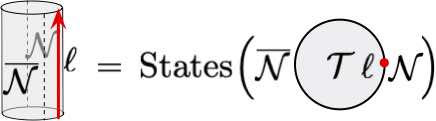}} \ee
We can glue this solid $M_\ell$ cylinder to a bounded $\kDk$ strip on a separate $\CN$ boundary, using a strip gluing:
\be \raisebox{-.5in}{\includegraphics[width=5in]{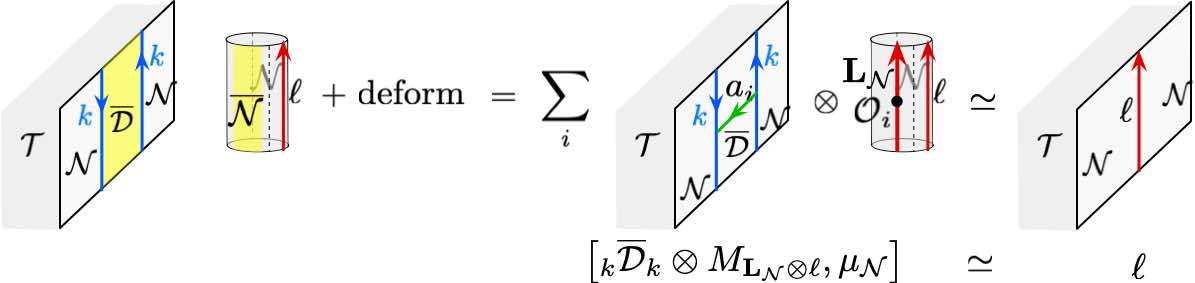}}  \label{invert-gen} \ee
The result, of course, is the original line operator $\ell$.

The analysis of the strip gluing from the previous section tells us what form the operation \eqref{invert-gen} can take. Gluing is done by 1) inserting a line operator $\mb L_{\CN}$ on the side of the solid $M_\ell$ cylinder, which simply produces a new vector space
\be V_\ell := M_{ \mb L_{\CN}\otimes \ell}  \ee
and 2) deforming the product of the $\kDk$ strip and $V_\ell$ with an insertion $\mu_{\CN} = \sum a^i \otimes \CO_i$ (swapping the factors due to our reversed orientation here). Thus, the gluing says that
\be [\kDk \otimes V_\ell,\mu_\CN] \simeq \ell \label{gen-l} \ee
where the LHS is interpreted as the image of $\mu_\CN$ acting on $\kDk\otimes V_\ell$ in the abelian case, and as a deformation by the MC element $\mu_\CN$ in the dg/infinity case. This precisely states that $\kDk$ is a generator of $\CC_\CN$.

Showing that $\kNk$ is a generator of $\CC_\CD$ is done the same way.

The bulk category is slightly trickier. Let's twist $\CDCN$ by 180$^\circ$, so that it looks like an empty tube with $\CN$ and $\CD$ boundaries (rather than $\ol\CD$ and $\ol \CN$). Then, for any $\ell\in \CC_\CT$, let $\CF_\CT(\ell)\in \text{Vect}$ be its image under the bulk fiber functor, represented by the solid cylinder in \eqref{defFT}. We can strip-glue $\CF_\CT(\ell)$ to $\CDCN$ using a bounded $\CD$ strip on $\CDCN$ and an unbounded $\CN$ strip on $\CF_\CT(\ell)$, as shown here in spatial cross-section:
\be \raisebox{-.4in}{\includegraphics[width=4.8in]{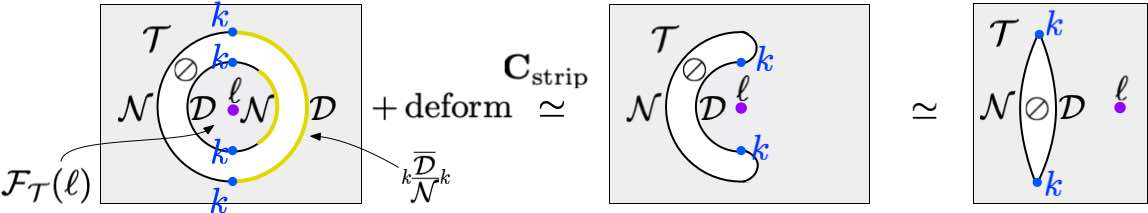}} \label{bulk-gen-0} \ee
The gluing is done by 1) inserting the line operator $\mb L_\CN$ on the boundary of $\CF_\CT(\ell)$, which just modifies $\CF_\CT(\ell)$ to a new vector space $W_\ell$; and 2) deforming the product $\CDCN\otimes W_\ell$ by the insertion of $\mu_\CN$. 

However, as seen on the RHS of \eqref{bulk-gen-0}, we are not finished. We've actually just shown that $[\CDCN\otimes W_\ell,\mu_\CN]\simeq  \CDCN\otimes \ell$. The final step is to ``close up'' the hole in the extra factor of $\CDCN$ in order to remove it entirely. This is done by another, slightly modified, strip gluing. Insertions in the two bounded $\CN$ and $\CD$ strips appearing here (with common $k$ boundaries) can be reduced to sparks acting on the product of $\CDCN$ and some vector space $W'$ --- the vector space left over from the RHS of \eqref{strip-web}. The sparks induce some new deformation $\mu'$ such that
\be [\CDCN\otimes W',\mu'] \simeq \id_\CT\,. \ee
Altogether, we find that
\be [\CDCN\otimes W_\ell\otimes W',\mu_\CN+\mu']\simeq \id_\CT \otimes \ell\simeq \ell\,. \ee
This is the statement of $\CDCN$ being a generator that we were after.

Note that, if one is careful, combining tensor products and deformations freely like this requires the technical assumptions from Section \ref{sec:cts} (rephrased as in Section \ref{sec:FF-Hom}): that $\CDCN$ is compact (guaranteed by \AD$_k$) and, in an abelian setting, projective.

\subsubsection{Inverting fiber functors}

We've now got all we need to invert the fiber functors.

Mathematically, this is completely standard, by the Yoneda lemma: if $X$ is a compact (and in the abelian case, projective) generator of a category $\CC$, then it's well known that $\text{Hom}_\CC(X,-):\CC\to \text{End}(X)\text{-mod}$ is an equivalence \cite{CPGenerator}. It's instructive, however, to explain why it's true in our case, and to associate some pictures to it.

Consider the lifted fiber functor $\wt\CF_\CN:\CC_\CN\to \CH_\CD\text{-mod}$, given by $\wt\CF_\CN(\ell) = \text{Hom}_{\CC_\CN}(\kDk,\ell)$, with $\CH_\CD=\text{End}_{\CC_\CN}(\kDk)^{\rm op}$.  We can construct its inverse by defining
\be \wt \CF_\CN^\vee:\CH_\CD\text{-mod}\to \CC_\CN\,,\qquad \wt \CF_\CN^\vee(M) :=  \kDk\otimes_{\CH_\CD} M\,. \ee
On the RHS here, $M$ is an arbitrary left $\CH_\CD$ module. To define $\kDk\otimes_{\CH_\CD} M$, we are taking the product of the object $\kDk\in \CC_\CN$ and the vector space $M$, then quotienting out by an equivalence relation that any element of $\CH_\CD$ acts the same way as an endomorphism of $\kDk$ as it does on $M$. In the dg/infinity setting, the relative tensor product is derived, and in particular implemented by a complex that resolves the quotient operation.

To check that $\wt\CF_\CN^\vee$ is the inverse of $\wt\CF_\CN$, first observe that for all $M$ we have
\begin{align} \wt\CF_\CN(\wt\CF_\CN^\vee(M)) &= \text{Hom}_{\CC_\CN}(\kDk,\kDk\otimes_{\CH_\CD} M) \notag \\
&\simeq  \text{Hom}_{\CC_\CN}(\kDk,\kDk)\otimes_{\CH_\CD} M \label{comm-hom-tensor} \\
&  = \CH_\CD\otimes_{\CH_\CD} M \notag \\
& \simeq M 
 \end{align}
The step \eqref{comm-hom-tensor}, commuting Hom with tensor product, requires the assumptions from Section \ref{sec:cts} (rephrased as in Section \ref{sec:FF-Hom}), that $\kDk$ is compact and (in an abelian setting) projective.

In the other direction, let $\ell\in \CC_\CN$ be any object, and use the fact that $\kDk$ generates to write $\ell\simeq [\kDk\otimes V_\ell,\mu_\CN]$, a deformation of $\kDk$ tensored with a vector space. Then
\begin{align} \wt \CF_\CN^\vee(\wt\CF_\CN(\ell)) &= \kDk\otimes_{\CH_\CD} \text{Hom}_{\CC_\CN}(\kDk,[\kDk\otimes V_\ell,\mu_\CN]) \notag \\
&\simeq   \kDk\otimes_{\CH_\CD} \big[\text{Hom}_{\CC_\CN}(\kDk,\kDk)\otimes V_\ell,\mu_\CN\big] \label{comm-def} \\
&= \kDk\otimes_{\CH_\CD} [\CH_\CD^{\rm op}\otimes V_\ell,\mu_\CN] \notag \\
&\simeq  \big[ \kDk\otimes_{\CH_\CD} \CH_\CD^{\rm op}\otimes V_\ell,\mu_\CN\big] \label{comm-def2} \\
&\simeq \big[\kDk \otimes V_\ell,\mu_\CN\big]  \simeq \ell\,.
\end{align}
Here it's commuting Hom and tensor product with deformations in \eqref{comm-def}, \eqref{comm-def2} that requires $\kDk$ to be compact and (if abelian) projective.

We've established both that $\wt\CF_\CN$ is an equivalence, and that its inverse is defined by tensoring with $\kDk$. The same of course will be true for $\wt\CF_\CD$ and $\wt\CF_\CT$, with inverses given by tensoring with $\kNk$ and $\CDCN$, respectively.

There is a nice way to illustrate the inverse functors physically, or in TQFT.  For $\CC_\CN$ (say), the inverse functor sends any $\CH_\CD$-module $M$ to a tensor product with $\kDk$ (a bounded $\ol\CD$ strip), with a relation that sparks on the $\ol \CD$ strip are equivalent to the $\CH_\CD$ action on $M$:
\be  \raisebox{-.35in}{\includegraphics[width=5.2in]{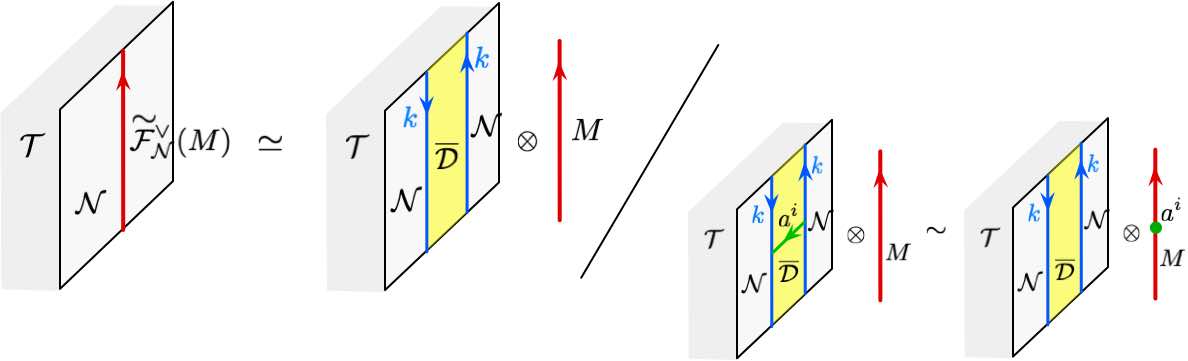}} \ee
In particular, applying this to $\wt\CF_\CN(\ell)$ for any $\ell$, we can re-glue the half-cylinder representing the fiber functor into a boundary by requiring that sparks act the same way on both sides:
\be \raisebox{-.5in}{\includegraphics[width=5.7in]{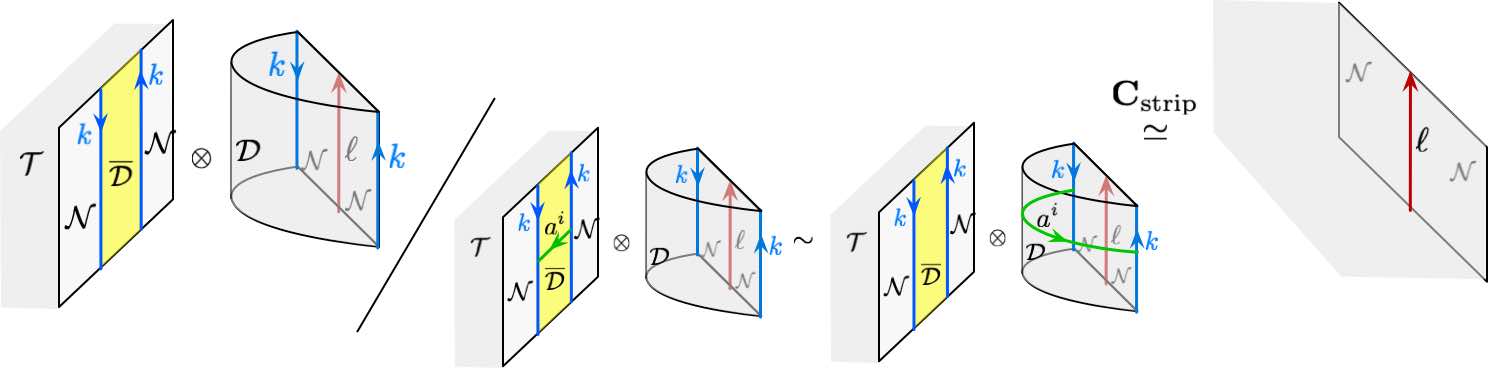}} \label{Invert-F} \ee
More generally, one can argue (though we won't need it here) that the strip gluing itself can be reformulated as a gluing of two $\kDk$ strips, with sparks acting the same way on both sides:
\be \raisebox{-.35in}{\includegraphics[width=5.7in]{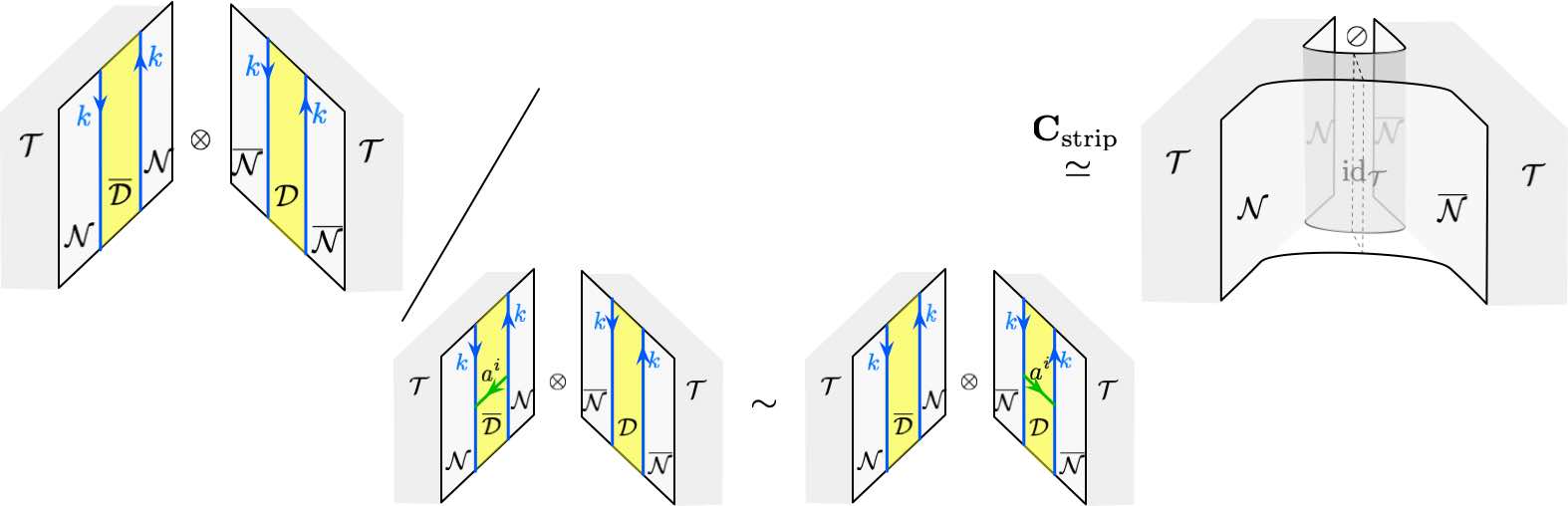}} \label{stripglue-tens} \ee

\section{Spark algebras are Hopf algebras}
\label{sec:Hopf-bdy}

In this section, we will begin to endow the spark algebras $\CH_\CD,\CH_\CN,U$ from Section \ref{sec:linefunctor} with additional operations.

As we review in Section  \ref{sec:Hopf-axioms}, a Hopf algebra is an associative algebra with a multiplicative unit, together with a coproduct $\Delta:\CH\to\CH\otimes \CH$, an antipode $S:\CH\to\CH^{\rm op}$, and a counit $\varepsilon:\CH\to\C$, satisfying various axioms. The axioms are precisely arranged so that the module category of a Hopf algebra is a monoidal category with duals and a tensor unit, \emph{cf.} \cite[Ch. 5]{egno}. 

Conversely, if $\CC$ is a monoidal category and $\CF:\CC\to \text{Vect}$ a well-behaved monoidal functor, with endomorphism algebra $\CH=\text{End}(\CF)$ such that the functor lifts
\be \wt \CF:\CC \to \CH\text{-mod}\,, \label{Flift4} \ee
then one can \emph{reconstruct} a coproduct $\Delta$ on $\CH$ such that the lift \eqref{Flift4} preserves monoidal structures. Similarly, if $\CC$ has a monoidal unit that's preserved by $\CF$, then one can reconstruct a counit $\varepsilon$ on $\CH$, so that $\wt \CF$ preserves the unit. And if $\CC$ has enough duals that are preserved by $\CF$, one can construct an antipode $S$ on $\CH$, so that $\wt\CF$ preserves duals. 

This reconstruction of structure on $\CH$ is unique and also explicit, at an algebraic level. We will explain in Section \ref{sec:Hopf-def} how to apply it to our categories $\CC_\CN,\CC_\CD,\CC_\CT$ and fiber functors. Further mathematical details are collected in Appendix \ref{sec:math}.

Our main goal in Section \ref{sec:Hopf-def}, in fact, will be to translate the Hopf structures on $\CH_\CD,\CH_\CN,U$ induced algebraically to \emph{topology}, thereby making them intrinsically defined and computable in terms of operators in topological QFT. We will connect to work of \cite{Reutter-slides,FreedTeleman-Ising}. In Section \ref{sec:Hopf-verify}, we will then verify topologically, using topological invariance and transversality of our QFT, that the topologically-defined $\Delta,\varepsilon,S$ do satisfy all the axioms of Hopf algebras. This is automatic algebraically; but if one is approaching from a QFT perspective, constructing $\CH_\CD,\CH_\CN,U$ as sparks, it is illuminating to see directly why the Hopf axioms are obeyed.

Finally, in Section \ref{sec:Hopf-pair}, we consider the interplay between $\CH_\CD$ and $\CH_\CN$. We already know from Prop. \ref{Prop:decomp} that there is a non-degenerate pairing 
\be h:\CH_\CD\otimes \CH_\CN \to \C\,, \ee
defined by inserting sparks along the boundary of a 3-ball. We will show that $h$ is a \emph{Hopf pairing}, roughly meaning that the adjoints of Hopf operations in $\CH_\CD$ are dual Hopf operations in $\CH_\CN$, \eg\ $h(a,\alpha\beta)=h(\Delta(a),\alpha\otimes\beta)$.

Altogether, we establish

\begin{Prop} \label{Prop:Hopf} Assuming topological invariance (\AD, Section \ref{sec:top}) and transversality (\AT, Section \ref{sec:trans}) of a bulk theory and its boundary conditions, as well as exactness and continuity of tensor products and fiber functors (Section \ref{sec:lines}, \ref{sec:cts}), the three spark algebras  $\CH_\CD,\CH_\CN,U$ are \emph{bilagebras}, with unit and counit. Further assuming control over dualizable objects as in \AD$_{\rm line}$ (Section \ref{sec:lines}), the three spark algebras become involutive Hopf algebras. The two boundary spark algebras $\CH_\CD,\CH_\CN$ are Hopf-paired by $h$.
\end{Prop}

When $\CH_\CD,\CH_\CN,U$ are infinite dimensional, topological completions may be required for various Hopf operations to be defined. We take this for granted throughout this section (and the next); see Appendix \ref{sec:math} for a mathematical discussion. We work at a level of generality that is appropriate for either a dg/infinity setting or an purely abelian one. In the dg/infinity setting, we in general expect to encounter dg/infinity analogues of Hopf (super)algebras; we won't consider higher Hopf operations in this paper.

\emph{Completeness} does not play an essential role in this section; nothing here requires fiber functors to be faithful or induce equivalences. If one does assume completeness, then Thm. \ref{Thm:faithful} combines with Prop. \ref{Prop:Hopf} to show that lifted fiber functors
\be \wt\CF_\CN:\CC_\CN\overset\sim\to \CH_\CD\text{-mod}\,,\qquad \wt\CF_\CD:\CC_\CD\overset\sim\to \CH_\CN\text{-mod}\,,\qquad \wt\CF_\CT:\CC_\CT\overset\sim\to U\text{-mod} \ee
are now equivalences of monoidal categories, mapping $\id$ to $\C$, and preserving duals.

\subsection{Axioms for Hopf algebras}
\label{sec:Hopf-axioms}

Axioms for Hopf algebras can now be found in many classic texts, \cf\ \cite[Sec 5.3]{egno} and references therein. We briefly recall them.

A Hopf algebra $\CH$ (over $\C$) is an associative algebra (over $\C$) that has a two-sided multiplicative unit `$1$' and
\begin{itemize}
\item[1)] a `coproduct' $\Delta:\CH\to\CH\otimes \CH$ that is an algebra morphism and is co-associative:
\be \Delta(ab) = \Delta(a)\Delta(b)\,,\qquad (\Delta\otimes\text{id})\circ\Delta(a) = (\text{id}\otimes \Delta)\circ\Delta(a) \qquad \forall\;a,b\in \CH \label{Delta-ax} \ee
\item[2)] a `counit' $\varepsilon:\CH\to\C$ that is an algebra morphism and a counit for the coproduct:
\be \varepsilon(ab)=\varepsilon(a)\varepsilon(b)\,,\qquad (\varepsilon\otimes \text{id})\circ\Delta(a)=( \text{id}\otimes\varepsilon)\circ\Delta(a) = a \qquad \forall\;a,b\in \CH \label{e-ax} \ee
\item[3)] an `antipode' $S:\CH\to \CH$ that is an anti-morphism and `cancels' the coproduct:
\be S(ab) = S(b)S(a)\,,\qquad m\circ (\text{id}\otimes S)\circ \Delta(a) = m\circ (S\otimes \text{id})\circ \Delta(a) = 1\varepsilon(a)\qquad \forall\;a,b\in \CH\,, \label{S-ax}\ee
where $m(a\otimes b) = ab$ is multiplication.
\end{itemize}

It is often convenient to introduce so-called Sweedler notation, writing
\be \Delta(a) = \sum a_{(1)}\otimes a_{(2)}  := \sum_i a_{(1)}^i \otimes a_{(2)}^i\,,\ee
where $(a_{(1)})$ and $(a_{(2)})$ are the two sequences of elements in $\CH$ that appear in the coproduct of a given element $a\in \CH$. Since $\Delta$ is co-associative, its powers are defined unambiguously, and one also writes (\emph{e.g.})
\be \Delta^2(a) :=  (\Delta\otimes\text{id})\circ\Delta(a) = (\text{id}\otimes \Delta)\circ\Delta(a) = \sum a_{(1)}\otimes a_{(2)}\otimes a_{(3)} \in \CH^{\otimes 3} \label{doubleD} \ee
for three sequences $(a_{(1)}),(a_{(2)}),(a_{(3)})$. In Sweedler notation some axioms become simpler to write, \emph{e.g.} the RHS of \eqref{e-ax} and \eqref{S-ax} read
\be  \sum \varepsilon(a_{(1)}) a_{(2)} =  \sum a_{(1)} \varepsilon(a_{(2)}) = a\,,\qquad \sum a_{(1)}S(a_{(2)}) = \sum S(a_{(1)})a_{(2)} = \varepsilon(a)\,. \ee

Finally, we remark that in the case of cohomological TQFT one will most generally encounter dg/infinity generalizations of Hopf algebras, which have an additional differential or BRST operator $Q$, and the above axioms may only hold up to $Q$-exact terms, which are in turn controlled by higher operations. In the examples we'll be interested in, we'll (conjecturally) be able to pass to $Q$-cohomology, setting the differential to zero. Even so, this lands us in the world of Hopf \emph{super}algebras, due to $\Z_2$ fermion-number gradings. All the above axioms are easily adapted to the case of superalgebras, by modifying the algebra structure on $\CH\otimes \CH$ to account for fermions --- see \cite[Rmk. 9.11.10]{egno} or \cite{andruskiewitsch2011pointed}. Explicitly, the axioms that $\Delta$ is a morphism and $S$ an anti-morphism become
\be \Delta(ab) = \sum (ab)_{(1)}\otimes (ab)_{(2)} =  \sum (-1)^{|a_{(2)}||b_{(1)}|} (a_{(1)}b_{(1)}\otimes a_{(2)}b_{(2)})=\Delta (a)\Delta (b)\,, \ee
\be S(ab) = (-1)^{|a||b|}S(b)S(a)\,, \ee
where $|a|\in \Z/2$ denotes the fermion number of $a\in \CH$. The other axioms are unchanged.

\subsection{Hopf operations on sparks}
\label{sec:Hopf-def}

We'll now define the Hopf-algebra operations on sparks topologically, guided by algebraic reconstruction.

\subsubsection{Multiplication and unit}

First, recall that spark algebras $\CH_\CD,\CH_\CN,U$ are in fact associative algebras. They each have a product defined by `vertical' collision of sparks, as in \eqref{spark-prod}, \eqref{spark-prodN}, \eqref{spark-prodU}. In this section, we'll also try to describe Hopf operations via 3d cobordisms. From this perspective, the products of boundary sparks come from including the neighborhoods of two intervals into a larger neighborhood:
\be \raisebox{-.5in}{\includegraphics[width=5.8in]{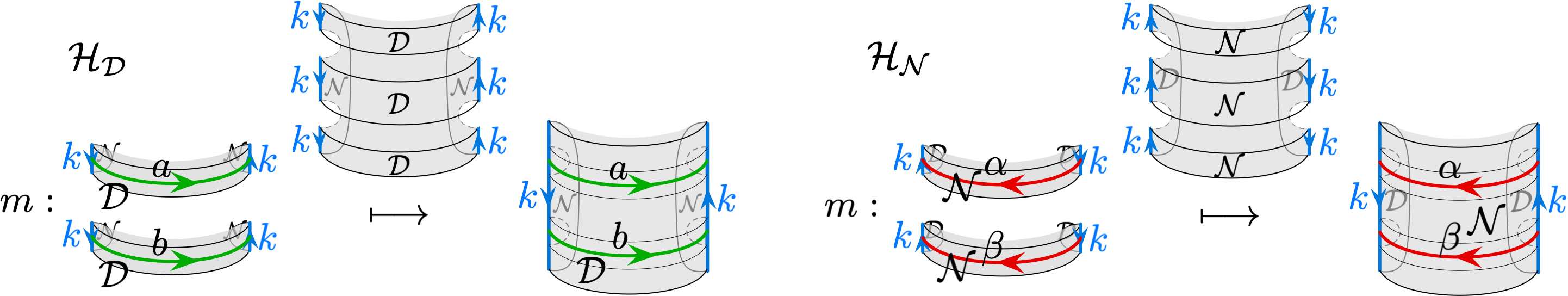}} \label{Hopf-m} \ee
The corresponding cobordisms are depicted above the `$\mapsto$' arrows in \eqref{Hopf-m}. For example, the cobordism governing multiplication $m:\CH_\CD^{\otimes 2}\to \CH_\CD$ has two incoming boundaries (in the front) where sparks can attach, and a single outgoing boundary (in the back) supporting the product. All the boundaries have the topology of~the~rectangle from~\eqref{Rectangles}. 

The product in $U$ from \eqref{spark-prodU} can be described in a similar way:
\be  \raisebox{-.5in}{\includegraphics[width=2.8in]{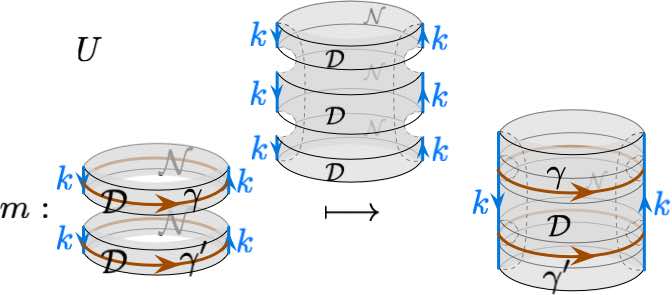}} \label{Hopf-mU} \ee
Its associated cobordism is a union of the $\CH_\CD$ and $\CH_\CN$ cobordisms from \eqref{Hopf-m}; it has two incoming annular boundaries and one big outgoing annular boundary (on the inside).

Each algebra $\CH_\CD,\CH_\CN,U$ also has a unit $1$, given by the trivial/empty spark. It may be thought of as an inclusion:
\be  \raisebox{-.3in}{\includegraphics[width=5in]{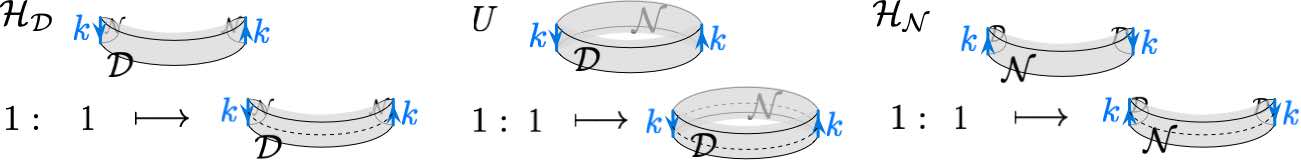}} \label{Hopf-u} \ee

\subsubsection{Coproduct}

Now consider coproducts. Algebraic reconstruction (see Appendix \ref{sec:math}) says that if we think of $\CH_\CD = \text{End}(\CF_\CN)$ as the algebra of natural transformations of the fiber functor $\CF_\CN:\CC_\CN\to\text{Vect}$ from \eqref{FiberBdy-N}, then we can construct $\Delta:\CH_\CD\to \CH_\CD\otimes \CH_\CD$ from the monoidal structure of the fiber functor. Explicitly, letting $J:\CF_\CN\otimes\CF_\CN\to\CF_\CN$ and $J^{-1}:\CF_\CN\to\CF_\CN\otimes\CF_\CN$ be the natural transformations from \eqref{defJ}, we are to define
\be \text{for $a\in \text{End}(\CF_\CN)$}\,,\qquad  \Delta(a):= J^{-1}\circ a \circ J\,. \ee

Translating this to topology, we first remove a neighborhood of $\ell$ and $\ell'$ from \eqref{defJ} (since we're thinking of $J,J^{-1}$ as natural transformations, rather than morphisms for fixed $\ell,\ell'$).  Then we define $\Delta(a)$
\be \hspace{-.2in} \includegraphics[width=6in]{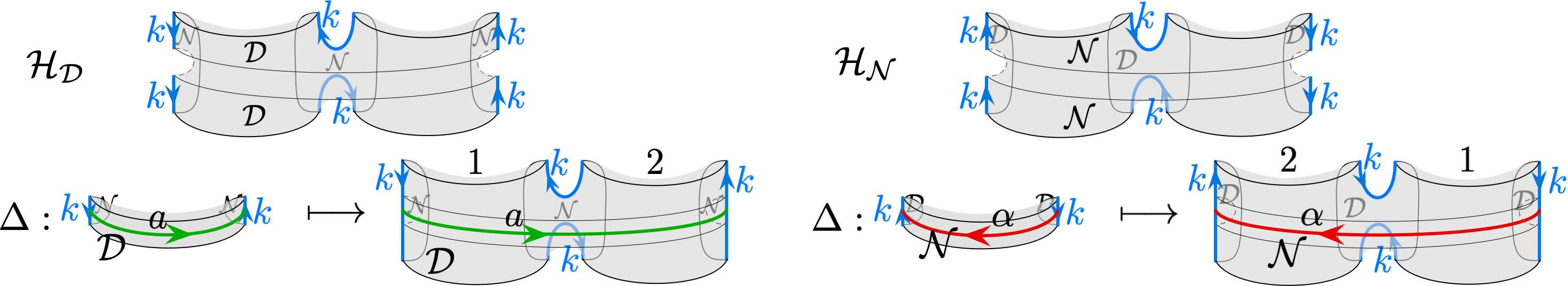} \label{Hopf-D} \ee
by \emph{stacking} the $J$ and $J^{-1}$ cobordisms on the bottom and top of a neighborhood of the spark $a$, to get the LHS of \eqref{Hopf-D}.
The RHS is the corresponding operation for $\CH_\CN$. For $U$, it looks like
\be \raisebox{-.6in}{\includegraphics[width=3in]{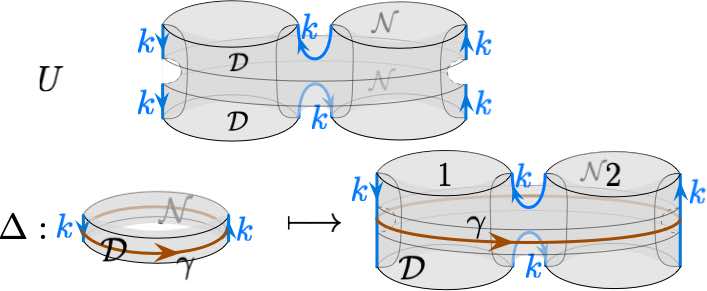}} \label{Hopf-DU} \ee

As before, we have drawn the 3d cobordisms governing $\Delta:\CH_\CD\to \CH_\CD\otimes \CH_\CD$ and $\Delta:\CH_\CN\to \CH_\CN\otimes \CH_\CN$. (The cobordism for bulk sparks is the union of these two.) The cobordisms now have one incoming rectangle (or bulk annulus) and two outgoing rectangles (or bulk annuli).  We indicate with `$1$' and `$2$' how the outgoing boundaries are meant to correspond to the ordered tensor factors in the coproduct. This does not make any difference for verifying Hopf-algebra axioms; it is ultimately done to match our conventions \eqref{TensorN} for the orders of tensor products in boundary line-operator categories. In Sweedler notation, we would have (for example)
\be \includegraphics[width=6in]{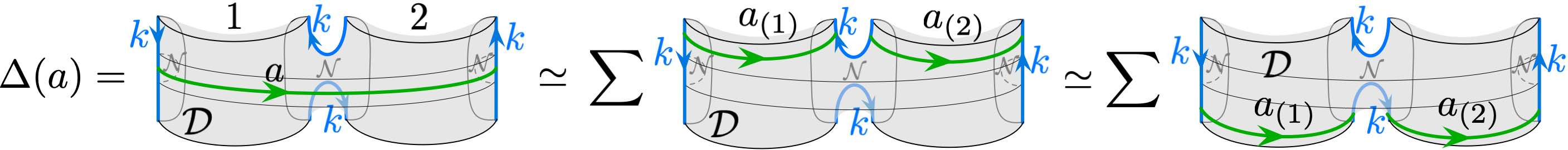} \label{Hopf-D-Sweedler} \ee
Thus, heuristically, the coproduct tells us how a spark decomposes as it traverses a pair of pants.

\subsubsection{Counit}

Algebraically, the counit in $\CH_\CD=\text{End}(\CF_\CN)$ (say) is defined by using the isomorphism $\CF_\CN(\id)\simeq \C$. Recall that this is implemented explicitly by an isomorphism $e:\CF_\CN(\id)\to\C$ and its inverse $e^*:\C\to \CF_\CN(\id)$, which topologically come from caps and cups formed from the transversality interface, as in \eqref{hole-a} and \eqref{e-counit}. Then  $\varepsilon:\CH_\CD\to \C$ is given by
\be \varepsilon(a) = e\circ a(\id) \circ e^*\,. \ee

Translating this to topology, using the cups and caps from \eqref{e-counit}, we find that the counit simply includes a spark into a solid ball (with boundary split into $\CD$ and $\CN$), evaluating the corresponding partition function:
\be \raisebox{-.8in}{\includegraphics[width=5in]{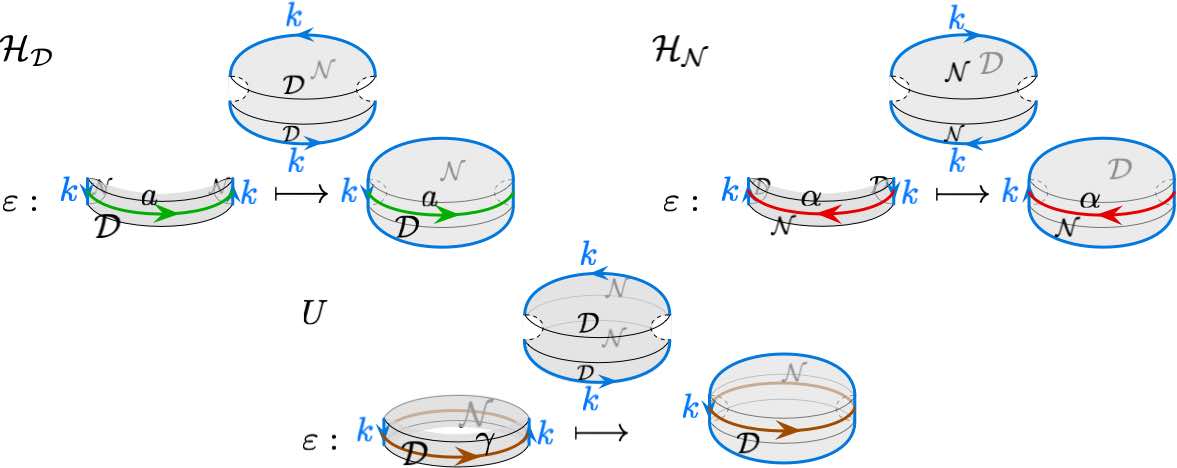}} \label{Hopf-eU} \ee

\subsubsection{Antipode}

Finally, let's consider duals. As we saw in Section \ref{sec:mon}, transversality guarantees that our fiber functors preserve dualizable objects. That means that for (say) $\CC_\CN$ and all dualizable $\ell\in \CC_\CN$ there are natural morphisms
\be
\begin{array}{c} \text{Ev}_\ell:\CF_\CN(\ell^*)\otimes \CF_\CN(\ell)\to \C\,,\qquad \text{Coev}_\ell:\C\to\CF_\CN(\ell)\otimes\CF_\CN(\ell^*)\,, \\
\text{Ev}'_\ell:\CF_\CN(\ell)\otimes \CF_\CN(\ell^*)\to \C\,,\qquad \text{Coev}'_\ell:\C\to\CF_\CN(\ell^*)\otimes\CF_\CN(\ell)\,, \end{array}
 \ee
obeying S-moves. (We also recall that left and right duals of $\ell$ are canonically isomorphic due to the assumed absence of framing anomaly.) A putative antipode $S:\CH_\CD\to\CH_\CD^{\rm op}$ is then constructed algebraically on $a\in \text{End}(\CF_\CN)$ by
\begin{align} S(a) &:= (\text{id}\otimes \text{Ev})\circ  (\text{id}\otimes a \otimes\text{id})  \circ  (\text{Coev}\otimes\text{id}) \notag \\
 &\simeq (\text{Ev}'\otimes\text{id} )\circ  (\text{id}\otimes a \otimes\text{id})  \circ  (\text{id}\otimes \text{Coev}')\,. \label{S-alg} \end{align}
The two lines here are equivalent due to the isomorphism between left and right duals; otherwise, one would define $S$ and the other would define $S^{-1}$.

There is a slight complication here, discussed further in Appendix \ref{sec:math}, because we have not assumed that all line operators are dualizable. (As discussed in Section \ref{sec:lines}, some lines may be too massive or infinite to bend 180$^\circ$ in a topologically invariant way.) Strictly speaking, the RHS of \eqref{S-alg} only makes sense as a natural transformation of the fiber functor
\be \CF_\CN^{fd}:\CC_\CN^{fd}\to \text{Vect}^{fd} \ee
that restricts $\CF_\CN$ to the subcategory of dualizable objects $\CC_\CN^{fd}\subseteq \CC_\CN$. However, if dualizable objects are sufficiently ``dense''  and the fiber functor is sufficiently well behaved, then a natural transformation of $\CF_\CN^{fd}$ should induce a natural transformation of $\CF_\CN$.

Concretely, we assume \AD$_{\rm line}$ from Section \ref{sec:lines}, saying that our line-operator categories are generated (in the sense of ind-completions) by compact, dualizable objects. Moreover, we assume that fiber functors are continuous as in Section \ref{sec:cts}, ensuring that they preserve the sorts of limits required to build arbitrary objects from dualizable ones. Then a natural transformation of $\CF_\CN^{fd}$ extends to a unique natural transformation of $\CF_\CN$, and \eqref{S-alg} indeed defines an antipode algebraically on $\CH_\CD=\text{End}(\CF_\CN)$. The antipodes for $\CH_\CN$ and $U$ are constructed the same way. 

Now let's translate \eqref{S-alg} to topology. Composing a spark $a\in \CH_\CD$ with the Ev and Coev morphisms from \eqref{defEv}, we get
\be  \raisebox{-.8in}{\includegraphics[width=5in]{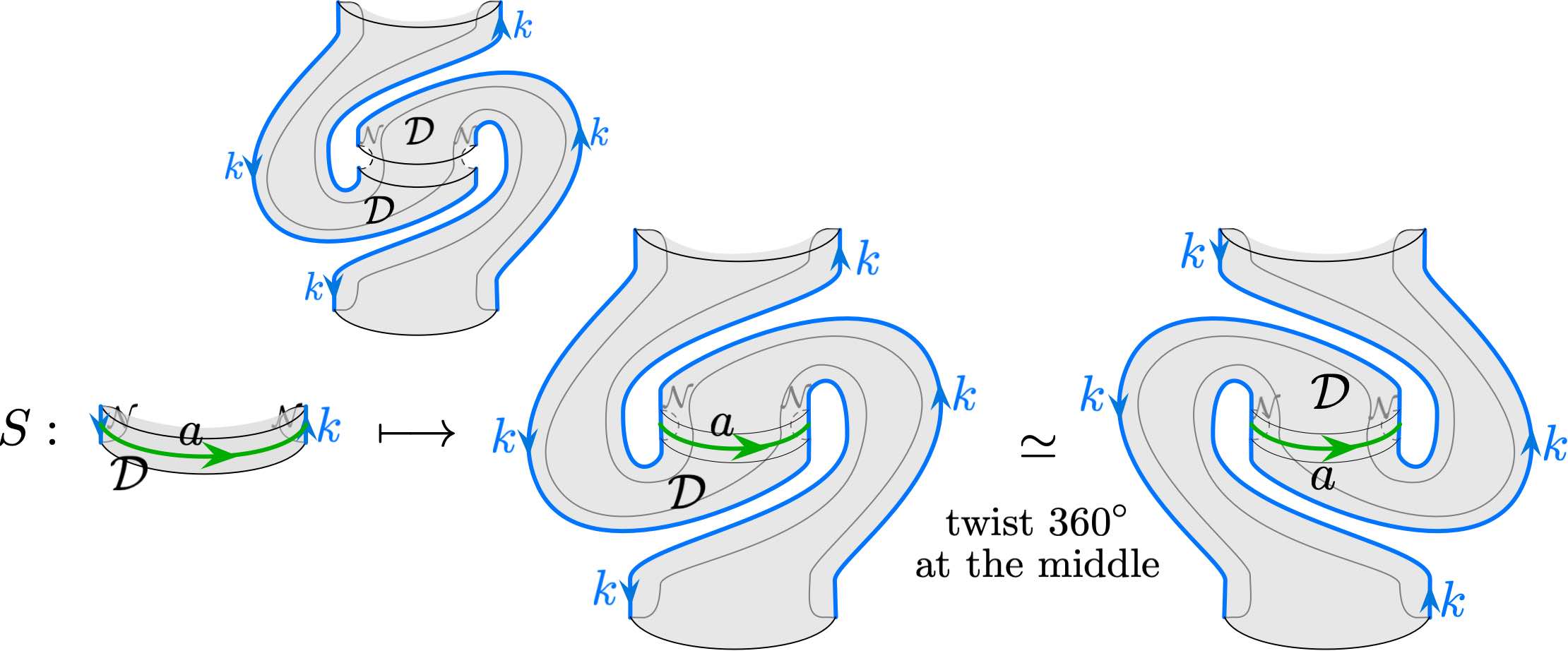}} \label{Hopf-S} \ee
Being able to twist by 360$^\circ$ in the middle due to absence of framing anomaly gives us the additional property $S=S^{-1}$, or $S^2=\text{id}$.

Note that by sliding the spark $a$ to either the top or bottom of the configuration in \eqref{Hopf-S}, it simply gets rotated by 180$^\circ$. Thus we can most efficiently represent $S$ by 180$^\circ$ rotation:
\be \raisebox{-.6in}{\includegraphics[width=4.8in]{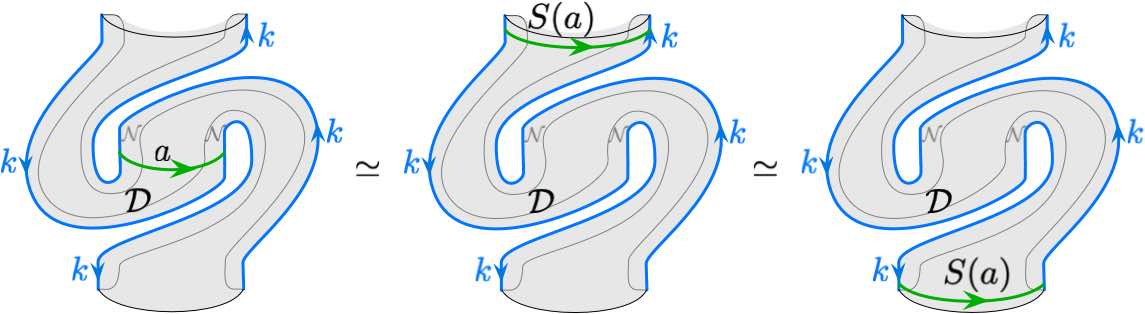}} \label{Hopf-S-S} \ee

For $\CH_\CN$ the picture is the same, with $\CD$ and $\CN$ boundaries swapped. For $U$, we have the analogous
\be  \raisebox{-.6in}{\includegraphics[width=2in]{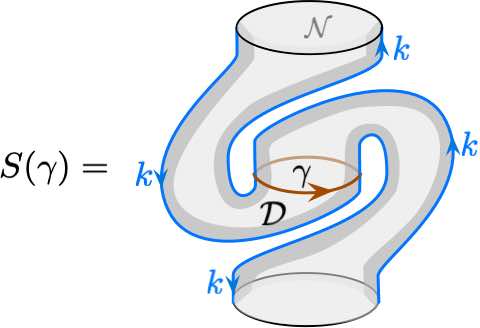}} \label{Hopf-SU} \ee

\subsubsection{A closer look at cobordisms}
\label{sec:Hopf-cob}

If one assumes the stronger dualizability condition \AD$_k$ from Section \ref{sec:trans} (that $k$ is dualizable as an interface between $\CN$ and $\ol \CD$, or between $\CD$ and $\ol\CN$) then the 3d cobordisms appearing above for $\CH_\CD,\CH_\CN$ can all be drawn in a way that makes their rectangular incoming and outgoing boundaries look more regular. This is quite revealing, in that it makes manifest the duality of $\CH_\CD$ and $\CH_\CN$ as Hopf algebras. Moreover, this re-drawing connects our constructions directly to the work of D. Reutter \cite{Reutter-slides}, \cite[Sec. 8.2]{FreedTeleman-Ising} and \cite{Hopkins-slides}, where the same cobordisms appear.

We recall, though, that \AD$_k$ forces spark algebras to be finite dimensional (Sec.~\ref{sec:spark-finite}).

Assuming \AD$_k$, the cobordisms appearing in \eqref{Hopf-m} and \eqref{Hopf-D}, which control products and coproducts are homeomorphic to
\begin{subequations} \label{Hopf-cob}
\be \hspace{-.2in} \raisebox{-.4in}{\includegraphics[width=5.3in]{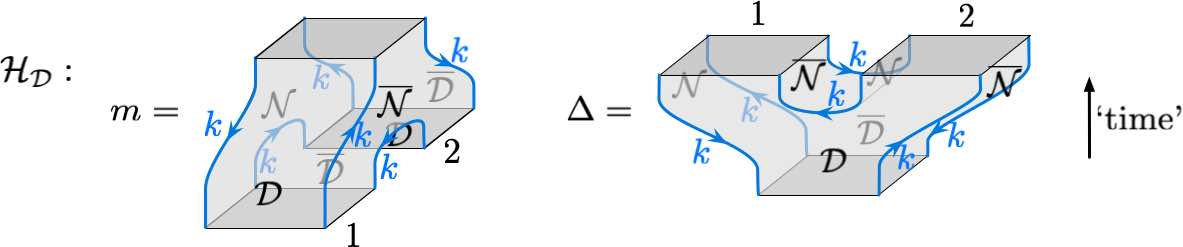}} \ee
\be \hspace{-.2in} \raisebox{-.4in}{\includegraphics[width=5.3in]{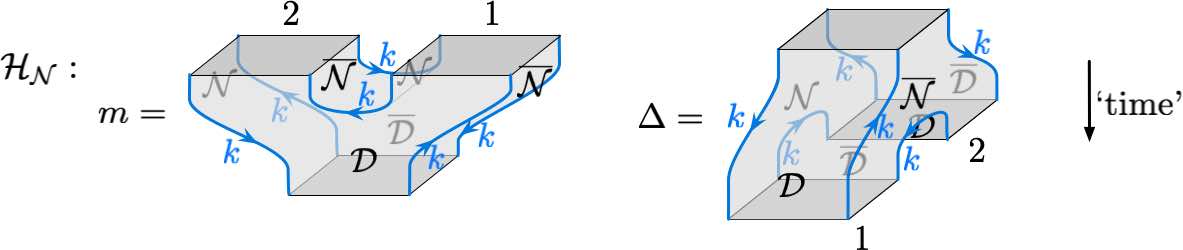}} \ee
\end{subequations}
For $\CH_\CD$, in these pictures, the incoming boundaries are at the bottom and the outgoing boundaries are at the top. Thus, $m$ joins two rectangles along $(\CD$,$\ol\CD)$ boundaries and $\Delta$ splits a rectangle, creating a new $(\ol\CN$,$\CN)$ boundary pair. For $\CH_\CN$, the cobordisms are \emph{identical} up to 1) the swap of incoming and outgoing boundaries (indicated by a reversed direction of `time'); 2) the swap of roles of $m$ and $\Delta$; 3) a partial swap of the orders of tensor factors.

Similarly, the cobordisms in \eqref{Hopf-u}, \eqref{Hopf-eU} for units and counits are homeomorphic to
\be \raisebox{-.5in}{\includegraphics[width=4.2in]{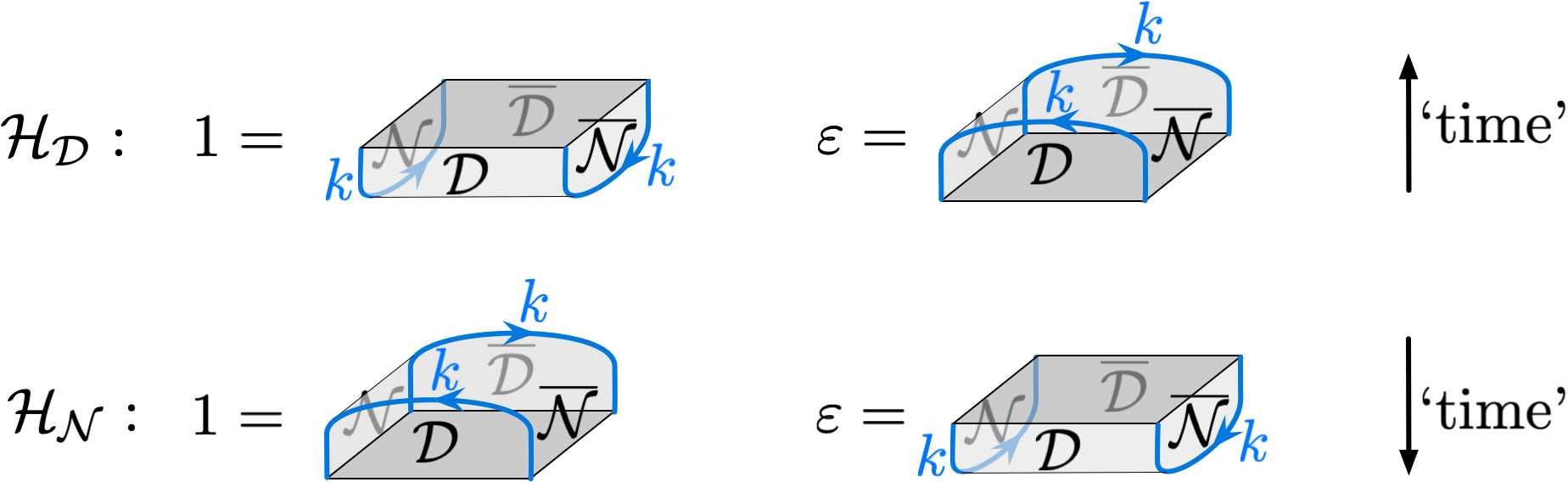}} \label{Hopf-cobe} \ee
and are identical for $\CH_\CD$ and $\CH_\CN$ up to 1) a swap of incoming and outgoing boundaries; and 2) a swap of the roles of $1$ and $\varepsilon$.

Finally, the antipode for both $\CH_\CD$ and $\CH_\CN$ is encoded in the cobordism
\be  \raisebox{-.5in}{\includegraphics[width=3.3in]{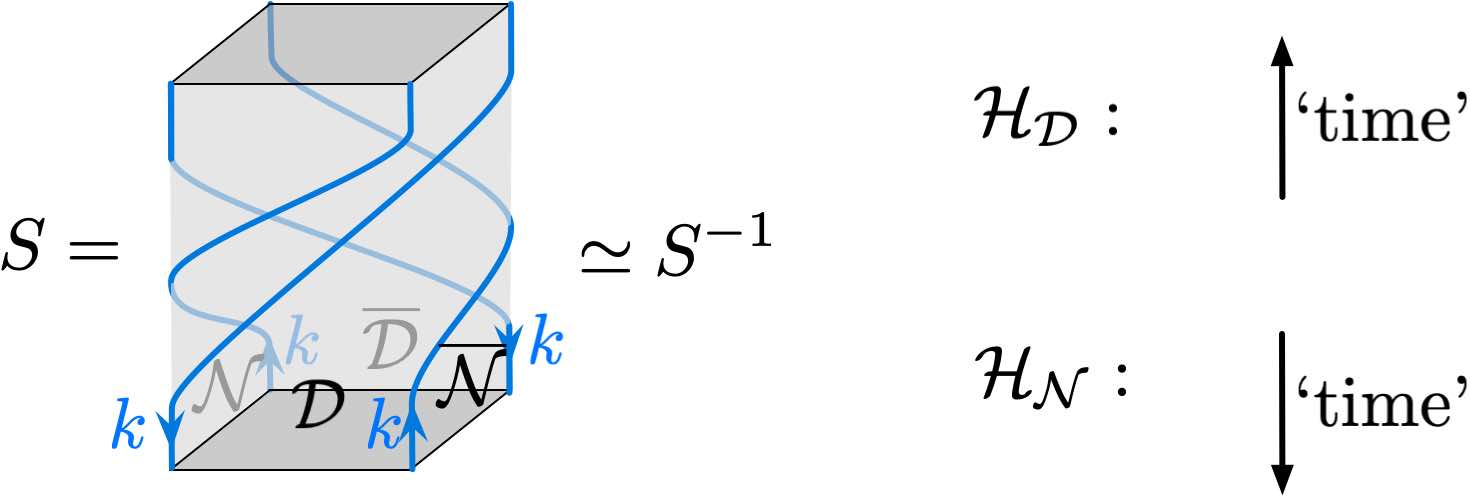}} \label{Hopf-cobS} \ee
with the only difference between $\CH_\CD$ and $\CH_\CN$ given by swapping incoming and outgoing boundaries.

The above relations precisely imply that $\CH_\CD$ and $\CH_\CN$ are not just dual vector spaces as in \eqref{DN-dualspaces}, but dual Hopf algebras. If they are finite dimensional and we choose a basis $\{a^i\}$ for $\CH_\CD$ and a dual basis $\{\alpha_i\}$ for $\CH_\CN$, with respective structure constants
\be \begin{array}{l} a^ia^j = m^{ij}_k a^k\,,\quad \Delta(a^i) = \Delta^i_{jk}a^j\otimes a^k\,,\quad 1 = \eta_ia^i\,,\quad  \varepsilon(a^i) = \varepsilon^i\,,\quad S(a^i) = S^i{}_j a^j\,, \\[.2cm] 
\alpha_i\alpha_j = m_{ij}^k \alpha_k\,,\quad \Delta(\alpha_i) =  \Delta_i^{jk}\alpha_j\otimes \alpha_k\,,\quad 1=\eta^i\alpha_i\,,\quad \varepsilon(\alpha_i) = \varepsilon_i\,,\quad S(\alpha_i) = S_i{}^j \alpha_j\,, \end{array}
\ee
we must have that
\be m^{ij}_k = \Delta^{ji}_j\,,\qquad \Delta^i_{jk} = m^i_{jk}\,,\qquad \eta_i=\varepsilon_i\,,\qquad \varepsilon^i=\eta_i\,,\qquad S^i{}_j = S_j{}^i\,. \label{mD-dual} \ee

\subsection{Topological proofs of the Hopf axioms}
\label{sec:Hopf-verify}

We'll now verify that the topological operations proposed above actually satisfy the required axioms of a Hopf algebra. This should be automatic, as the topological operations are just translations of algebraic reconstruction. Proofs were also already 
sketched by Reutter \cite{Reutter-slides}, assuming \AD$_k$ and using cobordisms in the form \eqref{Hopf-cob}, \eqref{Hopf-cobe}, \eqref{Hopf-cobS}. We'll present arguments that are more intrinsic to the spark pictures, and highlight where various assumptions are used.

It suffices to work with the spark algebra $\CH_\CD$. The arguments for $\CH_\CN$ and $U$ are identical, up to swapping $\CD/\CN$ boundaries or doubling the pictures.

Following Section \ref{sec:Hopf-axioms}, we first need to show that the coproduct is an algebra morphism and is co-associative. Co-associativity is easy, as it follows from simple topological invariance:
\be \includegraphics[width=5.5in]{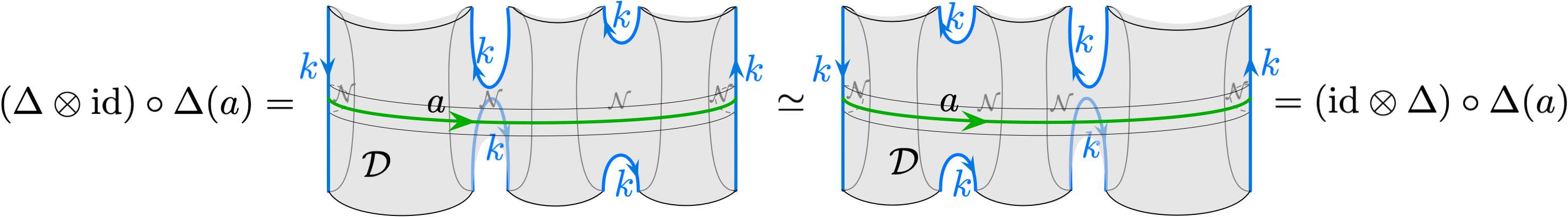} \ee
On the other hand, showing that $\Delta$ is an algebra morphism requires using transversality to create a ``hole'' between a ($\CD$,$\CN$) pair as on the LHS of \eqref{hole2}. Creating such a hole allows us to separate $\Delta(ab)$ into a composition (vertical stacking) of $\Delta(a)$ and~$\Delta(b)$:
\be \hspace{-.1in} \includegraphics[width=6in]{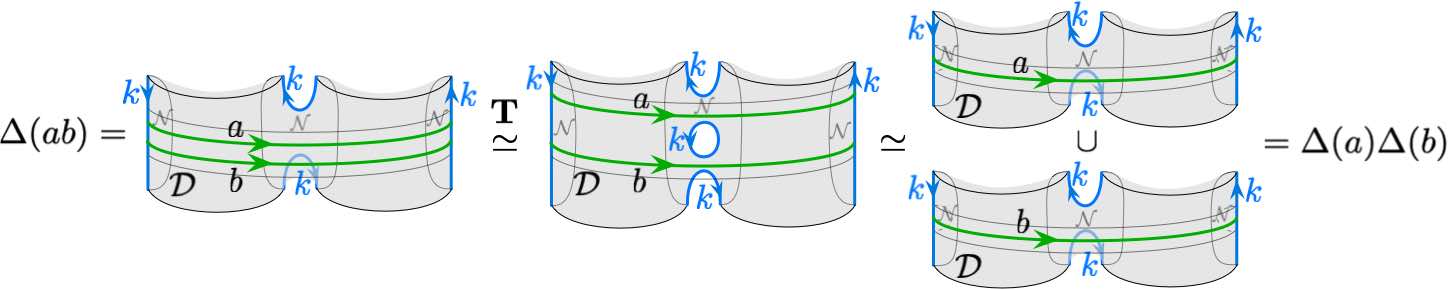} \ee

Next, consider the counit $\varepsilon$. Showing that it really is a counit for the coproduct follows from topological invariance
\be \raisebox{-.4in}{\includegraphics[width=4.3in]{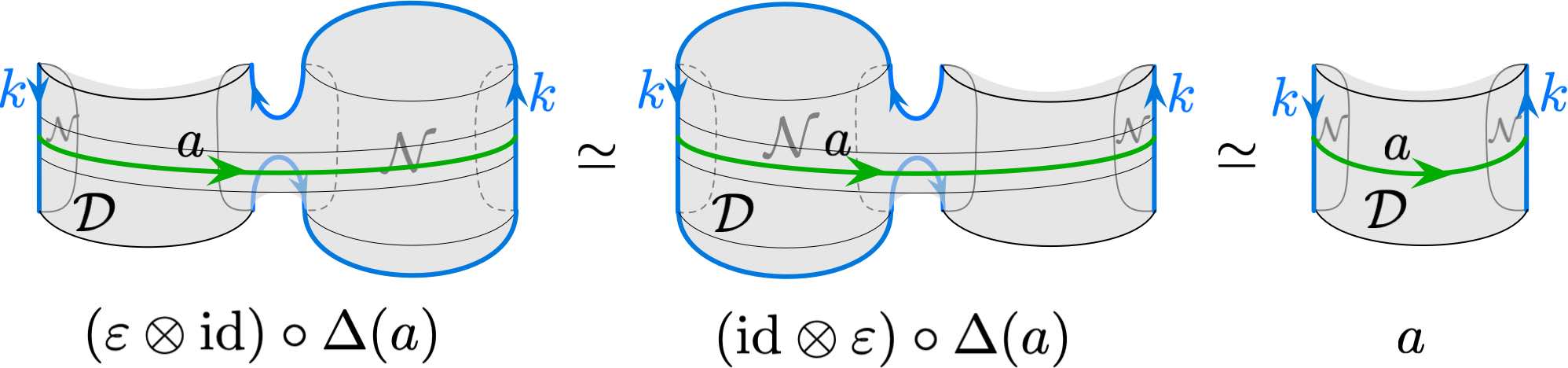}} \ee
whereas showing that it is an algebra morphism requires transversality \eqref{hole1}
\be \raisebox{-.4in}{\includegraphics[width=4in]{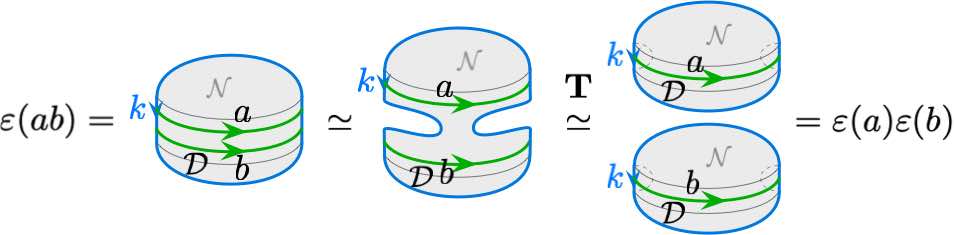}} \ee

Finally, consider the antipode. Being an algebra anti-morphism $S(ab)=S(b)S(a)$ is a straightforward consequence of topological invariance. In the presence of fermions, an extra sign $(-1)^{|a||b|}$ appears simply due to swapping the linear order in which operators are represented in correlation functions.
On the other hand, `canceling' the coproduct as on the RHS of \eqref{S-ax} uses transversality. This looks like
\be \hspace{-.2in} \includegraphics[width=6.2in]{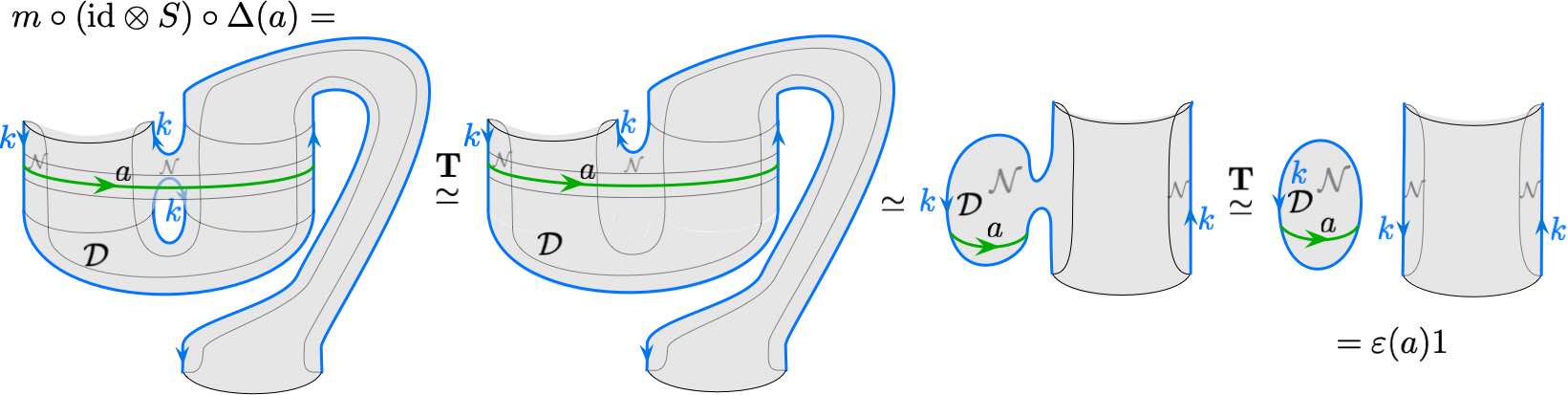} \notag \ee
\be \hspace{-.2in} \includegraphics[width=6.2in]{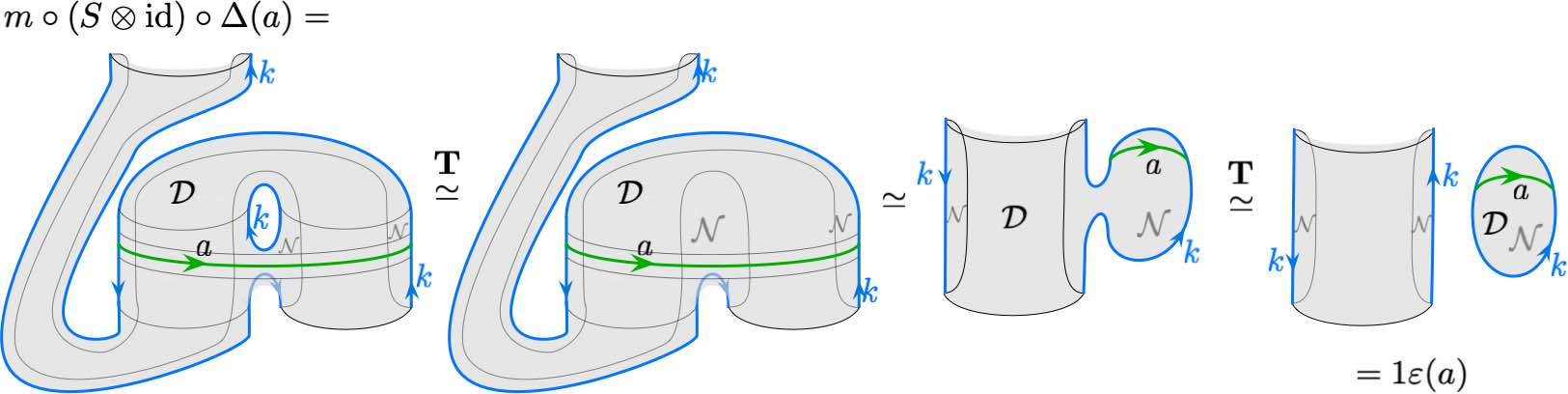} \ee
where transversality is first used to undo a hole, and then to separate off a half-ball containing the spark $a$.

Altogether, we've established the bulk of Proposition \ref{Prop:Hopf}: given topological invariance, transversality, enough dualizable objects (\AD$_{\rm line}$), and continuous fiber functors, the spark algebras $\CH_\CD,\CH_\CN,U$ are all Hopf algebras.

\subsection{The Hopf pairing}
\label{sec:Hopf-pair}

We already saw in Section \ref{sec:Hopf-cob} that if \AD$_k$ holds (and thus spark algebras are finite dimensional), the boundary algebras $\CH_\CD$ and $\CH_\CN$ are Hopf-dual to each other. We'd now like to rephrase this in terms of the nondegenerate pairing defined in \eqref{link-Hopf}, which places a pair of sparks on opposite hemispheres of a solid 3-ball and evaluates the resulting partition function:
\be h:\CH_\CD\otimes \CH_\CN \to \C\,,  \qquad \raisebox{-.47in}{\includegraphics[width=1.7in]{Pair.jpg}} \label{def-h} \ee
This pairing makes sense, as a continuous bilinear form, even when spark algebras are infinite dimensional. We'd like to show that $h$ is a \emph{Hopf pairing}, meaning that that it satisfies
\be \begin{array}{c} h(ab,\alpha) = \sum h(a\otimes b,\Delta^{\rm op}(\alpha)) \\
h(a,\alpha\beta) = \sum h(\Delta (a),\alpha\otimes \beta) \end{array}\,,\quad 
\begin{array}{c} h(a,1) =\varepsilon(a) \\
 h(1,\alpha) = \varepsilon(\alpha) \end{array}\,,\quad
  h(S(a),\alpha) = h(a,S^{-1}(\alpha))\,. \label{h-pair-def}   \ee
Here $\Delta^{\rm op}(a)=\sum (-1)^{|a^{(1)}||a^{(2)}|}a^{(2)}\otimes a^{(1)}$ means the coproduct where the two tensor factors are swapped. These properties imply that we can identify $\CH_\CD$ as the continuous, topological Hopf-dual of $\CH_\CN$ even in the infinite-dimensional case.

The relations on the RHS of \eqref{h-pair-def}, intertwining product and coproduct, come from combining the coproduct in the form \eqref{Hopf-D-Sweedler} to split one of the sparks, and then using transversality to fully separate the ball into two balls:
\be \raisebox{-.5in}{\includegraphics[width=5in]{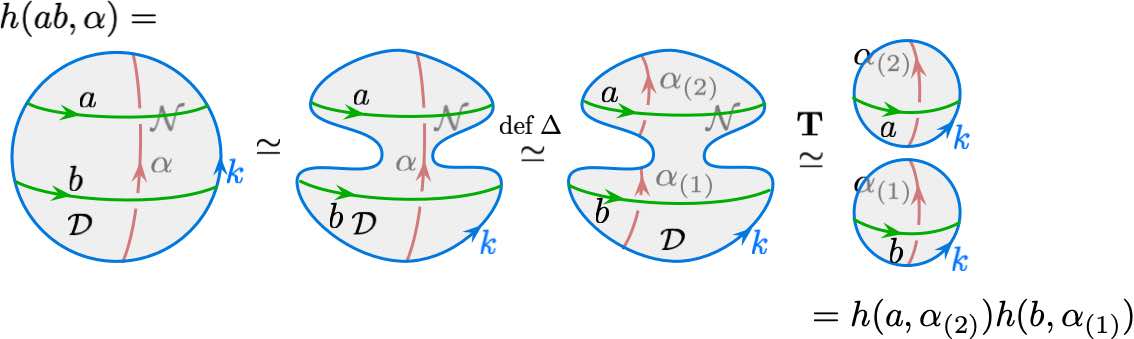}} \label{habalpha} \ee \vspace{-.1in}
\be \raisebox{-.5in}{\includegraphics[width=5in]{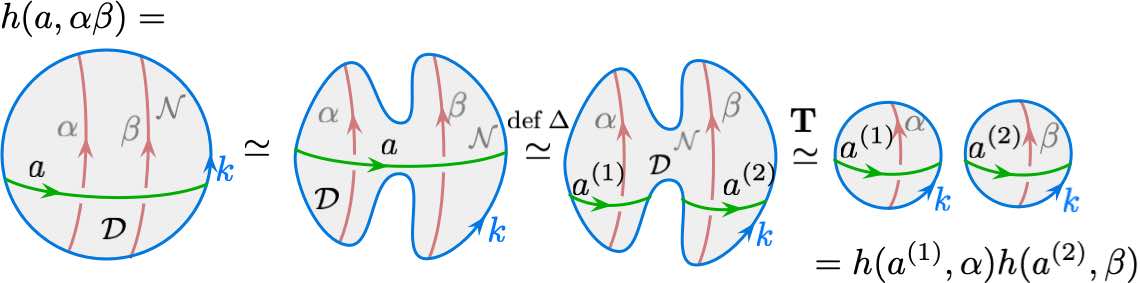}} \ee
If sparks are fermionic, extra signs appear whenever the linear order of operators is swapped, \emph{e.g.} \eqref{habalpha} reads $h(ab,\alpha) = \sum (-1)^{(|b|+|\alpha_{(1)}|)|\alpha_{(2)}|} h(a,\alpha_{(2)})h(b,\alpha_{(1)})$.

The relations in the middle of \eqref{h-pair-def}, intertwining unit and counint, result immediately by comparing the definition of $h$ in \eqref{def-h} and the counits in \eqref{Hopf-eU}: the counits are just the Hopf pairing with the identity spark inserted on one side or the other. The final relation in the RHS of \eqref{h-pair-def}, involving the antipode, reflects the fact that twisting one hemisphere by 180$^\circ$ is equivalent to un-twisting the other hemisphere by 180$^\circ$; in pictures,
\be \raisebox{-.4in}{\includegraphics[width=5.2in]{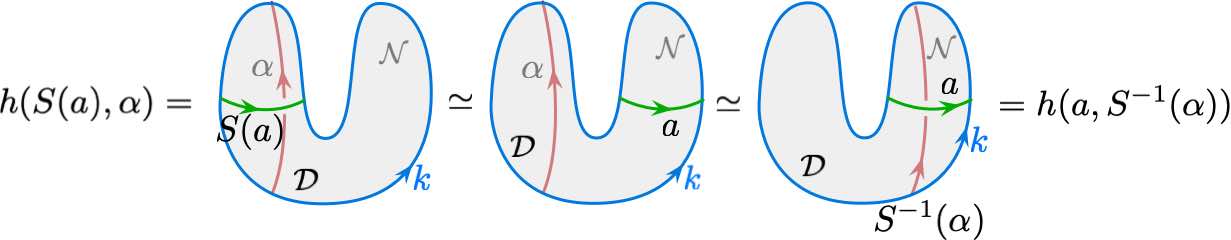}} \ee

This completes our argument for Proposition \ref{Prop:Hopf}.

\section{Bulk sparks}
\label{sec:bulk}

In this final theoretical section of the paper, we consider further the structure of the bulk spark algebra $U$. We have already showed that, under suitable conditions, $U$ will be a Hopf algebra. However, since $U$ should represent bulk line operators, which form a ribbon category (a braided monoidal category, with ribbon twists), we expect $U$ to be a ribbon Hopf algebra (a quasitriangular Hopf algebra, with R-matrix, and ribbon element). After reviewing the definitions of quasitriangular and ribbon Hopf algebras in Section \ref{sec:QT-ax}, we will show in Section \ref{sec:U-R} that

\begin{Prop} \label{Prop:R} Assuming topological invariance (\AD, Section \ref{sec:top}), transversality (\AT, Section \ref{sec:trans}), enough dualizable objects (\AD$_{\rm line}$, Section \ref{sec:lines}), and exact and continuous fiber functors (Section \ref{sec:cts}), the bulk spark algebra $U$ is an involutive ribbon Hopf algebra, with an R-matrix $\CR$ and ribbon element $v$ satisfying $v=m\circ (S\otimes \text{id})(\CR_{21})$.

When \AD$_k$ (Section \ref{sec:trans}) holds, $U$ is finite dimensional and $\CR$ is a genuine element of $U\otimes U$. Otherwise, $\CR$ belongs to a topologically completed tensor product.
\end{Prop}

Our perspective shall be the same as in Section \ref{sec:Hopf-bdy}. Namely, the braiding and twist on the bulk category $\CC_\CT$ can be used to reconstruct an R-matrix and ribbon element on the endomorphism algebra $U$ of a fiber functor $\CF_\CT:\CC_\CT\to \text{Vect}$, so that the lift
\be \wt\CF_\CT:\CC_\CT\to U\text{-mod} \ee
becomes a functor of ribbon categories. We translate this algebraic statement to topology, and then give independent topological proofs of the quasitriangularity/ribbon axioms --- which would make sense even if one did not know about $\CC_\CT$.

Our last major development is to extend the result that $U\simeq \CH_\CD\otimes\CH_\CN$ as vector spaces (Section \ref{sec:linefunctor}), and the result that $\CH_\CD$ and $\CH_\CN$ are Hopf-paired (Section \ref{sec:Hopf-bdy}) to

\begin{Thm} \label{Thm:double} Under the same assumptions as for Prop. \ref{Prop:R}, the bulk spark algebra $U$ is the Drinfeld double of the pair $(\CH_\CD,\CH_\CN)$, paired by the nondegenerate Hopf pairing $h:\CH_\CD\otimes\CH_\CN\to \C$ from \eqref{def-h}. In particular, when \AD$_k$ holds and $\CH_\CD,\CH_\CN$ are finite-dimensional, with bases $\{a^i\},\{\alpha_i\}$ that are dual under $h$, the R-matrix is $\CR = \sum_i (-1)^{|a^i||\alpha_i|}a^i\otimes\alpha_i$.
\end{Thm}

\noindent We verify all defining properties of the Drinfeld double topologically. A key result we derive and use is that the box gluing of \eqref{boxglue} (which does not actually require completeness) is performed by inserting sparks $\sum_i \alpha_i\otimes a^i$.

We do not use completeness anywhere in this section. Combining Prop. \ref{Prop:R} and Prop. \ref{Prop:Hopf} we obtain not-necessarily-faithful lifted functors
\be \wt\CF_\CN:\CC_\CN\to \CH_\CD\text{-mod}\,,\qquad  \wt\CF_\CT:\CC_\CT\to U\text{-mod}\,,\qquad  \wt\CF_\CD:\CC_\CD\to \CH_\CN\text{-mod} \label{Flift-R} \ee
that preserve monoidal structure, units, duals, and (for $U$) braiding and twists. Moreover, at the level of module categories, Theorem \ref{Thm:double} implies that there are equivalences of braided tensor categories
\be U\text{-mod} \simeq \CZ_{\rm Drin}(\CH_\CD\text{-mod})  \simeq \CZ_{\rm Drin}(\CH_\CN\text{-mod})\,,  \ee
where $\CZ_{\rm Drin}$ denotes the Drinfeld center. (In the abelian case, the Drinfeld double is engineered precisely so that Drinfeld centers have this property, \cf\ \cite[Prop. 7.14.6]{egno}; in the dg/infinity setting, a derived Drinfeld center as in \cite{Ben-Zvi:2008vtm} should be used.) If we do happen to assume completeness (Thm. \ref{Thm:faithful}), then an immediate corollary of Prop. \ref{Prop:Hopf}, Prop. \ref{Prop:R}, and Thm. \ref{Thm:double} is

\begin{Thm} \label{Thm:Rep} Assuming completeness (\AC$_{\rm strip}$ as well as \AD$_k$) and the same conditions as in Prop. \ref{Prop:R}, the lifted functors $\wt\CF_\CN,\wt\CF_\CD$ are equivalences of monoidal categories (preserving duals) and $\wt\CF_\CT$ is an equivalence of ribbon categories (preserving duals). In particular, $\CC_\CT\simeq \CZ_{\rm Drin}(\CC_\CN)\simeq \CZ_{\rm Drin}(\CC_\CD)$.
\end{Thm}

\noindent The final statement about Drinfeld centers is completely expected for a 3d topological QFT with boundary conditions $\CD,\CN$ that are generators of their 2-categories.
Mathematically, $\CC_\CT\simeq \CZ_{\rm Drin}(\CC_\CN)\simeq \CZ_{\rm Drin}(\CC_\CD)$ is sometimes taken as the definition of $\CD,\CN$ being generators, \cf\ Appendix \ref{app:line-setup}.

\subsection{Axioms of ribbon, quasitriangular Hopf algebras and doubles}
\label{sec:QT-ax}

We recall some standard definitions, mainly following \cite[Ch. 8]{egno}.

A quasitriangular Hopf algebra is a Hopf algebra $U$ together with an ``R-matrix,'' an invertible element $\CR\in U\otimes U$ that satisfies
\be \CR\Delta(a)\CR^{-1} =\Delta^\text{op}(a)\qquad\forall\; a\in A\,, \label{Rax-1} \ee \vspace{-.3in}
\be (\Delta\otimes\text{id})(\CR) = \CR_{13} \CR_{23}\,,\qquad (\text{id}\otimes \Delta)(\CR) = \CR_{13}\CR_{12} \label{Rax-2} \ee
where $\Delta^{\rm op}(a)$ is the coproduct with tensor factors swapped; and if $\CR = \sum_i a^i\otimes b^i$ for some $a^i,b^i\in U$ then $\CR_{12}$, $\CR_{12}$, $\CR_{23}$ are the elements of $U\otimes U\otimes U$ defined by
\be \CR_{12} = \sum_i a^i \otimes b^i \otimes 1 \,,\quad  \CR_{13}= \sum_i a^i\otimes 1 \otimes b^i\,,\quad \CR_{23} = \sum_i 1\otimes a^i\otimes b^i\,. \ee
Note that \eqref{Rax-2} implies the Yang-Baxter equation
\be \CR_{12}\CR_{13}\CR_{23}=\CR_{23}\CR_{13}\CR_{12}\,. \label{YB} \ee

A ribbon Hopf algebra $U$ is a quasitriangular Hopf algebra with a ``ribbon element,'' an invertible central element $v\in U$ that satisfies 
\be
\Delta(v)=(\CR_{21}R)^{-1}v\otimes v\,, \qquad S(v)=v\,. \label{defv}
\ee
When a ribbon element exists, it obeys $v^2 = uS(u)$, where $u\in U$, known as the Drinfeld element, is canonically built out of the R-matrix as
\be u := m\circ (S\otimes \text{id})(\CR_{21}) = \sum_i S(b^i)a^i\,. \label{defu} \ee
In our setting, we'll eventually obtain ribbon Hopf algebras where $v=u$.

A common source of quasitriangular Hopf algebras is Drinfeld's quantum double construction, reviewed in \cite[Ch. 7]{egno}; see also \cite[Sec I.5]{Ram} for an algebraic summary that's well adapted for our purposes. 

Suppose that $\CH_L$ and $\CH_R$ are both Hopf algebras, with a nondegenerate Hopf pairing $h:\CH_L\otimes \CH_R\to \C$, satisfying the Hopf-pairing axioms \eqref{h-pair-def}. Then the Drinfeld double is defined as the vector space $U:=\CH_L\otimes \CH_R$, with elements denoted $a\alpha$ for $a\in \CH_L$, $\alpha\in \CH_R$. It has an associative product given by
\be  \begin{array}{l} a\cdot\alpha = (a1_R)\cdot (1_L\alpha) := a\alpha \\[.2cm]
   \alpha \cdot a = (1_L\alpha) \cdot (a1_R) := \sum (-1)^{\xi(\alpha, a)} h\big(S^{-1}(a^{(1)}),\alpha_{(1)}\big)\; a^{(2)}\alpha_{(2)}\; h\big(a^{(3)},\alpha_{(3)}\big) \,. \end{array} \label{double-prod} \ee
Here we are using Sweedler notation for the double coproduct $\Delta^2$ in both $\CH_L$ and $\CH_R$, \eg\ $\Delta^2(a):= (\Delta\otimes \text{id})\circ\Delta(a) = (\text{id}\otimes\Delta)\circ\Delta(a) = \sum_i a^{(1)}_i\otimes a^{(2)}_i \otimes a^{(3)}_i$ for some sequences $(a^{(1)}),(a^{(2)}),(a^{(3)})$ in $\CH_L$.
Moreover, when fermions are present, $\xi(\alpha, a)$ is an element in $\Z_2$ defined by 
\begin{align}  
\xi(\alpha, a) &=(|\alpha_{(2)}|+|\alpha_{(3)}|)(|a^{(1)}|+|a^{(2)}|)+|\alpha_{(1)}||a^{(1)}|+|\alpha_{(3)}||a^{(3)}| \\
 &=  (|\alpha_{(1)}|+|\alpha_{(2)}|+|\alpha_{(3)}|)|a^{(1)}|+(|\alpha_{(2)}|+|\alpha_{(3)}|)|a^{(2)}| + |\alpha_{(3)}||a^{(3)}| \label{D-sign}
\end{align}
 The formula for this sign can be found in \cite{gould1993quantum}.\footnote{To exactly match our expression with the sign of \cite{gould1993quantum}, we first use that the pairing $h$ is even, and therefore $|a_{(1)}|=|\alpha_{(1)}|$; and we add an extra $|\alpha_{(1)}||a_{(1)}|+|\alpha_{(3)}||a_{(3)}|$ due to the change of order of $\alpha$ and $a$ in the pairing $h$. We find \eqref{D-sign} to be a bit more intuitive.} Equation \eqref{double-prod} implies on general elements that
\be (a\alpha)\cdot (b\beta) = \sum (-1)^{\xi(\alpha, b) }h\big(S^{-1}(b^{(1)}),\alpha_{(1)}\big)\; ab^{(2)}\alpha_{(2)}\beta\; h\big(b^{(3)},\alpha_{(3)}\big)\,. \label{double-prod-gen} \ee

Given \eqref{double-prod}, there is a unique Hopf-algebra structure on $U$ such that the maps $\CH_L\to \CH_L\otimes 1\hookrightarrow U$ and $\CH_R\to  1\otimes  \CH_R\hookrightarrow U$ are both Hopf-algebra morphisms. Its Hopf operations are simply given by (again with appropriate signs in the dg/super case)
\be \Delta(a\alpha) = \Delta(a)\Delta(\alpha)\,,\qquad \varepsilon(a\alpha) = \varepsilon(a)\varepsilon(\alpha)\,,\qquad S(a\alpha)=S(\alpha)S(a)\,. \label{double-Hopf}\ee

Most famously, the Drinfeld double has a canonical R-matrix that makes it quasitriangular, at least when $\CH_L,\CH_R$ are finite dimensional. Let $\{a^i\}$, $\{\alpha_i\}$ be dual bases of $\CH_L,\CH_R$, with respect to the Hopf pairing $h$. Then the R-matrix is given by
\be \CR = \sum_{i}(-1)^{|a^i||\alpha_i|} a^i\otimes \alpha_i\,. \label{double-R} \ee
In the infinite-dimensional case, the R-matrix has been constructed with the help of filtrations on $\CH_L,\CH_R$, and topological bases preserving the filtrations (\cf\ the discussion and references in \cite[Sec I.5.4]{Ram}). In general, the element $\CR$ lives in a completion of the tensor product $\CH_L\otimes \CH_R$.

\subsection{Bulk sparks are quasitriangular and ribbon}
\label{sec:U-R}

We'll now use categorical reconstruction to endow $U$ with an R-matrix and ribbon element, by translating algebra to topology, and establish Proposition \ref{Prop:R}. First, though, let's recall what the Hopf operations are, from Section \ref{sec:Hopf-def}.

\subsubsection{Reminder of the Hopf operations}
\label{sec:U-Hopf}

Using transversality to establish an isomorphism of vector spaces $U \simeq \CH_\CD\otimes \CH_\CN$ as in \eqref{T-DN-proof}, we may represent any bulk spark uniquely as `$a\alpha$', for some $a\in \CH_\CD$ and $\alpha\in \CD_\CN$. In pictures, this spark looks like
\be \raisebox{-.2in}{\includegraphics[width=3in]{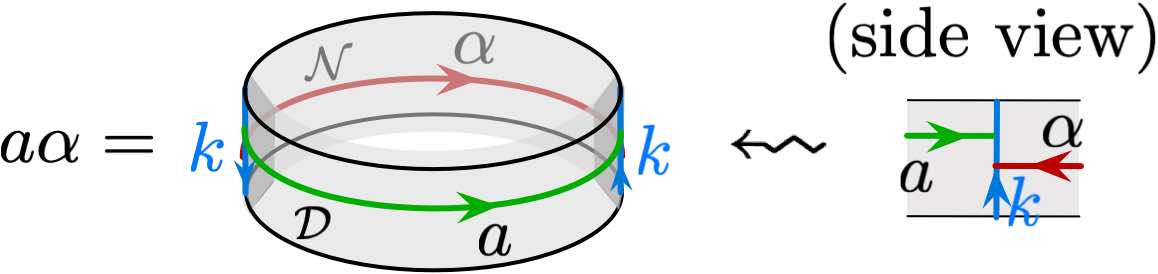}} \label{fig:Ubasic} \ee
Note that we must \emph{choose} a vertical order in which to place the $\CH_\CD$ and $\CH_\CN$ sparks, as the mutual insertion is only well defined when their endpoints lie at distinct points on the $k$ boundaries. Different orders correspond to different isomorphisms $U \simeq \CH_\CD\otimes \CH_\CN$, and we must fix one. We choose to place the $\CH_\CD$ spark above the $\CH_\CN$ spark, as indicated by writing `$a\alpha$' (as opposed to `$\alpha a$').

Representing bulk sparks as $a\alpha$, the Hopf operations from Section \ref{sec:Hopf-def} now look like 
\be  \raisebox{-.8in}{\includegraphics[width=5.5in]{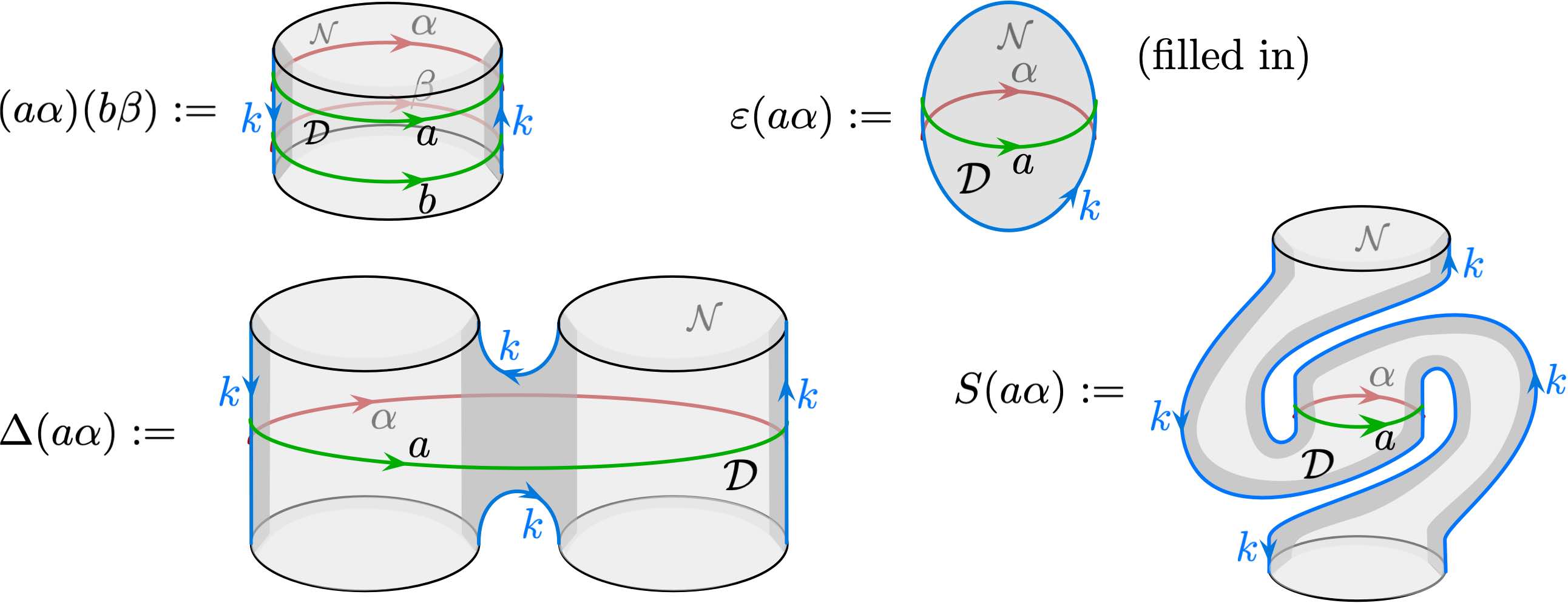}} \label{fig:UHopf} \ee
Note that in the product, the core of the cylinder is empty. In the coproduct, there are two distinct empty cores on either side (1 and 2), but the central region between the legs is filled in with bulk theory $\CT$. The counit is insertion in a solid ball with $\CD$ and $\CN$ boundaries (with endpoints of sparks in a different configuration than in the Hopf pairing \eqref{def-h}). The antipode is one long, twisted cylinder with the core drilled out; applying the antipode is equivalent to rotating a spark by 180$^\circ$.

\subsubsection{The R-matrix}
\label{sec:U-R2}

Now consider our fiber functor on bulk line operators $\CF_\CT:\CC_\CT\to \text{Vect}$. The bulk category $\CC_\CT$ is braided, meaning that there are isomorphisms
\be c_{\ell,\ell'} : \ell\otimes \ell'\to\ell'\otimes \ell\,,\qquad \forall\;\ell,\ell'\in\CC_\CT \ee
that obey certain relations with respect to the tensor product (\cite[Sec. 8.1]{egno}). The braiding isomorphisms, composed with a permutation $P$ to undo the swap in order of $\ell,\ell'$, can be collected into a single natural transformation of the functor $\CF_\CT\otimes \CF_\CT$,
\be \CR := Pc \in \text{End}(\CF_\CT\otimes \CF_\CT) \ee
When $U=\text{End}(\CF_\CT)$ is finite dimensional, we have $\text{End}(\CF_\CT\otimes \CF_\CT)\simeq \text{End}(\CF_\CT)\otimes \text{End}(\CF_\CT)$, so $\CR \in U\otimes U$. If $U$ is infinite dimensional, then $\CR$ generally lives in a completion of $U\otimes U$.

Translating this to topology, we find that $\CR$ is defined by the 3-manifold on the LHS of \eqref{fig:R-def}: a double pair of pants, whose outer boundary is split into $\CD$ (front) and $\CN$ (back), and whose interior is formed by removing two crossed (non-intersecting) thin solid cylinders.
The inner boundary is two annuli (as in \eqref{Annulus}), so this defines a state in $U\otimes U$. To keep track of the two tensor factors it can be useful to straighten out the inner cylinders. We do this by twisting the top as indicated on the RHS:
\be   \raisebox{-.5in}{\includegraphics[width=4.4in]{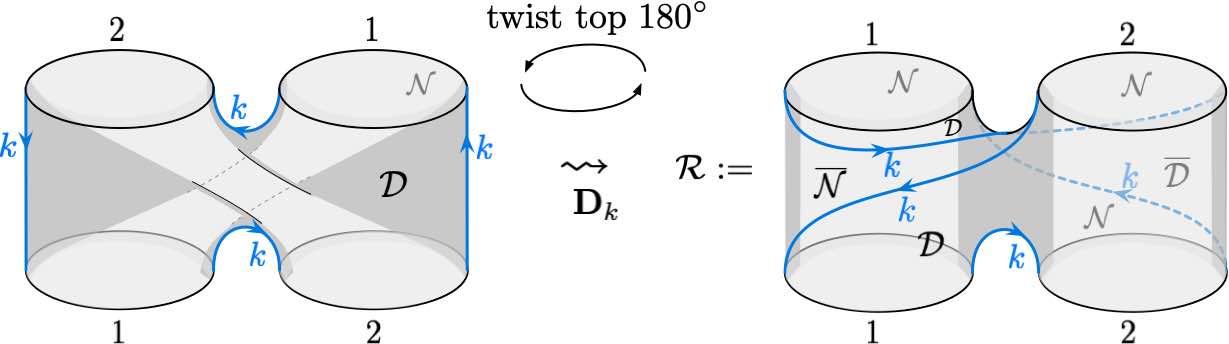}} \label{fig:R-def} \ee

If one is careful, the point where a completion of $U\otimes U$ arises, as opposed to an honest state, is in the twisting of $k$ interfaces on the boundary on the RHS of \eqref{fig:R-def}. For the RHS to be honestly defined, we should impose the dualizability condition \AD$_k$ (Sec. \ref{sec:trans}), which we know forces spark algebras to be finite dimensional. Without \AD$_k$, we use the LHS and expect to find a generalized state.

There are several other closely related elements of $U\otimes U$ that can be introduced at the same time. Assuming \AD$_k$, they look like
\be \hspace{-.2in} \raisebox{-.5in}{\includegraphics[width=6.2in]{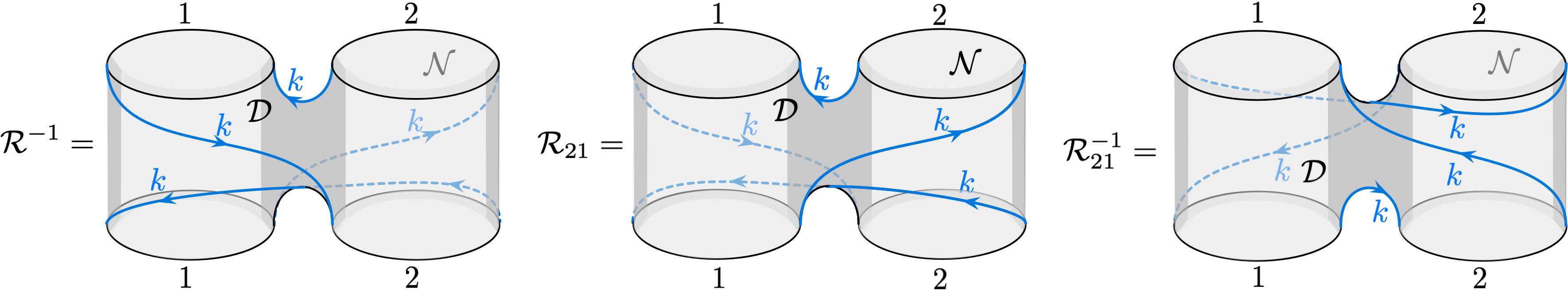}} \label{fig:R-alt} \ee
The inverse $\CR^{-1}$ is obtained by taking a vertical reflection of the crossed tubes on the LHS of \eqref{fig:R-def}, and then untwisting to straighten the tubes. Alternatively, we obtain an element  $\CR_{12}\in U\otimes U$ (with tensor factors swapped) by swapping `$1$' and `$2$' labels on the LHS of \eqref{fig:R-def}, and then untwisting. We get  $\CR_{12}^{-1}$ by swapping labels and reflecting vertically, and then untwisting. (Without \AD$_k$, we should keep these twisted, as on the LHS of \eqref{fig:R-def}.)

Note that choices made in this paper have led us most naturally to a left-handed braided convention, which is opposite that used in much of the literature. We've chosen the R-matrix to be \eqref{fig:R-def} ultimately so that its expression in terms of the Hopf pairing will be given by the standard Drinfeld-double formula in \eqref{double-R} (see below). An alternative option would be to take $\CR_{21}^{-1}$ as ``the'' R-matrix. It represents a right-handed braiding, and satisfies all the same quasitriangularity axioms as $\CR$; but it has a more complicated relation to the Hopf pairing, given in \eqref{fig:Rglue-alt} below.

\subsubsection{Topological proofs of quasitriangularity}
\label{sec:U-R-proofs}

The R-matrix defined in \eqref{fig:R-def} is guaranteed to satisfy the axioms \eqref{Rax-1}--\eqref{Rax-2} of a quasi-triangular Hopf algebra, due to the relations satisfied by the braiding isomorphisms $c_{\ell,\ell'}$ in the category $\CC_\CT$. However, one can also verify the axioms directly from topology.

Consider the first axiom \eqref{Rax-1}, stating that conjugation by $\CR$ swaps the factors in the coproduct. The topological proof looks like this:
\be   \raisebox{-.5in}{\includegraphics[width=5.5in]{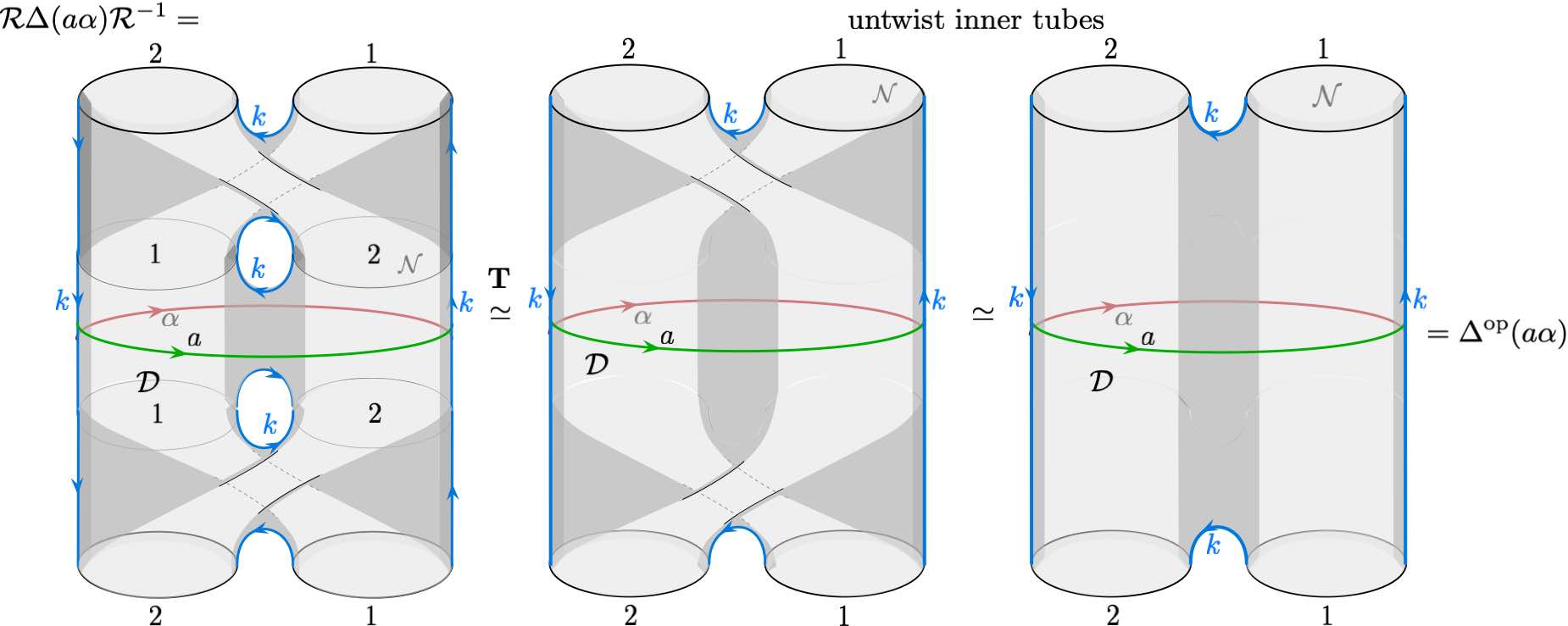}} \hspace{-.5in} \label{fig:RDop}  \ee
Here we've used the representations of $\CR$ and $\CR^{-1}$ as on the LHS of \eqref{fig:R-def}, with crossed tubes.
After using transversality to fill in the holes with the bulk theory $\CT$, we untwist the inner tubes, arriving at a version of the coproduct with its `$1$' and `$2$' legs swapped.

The first identity in \eqref{Rax-2}, $(\Delta\otimes \text{id})(\CR)=\CR_{13}\CR_{23}$ follows entirely from a topological deformation shown in \eqref{fig:RD}, with no need even for transversality. In the top-left figure here, we have rotated the top `2' hole toward the back, allowing us to unravel several of the $k$ seams going around it. In bottom figure, we moved the `2' hole along the back of the manifold all the way to the top. 
\be  \raisebox{-.7in}{\includegraphics[width=5.5in]{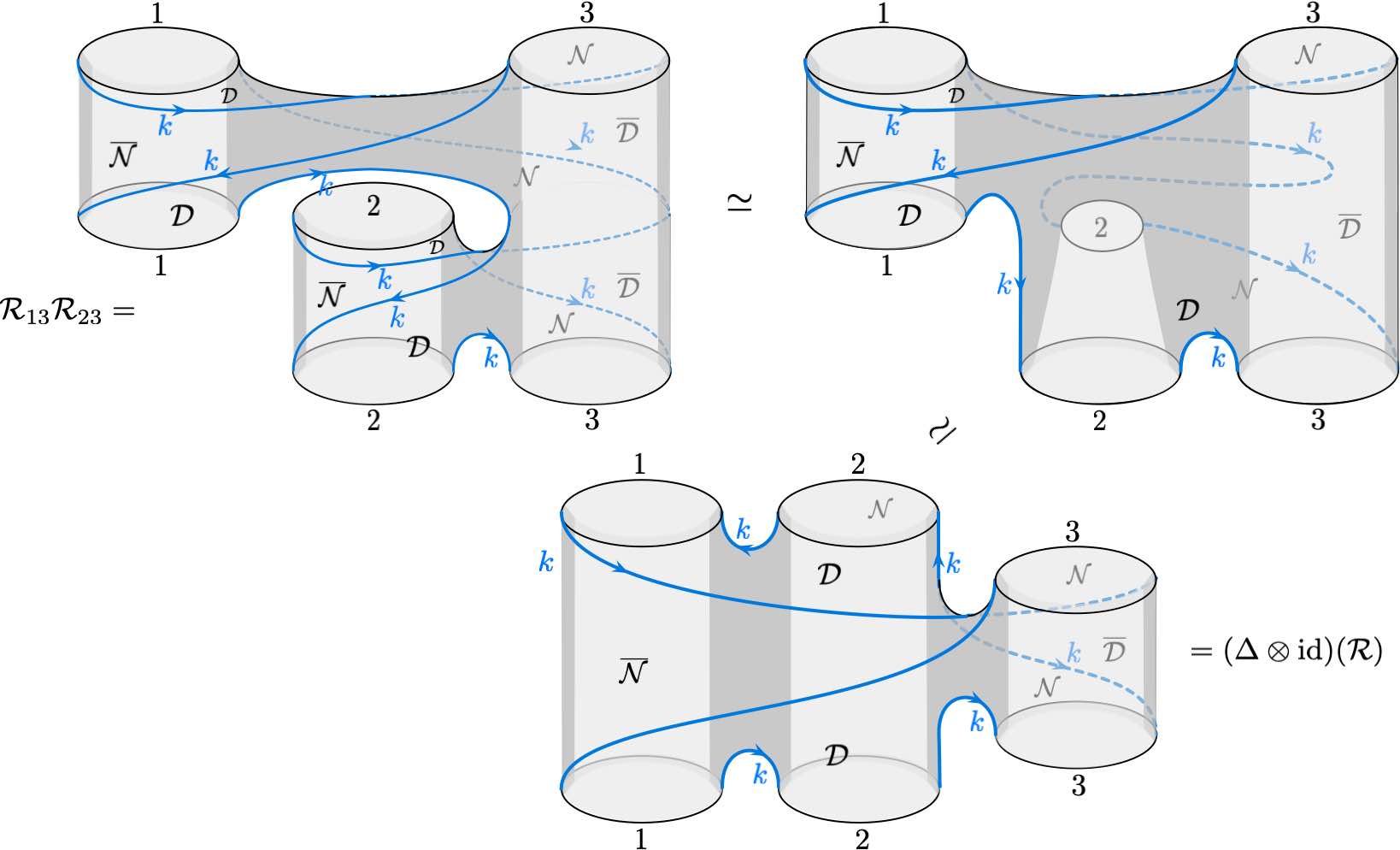}} \label{fig:RD}   \ee
The second identity $(\text{id}\otimes\Delta)(\CR)=\CR_{13}\CR_{12}$ works essentially the same way.

\subsubsection{The ribbon element}
\label{sec:ribbon}

Finally, the ribbon twists in the category $\CC_\CT$, isomorphisms $\tau_\ell:\ell\to\ell$ for each object $\ell$, can be assembled into a natural transformation of the fiber functor that defines a ribbon element
\be v := \tau \in \text{End}(\CF_\CT)\,. \ee
Translating to topology, the ribbon element $v\in U$ is given by an empty cylinder with twisting $k$ interfaces on its outer boundary:
\be  \raisebox{-.7in}{\includegraphics[width=2.8in]{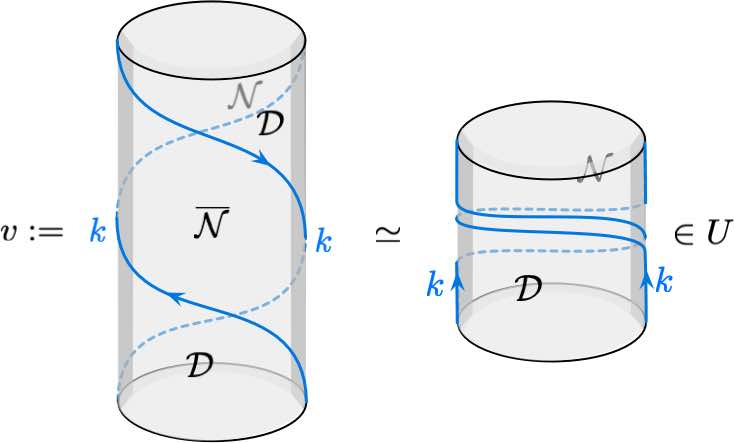}} \label{fig:Ribbon}  \ee

If we identify bulk sparks with states on an annulus as in \eqref{Annulus}, then the identity spark and $v$ are related by a Dehn twist of the annulus.

Our ribbon element $v$ turns out to be identical to the Drinfeld element $u=m\circ(S\otimes \text{id})(\CR_{21})$ constructed from the R-matrix in \eqref{defu}. A topological proof of this uses transversality and a topological deformation:
\be   \raisebox{-.7in}{\includegraphics[width=5.7in]{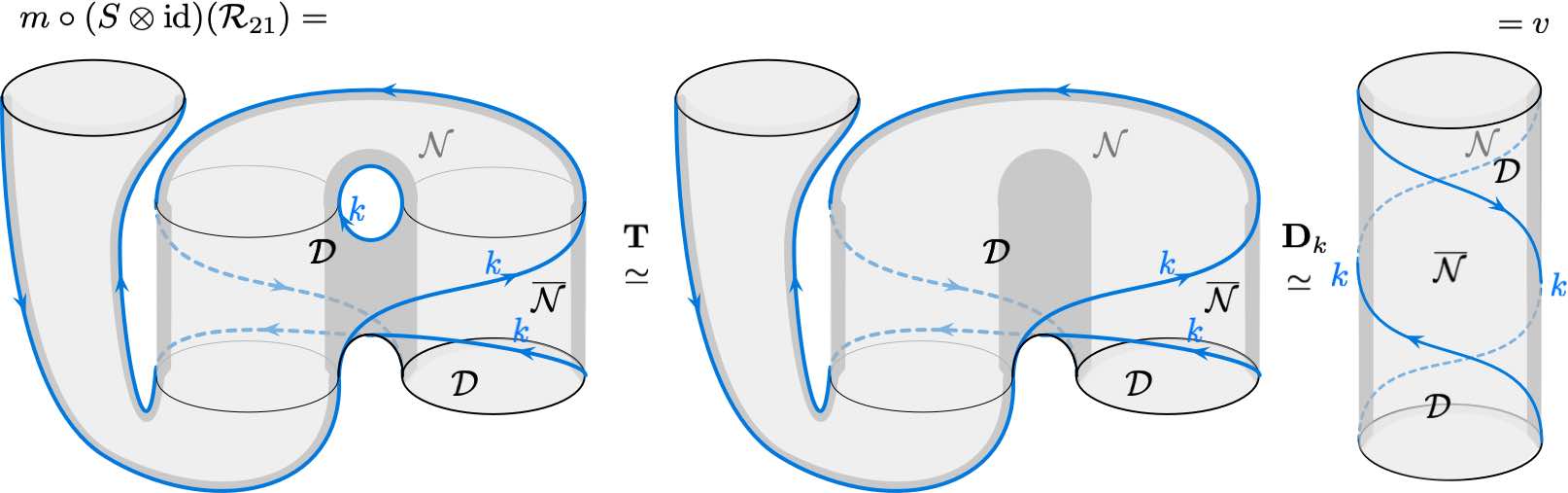}} \label{fig:R-Ribbon}   \ee
In the final step, we need \AD$_k$, dualizability of $k$ as an interface between $\CD$ and $\ol \CN$. Without this, we would still expect to have $v=m\circ(S\otimes \text{id})(\CR_{21})$ in a generalized sense, with $\CR$ defined in a completion of $U\otimes U$.

To finish the argument that $v=u$ given by \eqref{fig:Ribbon} really is a ribbon element, we note that 1) it follows automatically from the axioms of quasitriangular Hopf algebra that $\Delta(u)=(\CR_{21}\CR)^{-1}(u\otimes u)$ \cite[Prop. 8.9.6]{egno}, satisfying the LHS of \eqref{defv}; 2) rotating the cylinder in \eqref{fig:R-Ribbon} by 180$^\circ$ in the plane of the page leaves it invariant, so $S(u)=u$, satisfying the RHS of \eqref{defu}; 3) $u$ is invertible (its inverse is given by the cylinder in \eqref{fig:R-Ribbon} with opposite twist); and 4) $u$ is central, because passing any spark from the top to the bottom of the cylinder in \eqref{fig:R-Ribbon} twists it by 360$^\circ$ (along the axis of the cylinder), which is trivial due to absence of framing anomaly.  (In general, $u\gamma u^{-1} = S^2(\gamma)$ for all $\gamma$ in a quasitriangular Hopf algebra. For us, $S^2=\text{id}$.)

\subsection{Bulk sparks are the double of boundary sparks}
\label{sec:U-double}

So far, we have independently shown that boundary sparks are Hopf-paired Hopf algebras $\CH_\CD,\CH_\CN$ (Section \ref{sec:Hopf-bdy}), and that bulk sparks are a quasitriangular Hopf algebra $U$. We will now argue that bulk sparks are in fact the Drinfeld double formed from the paper $(\CH_\CD,\CH_\CN)$ with their Hopf pairing $h$, establishing Theorem \ref{Thm:double}.

Our argument will be entirely based on topology/QFT. We'll go through all the axioms/properties of a Drinfeld double, from Section \ref{sec:QT-ax}. First we'll show that the associative algebra structure and the Hopf algebra structure in $U$ is the one expected for the double. Then, after a detour into the topology of the ``box gluing,'' we'll show that the R-matrix is given by the inverse of the Hopf pairing.

\subsubsection{The double algebra}
\label{sec:swap}

Let's write $U\simeq \CH_\CD\otimes \CH_\CN$ as a vector space, representing bulk sparks as an ordered product $a\alpha$ of boundary sparks, as shown in \eqref{fig:Ubasic}. 

The first, and most nontrivial, property of the Drinfeld double to check is the second formula in \eqref{double-prod}, namely
\be \alpha \cdot a = \sum (-1)^{\xi(\alpha, a)}h\big(S^{-1}(a^{(1)}),\alpha_{(1)}\big)\; a^{(2)}\alpha_{(2)}\; h\big(a^{(3)},\alpha_{(3)}\big)\,. \label{double-prod2} \ee
This formula captures how a product of elements in a non-standard order can be reversed to match the standard one.
Let's check that it holds for us.

The LHS of \eqref{double-prod2} translates topologically to $\alpha\in \CH_\CN$ placed above $a\in \CH_\CD$:
\be \hspace{-.4in} \raisebox{-.6in}{\includegraphics[width=6.4in]{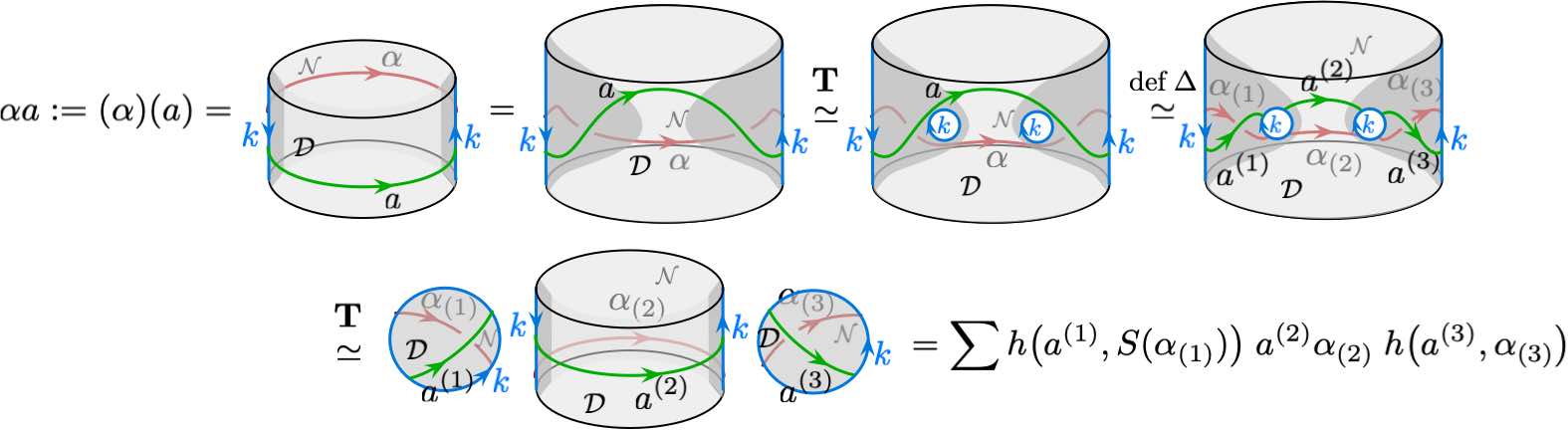}} \label{Dproduct-top} \ee
The order can be almost reversed by pulling the $a$ spark above the $\alpha$ spark, as shown in the first step. Recalling that the sparks are on the boundary of an empty cylinder (with the core drilled out), we also narrow the diameter of the core --- pushing the bulk $\CT$ regions (shaded) closer into the center of the cylinder. In the second step, we use transversality to punch two holes through the cylinder, from front to back, through the bulk $\CT$ regions. In the third step, we use the coproduct twice to reconnect both $a$ and $\alpha$ sparks to the new holes. In fourth step, we use transversality again to split off two solid balls containing the crossings near the sparks' endpoints.

Finally, we may use $S^2=\text{id}$ and properties of the Hopf pairing to rewrite $h\big(a^{(1)},S(\alpha_{(1)})\big) = h\big(a^{(1)},S^{-1}(\alpha_{(1)})\big) = h\big(S(a^{(1)}),\alpha_{(1)}\big) =  h\big(S^{-1}(a^{(1)}),\alpha_{(1)}\big)$. We then precisely recover \eqref{double-prod2}! More so, we have found its topological meaning: the two Hopf pairings that appear are merely correcting for a reversal of the order of endpoints of sparks.

When sparks are fermionic, an sign $\xi(\alpha,a)$ appears in \eqref{Dproduct-top}, matching \eqref{D-sign}. The sign simply accounts for swapping the orders of operators representing sparks, as they would be written linearly in a physical correlation function.

There is a similar formula/manipulation for writing the original product $a\alpha$ in the reverse order. We include it for completeness:
\be \hspace{-.2in} \raisebox{-.3in}{\includegraphics[width=6.4in]{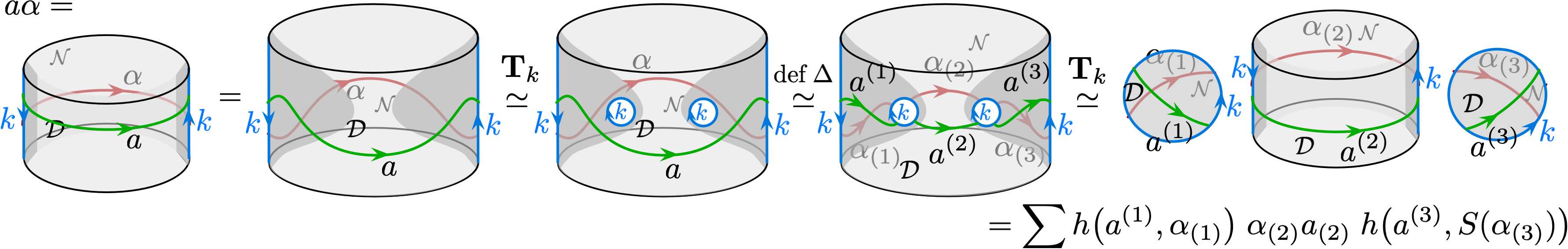}} \ee

We also check that the coproduct, counit, and antipode in $U$ can be written simply in terms of the corresponding operations in $\CH_\CD$ and $\CH_\CN$, as required by \eqref{double-Hopf}. This is remarkably straightforward. For the coproduct, we have:
\be \hspace{-.3in} \raisebox{-.4in}{\includegraphics[width=6.3in]{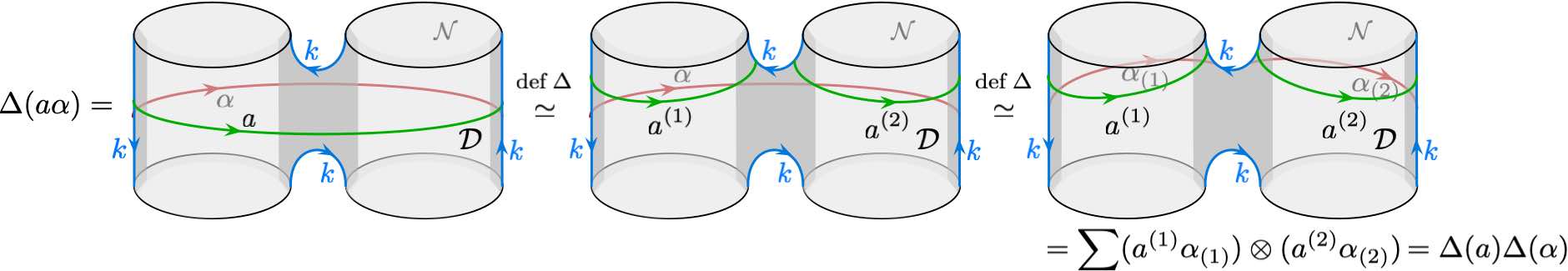}}  \label{double-cop} \ee
where first apply the coproduct to the $\CH_\CD$ spark and then the $\CH_\CN$ spark to get $\Delta(a\alpha)=\Delta(a)\Delta(\alpha)$. For the counit, we separate the sparks vertically and use transversality to split them apart, getting $\varepsilon(a\alpha) = \varepsilon(a)\varepsilon(\alpha)$:
\be\raisebox{-.4in}{\includegraphics[width=4.3in]{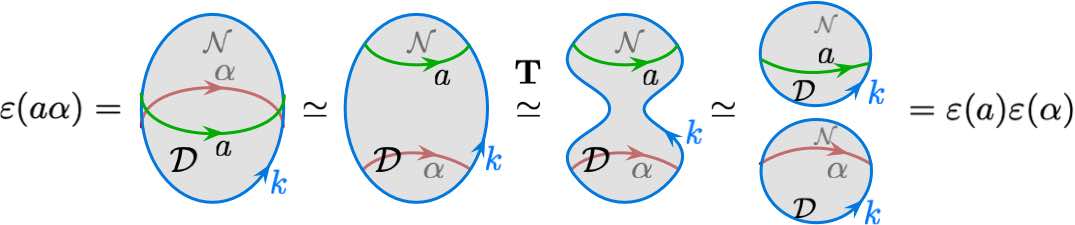}} \label{double-counit} \ee
For the antipode, we observe that rotating a bulk spark by 180$^\circ$ swaps the vertical orders of its $\CH_\CD$ and $\CH_\CN$ factors, and rotates them both, so $S(a\alpha) = S(\alpha)S(a)$:
\be\raisebox{-.4in}{\includegraphics[width=4in]{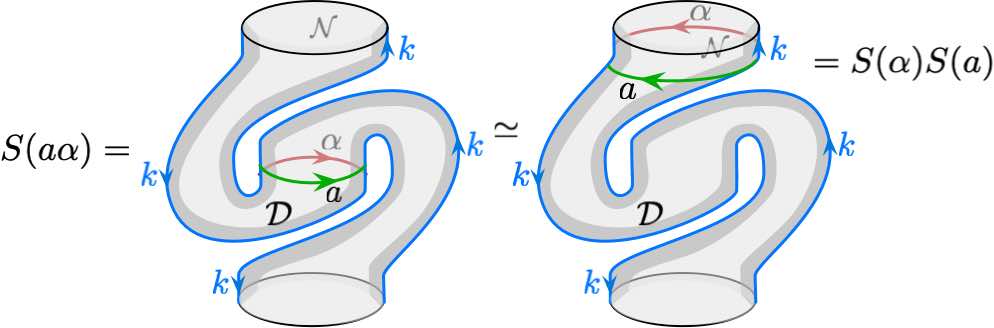}} \label{double-S}  \ee

\subsubsection{The Hopf pairing and the box gluing}
\label{sec:h-box}

Next, we prove a key result about cutting and gluing along rectangles that we will use to relate the R-matrix in $U$ to sparks. For this section, we will initially assume that \AD$_k$ holds: $k$ is dualizable as an interface between $\CN$ and $\ol\CD$ (and $\CD$ and $\ol \CN$), and thus spark algebras are finite dimensional.

Suppose that our topological QFT $\CT$ is placed on a 3-manifold that contains a region of the form (rectangle)$\times\R$, as on the LHS of \eqref{boxglue-sparks}:
\be \raisebox{-.6in}{\includegraphics[width=5.2in]{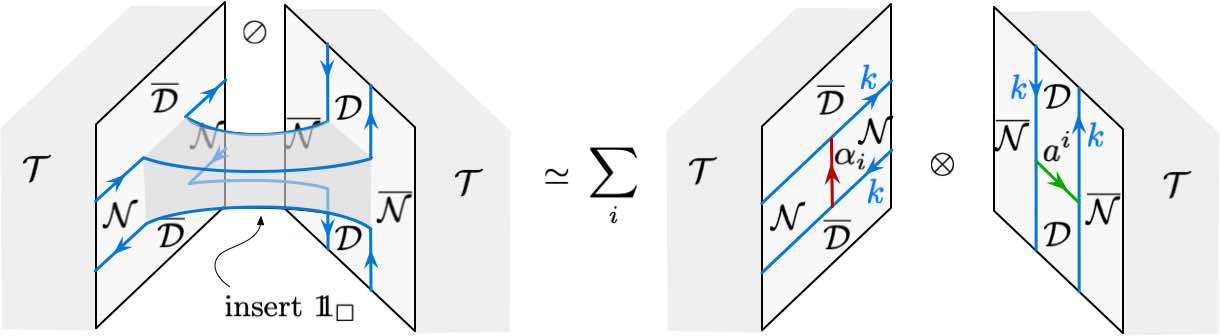}} \label{boxglue-sparks} \ee
If we assume \AD$_k$, the space of states in the tube is finite dimensional, and isomorphic to either $\CH_\CD$ or $\CH_\CN$, up to taking duals. Let's imagine the local direction of time in the tube as propagating from right to left, with incoming states (from the right) identified with $\CH_\CN$. Then we can cut the tube in half by inserting a copy of the identity operator $\id_\square \in \text{End}(\CH_\CN) \simeq \CH_\CN\otimes \CH_\CN^* = \CH_\CN\otimes \CH_\CD$. Choosing any basis $\{\alpha_i\}$ of $\CH_\CN$ and a dual basis $\{a^i\}$ of $\CH_\CD$ such that $h(a^i,\alpha_j)=\delta^i{}_j$, the identity operator looks like
\be \id_\square  = \sum_i \alpha_i \otimes a^i \label{id-res-fin}  \ee
Explicitly, it obeys $\big(\sum_i \alpha_i\otimes a^i\big) \cdot \beta_j = \sum_i\alpha_i\otimes h(a^i,\beta_j) = \beta_j$, as well as $b^j\cdot \big(\sum_i \alpha_i\otimes a^i\big) = \sum_i h(b^j,\alpha_i)\otimes a^i = b^j$. (The latter is the statement that `$\id$' is the identity for states propagating from the right as well.) Inserting the identity operator in the form \eqref{id-res-fin} is equivalent to inserting the corresponding sparks $\alpha_i$ and $a^i$ on the left and right sides of the cut tube, as shown on the RHS of \eqref{boxglue-sparks}.

There is some care involved in choosing the correct orientations of sparks to place on the RHS of \eqref{boxglue-sparks}. The choice is fixed by our convention for the Hopf pairing. The convention in \eqref{boxglue-sparks} ensures that we match the required identities
\begin{subequations}
\be \raisebox{-.5in}{\includegraphics[width=4.2in]{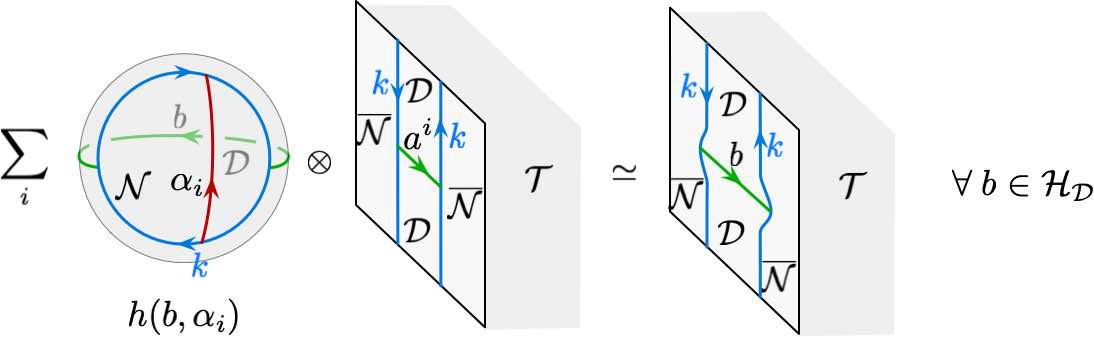}} \ee
\be \raisebox{-.5in}{\includegraphics[width=4.2in]{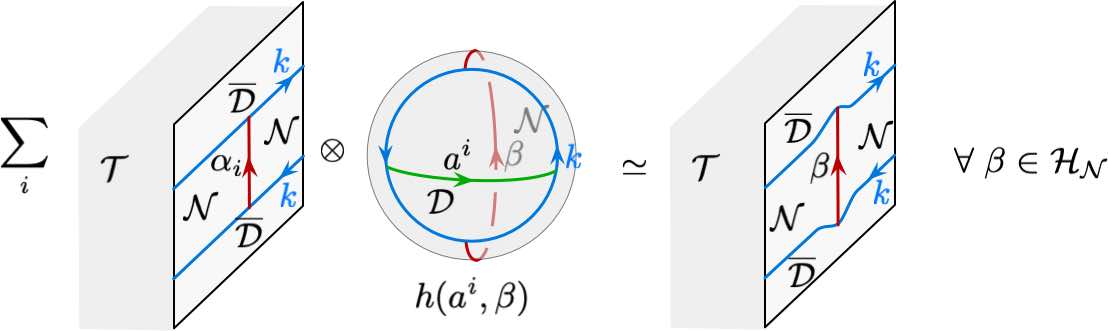}} \ee
\end{subequations}

It is sometimes convenient to work with other bases for the spark algebras. If we choose any basis $\{b^i\}$ for $\CH_\CD$ and $\{\beta^i\}$ for $\CH_\CN$, with $h(b^i,\beta_j)=h^i{}_j$ (a nondegenerate square matrix), then the identity operator instead gets expressed as
\be \id_\square = \sum_{i,j} (h^{-1})^i{}_j\, \beta_i\otimes b^j\,, \label{box-hb} \ee
using the inverse of the Hopf-pairing matrix.

If we relax condition \AD$_k$ and have infinite-dimensional spark algebras, we expect that a generalization of the decomposition \eqref{boxglue-sparks} still exists, in a suitable completion. Namely, we can still break the tube on the LHS of \eqref{boxglue-sparks} by inserting the identity $\id_\square \in \text{End}(\CH_\CN)$. However, we will only have $\text{End}(\CH_\CN)\supset\CH_\CN\otimes \CH_\CD$, and the identity operator $\id_\square$ may not strictly be an element of $\CH_\CN\otimes\CH_\CD$ but rather a completion of this tensor product.

We finally remark that decomposition we have found in \eqref{boxglue-sparks} is the same as the box gluing \AC$_{\rm box}$ we had anticipated back in \eqref{boxglue}, as a consequence of completeness. We have identified the specific \emph{deformation} required for the box gluing, as an insertion of sparks:
\be \raisebox{-.6in}{\includegraphics[width=5in]{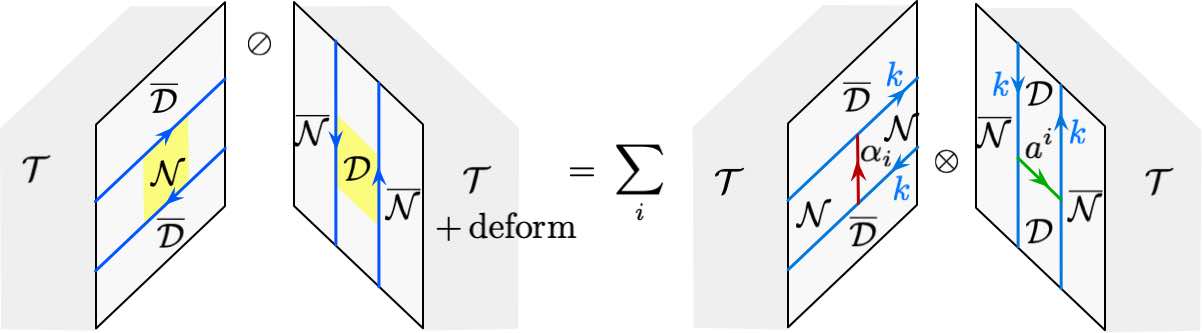}} \label{boxglue-C} \ee
More so, we see that the full power of completeness is not required. Since the box/rectangle is compact, box gluing just amounts to a decomposition of the rectangle state space.

\subsubsection{The R-matrix inverts the Hopf pairing}
\label{sec:R-double}

Finally, we show that our topological R-matrix \eqref{fig:R-def} is in fact given by the canonical formula in the Drinfeld double \eqref{double-R}, as the inverse of the Hopf pairing.

We observe that the 3-manifold \eqref{fig:R-def} defining the R-matrix is obtained by taking two trivial empty cylinders, and gluing them together across a rectangular region:
\be  \raisebox{-.5in}{\includegraphics[width=5.8in]{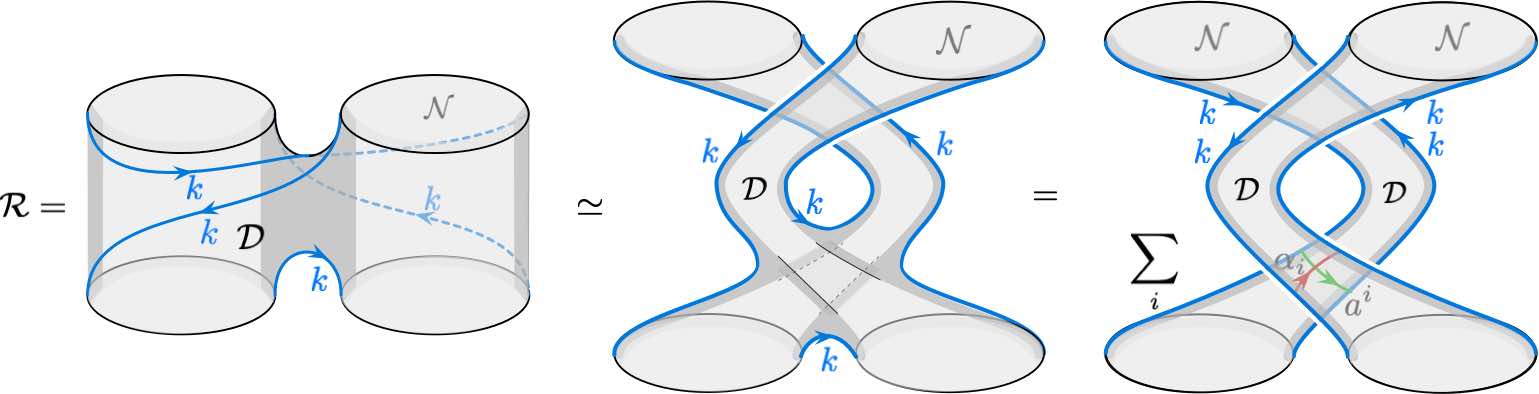}} \hspace{-.0in} \label{fig:R-glue3} \ee
Using the decomposition formula \eqref{boxglue-sparks}, we can break apart the gluing by inserting an identity operator $\id_\square = \sum_i \alpha_i\otimes a^i \in \CH_\CN\otimes \CH_\CD$, as a product of sparks and dual sparks. Splitting apart the two cylinders, and keeping track of orders and orientations of the inserted sparks, we find that the R-matrix becomes
\be  \raisebox{-.5in}{\includegraphics[width=5.2in]{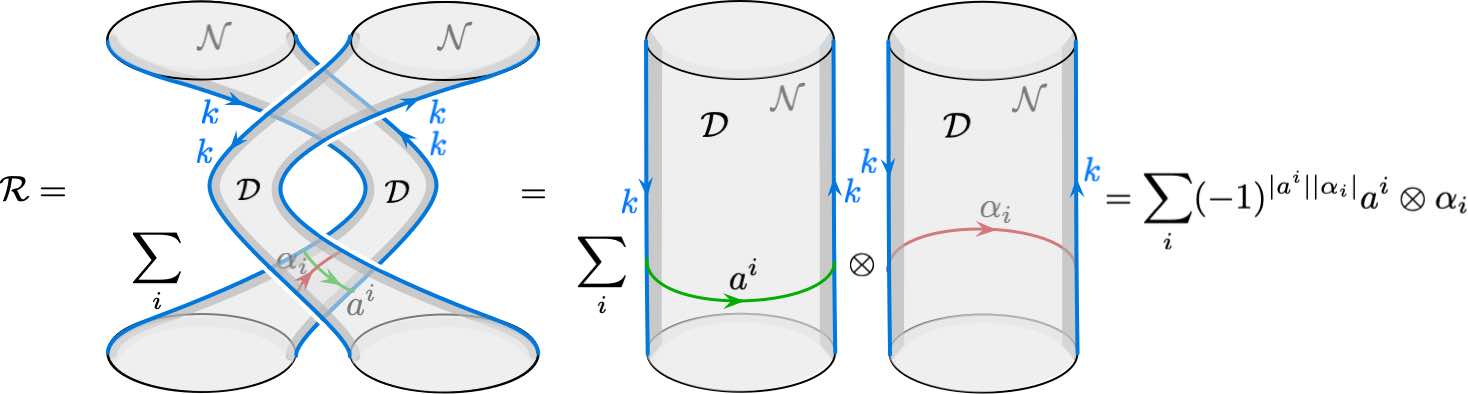}} \hspace{-.0in} \label{fig:R-glue4} \ee
In other words, $\CR$ is the permutation of the resolution of the identity
\begin{align} \label{R-double}  \CR = (\id_\square)_{21} &= \sum_i (-1)^{|a^i||\alpha_i|}a^i\otimes \alpha_i \quad\text{(in dual bases)} \\  &= \sum_{ij}(-1)^{|b^j||\beta_i|} (h^{-1})^i{}_j \,b^j\otimes \beta_i \quad \text{(in any basis)}\,. \notag  \end{align}
The sign $(-1)^{|a^i||\alpha_i|}$ comes from swapping the orders of the operators representing sparks in $\id_\square$, as they would be written linearly in a physical correlation function.
This beautifully reproduces \eqref{double-R}.

As remarked previously, for both the $\CR$ matrix to make sense as an honest element of $\CH_\CD\otimes \CH_\CN$, and for the decomposition \eqref{boxglue-sparks} to work with the identity $\id_\square\in \CH_\CN\otimes\CH_\CD$, we need to assume \AD$_k$, making spark algebras finite dimensional. In general, both $\CR$ and $\id_\square$ will live in completed tensor products.

By considering other box gluings of two crossed cylinders, we can give similar formulas for $\CR^{-1}$, $\CR_{21}$, and $\CR_{21}^{-1}$. They are summarized here:
\begin{subequations}\label{fig:Rglue-alt}
\be \raisebox{-.6in}{\includegraphics[width=5.2in]{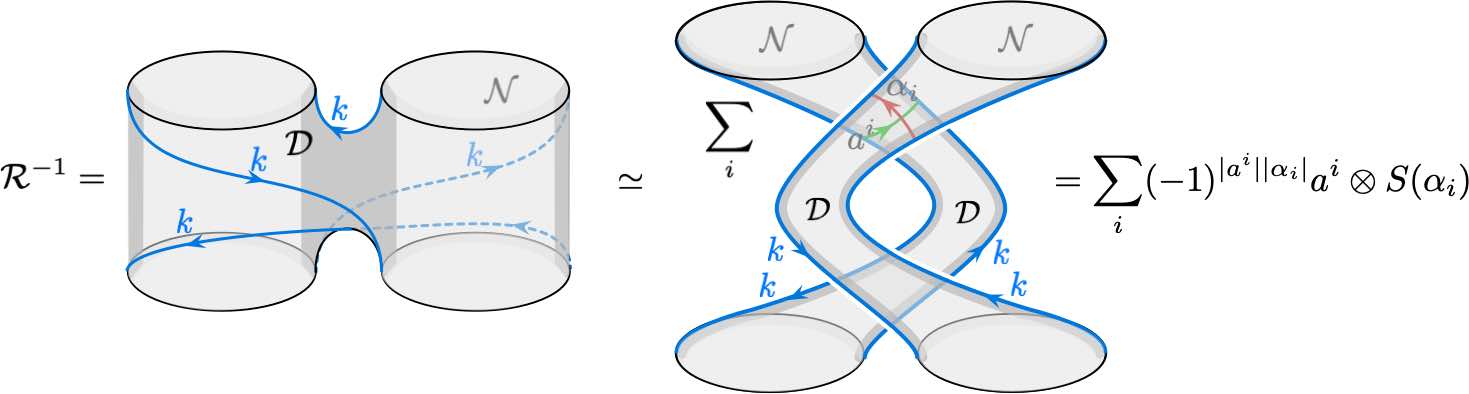}} \ee
\be \raisebox{-.6in}{\includegraphics[width=4.6in]{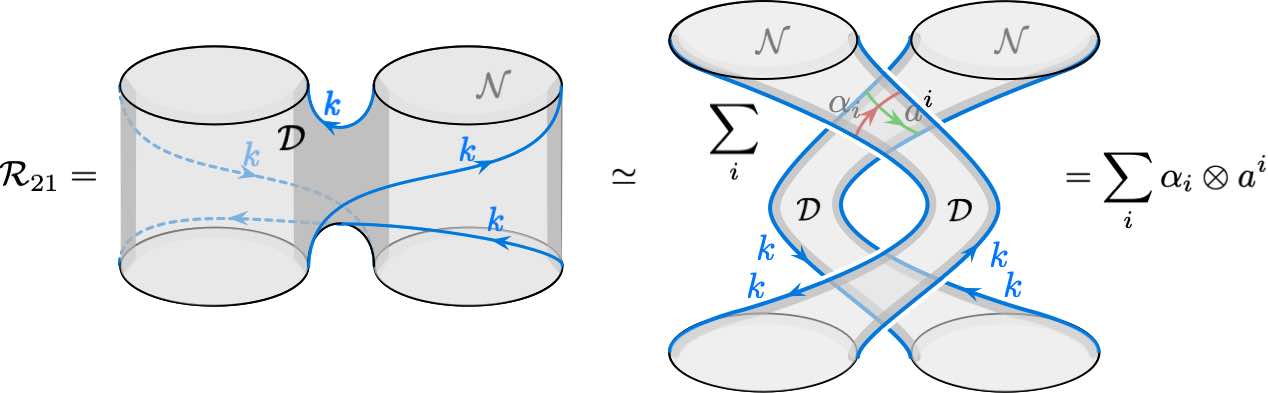}} \ee
\be \raisebox{-.6in}{\includegraphics[width=4.6in]{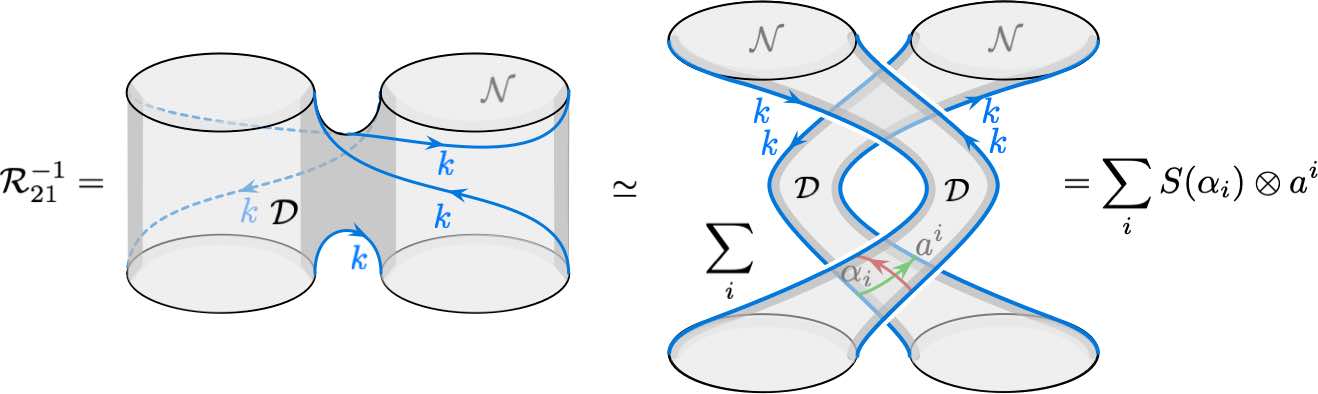}} \ee
\end{subequations}
The antipode makes an appearance in these formulas to account for the reversed orientation of some of the sparks that appear, compared to their standard (right-pointing) orientation.

\section{Finite-group gauge theory}
\label{sec:DW}

We'll now apply spark-algebra techniques to find the Hopf algebras representing line operators in a variety of 3d topological QFT's. Our main examples of interest are topological twists of 3d $\CN=4$ gauge theories. As a warmup, though, we'll first discuss a much simpler theory: pure gauge theory with finite gauge group $G$, \emph{a.k.a.} Dijkgraaf-Witten theory \cite{DW}. 
We will illustrate the constructions of Sections \ref{sec:linefunctor}--\ref{sec:bulk} by rederiving \emph{well-known} structures in Dijkgraaf-Witten theory, in the language of sparks.

Dijkgraaf-Witten (DW) theory has been rigorously defined as a fully extended 3d TQFT \cite{FreedQuinn,Freed-DW,FHLT}.
All observables in Dijkgraaf-Witten theory are explicitly calculable, and closely tied to the representation theory of finite groups.
 Tannakian ideas were applied early on to analyze categories of line operators \cite{RPD-quasiHopf,Freed-DW}, though with a slightly different setup than ours, constructing quasi-fiber functors from a single boundary condition. DW theory and its generalizations have played a central role in the rapid recent developments in understanding the structure of global symmetries in QFT, \cf\ \cite{GKSW,FMT}, and we will frequently use the language of global symmetries in our analysis.

DW theory in three dimensions depends on a choice of finite group $G$ and a 3-cocycle $\omega\in C^3(G,U(1))$, a discrete version of a Chern-Simons term. The theory always admits a topological Dirichlet boundary condition $\CD$, on which gauge bundles are trivialized. The natural candidate for a transverse boundary condition is Neumann $\CN$, on which gauge bundles are freely summed over. However, Neumann is only gauge invariant when the 3-cocycle is trivial. Thus, to apply our formalism, we will assume
\be \omega  = 1\,. \ee

For the remainder of this section, we fix an arbitrary finite group $G$ and denote
\begin{itemize}
\item[$\CT$:] Bulk DW theory with group $G$ and trivial $\omega$, restricted to orientable 3-manifolds.

Recall that the path integral of this theory is a sum over all $G$-bundles on a given manifold.
\item[$\CD$:] Dirichlet boundary condition, fixing a trivialization of $G$ bundles at the boundary, and thus breaking $G$ gauge symmetry to a global $G_\pd$ symmetry there.

(Locally, the trivialization of a $G$ bundle depends on a choice of $g\in G$. The boundary global symmetry $G_\pd$ acts on these choices.)

\item[$\CN$:] Neumann boundary condition, which in DW theory is essentially a free boundary condition: the path integral freely sums over $G$ bundles at the boundary.
\end{itemize}

We'll argue in Section \ref{sec:DW-setup} that DW theory satisfies every single axiom of Tannakian QFT from Section~\ref{sec:setup}. Moreover, at a technical level, since categories of line operators in DW theory are all finite, abelian, and \emph{semisimple}, it is automatic that tensor products are exact, and that fiber functors are exact and continuous. Thus, we expect spark algebras to be finite dimensional and for all from Sections \ref{sec:linefunctor}--\ref{sec:bulk} to hold. 

In Section \ref{sec:DW-spark} we'll explain how to derive spark algebras and their Hopf structures, Hopf pairing, R-matrix, and ribbon element directly from intuitive physical constructions. We'll find that $\CH_\CD$ is the group algebra of $G$, while $\CH_\CN$ is the algebra of functions on $G$, with their classic (dual) Hopf structures:
\be  \CH_\CD \simeq \C G\,,\qquad \CH_\CN\simeq \CO(G)\,,\qquad U \simeq \C G\otimes \CO(G) \quad \Big({\textstyle \CR = \sum_g g\otimes \delta_g}\Big)\,.\hspace{-.3in} \ee
In Section \ref{sec:DW-cats} we'll then look at the module categories of $\CH_\CD,\CH_\CN,U$, recovering the well-known categories of line operators in DW theory. We'll also spell out what our fiber functors were, and how they are represented by the special algebra objects $\kDk,\kNk,\CDCN$.

In Section \ref{sec:DW-C}, we revisit completeness, and describe explicitly how regions on opposing $\CD$ and $\CN$ boundaries may be glued together by means of ``condensation'' --- in this case, inserting coupled webs of defects, formed out of both $\kDk$ and $\kNk$.

\subsection{DW theory as a Tannakian QFT}
\label{sec:DW-setup}

In DW theory, the bulk theory and boundary conditions are all fully dualizable and orientable, satisfying \AD\ of Section \ref{sec:top} and more. Moreover, line operators are well known to form finite, abelian categories, where all objects are dualizable, so $\CC=\CC^{fd}$ for all categories and \AD$_{\rm line}$ (Section \ref{sec:lines}) holds trivially.

Let's take a look at transversality.  If we form a sandwich $\CD\circ \CN$, then we expect to get a trivial 2d theory (so transversality holds) because there will be no degrees of freedom left: the trivialization on $\CD$ leaves nothing to sum over in the path integral.

A more careful analysis could be done as follows. Note that the $\CD$ boundary condition can be obtained starting from an $\ol\CN$ boundary condition by coupling to $G$-valued fields on the boundary, explicitly implementing the choice of trivialization. Compactification on a $\ol\CN\circ \CN$ sandwich produces 2d DW theory with gauge group $G$. Therefore, compactification on a $\CD\circ\CN$ sandwich produces a 2d theory that's a topological sigma-model to $G$ (due to the extra fields on the $\CD$ boundary), coupled to the DW theory:
\be \raisebox{-.6in}{\includegraphics[width=4.3in]{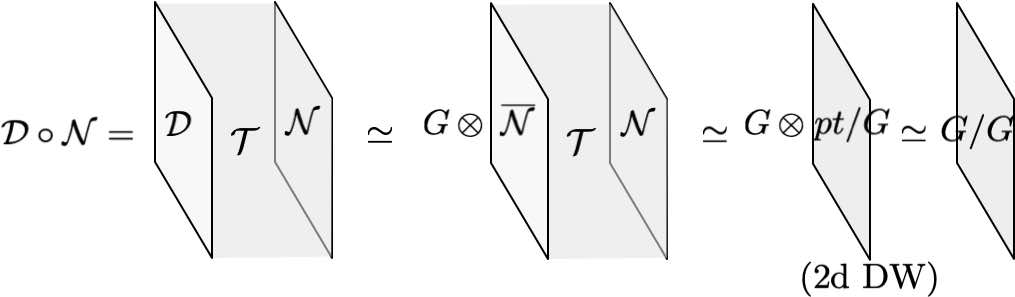}} \label{DW-sand} \ee 
In other words $\CD\circ\CN$ is a 2d topological gauged sigma-model (a finite homotopy theory) with target $G/G$. Here the gauge group is acting via its right action; but since the right action of $G$ on $G$ is free, one can just perform the the gauge quotient to find that $\CD\circ\CN$ should be \emph{equivalent} to a 2d sigma-model to a point $G/G=pt$. The sigma-model to a point is the trivial 2d TQFT ``id$_\oslash$''.

The isomorphism between the 2d $G/G$ theory and the trivial $pt$ theory can be implemented by an interface $k$ --- the transversality interface from \eqref{trans-loc}. From this perspective, $k$ is a boundary condition for the $G/G$ theory. Explicitly, one can check that it is given by a Neumann boundary condition: both the 2d $G$ ``matter fields'' and the $G$ ``gauge fields'' are unconstrained at $k$.

Another perspective comes from considering junctions between $\CN$ and $\ol\CD$, as in \eqref{k-bdy}. The category of such junctions in DW theory can be computed by 1) taking the link of the interface, which is a semicircle ``$C$'' with one end on $\CN$ and one end on $\CD$ as in \eqref{k-bdy-link}; 2) finding the moduli stack $\text{Bun}_G(C)$ of $G$-bundles on this semicircle; and 3) taking the category of finite-dimensional sheaves on the moduli stack:
\be  \raisebox{-.4in}{\includegraphics[width=3.3in]{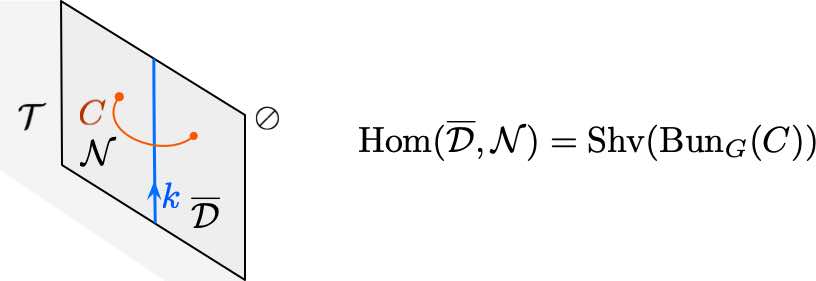}}  \label{k-bdy-link}\,. \ee 
The stack of $G$-bundles on $C$ is $\text{Bun}_G(C) \simeq G\backslash G$, since there are $G$ worth of choices of trivialization at the $\ol\CD$ end, and the gauge group on the rest of $C$ acts on this by quotienting on the left. This gives
\be \text{Bun}_G(C) = G\backslash G = pt\qquad \Rightarrow\qquad \text{Hom}(\ol\CD,\CN) \simeq \text{Shv}(pt) =\text{Vect}^{fd} \,. \ee
Then $k\in \text{Hom}(\CN,\ol\CD)$ is the image of $\C\in \text{Vect}^{fd}$ under the equivalence on the RHS. It is a dualizable interface --- like all interfaces in DW theory --- so condition \AD$_k$ is satisfied.

Finally, completeness of the $(\CD,\CN)$ pair follows heuristically from the fact that one can couple $\CN$ to $\CD$ by using the $G$ gauge symmetry on $\CN$ to gauge the $G_\pd$ global symmetry on $\CD$:
\be \raisebox{-.5in}{\includegraphics[width=2.7in]{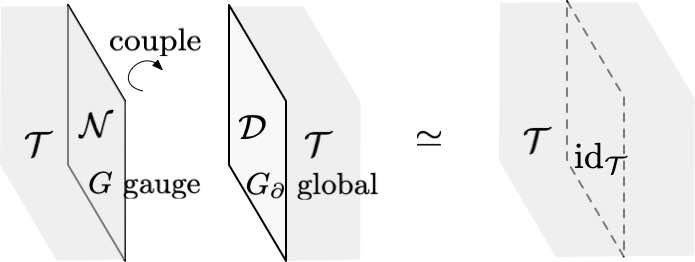}} \ee
The result is a theory that's been smoothly glued together.
We'll revisit completeness in Section \ref{sec:DW-C}, explaining how the gauging operation can be implemented explicitly by inserting coupled webs of defects on both $\CN$ and $\CD$.

\subsection{Spark algebras}
\label{sec:DW-spark}

Now that we've verified the assumptions of ``Tannakian QFT,'' let's do a direct computation of the spark algebras, and their various structures.

\subsubsection{Vector spaces}
\label{sec:DW-space}

We know, since \AD$_k$ holds, that spark algebras will be finite-dimensional (Section \ref{sec:spark-finite}) and that $\CH_\CD\simeq \CH_\CN^*$ on the nose. Let's begin by computing these vector spaces, as state spaces on rectangles \eqref{Rectangles}.

The space of states on a surface $\Sigma$ in DW theory is given by functions on the space of $G$-bundles, $\text{States}(\Sigma) = \C[\text{Bun}_G(\Sigma)]$. In the case of a rectangle with $\CD$ (or $\ol \CD$) on two sides, and $\CN$ (or $\ol\CN$) on the other two sides, $G$-bundles are uniquely determined by the holonomy along the path from one $\CD$ side to the other. The holonomy makes sense, and is gauge invariant, because bundles are trivialized on both of these sides. Thus,
\be \text{Bun}_G(\text{rectangle}) = \{\text{holonomy from $\CD$ to $\ol \CD$}\} = \{g\in G\}  \ee
and 
\be \CH_\CD \simeq \; \raisebox{-.3in}{\includegraphics[width=1.55in]{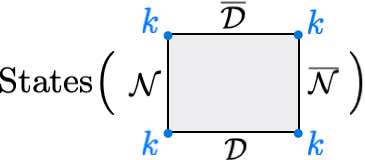}}  \simeq \C\langle G\rangle\,. \label{DW-HD-G} \ee
is isomorphic, as a vector space, to the space of functions on $G$.
The same calculation gives
\be \CH_\CN \simeq \C\langle G\rangle\,. \label{DW-HN-G} \ee

The bulk spark algebra has an underlying vector space
\begin{align} U &\simeq \; \raisebox{-.5in}{\includegraphics[width=1.7in]{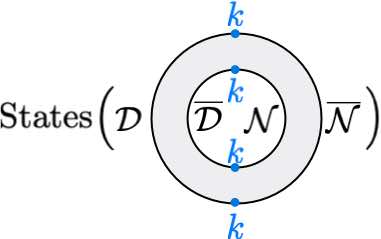}} \notag \\
 & =\C\big\langle \text{\{$G$-bundles on an annulus, trivialized on the two $\CD$ segments\}}\big\rangle 
 \notag \\
 & \simeq \C\langle G\times G \rangle\,.
\end{align}
Abstractly, this is simply a space of dimension $|G|\times |G|$. The first factor of $G$ can be thought of as the holonomy from one $\CD$ edge to the other; while the second factor measures the holonomy around a closed loop starting at one $\CD$ edge and winding once around the annulus.

\subsubsection{$\CH_\CD$ as a Hopf algebra}
\label{sec:DW-HD}

Next, let's determine $\CH_\CD$ as a Hopf algebra. There are multiple ways to do this, at different levels of mathematical/TQFT formalism. Since we would like to emphasize QFT computations of operator algebras, and generalize them in later sections, we will use somewhat more physical language, intuition from the study of symmetries.  Alternatively, can do a direct mathematical computation using the cobordisms of Section \ref{sec:Hopf-cob}.

As we've already noted, a Dirichlet boundary condition in DW theory has global $G_\pd$ symmetry. Thus it carries symmetry defects $V_g$ labelled by elements $g\in G$:
\be \raisebox{-.4in}{\includegraphics[width=.9in]{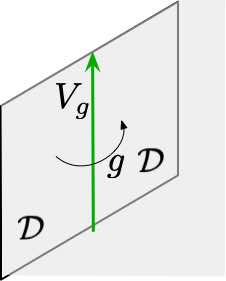}} \ee
As one crosses $V_g$, the boundary trivialization changes by $g$. 
If we place $V_g$ horizontally on a strip of $\CD$ b.c. sandwiched by $k$ interfaces, we get a spark that we'll simply call `$g$'
\be  \raisebox{-.35in}{\includegraphics[width=4.6in]{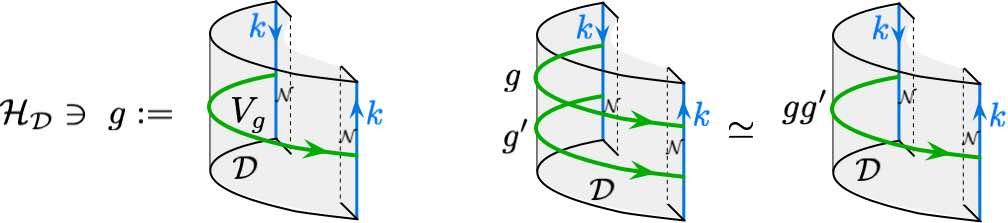}} \ee
Moreover, symmetry defects like this must span the entire space of sparks, since there are $|G|$ independent defects and this matches the dimension computed in \eqref{DW-HD-G}.

The associative product of sparks, coming from vertical stacking of symmetry defects, is just group multiplication, $m(g,g')=gg'$.  Thus, as an associative algebra,
\be \CH_\CD \simeq \C G = \C\langle g\rangle_{g\in G} \label{DW-HD-alg} \ee
is the group algebra of $G$. The unit in $\CH_\CD$ corresponds to the group identity, $1=e$.

To get the coproduct, we insert a symmetry defect along the `waist' of a pair of pants:
\be \Delta(g)=\;\; \raisebox{-.25in}{\includegraphics[width=3.3in]{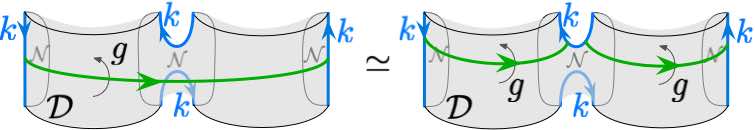}}\;\;= g\otimes g\,. \ee
Changing the boundary trivialization by $g$ along the waist is equivalent to changing it by $g$ simultaneously along the two cuffs. Therefore, as shown, $\Delta(g) = g\otimes g$ for all $g\in G$.

The counit $\varepsilon(g)$ is given by the partition function of a half-ball with a defect
\be \raisebox{-.2in}{\includegraphics[width=1.6in]{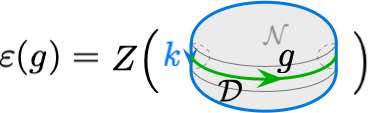}} \;= 1\quad \forall\; g\in G\,. \label{DW-counit} \ee
Thus partition function is computed by summing all the $G$-bundles on the ball, with the inserted change of trivialization on the $\CD$ side (and dividing by the order of the automorphism group of each bundle). However, for any $g$, there is a \emph{unique} $G$-bundle on the ball \eqref{DW-counit}, with trivial automorphism group. Thus $\varepsilon(g)\equiv 1$.

Finally, the antipode is given by rotating a spark by $180^\circ$, changing its orientation. This sends a $g$ symmetry defect to a $g^{-1}$ defect, whence $S(g)=g^{-1}$. In summary,
\be \CH_\CD= \C G\,:\quad 1=e\,,\qquad \begin{array}{c} \Delta(g) = g\otimes g \\[.1cm] \varepsilon(g) \equiv 1 \end{array}\,,\qquad S(g)=g^{-1}\,.  \label{DW-HD-Hopf} \ee
We've found the classic Hopf structure on the group algebra of a finite group.

\subsubsection{$\CH_\CN$ as a Hopf algebra}
\label{sec:DW-HN}

In a similar way, to construct $\CH_\CN$ we use the observation that the Neumann boundary condition supports Wilson lines $W_\rho$, labelled by finite-dimensional representations $\rho$ of~$G$.

Closed Wilson lines inserted in the path integral are observables that measure the trace of the holonomy of the gauge bundle, in representation $\rho$. On a strip of $\CN$ bounded by $k$ on either side, we can produce a gauge-invariant spark by stretching a Wilson line $W_\rho$ from one boundary to the other, as well as choosing a vector $v\in \rho$ at its starting point and a covector $w\in \rho^*$ at its endpoint:
\be   \raisebox{-.4in}{\includegraphics[width=2.8in]{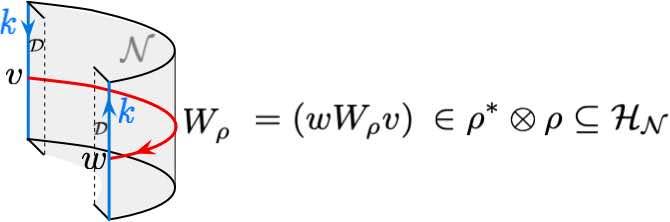}} \label{DW-Nsparks} \ee
Let $\varphi_\rho:G\to \text{End}(\rho)$ denote the representation map. The insertion of this spark in the gauge-theory path integral measures the quantity $w\cdot\varphi_\rho(g)v$
where $g\in G$ is the holonomy of a given $G$-bundle from one $k$ boundary to the other. Put differently, the insertion of the framed $W_\rho$ segment computes a particular matrix element~of~$\varphi_\rho(g)$.

Altogether, the linearly independent sparks that we can construct from Wilson lines span the space
\be \bigoplus_{\rho\,\in\,\text{irrep}(G)} \rho^*\otimes \rho\,. \label{Wrr} \ee 
It is a basic result of representation theory that \eqref{Wrr} is isomorphic to the space of \emph{complex-valued functions} on $G$, which we denote $\CO(G)$. In particular, under the left action of $G$ on itself, functions on $G$ form what is known as the regular representation, which decomposes into irreducibles $\rho$ with multiplicities $\rho^*$, just as in \eqref{Wrr}.

Since $\text{dim}_\C \,\CO(G) = |G|$, we compare with \eqref{DW-HN-G} to verify that we've recovered all elements of $\CH_\CN$. Let us choose a basis of $\CO(G)$ given by delta functions $\delta_g$, so that
\be \CH_\CN \simeq \CO(G) = \C\langle \delta_g\rangle_{g\in G}\,. \ee
The spark $\delta_g$ represents a projection: it is a superposition of framed Wilson lines with the property that its insertion in the path integral equals $1$ for any bundle whose holonomy across the $\CN$ strip is $g$, and equals $0$ otherwise.

Now consider the Hopf-algebra operations. Since $\delta_g$ is a projection to bundles with holonomy $g$ across the strip, we easily have that $\forall\;g,g'$
\be \delta_g \delta_{g'} = \delta_{g'} \delta_g= \begin{cases} \delta_g & g=g' \\ 0 & g\neq g' \end{cases}\,. \ee
In particular, $\CH_\CN$ is commutative. The identity, given by the trivial Wilson line, is represented in the delta-function basis by $1= \sum_{g\in G} \delta_g\,.$

To describe the coproduct, we note that a Wilson line $W_\rho$ can be broken by inserting a complete set of states in $\rho$. In other words, given a basis $v^i\in\rho$ and a dual basis $w_i\in \rho^*$, we have
\be \Delta(w W_\rho v) = \raisebox{-.22in}{\includegraphics[width=3.5in]{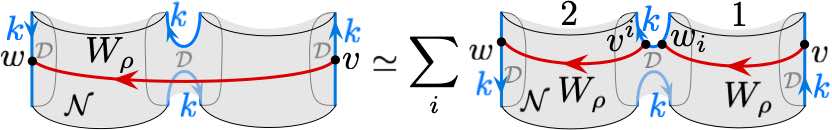}}  = \sum_i  (w_i W_\rho v)\otimes (w W_\rho v^i)\,.  \ee
Note the order in the two factors of the coproduct, matching our reversed conventions for $\CH_\CN$. Alternatively, in a delta-function basis, we observe that a projection $\delta_g$ along the waist of a pair of pants is equivalent to the sum of all products of projections $\delta_{k'},\delta_{k}$ along the cuffs, for $k'k=g$:
\be \Delta(\delta_g) = \raisebox{-.37in}{\includegraphics[width=3.5in]{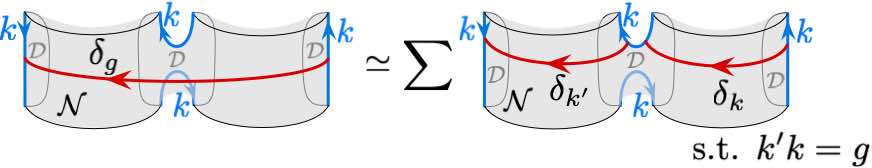}} \hspace{-.1in} =  \sum_{k,k'\;\text{s.t.}\;k'k=g} \delta_k\otimes \delta_{k'}\,. \label{HN-D-DW} \ee
This coproduct is known as convolution in $\CO(G)$, induced from multiplication on $G$.

Finally, the counit is the partition function
\be  \raisebox{-.2in}{\includegraphics[width=1.6in]{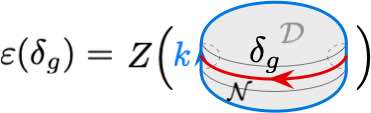}}   = \begin{cases} 1 & g = e \\ 0 & g\neq e\,, \end{cases} \ee
which takes this value because the trivialization on $\CD$ side forces $G$-bundles to have trivial holonomy all along the $\CN$ side as well. The antipode corresponds to swapping orientation, and thus inverting holonomy, so $S(\delta_g) = \delta_{g^{-1}}$. In summary:
\be \CH_\CN= \CO(G)\,:\quad
\begin{array}{c} \delta_g\delta_{g'} = \delta_{g=g'}\delta_g \\[.1cm] 1=\delta_e \end{array}\,,\quad \begin{array}{c} \Delta(\delta_g) = \sum_{k'k=g}\delta_k\otimes \delta_{k'} \\[.1cm] \varepsilon(\delta_g) =\delta_{g=e} \end{array}\,,\quad S(\delta_g)=\delta_{g^{-1}}\,.  \label{DW-HN-Hopf} \ee
This is the classic Hopf structure on the commutative algebra of functions on $G$.

\subsubsection{Hopf pairing and box gluing}
\label{sec:DW-box}

In preparation for deriving the bulk spark algebra $U$, we'll also spell out the Hopf pairing between $\CH_\CD$ and $\CH_\CN$, and check that the box gluing can be implemented by its inverse.

Consider the Hopf pairing $h:\CH_\CD\otimes \CH_\C\to \C$, defined as in Section \ref{sec:Hopf-pair} by inserting sparks on two $\CD,\CN$ hemispheres of a solid ball. If we insert a $g$ spark on the $\CD$ hemisphere, we force the holonomy of the $G$ bundle to jump by $g$ across the $\CN$ hemisphere as well. Thus, additionally inserting $\delta_k$ in the $\CN$ hemisphere we get
\be \raisebox{-.44in}{\includegraphics[width=1.6in]{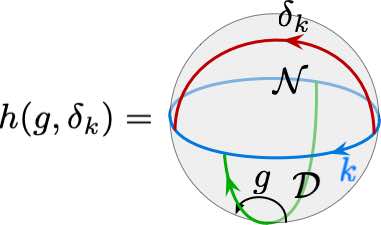}} \;\;= \begin{cases} 1 & g=k \\ 0 & g\neq k\,, \end{cases}   \label{DW-Hopf} \ee 
showing that $\{g\}_{g\in G}$ and $\{\delta_g\}_{g\in G}$ are dual bases of $\CH_\CD$ and $\CH_\CN$.  Indeed, it is well known that the group algebra $\C G$ and functions $\CO(G)$ are Hopf-dual Hopf algebras under the pairing \eqref{DW-Hopf}.

As for the box gluing, we now expect from Section \ref{sec:h-box} that 
\be \raisebox{-.6in}{\includegraphics[width=5in]{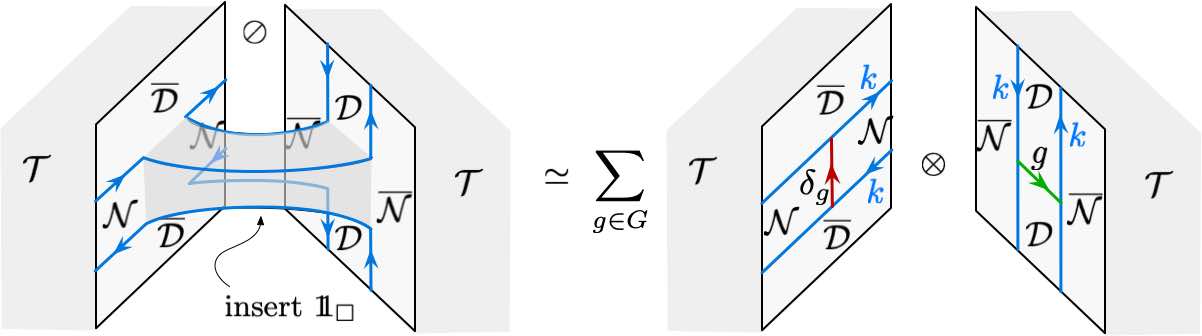}} \label{boxglue-sparks-DW} \ee
Let's interpret this physically. To glue a small rectangle on the $\CN$ strip to a small rectangle on the $\CD$ strip, we
\begin{enumerate}
\item Relax the trivialization of bundles on $\CD$, by inserting all possible $g$ sparks there. So now all possible bundles along the $\CD$ strip will be included in the path integral.
\item Match the bundles on $\CD$ with those that appear on the opposing $\CN$ strip, by inserting projection operators $\delta_g$ (for each $g$ insertion on $\CD$). Thus the path integral doesn't sum independently over bundles on the $\CN$ strip and bundles on the $\CD$ strip, but rather sums just once over bundles in the rectangular region that we want to identify across the two sides.
\end{enumerate}
It's in this way that the insertion of $\id_\square = \sum_{g\in G} \delta_g\otimes g$ implements the gluing.

\subsubsection{Bulk sparks}
\label{sec:DW-U}

We now follow Section \ref{sec:bulk} to determine the bulk spark algebra $U$. Transversality guarantees that, as a vector space, $U\simeq \CH_\CD\otimes \CH_\CN$. Then the constructions in Section \ref{sec:U-double} (Theorem \ref{Thm:double}), which depend only on the existence of the box gluing, together with transversality and topological invariance, identify $U$ as the Drinfeld double of $\CH_\CD$ and $\CH_\CN$ under the Hopf pairing.

Explicitly, $U$ has basis
\be U = \C\langle g\delta_k\rangle_{g,k\in G}\,, \ee
with multiplication on each factor inherited from $\CH_\CD,\CH_\CN$, along with
\be g\delta_k = \delta_{gkg^{-1}}g \label{DW-U-prod} \ee
due to \eqref{double-prod2}. Physically, this comes from the fact that the framings $v,w$ at the endpoints of sparks on $\CN$ transform nontrivially under the global $G_\pd$ symmetry on $\CD$: 
\be \raisebox{-.4in}{\includegraphics[width=5.7in]{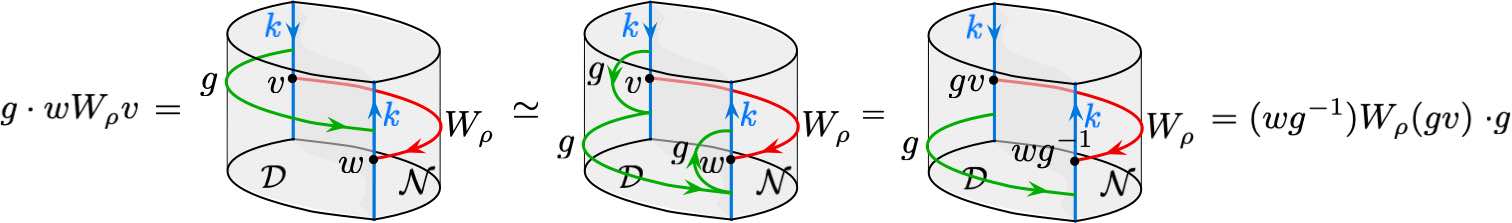}} \label{DW-U-comm} \ee
If we think of elements of $\CH_\CN$ as functions on $G$, then they get conjugated when a spark $g\in \CH_\CD$ is passed by them.

The coproduct and counit in $U$ are inherited directly from $\CH_\CD,\CH_\CN$. Now, however, there is an R-matrix induced from the gluing deformation \eqref{boxglue-sparks-DW}, and corresponding ribbon element
\be \CR = \sum_{g\in G} g\otimes \delta_g \; \in U\otimes U\,,\qquad  v = \sum_{g\in G} \delta_{g^{-1}} g \; \in U\,. \ee

\subsection{Categories and fiber functors}
\label{sec:DW-cats}

Assuming completeness works the way we claim, Theorem \ref{Thm:faithful} now guarantees that there are equivalences of tensor categories (ribbon in the case of $\CC_\CT$), 
\be \wt\CF_\CN: \CC_\CN \overset\sim\to \CH_\CD\text{-mod}^{fd}\,,\qquad  \wt\CF_\CD: \CC_\CD \overset\sim\to \CH_\CN\text{-mod}^{fd}\,,\qquad
   \wt\CF_\CT:\CC_\CT \overset\sim\to U\text{-mod}^{fd}\,. \label{DW-cats}\ee
(Note that we're just using finite-dimensional modules, in order to match the standard rigid categories of DW theory.) 
Let's spell this out, and check that the result is reasonable.

In DW theory, the category of lines on $\CN$ is $\CC_\CN=\text{Rep}(G)$, the category of finite-dimensional representations of $G$. It can be computed directly just as in \eqref{k-bdy-link}, as sheaves on the stack of $G$-bundles on a semicircle ``$C$'' that links a line on $\CN$:
\be   \raisebox{-.4in}{\includegraphics[width=1in]{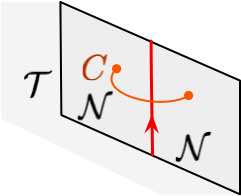}} \quad \CC_\CN = \text{Shv}(\text{Bun}_G(C)) = \text{Shv}(pt/G) = \text{Shv}^G(pt) = \text{Rep}(G) \label{DW-sandN}   \ee
The stack of bundles is a point (all bundles on $C$ with $\CN$ b.c. on the two ends are trivial), with a residual $G$ gauge action, whence $pt/G$. Sheaves on a point with a $G$ action is precisely $\text{Rep}(G)$. This agrees with the module category of the group algebra
\be \CH_\CD\text{-mod}^{fd} = \CG\text{-mod}^{fd} = \text{Rep}(G)\,.\ee
The Hopf structure on $\CH_\CD$ agrees with the standard tensor product of representations in $\text{Rep}(G)$.

Similarly, the category of lines on $\CD$ is computed directly from the stack of $G$-bundles on a semicircle with $\CD$ b.c. on both ends. Now this space is $G$ itself, measuring the holonomy of the bundle from one end to the other, so
\be   \raisebox{-.4in}{\includegraphics[width=1in]{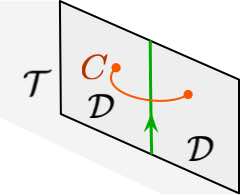}} \quad  \CC_\CD = \text{Shv}(\text{Bun}_G(C)) = \text{Shv}(G) \simeq \text{Vect}_G^{fd}  \label{DW-sandD}  \ee
The resulting category, sheaves on the discrete space $G$, is usually denoted $\text{Vect}_G$, and known as $G$-graded vector spaces. This again agrees with modules for the algebra of functions:
\be \CH_\CN\text{-mod}^{fd} = \CO(G)\text{-mod}^{fd} =\text{Vect}_G^{fd} \,.\ee

Finally the category of lines in the bulk $\CC_\CT$ is computed as sheaves on the stack of bundles on a circle. This stack is a copy of $G$ (the holonomy around the circle), modulo $G$ gauge transformations acting in the adjoint representation. Thus:
\be  \raisebox{-.3in}{\includegraphics[width=.6in]{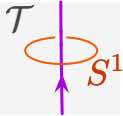}} \qquad  \CC_\CT = \text{Shv}(\text{Bun}_G(S^1)) = \text{Shv}(G/G_{ad}) = \text{Shv}^{G_{ad}}(G) \label{DW-bulkcat}  \ee
This agrees perfectly with modules for $U$, and the R-matrix of $U$ makes the braiding in $\CC_\CT$ completely explicit. It's also well known that the bulk category is the Drinfeld center of either boundary one, in line with Thm. \ref{Thm:Rep}:
\be \CZ_{\rm Drin}(\text{Shv}(G)) \simeq  \text{Shv}(G/G_{ad}) \simeq   \CZ_{\rm Drin}(\text{Shv}(pt/G))\,. \ee

\subsubsection{Algebra objects}
\label{sec:DW-alg}

We can also describe the fiber functors for our three categories more explicitly, and in the process identify the three special algebra objects $\kDk,\kNk,\CDCN$ that represent them, as in Section \ref{sec:FF-Hom}.

The functors for our categories $\CC_\CN\simeq \CH_\CD\text{-mod}^{fd}$, $\CC_\CD\simeq \CH_\CN\text{-mod}^{fd}$, $\CC_\CT\simeq U\text{-mod}^{fd}$ simply forget the actions of the respective sparks $\CC_\CD,\CC_\CN,U$ on modules, leaving behind the underlying vector spaces.

Thus, if we consider $\CC_\CN$ (say), the fiber functor
\be \CF_\CN : \text{Rep}(G)^{fd} \to \text{Vect}^{fd} \ee
must be the one that forgets the $\CH_\CD=\C G$ action. On Wilson lines, it must map each $W_\rho$ to its underlying $\rho$. 
We can represent $\CF_\CN=\text{Hom}(\kDk,-)$, by taking Hom with a Wilson line in the regular representation,
\be  {}_k\ol\CD_k := W_{\CO(G)} = \bigoplus_{\rho\in\text{irrep}(G)} \rho^*\otimes W_\rho\,. \label{kDk-DW} \ee
Indeed, for any irreducible $\rho'$, we get the desired result
\be \text{Hom}_{\CC_\CN}(\kDk,W_{\rho'}) = \bigoplus_{\rho\in\text{irrep}(G)} \rho \otimes \text{Hom}(W_\rho,W_{\rho'})  = \bigoplus_{\rho\in\text{irrep}(G)} \rho \otimes \C\delta_{\rho=\rho'} = \rho'\,.\ee

To understand more physically why $\kDk$ would be a Wilson line in the regular representation, note that a narrow strip of $\ol \CD$ on an $\CN$ boundary breaks gauge symmetry there. This same breaking can be achieved by coupling $\CN$ to a 1d topological quantum mechanics with target $G$, on which $G$ acts freely by (say) left multiplication. The state space of the quantum mechanics is precisely the functions $\CO(G)$, with an induced left action $g\cdot f(x)=f(g^{-1}x)$. Coupling to the gauge group on the $\CN$ boundary then constructs the Wilson line $W_{\CO(G)}$.

Similarly, the fiber functor $\CF_\CD:\text{Vect}_G^{fd}\to\text{Vect}^{fd}$ forgets the $G$-grading (which comes from the action of $\CH_\CN=\CO(G)$). It simply sends each boundary global symmetry defect $V_g$ to the one-dimensional space $\C$. We can represent $\CF_\CD = \text{Hom}(\kNk,-)$, with
\be \kNk := \bigoplus_{g\in G} V_g \label{kNk-DW} \ee
a sum of all the symmetry defects. Now for any $V_{g'}$ we check that
\be \text{Hom}_{\CC_\CD}(\kNk,V_{g'}) = \bigoplus_{g\in G}\text{Hom}(V_g,V_{g'}) = \bigoplus_{g\in G} \C \delta_{g=g'} = \C \,.\ee

Physically, the identification \eqref{kNk-DW} makes sense because placing a thin strip of $\CN$ on a $\CD$ boundary gauges the global $G_\pd$ symmetry on the strip. Gauging the symmetry acts like a projection operator, projecting boundary operators $G_\pd$-invariants as one traverses across the strip. The sum of defects \eqref{kNk-DW} has precisely this projection-to-invariants property.

Finally, the bulk fiber functor $\CF_\CT$ sends a sheaf $\CE\in \text{Shv}^{G_{ad}}(G)$ to the sum of its stalks at every point of $G$, forgetting the equivariant $G_{ad}$ structure. This may be represented as $\CF_\CT=\text{Hom}_{\CC_\CT}(\CDCN,-)$ for an object $\CDCN$ that's a sheaf supported on all of $G$, whose stalk at each point is the regular representation. 

\subsubsection{Sparks from algebra objects} 

To close the circle, we can check that we recover spark algebras as endomorphisms of the algebra objects $\kDk,\kNk$, as required by \eqref{EndCDC} in Section~\ref{sec:spark-Yo}. We'll just consider  boundary sparks.

Endomorphisms of $W_{\CO(G)}$ come from the right action of $G$ on $\CO(G)$, which commutes with the left action that was used to upgrade $\CO(G)$ to a Wilson line. Using the right action gives rise to the opposite group algebra
\be \text{End}_{\CC_\CN}(\kDk)= \text{End}_{\CC_\CN}(W_{\CO(G)}) \simeq \C G^{\rm op} \qquad\leftrightarrow\qquad \CH_\CD = \C G\,. \label{End-spark-DW1} \ee

On the other hand, each symmetry defect $V_g$ is a simple object of $\CC_\CD$, with no morphisms between $V_g$ and $V_{g'}$ possible when $g\neq g'$. Thus,
\be \text{End}_{\CC_\CD}(\kNk) = \text{End}_{\CC_\CD}(\oplus_g V_g) \simeq \oplus_g \C \simeq \CO(G)^{\rm op}\qquad\leftrightarrow\qquad \CH_\CN = \CO(G)  \label{End-spark-DW2}  \ee
is identified with functions on the finite set $G$. (Since $\CO(G)$ is commutative, we have $\CO(G)\simeq \CO(G)^{\rm op}$.) Explicitly, the delta function $\delta_k\in \CO(G)$ is identified with the projection morphism  \vspace{-.2in}
\be \delta_k:\oplus_gV_g\to \oplus_g V_g\,,\qquad \delta_k\big|_{V_g} = \begin{cases} \text{id}\big|_{V_g} & k=g \\ 0 & k\neq g \end{cases}\,. \label{delta-proj} \ee

\subsection{Completeness}
\label{sec:DW-C}

We argued heuristically in Section \ref{sec:DW-setup} that the $\CN$ and $\CD$ boundary conditions of DW theory obeyed completeness. We'd like to be more explicit about this, and actually describe the deformation that implements a gluing of regions on opposite $\CN$ and $\CD$ boundaries.

Suppose that $R$ is some region on $\CD$, and $\bar R$ its orientation-reversed image on $\CN$ (as in \eqref{stripglue}). To glue, we'd like to replicate the same steps described at the bottom of Section \ref{sec:DW-box} for the box/rectangle gluing. Namely:
\begin{enumerate}
\item Insert defects along $R$ to relax the $\CD$ trivialization there --- effectively inserting a sum over all boundary gauge bundles in the path integral.
\item Insert defects along $\bar R$ to perfectly measure the fluctuating $\CN$ bundles there, so that by taking sums of measurements we can project to any given bundle.
\item Couple the two sides: sum just once over bundles on $R$ and \emph{matched} projections on $\bar R$ to replicate a smooth gluing through the bulk.
\end{enumerate}
We present a general proposal for implementing this, in the spirit of condensation webs/foams that are used to implement gauging (see \eg\ \cite[Sec. 5]{FMT}, \cite{Carqueville-review} for a review%
\footnote{We suspect that what we describe here is closely related to Example 5.17 of \cite{FMT}, and that the object $x_{\rm reg}$ described there is the DW/Turaev-Viro analogue of the 2-Maurer-Cartan element that we use to control gluing in later examples of cohomological TQFT.}).
We will not be fully rigorous -- \eg\ we leave proofs of invariance of the webs in this proposal to future work.

\subsubsection{Junctions of $\kDk$ and $\kNk$}
\label{sec:DW-mm}

The defects that we'll insert on $\CD$ are combinations of $V_g$'s, and the defects that we insert on $\CN$ are combinations of $W_\rho$'s. These can all be written, respectively, in terms of  $\kNk$ and $\kDk$. (For example, a particular $V_g$ is obtained by starting with $\kNk$ and applying a projection.)

Since we'll ultimately want to create webs of defects, it helps to understand better the algebra structure of $\kNk$ and $\kDk$.

Both ${}_k\ol\CN_k$ and ${}_k\ol\CD_k$ are Frobenius algebra objects of their respective categories, by virtue of being constructed from strips, as depicted in \eqref{kDk-alg} of Section \ref{sec:FF-Hom}. In particular, there are canonical multiplication and comultiplication morphisms $\mb m:{}_k\ol\CN_k\otimes {}_k\ol\CN_k\to {}_k\ol\CN_k$, $\mb m^*:{}_k\ol\CN_k\to {}_k\ol\CN_k\otimes {}_k\ol\CN_k$ (and similarly for ${}_k\ol\CD_k$) defined by junctions of strips:
\be \raisebox{-.3in}{\includegraphics[width=4in]{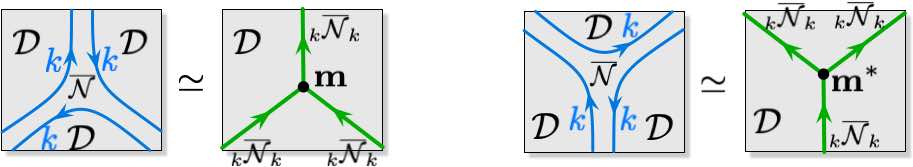}} \label{mm*-strip} \ee
These morphisms must be compatible with the coproducts in the spark algebras \eqref{End-spark-DW1}--\eqref{End-spark-DW2}; \emph{e.g.} for all $\alpha\in \CH_\CN = \text{End}_{\CC_\CD}({}_k\ol\CN_k)$, we have
\be \mb m\circ \Delta(\alpha) = \alpha\circ \mb m\,,\qquad \mb m^*\circ \alpha = \Delta(\alpha)\circ \mb m^*\,. \label{mm*-D}\ee

Explicitly, in DW theory, the (co)multiplication morphisms are given as follows. For ${}_k\ol\CD_k=W_{\CO(G)}$ we use the identification $\CO(G)\otimes \CO(G)\simeq \CO(G\times G)$ and the diagonal map $\text{diag}:G\hookrightarrow G\times G$, $\text{diag}(g):=(g,g)$ to pull back and push forward functions
\be \begin{array}{c@{}c@{}c} \CO(G\times G) & \overset{\text{diag}^*}\longrightarrow & \CO(G) \\ 
    \Big((g_1,g_2)\mapsto f(g_1,g_2)\Big) &\mapsto & \Big(g\mapsto f(g,g)\Big)
    \end{array},\quad
     \begin{array}{c@{}c@{}c} \CO(G) & \overset{\text{diag}_*}\longrightarrow & \CO(G\times G) \\ 
    \Big(g\mapsto f(g)\Big) &\mapsto & \Big((g_1,g_2)\mapsto \delta_{g_1=g_2}f(g_1)\Big)
    \end{array} \label{diag-push}
\ee
These maps intertwine the left  $G$-action on $G$ with the diagonal left $G$-action on $G\times G$. Thus they give us maps of Wilson lines
\be  \begin{array}{rl}\mb m^* = \text{diag}_*: & W_{\CO(G)}\to W_{\CO(G\times G)}\simeq W_{\CO(G)}\otimes W_{\CO(G)} \\[.1cm]
\mb m = \text{diag}^*: &  W_{\CO(G)}\otimes W_{\CO(G)}\simeq W_{\CO(G\times G)} \to W_{\CO(G)}\,. \end{array} \ee
They also intertwine the corresponding right $G$-actions, thus are compatible with the coproduct $\Delta(g)=g\otimes g$ in $\CH_\CD$. Moreover, it is easy to compute that $\mb m\circ \mb m^* = \text{id}\big|_{W_{\CO(G)}}$, implying that $W_{\CO(G)}$ is a \emph{separable} Frobenius algebra object.

For ${}_k\ol\CN_k=\oplus_gV_g$, we have $\big(\oplus_gV_g\big)\otimes \big(\oplus_gV_g\big) = \oplus_{g,g'} V_g\otimes V_{g'}$.
Recall that the tensor product identifies $V_g\otimes V_{g'}\simeq V_{g'g}$. Then multiplication is given on each summand by
\be \mb m : V_g\otimes V_{g'}\simeq V_{g'g} \overset{\varphi}\to V_{g'g} \hookrightarrow \oplus_g V_g\,,\qquad \varphi = \frac{1}{|G|} \text{id}\big|_{V_{g'g}}\,, \ee
while comultiplication is given on each summand by
\be \mb m^*: V_g \overset{\varphi^*}{\to} \bigoplus_{k\in G} V_{k^{-1}g} \otimes V_k \simeq \C^{|G|} \otimes V_g\,,\qquad \varphi^* = \left(\begin{smallmatrix} 1\\1 \\ \vdots \\ 1\end{smallmatrix}\right)\otimes \text{id}\big|_{V_g}\,. \ee
The correction by $\frac{1}{|G|}$ in $\varphi$ assures that $\mb m\circ \mb m^*=\text{id}_{\oplus_gV_g}$, making $\oplus_gV_g$ a separable Frobenius algebra object as well. We leave it to the reader to check that the coproduct \eqref{HN-D-DW} in $\CH_\CN$ is compatible with $\mb m,\mb m^*$ as in \eqref{mm*-D}, when $\delta_k$'s are interpreted as projections via \eqref{delta-proj}.

\subsubsection{General gluing proposal}
\label{sec:DW-glue}

Separable Frobenius algebra objects play a distinguished role in the formal TQFT construction of 2d generalized gauging (\cf\ \cite{Carqueville-review}). Roughly, the insertion of such algebra objects along webs (\emph{e.g.} the skeletons of triangulations of 2d spacetime) can be used to define a new set of modified correlation functions, corresponding to a gauged/condensed version of the theory.

For example, in DW theory, the insertion of ${}_k\ol\CN_k=\oplus_g V_g$ along webs on the $\CD$ boundary effectively gauges the boundary $G$ symmetry, converting the boundary condition to $\ol \CN$.
Conversely, the insertion of ${}_k\ol\CD_k=W_{\CO(G)}$ along webs on an $\CN$ boundary effectively converts the $\CN$ boundary to $\ol\CD$:
\be \raisebox{-.5in}{\includegraphics[width=5in]{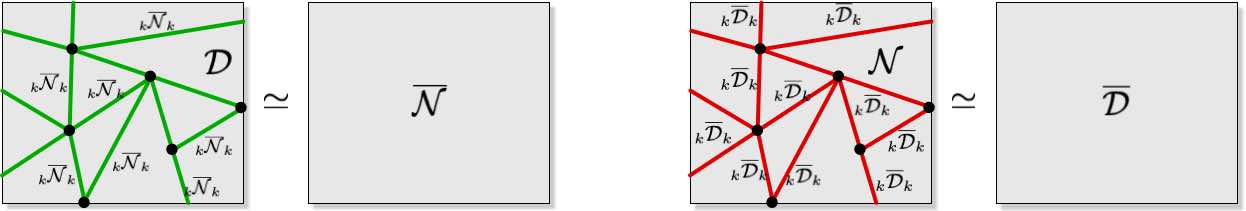}} \ee

In order to glue $\CN$ to $\CD$ as required for completeness, we propose to insert both ${}_k\ol\CD_k$ webs on $\CN$ and dual ${}_k\ol\CN_k$ webs on $\CD$; \emph{and} to couple them together with further insertions of local operators. We proceed, somewhat heuristically, as follows.

We assume that the regions $R$ (on $\CD$) and $\bar R$ (on $\CN$) that we wish to glue are well-bounded, in the sense of Definition \ref{def-bounded} from Section \ref{sec:comp}. Namely, if $\pd R$ is nonempty, we assume that it consists of a finite union of smooth curves, which are either
\begin{itemize}
\item $k$ junctions between $\CD$ and $\ol \CN$ (on the boundary of $R$); or
\item $k$ junctions between $\CN$ and $\ol\CD$ (on the boundary of $\bar R$)
\end{itemize}
(but not both).
Choose a triangulation $\tau$ of the interior of $R$, with oriented edges, such that the edges may end on junctions with $\ol\CN$ (at least one edge ends on every such junction) or be semi-infinite and shoot into asymptotic regions at infinity. Boundary cells of the `triangulation' need not be actual triangles. Let $\tau^\vee$ be the cell decomposition of $\bar R$ dual to $\tau$, with at least one edge ending at every junction with $\ol\CD$.

Two examples are depicted here, for $R$ a torus and $R$ a semi-infinite strip:
\be \raisebox{-1.2in}{\includegraphics[width=5in]{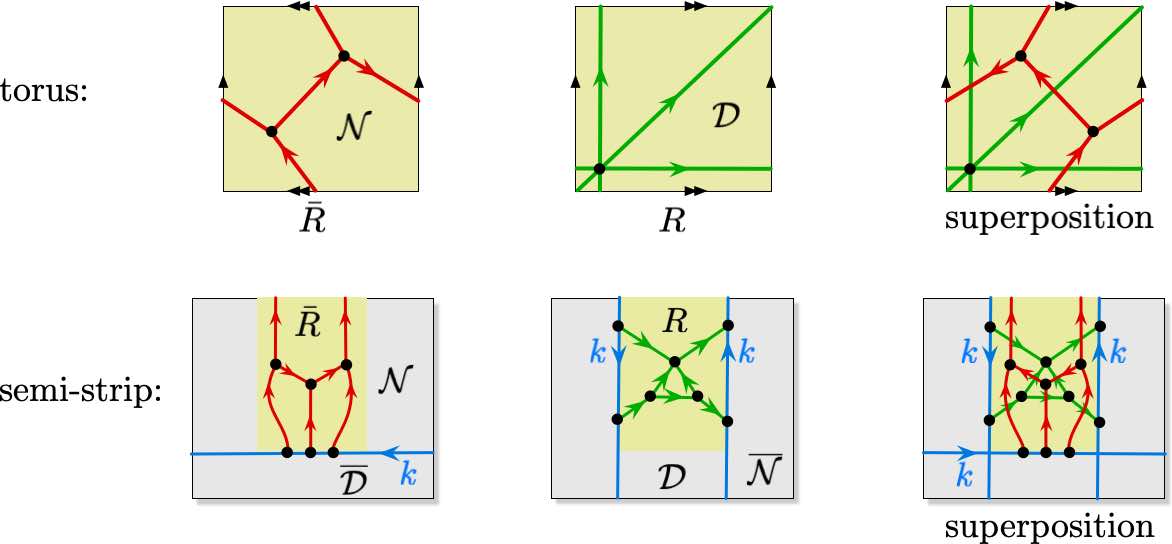}} \label{torus-strip} \ee
On the left we show $(\bar R,\tau^\vee)$ and $(R,\tau)$  separately, each viewed from `outside' the boundary of our 3d theory $\CT$. On the right we show their superposition, as viewed from inside the $\bar R$ boundary and outside the $R$ boundary.

Now, we will place ${}_k\ol\CN_k = \oplus_g V_g$ along the edges of $\tau$, and place ${}_k\ol\CD_k=W_{\CO(G)}$ along the edges of $\tau^\vee$. At the vertices we place the canonical morphisms ($\mb m,\mb m^*$, or compositions thereof) induced by resolving these objects into strips, just as in \eqref{mm*-strip}.

At this point, the procedure has deformed $R$ into an $\ol \CN$ boundary, and has deformed $\bar R$ into an $\ol \CD$ boundary. But we want to do more!

In addition, at each intersection of an edge of $\tau$ and an edge of $\tau^\vee$ (identified via the gluing map), we insert the gluing deformation
\be  o = \sum_{k\in G} k^{-1}\otimes \delta_k\,, \label{glue-def-DW} \ee
where the first factor `$k$' is interpreted as an endomorphism of $W_{\CO(G)}$ and the second factor as an endomorphism of $\oplus_g V_g$ (\cf\ \eqref{End-spark-DW1}--\eqref{End-spark-DW2}). For example, in the case of the torus and semi-strip of \eqref{torus-strip}, the insertions look like
\be \raisebox{-.5in}{\includegraphics[width=3.5in]{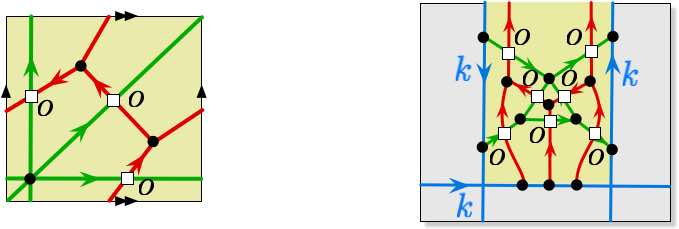}} \ee

Consider what this does. The insertion of $\delta_k$ along an edge of $\tau$ projects the line operator $\oplus_g V_g$ to $V_k$ along that edge. Insertions of $\delta_k$'s along all edges thus construct a network of trivialization-changing defects (or ``symmetry defects''). As the $\delta_k$'s on different edges are varied, these insertions construct all possible $G$-bundles in the $R$ region on the $\CD$ boundary. (Compatibility with $\mb m,\mb m^*$ ensures that the $G$-bundles extend over vertices.) Dually, the insertion of $k^{-1}$'s in the network of $W_{\CO(G)}$ Wilson lines \emph{projects} the sum of bundles on the $\CN$ boundary containing $\bar R$, to a single bundle that matches the one constructed on $R$. Altogether, inserting a sum \eqref{glue-def-DW} on every single pair of transverse $(\tau^\vee,\tau)$ edges restricts the path integral to a \emph{single}, aligned sum over all $G$-bundles along $\bar R\cup R$, thus gluing together the two regions.

\section{3d $\CN=4$ matter}
\label{sec:matter}

Our next example is slightly less well known. We'll consider the simplest 3d topological QFT $\CT$ of \emph{cohomological} type, which has several equivalent descriptions:
\begin{itemize}
\item The B-twist of $n$ free 3d $\CN=4$ hypermultiplets, \emph{a.k.a.} Rozansky-Witten theory \cite{RW} with target $T^*V$, where $V=\C^n$. (Our main perspective.)

Rozansky-Witten theory with cotangent target was developed as an extended TQFT in \cite{KRS, KapustinRozansky}, whose methods we'll use extensively here. For affine, linear target $T^*V$, the theory was recently revisited and defined as a functorial, extended, truncated 2-1-0 TQFT in \cite{BCR-extended,BCFR-defect}. 

\item Chern-Simons theory for superalgebra $\mathfrak{psl}(1|1)^n$, in BRST quantization, keeping fields/operators of all ghost numbers (where ghost number = cohomological degree) \cite{Mikhaylov}.
\item BF theory for odd superalgebra $\C^{0|n}$, in BRST quantization.
\end{itemize}

This example is especially nice because, while of cohomological/dg type, it still satisfies all the axioms of ``Tannakian QFT.'' 
The $\CN$ and $\CD$ boundary conditions will be associated to the transverse Lagrangian submanifolds $V\subset T^*V$ and $V^*\subset T^*V$. Moreover, this example will turn out to satisfy the strong dualizability condition \AD$_k$ for the transversality interface, which renders spark algebras finite-dimensional; and the condition \ADline\ that categories are generated by dualizable objects, which ensures antipodes make sense.

Moreover, this example serves as a toy model to help us illustrate characteristic features of a cohomological setting. In particular, a key role will be played by \emph{topological descent} \cite{Witten-Donaldson}. We'll show in Section \ref{sec:matter-sparks} that sparks arise as integrated descendants of local operators. In Section \ref{sec:matter-Hopf} we'll relate the Hopf pairing of boundary sparks to descendants of bulk local operators integrated along a Hopf link, which in turn defines the ``higher'' $E_3$/Poisson bracket of bulk local operators \cite{descent}. We'll also see concretely in Section \ref{sec:matter-comp} how completeness is governed by the second descendant of a degree-two local operator, a ``2-Maurer-Cartan'' element.

For concrete calculations, we'll work in a twisted BV formalism, which we briefly review in Section \ref{sec:matter-def}. This is not at all essential. The same calculations could be done, somewhat more clumsily, in standard B-twists of 3d $\CN=4$, 2d $\CN=(2,2)$, and 1d $\CN=2$ SUSY (all of which will play a role). We'll try to indicate throughout the section how BV formulas match more familiar SUSY expressions.

The ribbon Hopf algebra $U$ that we derive for bulk line operators looks like a free exterior algebra in $2n$ fermions $\psi_+^i$, $\psi^-_i$, $U = \C[\psi_+,\psi^-] = \Lambda^\bullet(T^*V),$ with
\be U:\quad \begin{array}{c} \Delta(\psi)=\psi\otimes 1+1\otimes \psi \\ \varepsilon(\psi)=0\,,\quad S(\psi)=-\psi\end{array}\quad \forall\;\psi\,,\qquad \CR = e^{-\frac1\hbar \psi_+\otimes \psi^-}\,,\quad v = e^{-\frac1\hbar \psi^-\cdot \psi_+} \ee
This is otherwise known as the quantum group $U_q(\mathfrak{psl}(1|1)^n)$, which is not surprising given the relation of our theory to $\mathfrak{psl}(1|1)^n$ Chern-Simons, \cf\ \cite{Mikhaylov, CDGG}. In Section \ref{sec:matter-cat}, we'll relate $U$-mod to more familiar categories of lines in Rozansky-Witten theory, and indicate how our methods produce a fiber functor for the more familiar braided tensor category that Roberts-Willerton defined therein \cite{RobertsWillerton}; we'll also briefly relate $U$-mod to modules for a boundary VOA, given by symplectic fermions.

\subsection{Fields and action}
\label{sec:matter-def}

It is insightful (and highly simplifying) to recast the B twist of $n$ free hypermultiplets in a twisted BV-BRST formalism, as reviewed in \cite{CDGG, Garner-formalism}. Let $V=\C^n$. The fields of the theory on a 3-manifold $M$ are multi-forms
\be \mb X \in V\otimes \Omega^\bullet(M)[1]\,,\qquad \mb Y \in V^* \otimes \Omega^\bullet(M)[1]\,, \label{matter-XY} \ee
with action and BRST operator
\be S = \frac1\hbar\int_M \mb Yd \mb X = -\frac1\hbar\int \mb Xd\mb Y\,,\qquad\quad \left\{\begin{array}{l} Q(\mb X) = d\mb X\,, \\[.1cm] Q(\mb Y) = d\mb Y\,. \end{array}  \right.\label{matter-S} \ee
Given a basis $e^i$ for $V$ and a dual basis $e_i$ for $V^*$ $(i=1,...,n)$, we might also write the fields in components as $\mb X_i,\mb Y^i\in \C\otimes \Omega^\bullet(M)[1]$, and the action as $S = \int_M \sum_{i=1}^n \mb X_i d\mb Y^i$.

A nice feature of this formalism is that the BRST differential $Q$ is defined by BV bracket with the action, $Q = \{S,-\}_{\rm BV}$, where in turn the bracket is deduced entirely from the kinetic term, here $\{\mb X(x),\mb Y(x')\}_{\rm BV} = \hbar \,\delta^{(3)}(x-x')$.

Let's compare this to the more standard field content of 3d $\CN=4$ hypermultiplets. The leading (0-form) components $X,Y$ of $\mb X$ and $\mb Y$ are the complex bosonic scalars of $N$ hypermultiplets valued in $T^*V$. The shift `$[1]$' in \eqref{matter-XY} is a homological shift. It indicates that the fields $X,Y$ have degree $-1$ but that operators (\eg\ local operators) formed out of $X,Y$, which are functions of the fields and what we usually care about in physics, have degree $+1$.
Physically, this homological degree comes from charge under the maximal torus of the $SU(2)_H$ R-symmetry acting on Higgs branch. Conventions are also such that the exterior derivative also has cohomological degree $+1$ (consistent with $|Q|=1$). Thus at higher form degrees we encounter:
\begin{itemize}
\item A 1-form $X^{(1)} = \chi_\mu dx^\mu$ whose components are a triplet of hypermultiplet fermions, of cohomological degree zero (and similar for $Y^{(1)})$.
\item A 2-form $X^{(2)}$ that's $Q$-cohomologous to $*d\ol Y$; similarly $Y^{(2)}$ is cohomologous to $-*d\ol X$. Operators formed from their components are bosons of degree $-1$.
\item 3-forms $X^{(3)}$ and $Y^{(3)}$, such that $*X^{(3)},*Y^{(3)}$ are cohomologous to a final pair of physical hypermultiplet fermion, of degree $-2$\,.
\end{itemize}
This matches the field content and cohomological degrees (described as ghost numbers) from \cite{RW}.\label{matter-field-gradings}

We emphasize that fermion number, controlling parity, is distinct from cohomological degree in this theory. There are cohomologically-odd bosons (which commute) and cohomologically-even fermions (which anti-commute). The BRST operator $Q$ is \emph{both} fermionic and of cohomological degree +1.

In relating the theory to (say) Chern-Simons for superalgebra $\mathfrak{psl}(1|1)^n$, the fermionic 1-forms $X^{(1)}$, $Y^{(1)}$ play the role of the connections. The bosonic 0-forms $X,Y$ play the role of ghosts. Cohomological degree translates to ghost number. When saying that the B twist of hypermultiplets is a ``derived enhancement'' of Chern-Simons theory, we mean that we quantize in the BV-BRST formalism and keep states/operators of all ghost number, rather than truncating to ghost-number zero, as is sometimes done.

\subsubsection{Local operators and descendants}
\label{sec:matter-descent}

In $Q$-cohomology, the bulk local operators of the theory consist of polynomials in the $2n$ bosonic scalars $X_i,Y^i$,
\be  A = \C[X_i,Y^i]_{i=1}^n = \C[T^*V] \label{matter-bulkops} \ee
This is a commutative algebra, with no quantum corrections to the classical product of polynomials.

For what follows, it will be useful to recall that in any topological theory of cohomological type, local operators come with towers of ``topological descendants'' that manifest its topological invariance \cite{Witten-Donaldson}. Given a local operator $\CO=\CO^{(0)}$ satisfying $Q\CO=0$, the descendants $\CO^{(k)}$ ($k\leq 3$ in dimension $d=3$) satisfy 
\be  d\CO^{(k)} = Q\CO^{(k+1)}\,, \label{descent-eq} \ee
or simply
\be (d-Q)(\CO+\CO^{(1)}+\CO^{(2)}+\CO^{(3)})=0\,. \label{descent-form}  \ee
The descendants guarantee (for example) that insertions of $\CO$ at two different spacetime points $x,y$ that can be connected by a path $\gamma$
 are equivalent up to a $Q$-exact term, built from the integral along $\gamma$ of the non-closed operator $\CO^{(1)}$,
\be \CO(y) - \CO(x) = Q\int_x^y \CO^{(1)}\,. \label{descent-top} \ee
Thus, in $Q$-cohomology, $\CO$ becomes independent of insertion point.

The twisted BV formalism \eqref{matter-XY}--\eqref{matter-S} precisely organizes fields by grouping together their descendants into multiform-valued ``superfields.'' It aligns form degree with descent degree. In particular, given a polynomial $f(X,Y) \in A$ in the bosonic scalars, we have
\be f(X,Y)^{\text{($k$-th descendant)}} = f(\mb X,\mb Y)^{(k\text{-form})} \ee

While the integrals of descendants on open cycles $\gamma$ manifest the topological invariance of $Q$-closed operators on $\pd\gamma$, as in \eqref{descent-top}, the integrals of descendants on closed cycles $\gamma$ are themselves $Q$-closed (and depend only on the homology class of $\gamma$). By using configurations of multiple closed cycles that are mutually linked, one can use this idea to define higher operations on the algebra of local operators \cite{descent}.%
\footnote{This idea was explored in two-dimensional physics in \cite{WittenZwiebach,LianZuckerman,Getzler,PenkavaSchwarz}. Mathematically, it reflects the general expectation that local operators in a $d$-dimensional TQFT form an $E_d$ algebra, which for $d\geq 2$ is quasi-isomorphic to $P_d$, or a shifted-Poisson algebra.} %
In $d=3$, there is only one nontrivial higher operation, which ultimately endows the algebra $A$ of local operators with a (-2)-shifted Poisson bracket. The bracket of two local operators $\CO_1,\CO_2$ can be defined by integrating their first descendants along the loops of a Hopf link:
\be \big\{\!\!\big\{\CO_1,\CO_2\big\}\!\!\big\}:= \oint_{\gamma_1}\CO_1^{(1)} \oint_{\gamma_2}\CO_2^{(1)} =\quad \raisebox{-.3in}{\includegraphics[width=1.3in]{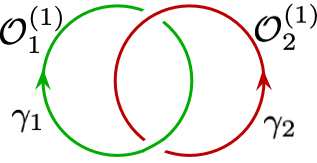}}
\label{def-PB}  \ee   
The ``shift'' refers to the fact that, due to taking descendants, the bracket decreases homological degree by 2.

It is a straightforward calculation in the B-twist of hypermultiplets \cite{descent} to show that
\be \{\!\!\{Y^i,X_j\}\!\!\} = \hbar\, \delta^i{}_j\,, \label{XY-PB}  \ee
up to a universal constant.
This comes from using the propagator between $\mb X,\mb Y$ in \eqref{matter-S} to evaluate $\langle Y^{(1)}(x)X^{(1)}(y)\rangle$ on the RHS of \eqref{def-PB}, leading to a Gauss integral for the linking number between the two components of the Hopf link. This is a highly simplified analgue of Gauss integrals in abelian Chern-Simons theory \cite{Witten-Jones}.

\subsection{Boundaries, transversality, and completeness}
\label{sec:matter-bdy}

In \cite{KRS}, it was explained that basic topological boundary conditions in a 3d B model are labelled by smooth holomorphic Lagrangian support in the target, in this case $T^*V = V\times V^*$. For our boundary conditions, we take those corresponding to the obvious Lagrangians $\CN = V\times \{0\}$, $\CD = \{0\}\times V^*$.

They have several equivalent descriptions. In 3d $\CN=4$ terms, these are boundary conditions that preserve 2d $\CN=(2,2)$ boundary supersymmetry, as discussed in \cite{BDGH,ChungOkazaki}. Namely, $\CD,\CN$ are the unique supersymmetric completions of boundary conditions that impose Dirichlet/Neumann (or vice versa) on the complex scalars,
\be \CD = \big\{X\big|_\pd = 0\,,\; \pd_\perp Y\big|_\pd = 0\}\,,\qquad \CN = \big\{Y\big|_\pd = 0\,,\; \pd_\perp X\big|_\pd = 0\}\,. \ee
In twisted BV formalism, the boundary conditions are simply
\be \label{matter-DN}
 \CD = \big\{ \mb X\big|_\pd = 0\big\}\,,\qquad  \CN = \big\{ \mb Y\big|_\pd = 0\big\}\,,
\ee
where  ``$\big|_\pd$'' denotes pullback of a form to the boundary. They may be supplemented by $\iota_\perp\mb Y\big|_\pd = 0,\pd_\perp \mb Y\big|_0=0$ (for $\CD$) and $\iota_\perp\mb X\big|_\pd = 0,\pd_\perp \mb X\big|_0=0$ (for $\CD$), where $\iota_\perp$ is contraction with a vector field normal to the boundary; though these additional conditions may just be viewed as a gauge choice in the BV formalism. 

Classically, these boundary conditions are fully topological. At a quantum level, they may have a framing anomaly, which would invalidate our basic construction. A putative framing anomaly can be analyzed by going back to the untwisted 3d $\CN=4$ theory, with its $SU(2)_C$ (Coulomb) and $SU(2)_H$ (Higgs) R-symmetries. The B twist introduces an $SU(2)_C$ background aligned with spacetime curvature. An $\CN=(2,2)$ boundary condition preserves $U(1)_C\times U(1)_H$ subgroups of the bulk R-symmetry, and at the boundary the B twist aligns $U(1)_C$ backgrounds with spacetime curvature. Thus the boundary is unframed as long as $U(1)_C$ is unbroken. (Conversely, the boundary has a framing anomaly if $U(1)_C$ has a boundary anomaly.) 

Using the analysis of boundary anomalies in \cite{BDGH} or \cite{dimofte2018dual}, we find boundary anomaly polynomials%
\footnote{An easy shortcut to calculating \eqref{matter-anom} is to find the anomaly polynomial for the $\CN=(2,2)$ sigma model to either $V^*$ or $V$ as appropriate, \emph{cf.}  \cite{Witten-phases}, and then divide by two. The $U(1)_H$ and $U(1)_C$ R-symmetries discussed here are vector and axial symmetries, from a $\CN=(2,2)$ perspective.}
\be \CD: \quad - \text{Tr}(\mb f)\mb c\,,\qquad  \CN:\quad  + \text{Tr}(\mb f) \mb c\,, \label{matter-anom} \ee
where `$\mb f$' is a field strength for the $U(n)$ flavor symmetry that acts on $X,Y$ in fundamental/antifundamental representations, and `$\mb c$' is a field strength for $U(1)_C$. These are 't Hooft anomalies for global symmetries. As long as we don't turn on a curved flavor background, $U(1)_C$ will remain unbroken, and there is no boundary framing anomaly.

We also note that there is no problem with cohomological degree being preserved by the boundary conditions. Cohomological degree comes from $U(1)_H$, which simply rotates $(\mb X,\mb Y)$ with charges $(+1,+1)$. This is preserved classically by \eqref{matter-DN} and there is never a boundary anomaly for $U(1)_H$.

As required for 1-dualizability of $(\CD,\CN)$ as in \AD\ of Section \ref{sec:top}, all cups, caps, S-bends, and pinches of these boundary conditions are well defined. They were initially described in \cite{KRS}, and then revisited systematically from the perspective of functorial extended TQFT in \cite{BCR-extended}. 

However, the boundary conditions are not fully dualizable (and do not need to be). For example, the space of states on a 2-disc with $\CD$ or $\CN$ boundary conditions are both infinite-dimensional. By a state-operator correspondence, the space of states on a disc with $\CD$ boundary conditions is equivalent to the space of local operators on the $\CD$ boundary itself, which is the commutative algebra $\C[Y^i]_{i=1}^n$ in bosonic scalars that survive on $\CD$. Similarly,
\be \CZ(D^2,\CD) \simeq \C[Y^i]_{i=1}^n = \text{Sym}^\bullet (V[-1])\,,\qquad \CZ(D^2,\CN) \simeq \C[X_i]_{i=1}^n = \text{Sym}^\bullet (V^*[-1])\,.\ee
(The notation on the RHS indicates symmetric algebras formed from elements of a vector space $V$ generated in homological degree $+1$ and even fermion number; see p. \pageref{matter-field-gradings}.)

\subsubsection{Transversality}
\label{sec:matter-trans}

Transversality \AT\ of $(\CD,\CN)$ now follows literally from the fact that $V^*,V$ are transverse holomorphic Lagrangians inside $T^*V$.

For a concrete calculation, we may follow now-standard techniques from \cite{KRS}.
We begin with the 3d B model on an interval $[0,\epsilon]\times \R^2$, with $\CN$ b.c. on both ends. This theory is equivalent to a 2d B model with target $V$ \cite{KRS}, parameterized by the scalar $X$. The $\CN$ b.c. on the left may be deformed, or flipped, to a $\CD$ b.c. by 1) tensoring it with a 2d B-twisted chiral multiplet $y \in V^*$ and 2) coupling $y$ to the restriction of $X$ to the left boundary with a boundary superpotential $W=Xy$. The upshot is that the 3d theory on a sandwich with $(\CD,\CN)$ b.c. on the two ends becomes equivalent to a 2d B model with target $V\times V^*$ (parameterized by $X,y$), with superpotential $W=Xy$.
 
To implement this in twisted BV formalism, we write the action on the interval with $\CN$ on both sides as
\be S_{\ol\CN\circ\CN} = \frac1\hbar \int _{[0,\epsilon]\times\R^2} \mb Yd\mb X\,. \label{matter-SNN} \ee
Note that when there are boundaries it matters how the bulk action is written, to avoid boundary terms in the BRST transformations of bulk fields.%
\footnote{Similar arguments can be made from a more traditional physical perspective, by first allowing free boundary conditions and then using boundary equations of motion (boundary terms in the variation of the bulk action) to dynamically impose boundary conditions.} %
 For example, letting $s$ be a coordinate on $[0,\epsilon]$, observe that \eqref{matter-SNN} induces $Q\mb X(s)= \{S_{\ol\CN\circ\CN},\mb X(s)\}_{\rm BV} = d\mb X(s)$ but $Q\mb Y(s)= \{S_{\ol\CN\circ\CN},\mb Y(s)\}_{\rm BV} = d\mb Y(s) + \delta(s-\epsilon) \mb Y|_\epsilon - \delta(s) \mb Y|_0$, with boundary terms vanishing precisely when $\CN$ b.c. are imposed on both sides. Adding a boundary superpotential $\mb X\mb Y$ at $s=0$, we get
\be S_{\CD\circ\CN} = \frac1\hbar\int_{\R^2} \mb X|_0\mb Y|_0 + \frac1\hbar \int _{[0,\epsilon]\times\R^2} \mb Yd\mb X = \frac1\hbar \int_{\R^2} \mb X|_\epsilon \mb Y|_\epsilon -\frac1\hbar\int_{[0,\epsilon]\times\R^2} \mb Xd\mb Y\,, \ee
modifying the BRST variations to $Q\mb X(s) = d\mb X(s)-\delta(s) \mb X|_0$, $Q\mb Y(s) = d\mb Y(s)+\delta(s-\epsilon) \mb Y|_\epsilon$. It is the correct action for the theory with $\CD$ b.c. at $s=0$ and $\CN$ b.c. at $s=\epsilon$.

Now, due essentially to topological invariance, we can take the limit $\epsilon\to 0$. All nonzero modes of $\mb X,\mb Y$ on the interval become massive and can be exactly integrated out. The zero-modes leave behind a 2d theory with action
\be S_{\CD\circ \CN} \simeq \frac1\hbar\int_{\R^2} \Big[\mb x d\bm \psi^x + \mb y d\bm \psi^y + \mb x\mb y\Big]\,, \label{matter-SDN} \ee
where $\mb x = \mb X|_0$,  $\bm\psi^x = \iota_{ds}\mb Y|_0$ and $\mb y = \mb Y|_0$,  $\bm\psi^y = -\iota_{ds}\mb X|_0$. This is the twisted BV form of a 2d B model with target $T^*V$, and superpotential $xy$ (to be identified with $W=Xy$ in the \cite{KRS} analysis above).

Transversality, globally, is now the statement that the 2d B-model with $xy$ superpotential is completely massive. The nondegenerate quadratic superpotential allows all fields to be integrated out exactly, leaving an empty, trivial 2d theory.

Describing transversality \emph{locally} requires us to identify an interface $k$ between the 2d B model \eqref{matter-SDN} that represents $\CD\circ\CN$ and the trivial 2d theory, which in turn satisfies $k\circ k^*,k^*\circ k\simeq \text{id}$ as in \eqref{k-ids}.  Note that interfaces between the $\CD\circ\CN$ and the trivial theory are the same as boundary conditions for $\CD\circ\CN$.

The requisite $k$ is well known in mathematics. The category of boundary conditions for the B model \eqref{matter-SDN} is a category of matrix factorizations \cite{Kapustin:2002bi}%
\footnote{Classic mathematical treatments of matrix factorizations in mathematics describe $\Z/2$ graded categories. The ``breaking'' of a putative $\Z$ homological grading to $\Z/2$ happens because objects in a matrix-factorization category are sheaves equipped with an endomorphism $D$ (a ``curved differential'') of degree 1 that must obey $D^2=W\cdot\text{id}$, where $W$ is the superpotential. In classic treatments, $W$ has degree 0, necessarily breaking the grading to $\Z/2$. 

In every 3d $\CN=4$ example in this paper, superpotentials will have degree 2, and matrix-factorization categories, where they appear, will be $\Z$ graded. The nontrivial degree of $W$ is comes from the nontrivial degree of bosons (such as the coordinate functions $X,Y$ on $T*V$, both of degree 1). Physically, this is ultimately a manifestation of unbroken $U(1)_H$ symmetry.\label{foot-MF}} %
\be \text{MF}(T^*V,W=xy)\,. \ee
The boundary condition $k$ is the so-called Koszul matrix factorization $\CK[x]$ associated to $x=0$ (or equivalently%
\footnote{In the standard mathematics setting, where $x,y$ are of degree 0, there is a slightly asymmetry between the $x=0$ and $y=0$ Koszul matrix factorizations. In our setting, where $x,y$ are of degree 1, the two are canonically equivalent.} %
 to $y=0$); it has the property that tensoring a dg vector space with it induces an equivalence of categories, known as Kn\"orrer periodicity \emph{cf.} \cite[Sec 2.2.3]{KapustinRozansky}
\be  \begin{array}{ccc} \text{Vect} &\overset{\sim}\longrightarrow & \text{MF}(T^*V,W=xy) \\
    M &  \longmapsto  & M\otimes   \CK[x]\,. \end{array}  \label{Knorrer} \ee
The properties $k\circ k^*,k^*\circ k\simeq \text{id}$ follow because the inverse of the Kn\"orrer functor \eqref{Knorrer} is given by its adjoint.

Finally, for those willing to do precise calculations with the $k$ interface, it may be useful to include an explicit formulation of it after ``unfolding'' the $\CD\circ\CN$ sandwich. Consider the bulk theory on $\R_t\times \R_p\times \R_{s\geq 0}$, with $\CD$ b.c. at $s=0,p<0$ and $\ol\CN=\CN$ b.c. at $s=0,p>0$. The BV action starts out as
\be S_{\CD_k\ol\CN} \overset{\text{?}}{=} \frac{1}{\hbar} \hspace{-.8cm}\int_{\R_t\times\R_p\times\{s\geq 0\}}  \hspace{-.5cm}\mb Yd\mb X +\frac1\hbar  \hspace{-.8cm} \int_{\R_t\times \{p<0\}\times\{s=0\}}  \hspace{-.5cm}\mb X\mb Y \label{SDkN-0} \ee
However, this action does not preserve the BRST symmetry it generates, as $Q S_{\CD_k\ol\CN} = \frac1\hbar \int_{\R_t\times \{p<0\}\times\{s=0\}} d(\mb X\mb Y) = \frac1\hbar \int_{\R_t\times \{p=0\}\times \{s=0\}}\mb X\mb Y$. This extra 1d interface term in the BRST variation can be cancelled by introducing a pair of 1d fermions (a B twisted 1d fermi multiplet). In twisted BV formalism, the 1d fermions sit in multiforms $\mb a,\bar{\mb a} \in \Pi (V\times V^*)\otimes \Omega^\bullet(\R_t)$, whose leading components are fermionic. The total action is 
\be S_{\CD_k\ol\CN} = \frac{1}{\hbar} \hspace{-.8cm} \int_{\R_t\times\R_p\times\{s\geq 0\}}\hspace{-.5cm}\mb Yd\mb X +\frac1\hbar  \hspace{-.8cm} \int_{\R_t\times \{p>0\}\times\{s=0\}} \hspace{-.5cm}\mb X\mb Y - \frac1{2\hbar}  \hspace{-.8cm} \int_{\R_t\times \{p=0\}\times \{s=0\}}\hspace{-.8cm}\Big[ \mb ad\bar{\mb a} + \mb a \mb Y +  \mb X\bar{\mb a}\Big]\,, \label{SDkN} \ee
where $Q(\mb a) = d\mb a+ \mb X\big|_{0,0}$, $Q(\bar{\mb a}) = d\bar{\mb a}+ \mb Y\big|_{0,0}$, and the variation of the final term in the action (interpreted physically as 1d ``$E$'' and ``$J$'' potential terms) is $Q\int \big(\mb a\mb Y+\mb X\bar{\mb a}\big) = \int 2 \mb X\mb Y$, which precisely cancels the 1d interface term in the variation of \eqref{SDkN-0}. Schematically:
\be \raisebox{-.5in}{\includegraphics[width=2.2in]{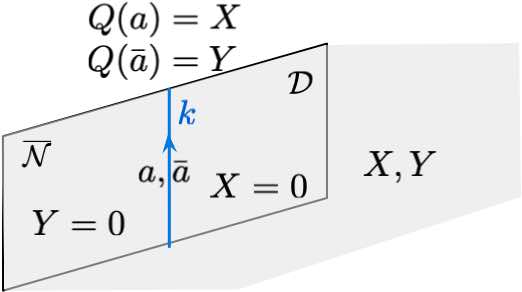}}  \label{matter-k-int} \ee

The new 1d fermions (leading components $a,\bar a$ of $\mb a,\bar{\mb a}$) satisfy an anti-commutation relation $[a^i,\bar a_j] = \hbar\, \delta^i_j$ when inserted at different positions along $\R_t$. The BRST operator acts on them as
\be Qa =  X\,,\qquad Q\bar a = Y\,. \label{trivXY} \ee
Thus, at the $k$ interface, both $X$ and $Y$ are \emph{effectively} set to zero in cohomology, by becoming $Q$-exact.

\subsubsection{Completeness}
\label{sec:matter-comp}

For completeness, we consider the bulk theory on two disjoint half-lines $\R^2\times \{s\leq 0\} \;\sqcup\; \R^2 \times  \{s\geq 0\}$, with an $\CN$ b.c. for the left half, and a $\CD$ b.c. for the right half. The appropriate BV action for this configuration is
\be S_{\rm split} = \frac{1}{\hbar} \int_{\R^2\times\{s\leq 0\}}^{(\CN)} \mb Y d\mb X -  \frac{1}{\hbar} \int_{\R^2\times\{s\geq 0\}}^{(\CD)} \mb X' d\mb Y'\,, \label{matter-Ssplit} \ee
using primes to distinguish the fields on the two sides. We claim that in order to glue these back together, it suffices to add a  boundary superpotential $-XY'$, or in BV terms
\be S_{\rm glue} = -\frac{1}{\hbar}\int_{\R^2} \mb X\big|_0 \mb Y'\big|_0\,,  \label{matter-Sglue} \ee
that couples the fields on the two sides.

The $-\mb X\mb Y'$ superpotential here, which has homological degree $+2$ (as required for a boundary superpotential) is the 2-Maurer-Cartan element we referred to back in the statement of completeness in Section \ref{sec:comp}. We suspect it is a continuous/dg analogue of the element in Example 5.17 of \cite{FMT}, in the analysis of global symmetries.

Following \cite{KRS}, a quick, geometric way to argue that the theory with combined action $S_{\rm split} + S_{\rm glue}$ is equivalent to a single, smooth 3d theory on $\R^2\times \R_s$ is to observe that the superpotential $-XY'$ is the generating function whose graph deforms the factorized Lagrangian $\big(V\times\{0\}\big)\times \big(\{0\}\times V^*\big) \subset T^*V\times T^*V$ (corresponding to separate $\CN$ and $\CD$ b.c. as in \eqref{matter-Ssplit}) to the diagonal $\Delta_{T^*V} \subset T^*V\times T^*V$.

More directly, we may consider the BRST variation of bulk fields (or the usual physical variation of the bulk+boundary action) $S_{\rm split} + S_{\rm glue}$. The variations with boundary terms are
\be Q\mb Y(s) = d\mb Y(s)+\delta(s)\big[\mb Y\big|_0 - \mb Y'\big|_0\big]\,,\qquad Q\mb X'(s) = d\mb X'(s)-\delta(s)\big[\mb X'\big|_0 - \mb X \big|_0\big]\,, \ee
which set $\mb X|_0 = \mb X'|_0$ and $\mb Y|_0=\mb Y'|_0$, thus gluing the theory back together.

Note that the deformation $S_{\rm glue}$ can be generalized so that it makes sense locally, in any region $R$ that's well bounded, in the sense of Section \ref{sec:comp} (Definition \ref{def-bounded}). Recall that such a region has boundary $\pd R$ consisting entirely of components $\bar\gamma_\alpha$ that look like a $\CN_k\ol\CD$ junction for the left half-theory, where $\mb X$ will be trivialized by some fermions $\mb a_\alpha$ (with $Q\mb a_\alpha=d\mb a_\alpha+\mb X$); and components $\gamma_\beta$ that look like a $\CD_k\ol\CN$ junction for the right half-theory, where $\mb Y'$ will be trivialized by some fermions $\bar{\mb b}_\beta$ (with $Q\bar {\mb b}_\beta=d\bar{\mb b}_\beta+\mb Y'$). To define the gluing deformation on $R$, we just dress it with the appropriate boundary fermions:
\be S_{\text{glue},R} = -\frac{1}{\hbar}\int_R \mb X\big|_0\mb Y'\big|_0 + \sum_{\bar\gamma_\alpha\in \pd R} \frac1\hbar\int_{\bar\gamma_\alpha} \mb a_\alpha \mb Y'\big|_{0,\pd R}+\sum_{\gamma_\beta\in\pd R} \frac1\hbar\int_{\gamma_\beta} \mb X\big|_{0,\pd R} \bar{\mb b}_\beta - \sum_{\bar\gamma_\alpha\cap\gamma_\beta}\frac{1}{\hbar} \big(\mb a_\alpha \bar {\mb b}_\beta\big)\big|_{\gamma_\alpha\cap\bar\gamma_\beta} \,. \label{matter-SglueR} \ee
See \eqref{XY-R} for an example. Then the BRST variation of the first term has a boundary term
\be Q\bigg( -\frac1\hbar \int_R  \mb X\big|_0\mb Y'\big|_0\bigg) = -\frac1\hbar \int_R d \big( \mb X\big|_0\mb Y'\big|_0\big) = -\frac1\hbar \int_{\pd R}  \mb X\big|_{0,\pd R}\mb Y'\big|_{0,\pd R}\,,  \ee
which is cancelled by the BRST variation of the middle terms, which may also have a boundary term --- which is finally cancelled by the BRST variation of the final term, an insertion of $a_\alpha\bar b_\beta$ at the vertices of $R$. Altogether, $Q(S_{\text{glue},R})=0$.

Heuristically, one could say that what's happening here is that either $\mb X$ or $\mb Y'$ is effectively being set to zero at each boundary component of $R$, so all boundary terms in $Q(S_{\text{glue},R})=0$ vanish. However, it's important that only one of $\mb X$ or $\mb Y'$ are set to zero (not both), or extra 1d degrees of freedom would be introduced and the gluing across $R$ would not be smooth. This is made clear by carefully keeping track of \emph{how} $\mb X,\mb Y'$ are set to zero at `$k$' interfaces.

\subsection{Spark algebras via descendants}
\label{sec:matter-sparks}

Let's now write down the spark algebras on $\CD$ and $\CN$.

On a $\CD$ boundary, the $Q$-cohomology of local operators is given by polynomials in the $Y^i$ scalars (zero-form components of $\mb Y$) that survive there. The $X_i$'s are set to zero. Similarly, local operators on $\CN$ are polynomials in $X_i$,
\be A_\CD = \C[Y^i]_{i=1}^n = \text{Sym}^\bullet(V[-1])\,,\qquad A_\CN = \C[X_i]_{i=1}^n = \text{Sym}^\bullet(V*[-1])\,. \label{matter-bdyops} \ee
Just as for bulk local operators \eqref{matter-bulkops}, the boundary operators are invariant under deformations of their insertion points, since if we take an open path $\gamma$ with endpoints on either $\CD$ or $\CN$  the 1-form components of $\mb X,\mb Y$ satisfy
\be Q\int_\gamma Y^{(1)} = Y\big|_{\pd\gamma} \quad\text{(on $\CD$)}\,,\qquad  Q\int_\gamma X^{(1)} = X\big|_{\pd\gamma} \quad\text{(on $\CN$)}\,.\ee

Now consider a strip of $\CD$ with coordinate $s\in [0,1]$ along its width, bounded by a $k$ interface on both sides. \emph{Both} $X$ and $Y$ are set to zero at $k$. Therefore, the integrals $\psi_+^i= \int_{[0,1]} Y^i{}^{(1)}$ define $Q$-closed operators.  In BV formalism, where additional fermions are used to set $Y=0$ at $k$, we should modify the integrals of $Y^{(1)}$ to
\be \psi_+^i:= \int_{[0,1]} Y^i{}^{(1)} - \bar a^i +\bar b^i = \quad \raisebox{-.3in}{\includegraphics[width=1.2in]{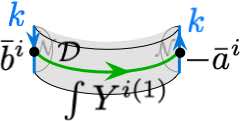}}\,. \label{def-psi-}\ee
Here $\bar a$ and $\bar b$ are 1d fermions inserted at $s=1$, $s=0$, respectively, to define $k, k^*$ as in \eqref{SDkN}; they satisfy $Q(\bar a) = Y(s=1)$ and $Q(\bar b)=Y(s=0)$, thus ensuring $Q\psi_+^i=0$.

We claim that the $\psi_+^i$ generate the spark algebra on $\CD$. We will verify this by (1) computing here the algebra structure on the $\psi_+^i$'s; and (2) arguing in Section \ref{sec:matter-spaces} that the algebra they generate in fact spans the space of operators on the $\CD$ strip.

To compute the algebra structure, we proceed as follows. Denote by $t$ the coordinate along the infinite length of the strip. Then observe that insertions of $\int_{[0,1]} Y^i{}^{(1)}$ at different $t$'s have non-singular correlation functions. This is easy to prove in the BV formalism. The bulk propagator of our theory couples (various form degrees of) $\mb X$ to $\mb Y$, as the kinetic term is simply $\mb Y d\mb X$; and there are no interaction vertices. Boundary conditions modify the functional form of the propagator, but not the basic fact that it couples $\mb X$ to $\mb Y$. Thus, there are no possible contractions of any number of insertions of integrals of $Y^i{}^{(1)}$.

Similarly, the fermions $\bar a^i, \bar b^i$ at the endpoints of sparks have nonsingular correlation functions, among themselves and with the integrated $Y^{(1)}$'s. The 1d propagator from \eqref{SDkN} couples $\bar{\mb a}$ to $\mb a$  (similarly for $\bar{\mb b}$ and $\mb b$), and the quadratic 1d vertices from \eqref{SDkN} do not allow any Feynman diagrams connecting $Y$'s with $\bar a$'s or $\bar b$'s.

Thus, perturbatively, correlation functions of the $\psi_+^i$ are nonsingular. There is also no room for nonperturbative corrections, as the equations of motion (\emph{a.k.a.} $Q$-fixed points) in the 3d B model are constant fields, and do not allow nontrivial instantons. We conclude that the algebra generated by the $\psi_+^i$, which are fermionic, is a \emph{free exterior algebra}, with anti-commutators
\be  [\psi_+^i,\psi_+^j]=0\quad\forall\;i,j=1,...,n\,. \ee
This algebra is also denoted  $\Lambda^\bullet V$.

Similarly, sparks on a strip of $\CN$ include integrated descendants
\be \psi^-_i := \int_{-[0,1]} X_i^{(1)} + a_i - b_i = \quad \raisebox{-.3in}{\includegraphics[width=1.2in]{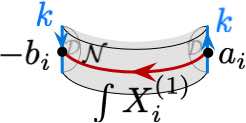}} \,.\label{matter-X-ferm} \ee
These generate a free exterior algebra $\Lambda^\bullet V$. Our claim is that
\be \CH_\CD \simeq \Lambda^\bullet V = \C[\psi_+^i]_{i=1}^n\,,\qquad  \CH_\CN \simeq \Lambda^\bullet V^* = \C[\psi^-_i]_{i=1}^n\,. \label{matter-HDN} \ee
Note that even though the $\psi_+,\psi^-$ are fermions, they lie in homological degree zero.

\subsubsection{As state spaces}
\label{sec:matter-spaces}

One way to argue that the exterior algebras in \eqref{matter-HDN} span the full spark algebras on $\CD$ and $\CN$ is to compare them with spaces of states on a rectangle, with two sides bounded by $\CN$ and two sided bounded by $\CD$, as in \eqref{Rectangles}. (The way we have set up the 3d B model, is no functional difference between $\CN,\CD$ and $\ol\CN,\ol\CD$.)

Let's use topological invariance to shrink the $\CD$ edges of the rectangle to an infinitessimal width:
\be \raisebox{-.3in}{\includegraphics[width=4.2in]{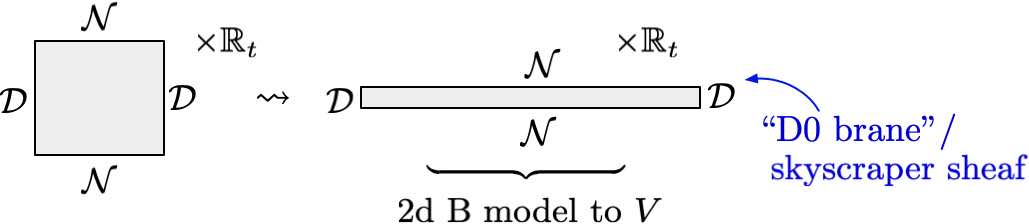}} \label{XY-boxQM} \ee
Then we end up with the 2d $(\CN,\CN)$ sandwich theory on a strip. The $(\CN,\CN)$ sandwich theory was described back in \ref{sec:matter-trans} as a B-model with target $V$. It has BV action
\be  \frac1\hbar \int_{\R_t\times[0,1]} \bm\psi^x d\mb x\,, \ee
where $\mb x = \mb X|_{\text{$\CN$ bdy}}$ and $\bm\psi^x= - \iota_\perp \mb Y|_{\text{$\CN$ bdy}}$. Moreover, the $\CD$ b.c. from 3d reduce to boundary conditions at the two ends of this strip; in 3d they set $\mb X|_{\text{$\CD$ bdy}} = 0$ and now in 2d they set $\mb x|_0=\mb x|_1=0$. The 2d B-model then reduces further to 1d B-model on $\R_t$, a B-twisted fermi multiplet valued in $V$. It has fields $\bm\psi^- = \bm\psi^x|_0 \in \Pi V \otimes \Omega^\bullet(\R_t)$ and $\bm \psi_+ = 2\iota_\perp \bm x|_0 \in \Pi V^* \otimes \Omega^\bullet(\R_t)$, with action
\be \frac1{2\hbar} \int_{\R_t} \bm\psi^- d\bm\psi_+\,. \ee

The quantized state space of this fermi multiplet depends on a choice of polarization. In $Q$-cohomology, the algebra of operators in the quantum mechanics is unambiguously a Clifford algebra, in the leading components of $\bm\psi^-,\bm\psi_+$\,:
\be  [\psi^-_i,\psi_+^j] = \hbar\, \delta^i{}_j\,. \ee
One natural polarization leads to a Hilbert space that's the underlying vector space of the exterior algebra $\C[\psi^-_i]_{i=1}^n$ (\ie\ a fermionic Fock space), with the operators $\psi^-_i$ acting as multiplication and $\psi_+^j$ as derivatives. A dual polarization leads to a Hilbert space $\C[\psi_+^i]_{i=1}^n$, with  $\psi^-_i$ acting as derivatives and $\psi_+^j$ as multiplication.

The space of states on the rectangle in these two dual polarizations clearly correspond, respectively, to the spark algebras on $\CN$ and on $\CD$ proposed in \eqref{matter-HDN}. So we have found all the possible sparks!

We note that the two polarizations are nearly isomorphic. They are distinguished only if one keeps track of the $U(n)$ flavor symmetry of the 3d theory, as states in $\C[\psi^-_i]_{i=1}$ and $\C[\psi_+^i]_{i=1}^n$ transform in dual representations.

\subsubsection{The antipode and counit}
\label{sec:matter-antipode}

Next, consider the antipode. From the topological definition \eqref{Hopf-S}, it follows that $S(\psi_+^i)$ is given by the integral of $Y^i{}^{(1)}$ on a strip of $\CD$, but with opposite orientation. Correcting by fermions on the $k$ interfaces to get a $Q$-closed operator, we get
\be S(\psi_+^i) = \int_{-[0,1]} Y^i{}^{(1)} +\bar a^i - \bar b^i = -\psi_+^i \,.\ee
By an identical argument, for sparks on $\CN$, we have
\be S(\psi^-_i) = \int_{[0,1]} X_i^{(1)} -a_i+b_i = -\psi^-_i\,.\ee

The unit `1' in our algebras is of course always the identity, or trivial spark. The counit on any generator is
\be \varepsilon(\psi_+^i) = \varepsilon(\psi^-_i) = 0 \qquad\forall\,i\,, \ee
because the loop representing the counit of a generator may be shrunk to zero size (up to $Q$-exact terms):
\be \raisebox{-.3in}{\includegraphics[width=3.7in]{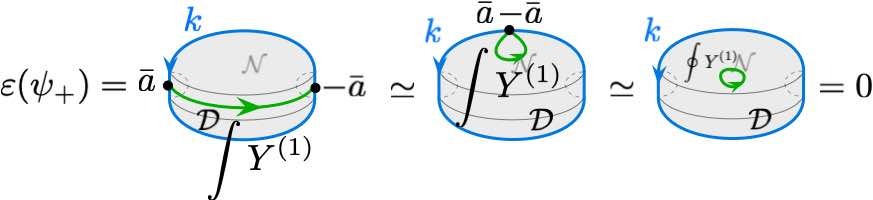}}  \ee

\subsubsection{The coproduct}
\label{sec:matter-coproduct}

Now we derive the coproduct on the two spark algebras $\CH_\CD$ and $\CH_\CN$.

Technically, we do not actually need to compute the coproduct at all. Prop. \ref{Prop:Hopf} \emph{guarantees} that $\CH_\CD$ and $\CH_\CN$ are Hopf-paired; and the dual of the graded-commutative product \eqref{matter-HDN} is
\be \Delta(\psi^-_i) = \psi^-_i\otimes 1 + 1 \otimes \psi^-_i\,,\qquad \Delta(\psi_+^i) = \psi_+^i\otimes 1 + 1\otimes \psi_+^i\,. \label{matter-coproduct} \ee

We will nevertheless see that it's easy to derive \eqref{matter-coproduct} from first principles. Consider a spark on $\CD$ given by $\psi_+^i = \int_{[0,1]} Y^i{}^{(1)} - \bar a^i +\bar b^i $ as in \eqref{def-psi-}. To compute its coproduct, we consider the double-pants configuration 
\be \raisebox{-.4in}{\includegraphics[width=5.7in]{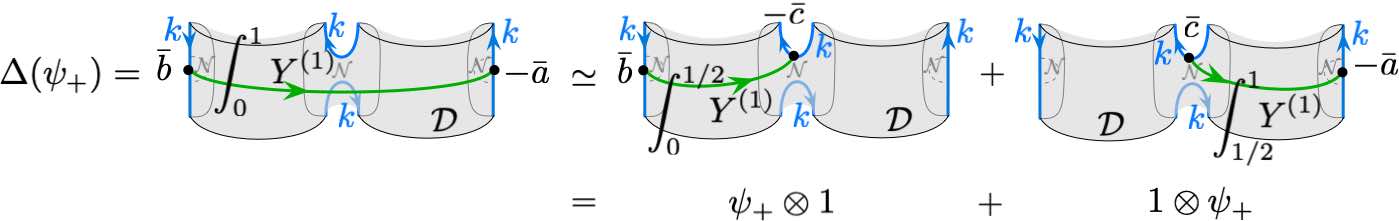}}\ee
from \eqref{Hopf-D-Sweedler}. Moving the integrated descendant upward, we can split the integral (using simple linearity) into two halves $\int_{[0,1]} Y^i{}^{(1)} = \int_{[0,1/2]} Y^i{}^{(1)} + \int_{[1/2,1]} Y^i{}^{(1)}$. We can also add and subtract the fermion $\bar c_i$ on the new $k$ interface to obtain
\be \Delta(\psi_+^i) = \bigg[\int_{[0,1/2]} Y^i{}^{(1)} - \bar c^i+\bar b^i \bigg] + \bigg[\int_{[1/2,1]} Y^i{}^{(1)} - \bar a^i+\bar c^i \bigg] \ee
The two terms are independently $Q$-closed, and (since they are each identical to the original spark integral, up to a rescaling of the interval) represent the sparks $\psi_+^i\otimes 1$ and $1\otimes \psi_+^i$.

The derivation of the coproduct for sparks on $\CN$ is identical.

\subsubsection{The Hopf pairing}
\label{sec:matter-Hopf}

The Hopf pairing $h:\CH_\CD\otimes \CH_\CN\to \C$ might be obvious, given the interpretation of $\CH_\CD$ and $\CH_\CN$ in Section \ref{sec:matter-spaces}, as quantizations of the space of states on a rectangle, in dual polarizations. That calculation immediately suggests 
\be h(\psi_+^i,\psi^-_j) = \hbar\, \delta^i{}_j\,.  \label{matter-Hopf} \ee
for generators, and more generally
\be h(\psi_+^{i_1}\psi_+^{i_2}...\psi_+^{i_k},\psi^-_{j_1}\psi^-_{j_2}...\psi^-_{j_{k'}}) = \hbar^k\frac{\pd}{\pd \psi^-_{i_1}}\frac{\pd}{\pd \psi^-_{i_2}}...\frac{\pd}{\pd \psi^-_{i_k}}(\psi^-_{j_1}\psi^-_{j_2}...\psi^-_{j_{k'}})\Big|_{\psi^-\equiv 0}  \label{matter-Hopf-gen} \ee
We can derive \eqref{matter-Hopf} in two other ways, each providing valuable insight.

First, recall the topological definition \eqref{link-Hopf}, \eqref{def-h} of the Hopf paring, as an insertion of two sparks in a solid ball with hemispheres bounded by $\CD$ and $\CN$. Consider an insertion of generators $\psi_+^i$ and $\psi_-^j$. There is now just a single pair of fermions $a^i,\bar a_i$ running along the $k$ interface between $\CD$ and $\CN$, which contributes to both spark integrals: \vspace{-.2in}
\be \hspace{-.2in}  \raisebox{-.4in}{\includegraphics[width=5.5in]{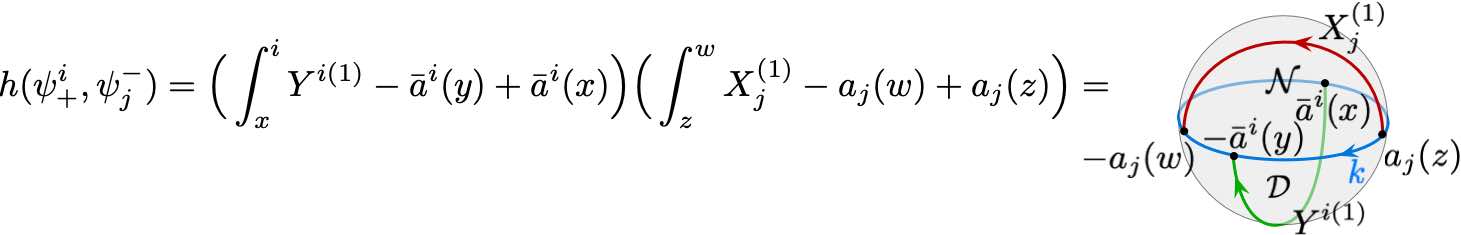}}   \label{matter-Hopfdef} \ee
We can use topological invariance to move (say) the rear endpoint of the $\psi_+^i$ spark (connected to an $\bar a_i$ fermion) to the right. However, if we try to pass it through the right endpoint of the $\psi^-_j$ spark (connected to an $a^j$ fermion), the Clifford commutation relation from \eqref{SDkN} implies
\be \raisebox{-.4in}{\includegraphics[width=4in]{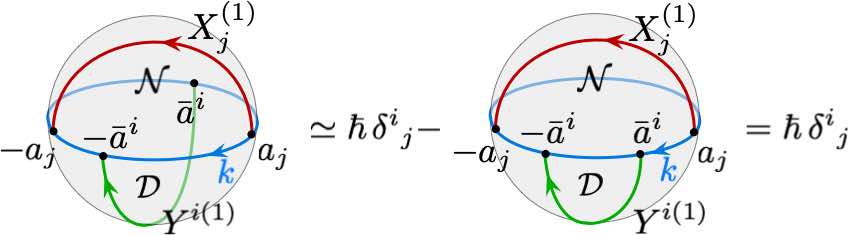}}  \ee
In the configuration on the right, both sparks can be further shrunken down to zero size, so the configuration evaluates to zero, leaving behind \eqref{matter-Hopf}.

Alternatively, we could start with \eqref{matter-Hopfdef} and move the two integrated descendants into the bulk. First, we can deform the $\psi^-_j$ spark off of the $\CN$ boundary:
\be  \raisebox{-.4in}{\includegraphics[width=5in]{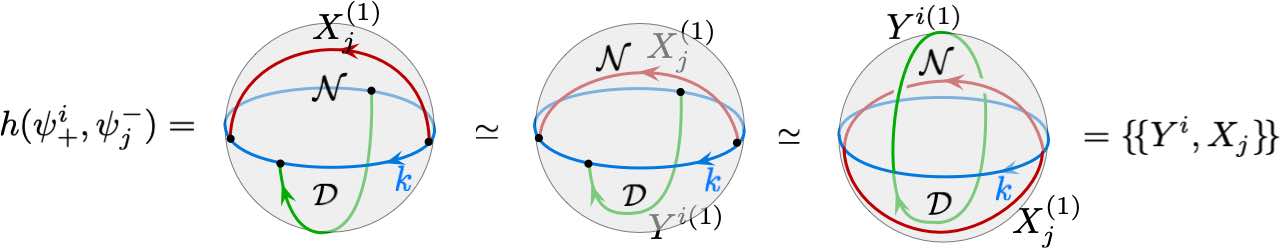}}  \ee
Then as shown, using the boundary condition $\mb Y|_{\text{$\CN$ bdy}}=0$, we can pull the $\psi_+^i$ spark entirely into the bulk, where it's represented by an integral of $Y^{i(1)}$ on a closed loop. Finally, we pull the $\psi^-_j$ spark into the bulk, where it's represented by an integral of $X_j^{(1)}$ closed loop. The two loops are linked once, in a Hopf link. Their correlation function precisely computes the Poisson bracket of bulk local operators as in \eqref{def-PB}, showing that the Hopf pairing is \emph{equivalent} to the Poisson bracket:
\be  h(\psi_+^i,\psi^-_j) = \{\!\{ Y^i,X_j \}\!\} = \hbar\,\delta^i{}_j \label{matter-HopfPoisson} \ee

\subsubsection{The bulk spark algebra}
\label{sec:matter-bulk}

The bulk spark algebra $U$ must be the quantum double formed from $\CH_\CD\otimes \CH_\CN$ with the Hopf pairing $h$. Algebraically, it is a straightforward computation to see that, as a Hopf algebra, $U$ is the free exterior algebra
\be U \simeq \Lambda^\bullet(V\oplus V^*) = \C[\psi_+^i,\psi^-_i]_{i=1}^n\,, \ee
with antipode $S(\psi) = -\psi$ on all generators and the same coproduct \eqref{matter-coproduct}.

Most of this is also easy to see directly from the definition of bulk sparks as in Section~\ref{sec:bulk}. In particular, the coproduct is derived from linearity of integrals of descendants, just as it was for boundary sparks. One aspect of $U$, however, deserves a special mention: the fact that generators $\psi_+^i$ and $\psi^-_j$ on opposite boundaries still commute with each other when they are considered as bulk sparks.

Physically, what is happening is the following. If we try to pass bulk sparks $\psi_+^i$ and $\psi^-_j$ through each other, the only potential failure of commutativity comes from their endpoints on the two $k$ interfaces:
\be \label{matter-pm-comm} \raisebox{-.3in}{\includegraphics[width=1.8in]{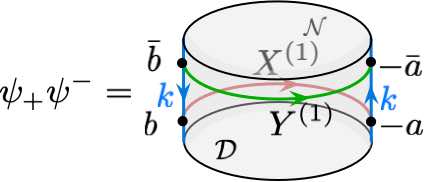}}\ee 
The interfaces support fermions $a,\bar a$ and $b,\bar b$, respectively, which do indeed have nontrivial Clifford commutation relations (that we already used in deriving the Hopf pairing). However, due to the opposite orientation of the two $k$'s, the contributions from the two sides cancel exactly. If we insert $\psi_+^i$ at time $t_2$ and $\psi^-_j$ at time $t_1$ into a correlation function, with $t_2>t_2$ and the times very close to each other, we get
\begin{align} \psi_+^i(t_2)\psi^-_j(t_1) &= \Big(\bar b^i(t_2) + \int Y^{i(1)}(t_2) - \bar a^i(t_2)\Big) \Big( b_j(t_1) + \int X_j^{(1)}(t_1) -  a_j(t_1)\Big)  \notag \\
&=  \bar b^i(t_2) b_j(t_1) +\bar a^i(t_2)a_j(t_1) + \text{terms regular at $t_2=t_1$} \notag \\ 
&= \hbar \delta^i{}_j + \bar b^i(t_1) b_j(t_2) -\hbar \delta^i{}_j + \bar a^i(t_1)a_j(t_2)  + \text{terms regular at $t_2=t_1$}  \notag\\
& = \Big(\bar b^i(t_1) + \int Y^{i(1)}(t_1) - \bar a^i(t_1)\Big) \Big( b_j(t_2) + \int X_j^{(1)}(t_2) -  a_j(t_2)\Big) \notag \\
&= - \Big( b_j(t_2) + \int X_j^{(1)}(t_2) -  a_j(t_2)\Big) \Big(\bar b^i(t_1) + \int Y^{i(1)}(t_1) - \bar a^i(t_1)\Big) \notag \\
&= - \psi^-_j(t_2)\psi_+^i(t_1)  \label{double-cancel}
 \end{align}
and we get our anti-commutation relation $[\psi_+^i,\psi^-_j]=0$ in $U$.

This edge contributions to the commutator, which cancelled here, are the same one that were captured by topology in Section \ref{sec:U-double}.

\subsubsection{The box gluing and the quantum group}
\label{sec:matter-R}

Finally, let's look at the R-matrix, and associated ribbon element, in $U$.

We know from Section \ref{sec:U-double} that the R-matrix must be the inverse of the Hopf pairing. To compute this, we should actually choose bases for the boundary spark algebras. Let $\big\{\psi_+^{i_1}\psi_+^{i_2}...\psi_+^{i_k}\big\}_{i_1<i_2<...<i_k}$ be a basis for $\CH_\CD$, which has dimension $2^n$. Then under the general pairing \eqref{matter-Hopf-gen}, a dual basis for $\CH_\CN$ is given by $\big\{\hbar^{-k}\psi^-_{i_k}...\psi^-_{i_2}\psi^-_{i_1}\big\}_{i_1<i_2<...<i_k}.$ The R-matrix then becomes
\be \CR \;= \hspace{-.4cm} \sum_{\scriptsize\begin{array}{c} i_1<i_2<...<i_k \\ 0 \leq k \leq n \end{array}} \hspace{-.2cm} (-1)^{k^2}\frac{1}{\hbar^k}\psi_+^{i_1}\psi_+^{i_2}...\psi_+^{i_k}\otimes \psi^-_{i_k}...\psi^-_{i_2}\psi^-_{i_1} \;=\; \exp \lp -\frac{1}{\hbar} \sum_{i=1}^n \psi_+^i \otimes \psi^-_i\rp \,.\label{matter-R} \ee
From Prop. \ref{Prop:R}, the associated ribbon element is the Drinfeld element
\be v  = \exp \bigg( -\frac{1}{\hbar} \sum_{i=1}^n  \psi^-_i \psi_+^i\bigg) \,. \label{matter-ribbon} \ee

Again, there is another useful perspective to give on the R-matrix. Recall from Section \ref{sec:U-double} that the R-matrix is essentially given by the linear combination of sparks used in the box gluing. In turn, the box gluing can \emph{either} be obtained from factoring the identity operator in the (box)$\times\R_t$ Hilbert space into (states)$\cdot$(dual states), leading directly to \eqref{matter-R}; or it can be obtained as a general consequence of completeness. We want to explain how to get the box gluing from completeness (\AC$_{\rm box}$).

Consider gluing along a box region $R$ as follows:
\be \raisebox{-.5in}{\includegraphics[width=5.5in]{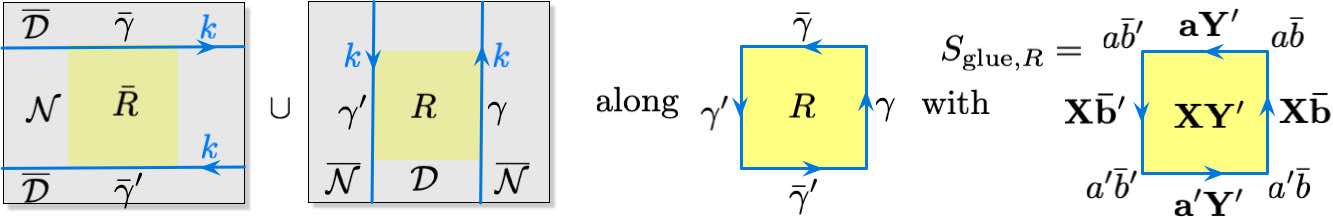}} \label{XY-R}\ee
The prescription of \eqref{matter-SglueR} tells us to glue by adding a boundary action $S_{\text{glue},R} = -\frac1\hbar\int_R\mb X\mb Y'+\ldots$, with correction terms, indicated in \eqref{XY-R}, involving fermions on $\pd R$ to ensure that $S_{\text{glue},R}$ is $Q$-closed.

We claim that, up to $Q$-exact terms, the boundary action is equivalent to
\be S_{\text{glue},R}\simeq   \Big( \int_{\bar\gamma'} X^{(1)}-a+a'\Big)\frac{1}{\hbar}\Big( \int_{\gamma} Y^{(1)}{}'-\bar b+\bar b'\Big) = \frac1\hbar \psi^-\psi_+{}'  \label{matter-Sglue-simp}\ee
so that after exponentiating and splitting apart the two sparks (as in \eqref{fig:R-glue4}), we recover the R-matrix \eqref{matter-R}. The identity \eqref{matter-Sglue-simp} follows from generalizing Riemann's bilinear identity to the multiforms $\mb X$ and $\mb Y'$, which aren't closed, but closed up to $Q$-exact terms, since they satisfy $(d-Q)=0$.

We'll give a simplified derivation that neglects the fermions $a,a',\bar b,\bar b'$ that trivialize $\mb X,\mb Y'$ at various parts of $\pd R$; we'll just assume that either $\mb X=0$ or $\mb Y'=0$ on the nose at the appropriate boundaries. (It's straightforward to add the fermions back in, but they make the derivation even more tedious.) The \emph{rough} idea is that we can expand
\be \int_R \mb X\mb Y' = \int_R X Y^{(2)}{}' + \int_R X^{(1)} Y^{(1)}{}' + \int_R X^{(2)} Y' \ee
and then we expect each term to factor: $\int_R X Y^{(2)}{}' \approx X \cdot \int_R Y^{(2)}{}'$, $\int_R X^{(2)} Y' \approx Y'\cdot \int_R X^{(2)}$, $\int_R X^{(1)} Y^{(1)}{}' \approx \int_{\gamma} X^{(1)} \int_{\bar\gamma'} Y^{(1)}{}'$.  If this were true, then we could just bring the local operators $X,Y'$ to any boundary where they vanish to kill their terms; and the surviving term involving 1-forms becomes the product of sparks $\psi^-\psi_+$. It is not true that each term factors separately; but they do factor all together.

Let's put coordinates $(s,t)\in [0,1]\times [0,1]$ on the box, introduce the contour $\eta(s,t)$: 
\be \raisebox{-.4in}{\includegraphics[width=1.8in]{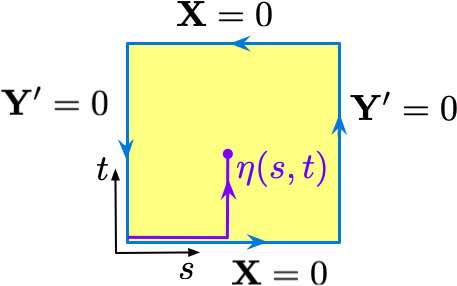}} \ee
and define $f(s,t) = \int_{\eta(s,t)} Y^{(1)}{}'$, noting that $Qf(s,t) = Y'(s,t) - Y'(0,0) = Y'(s,t)$. Then we rewrite
\begin{align}
\int_R  X^{(2)}Y'  &= \int_R X^{(2)} Qf =  - \int_R QX^{(2)} f \notag + Q(...) \notag \\
 &\simeq -\int_R d X^{(1)} f  = -\int_R d(X^{(1)} f) - \int_R X^{(1)} df \notag \\
 &= - \int_{\pd R} X^{(1)} f - \int_R X^{(1)} Y^{(1)}{}' - \int_R X^{(1)}\Big( \int_{(s,0)}^{(s,t)} \iota_{ds} dY^{(1)}{}' \Big) ds 
\end{align}
where we have $Q$-exact terms. The final term here simplifies further to
\begin{align}
 - \int_R X^{(1)}\Big( \int_{(s,0)}^{(s,t)} \iota_{ds} dY^{(1)}{}' \Big) &=  - \int_R X^{(1)}\Big( \int_{(s,0)}^{(s,t)} \iota_{ds} QY^{(2)}{}' \Big) ds   \simeq  -\int_R QX^{(1)} \Big( \int_{(s,0)}^{(s,t)} \iota_{ds} Y^{(2)}{}' \Big) ds \notag \\
 &=  -\int_R dX \Big( \int_{(s,0)}^{(s,t)} \iota_{ds} Y^{(2)}{}' \Big) ds 
 = -\int_{\pd R} X  \Big( \int_{(s,0)}^{(s,t)} \iota_{ds} Y^{(2)}{}' \Big) ds - \int_R X Y^{(2)}{}'\,. \notag
\end{align}
The boundary term in this last expression evaluates to zero: its $ds$ integration measure only allows it to get contributions from top and bottom edge of the box, where $X=0$. Putting everything together, we find $\int_R  X^{(2)}Y'  = - \int_{\pd R} X^{(1)} f - \int_R X^{(1)} Y^{(1)}{}'  - \int_R X Y^{(2)}{}'$ up to $Q$-exact terms, or in other words
\be -\frac1\hbar \int_R\mb X\mb Y' = \frac1\hbar \int_{\pd R} X^{(1)} f = \frac{1}{\hbar} \int_{\bar\gamma'} X^{(1)}\int_\gamma Y^{(1)}{}'  \underset{\text{no fermions}}{=} \frac1\hbar \psi^-\psi_+{}'\,. \label{Riem-bil} \ee
exactly as desired in the absence of boundary fermions to get \eqref{matter-Sglue-simp}.

\subsubsection{Summary}\label{sec:matterU}

Let's summarize the structure of boundary and bulk spark algebras that we've just found, through direct calculations.

The boundary spark algebras are exterior algebras, with co-commutative coproducts, and simple antipodes and counits:
\be \begin{array}{c} \CH_\CD = \Lambda^\bullet V = \C[\psi_+^i]_{i=1}^n\,,\qquad  \CH_\CN = \Lambda^\bullet V^* = \C[\psi^-_i]_{i=1}^n\,, \\[.2cm]
\ \Delta \psi = \psi\otimes 1+1\otimes \psi\,,\qquad S(\psi) = -\psi \,,\qquad \varepsilon(\psi)=0\qquad \forall\,\text{generators $\psi_+,\psi^-$}\,. \end{array} \label{XY-sum-HD} \ee
These are Hopf \emph{super}algebras. Just like the underlying B-twisted 3d theories, they have a $\Z_2$ fermion-number grading, which adds signs in commutators. They could have an additional cohomological $\Z$ grading, from $U(1)_H$ R-symmetry in the 3d theory, but they turn out to lie entirely in \emph{cohomological degree zero}.

The Hopf pairing acts on generators as
\be h(\psi_+^i,\psi^-_j) = \delta^i{}_j\,,\ee
leading to a bulk spark algebra given by the double $U \simeq \CH_\CD\otimes \CD_\CN$, which still turns out to be an exterior algebra
\be U = \Lambda^\bullet(V\oplus V^*) = \C[\psi_+^i,\psi^-_i]_{i=1}^n \ee
with the same coproduct and antipode as above, and R-matrix and ribbon element
\be \CR =  \exp \lp -\frac{1}{\hbar} \sum_{i=1}^n \psi_+^i \otimes \psi^-_i\rp  \,,\qquad  v  = \exp \lp -\frac{1}{\hbar} \sum_{i=1}^n  \psi^-_i\psi_+^i\rp \,. \ee

\subsection{Categories and fiber functors}
\label{sec:matter-cat}

Now let's look at dg module categories for the Hopf algebras we've found, and compare them to known categories of lines operators in Rozansky-Witten theory.

\subsubsection{Boundary lines and Koszul duality}
\label{sec:matter-bdycat}

We'll start at the boundary, say with the category $\CC_\CN$. This was identified by \cite{KRS} as the derived category of coherent sheaves on the affine space $V$
\be \CC_\CN^{fd} \simeq \text{Coh}(V)\,.  \label{CohV} \ee
It will actually correspond to the dualizable subcategory for us, and we anticipate this by writing `$fd$'.
This description comes from compactifying the 3d theory on the link `$C$' of a line on $\CN$:
\be \raisebox{-.3in}{\includegraphics[width=3.1in]{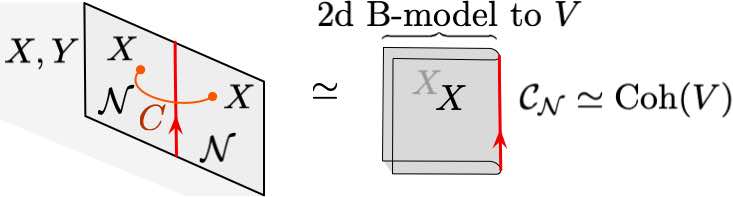}}  \label{matter-book} \ee
Just as in \eqref{XY-boxQM}, this results in a 2d B-model with target $V$. Then $\CC_\CN$ maps to its category of boundary conditions, which is given by \eqref{CohV}.

It's important to remember that the bosonic coordinates $X^i$ on $V$ have homological degree $+1$. This could be emphasized by writing \eqref{CohV} as $\text{Coh}(V[1])$. Alternatively, \eqref{CohV} means the dg category of finite-rank modules for the dg algebra $\C[X_i]_{i=1}^n = \text{Sym}^\bullet (V^*[-1])$, a polynomial algebra in $n$ commuting variables of degree 1:%
\footnote{We also emphasize once more, to avoid mathematical confusion: 3d B-twisted theories have a separate $\Z_2$ fermion-number grading that controls commutativity/signs, unrelated to the homological degree. The variables in \eqref{CohV} satisfy $x^ix^j=x^jx^i$. All the categories we're discussing here are $\Z\times\Z_2$ graded.}
\be \text{(more accurate)}\quad  \CC_\CN^{fd}  = \text{Coh}(V[1]) = \C[X]\text{-mod}^{fr}\,. \label{CohV-better} \ee

Due to completeness (Thm. \ref{Thm:faithful}), we expect an equivalence
\be \CF_\CN: \CC_\CN^{fd} \overset\sim\to \CH_\CD\text{-mod}^{fd} = \C[\psi_+]\text{-mod}^{fd}\,. \label{CohV-Koszul} \ee
The equivalence of dg categories \eqref{CohV-better} and \eqref{CohV-Koszul} is classic Koszul duality \cite{Priddy,beilinson1996koszul}. Our fiber functor $\CF_\CN$, which implements the equivalence, has the following explicit description. The strip $\kDk$ on an $\CN$ boundary sets $X^i =0$, and represents the module $\C[X]/(X^1,...,X^n)$ in $\C[X]\text{-mod}$, also known as the skyscraper sheaf $\CO_0\in\text{Coh}(V[1])$,
\be \kDk = \C[X]/(X^1,...,X^n)  = \CO_0 \;\in \CC_\CN\,. \ee
The derived endomorphism algebra of $\CO_0$ is the exterior algebra generated by tangent vectors at the origin, recovering our spark algebra (as expected from Section \ref{sec:spark-Yo}):
\be \text{End}_{\CC_\CN}(\kDk) = \text{End}_{\text{Coh}(V[1])}(\CO_0) = \C\Big[\tfrac{\pd}{\pd X_i}\Big]_{i=1}^n \simeq \Lambda^\bullet V = \CH_\CD^{(\rm op)}\,.\ee
Then $\CF_\CN$ sends any coherent sheaf to its derived Hom with the skyscraper sheaf,
\be \CF_\CN = \text{Hom}_{\CC_\CN}(\kDk,-) = \text{Hom}_{\text{Coh}(V[1])}(\CO_0,-)\,. \ee

Under Koszul duality, the tensor product in $\CH_\CD\text{-mod}$ coming from the coproduct \eqref{XY-sum-HD} in $\CH_\CD$ --- which is known as a convolution product --- maps to the ordinary tensor product of sheaves in $\text{Coh}(V[1])$. We have proven in the preceding physics derivations of products and dual coproducts that there are no quantum corrections to this. (This was expected from \cite{KRS}, given a flat target geometry $T^*V$).

Analogous statements hold for the opposite boundary category $\CC_\CD$. Namely, we have
\be \CC_\CD^{fd} = \text{Coh}(V^*[1])\,, \ee
with a fiber functor that implements Koszul duality
\be \CF_\CD: \CC_\CD^{fd} \overset\sim\to \CH_\CN\text{-mod}^{fd}\,,\qquad  \CF_\CD =\text{Hom}_{\CC_\CD}(\kNk,-)  =  \text{Hom}_{\text{Coh}(V^*[1])}(\CO_0,-)\,. \ee

\subsubsection{Dualizability}
\label{sec:matter-dual}

We note that the objects $\kDk$ and $\kNk$ are both dualizable in their respective categories. Indeed, though it was not  obvious from Section \ref{sec:matter-trans}, it can be checked easily now using \cite{KRS} methods that the interface $k$ between $\CD$ and $\ol \CN$ is dualizable, so our theory satisfies axiom \AD$_k$ (Section \ref{sec:trans}). This ensures that $\kDk$ and $\kNk$ are Frobenius algebra objects and (retroactively) explains why our Hopf algebras $\CH_\CN,\CH_\CD,U$ were finite dimensional.

We also note that line operators in this example have turned out to satisfy property \AD$_{\rm line}$ (Section \ref{sec:trans}). Namely, line operators are generated by compact, dualizable objects. As a warning, however, not all dualizable objects are compact. There are large categories $\CH_\CD\text{-mod}$,  $\CH_\CN\text{-mod}$ of all modules (possibly infinite-dimensional), which we might think of as all line operators in our theory. Inside them sit the finite-dimensional modules, which are the dualizable objects, and correspond to coherent sheaves above. Further inside sit the compact modules; they are the ``perfect'' modules, with finite projective resolutions by $\CH_\CD$ or $\CH_\CN$. In coherent sheaves, this amounts to being supported at the origin. Thus, \eg,
\be \hspace{-.1in} \begin{array}{ccccc}
\CC_\CN & \supset & \CC_\CN^{fd} & \supset & \CC_\CN^c \\
\rotatebox{90}{=} &&\rotatebox{90}{=}&&\rotatebox{90}{=} \\
\C[\psi_+]\text{-mod} &&  \C[\psi_+]\text{-mod}^{fd}\simeq \text{Coh}(V[1]) && \C[\psi_+]\text{-mod}^{\rm perf} \simeq \text{Coh}_0(V[1]) \end{array} \ee

\subsubsection{Bulk lines}
\label{sec:matter-bulkline}

The situation for bulk lines is very similar. From \cite{KRS}, we identify
\be \CC_\CT^{fd} \simeq \text{Coh}\big(T^*[2](V[1])\big)\,, \label{XY-CohTV} \ee
where $T^*[2](V[1])= (T^*V)[1]$ just denotes the ordinary bosonic cotangent bundle $T^*V$, with cohomological shifts such that the coordinates $X^i,Y_i$ each have degree 1. The line operator $\CDCN$ that represents the fiber functor here is a drilled-out cylinder bounded by a strip of $\ol\CD$ on one side and a strip of $\ol\CN$ on the other. It effectively sets \emph{both} $X$ and $Y$ to zero, and engineers a bulk line operator represented by the skyscraper sheaf at the origin of the cotangent bundle
\be \CDCN \simeq \CO_0\,,\qquad \text{End}_{\text{Coh}(T^*[2](V[1]))}(\CO_0) =\C\Big[\tfrac{\pd}{\pd X_i},\tfrac{\pd}{\pd Y^i}\Big]_{i=1}^n \simeq \C[\psi_+^i,\psi^-_i]_{i=1}^n = U^{(\rm op)} \ee
The fiber functor implements Koszul duality
\be \CF_\CT:\CC_\CT^{fd} \overset\sim\to U\text{-mod}^{fd}\,,\qquad \CF_\CT= \text{Hom}_{\text{Coh}(T^*[2](V[1]))}(\CO_0,-)\,.\ee

The coproduct in the Hopf algebra $U$ again matches the ordinary, uncorrected tensor product of coherent sheaves. Now, however, we've also got an R-matrix and ribbon element, encoding nontrivial braiding and twists of line operators. This might be surprising for line operators in a sigma-model with flat, linear target --- a free theory!

As an example of nontrivial braiding on the coherent-sheaf side, consider the line operators $\CO_V$ and $\CO_{V^*}$ (the structure sheaves of $V$ and $V^*$ inside $T^*V$). Their tensor product is $\CO_V\otimes \CO_{V^*} \simeq \CO_0$, the skyscraper at the origin. The R-matrix translates to the endomorphism
\be \CR = \exp \lp -\frac1\hbar \sum_i \frac{\pd}{\pd X_i}\wedge \frac{\pd}{\pd Y^i}\rp \;\in\;  \text{End}_{\text{Coh}(T^*[2](V[1]))}(\CO_0)\,. \label{XY-R-Coh} \ee
If we were working in an unshifted category $\text{Coh}(T^*V)$, the bi-vector $\frac{\pd}{\pd X_i}\wedge \frac{\pd}{\pd Y^i} $ would be an element of $\text{Ext}^2(\CO_0,\CO_0)$. However, our choice of cohomological degrees shifts it back down to a degree-zero endomorphism, so altogether $\CR$ defines a nontrivial degree-zero element in $\text{End}(\CO_0)$.

This R-matrix matches several known results on lines in Rozansky-Witten theory:
\begin{enumerate}[leftmargin=*]
\item Inspired by \cite{Kontsevich-RW, Kapranov-RW}, Roberts and Willerton \cite{RobertsWillerton} defined --- somewhat indirectly --- the structure of a braided tensor category on the derived category of coherent sheaves on an algebraic symplectic variety $\CX$. (Here $\CX$ is meant to be the target of a B-twisted 3d $\CN=4$ sigma model.) The basic idea was to think of the shifted (in cohomological degree and fermion number) tangent bundle $T[-1]\CX$ as a sheaf of metric Lie algebras. The Lie bracket is defined by the Atiyah class of $\CX$. The metric is define by the symplectic form $\Omega$. Then general coherent sheaves may be thought of as modules for $T[-1]\CX$, and acquire a perturbative quantization by solving a KZ equation \cite{drinfeld1990quasitriangular}. The result is a braided tensor category with simple braiding given by the inverse of the symplectic form, \emph{i.e.} a Poisson bivector
\be \CR\sim \exp \Omega^{-1} \label{RPBV} \ee
and a potentially complicated perturbative associator.

In our rather simple example, the affine space $\CX=T^*[2](V[1])$ is conical, so one might expect all information about the sheaf $T[-1]\CX$ to be captured by its stalk at the origin. The stalk at the origin is our exterior algebra:
\be T[-1]_0\CX = \C\big[\tfrac{\pd}{\pd X_i},\tfrac{\pd}{\pd Y^i}\big]_{i=1}^n \simeq \C[\psi_+^i,\psi^-_i] = U\,, \ee
where our cohomological shifts have put $\tfrac{\pd}{\pd X_i},\tfrac{\pd}{\pd Y^i}$ into degree zero (but the fermion-number shift remains and keeps them fermionic). Our symplectic form --- which controls the higher bracket of local operators \eqref{XY-PB} --- is $\hbar\sum_i dX_i\wedge dY^i$, and the Poisson bivector is $\Omega^{-1}= -\frac{1}{\hbar}\sum_i\tfrac{\pd}{\pd X_i}\wedge\tfrac{\pd}{\pd Y^i}$, whence \eqref{RPBV} agrees with our R-matrix $\CR= \exp\lp -\frac1\hbar\sum_i \psi_+^i\otimes \psi^-_i\rp$.

More precisely, in our example, we have found a monoidal fiber functor on the Roberts-Willerton category, by taking Hom with a skyscraper at the origin, which not only recaptures the expected R-matrix \eqref{RPBV} but \emph{trivializes the associator}. 

\item For any smooth stack $\CX$, one can attach to it a space $\CL\CX=\mathrm{Maps}(S^1, \CX)$, the so-called derived loop space of $\CX$. The loop in question is the link of bulk line operators, much as on the LHS of \eqref{DW-bulkcat}. It was shown in \cite{ben2012loop} that $\Coh (\CL \CX)$ has the structure of a braided tensor category, such that the infinitesimal twist is given by loop rotation on $S^1$. In \cite{riedel2019free}, it was shown that when $\CX$ is a smooth variety, there is an equivalence of categories (via Koszul duality)
\be
\Coh (\CL\CX)\simeq \text{Coh}\big(T^*[2]\CX\big),
\ee
under which the infinitesimal twist of the RHS is given by the Poisson bivector on $T^*[2]\CX$. Setting $\CX=V[1]$, this matches our expression for the twist, $\log v = -\frac1\hbar \psi_+\cdot\psi^-$.

\item There exists a holomorphic boundary condition for our B-twisted matter theory, supporting symplectic fermions valued in $V$, $SF(V)$ \cite{CG}. The VOA $SF(V)$ is one of the simplest examples of a logarithmic VOA:  its representation theory is  non-semisimple, and the definition of the braided tensor structure requires the theory of logarithmic intertwining operators (\cf\ \cite{huang2014logarithmic1}). In general, it is difficult to study the braided tensor category for logarithmic VOA's. Fortunately, in the case of $SF(V)$, the category $SF(V)\text{-mod}$ is well understood and was shown to have the structure of a rigid braided tensor category (see for instance, \cite{Abe:2011cd, runkel2014braided}). Moreover, it was shown in \cite{Farsad:2017eef} that the category (of generalized untwisted modules) is equivalent to modules of a quasi-triangular Hopf algebra. That quasi-triangular Hopf algebra is precisely our $U$, and
\be
SF(V)\text{-mod} \simeq U\text{-mod}\,,
\ee
as ribbon categories. The same is proved in \cite{creutzig2024kazhdan} using a different method.

None of these existing proofs make use of transverse boundary conditions, or the fiber functors associated with them.
\end{enumerate}

\section{3d $\CN=4$ gauge theory}
\label{sec:gauge}

In this section, we discuss an example that is both somewhat nontrivial, and slightly beyond the rigorous setting of the arguments from Sections \ref{sec:linefunctor}--\ref{sec:bulk}, due to a mild failure of dualizability. The example is the B-twist of 3d $\CN=4$ pure $G_c$ gauge theory, first introduced in \cite{BlauThompson}. (Throughout this section, we'll denote by $G_c$ a compact semisimple gauge group, and $G=(G_c)_\C$ its complexification, a reductive algebraic group.)
Our bulk theory $\CT$, the B-twist of $G_c$ gauge theory, may equivalently be thought of as
\begin{itemize}
\item $G$-equivariant Rozansky-Witten theory with target a point; or Rozansky-Witten theory to the symplectic stack $T^*(BG)$
\item 3d BF theory with gauge group $G_c$, quantized in the BRST formalism, keeping fields/operators of all ghost numbers (which becomes cohomological degree)
\item 3d $G_c\ltimes \fg$ Chern-Simons theory with off-diagonal level, in BRST quantization HERE
\item \emph{Roughly}, Dijkgraaf-Witten theory with infinite gauge group $G$ --- though as we will see, important features arise from keeping track of the topology of $G$, as opposed to just treating it as an infinite discrete group.
\end{itemize}

This example combines aspects of DW theory from Section \ref{sec:DW}, as well as B-twisted matter from Section \ref{sec:matter}. The constructions of spark algebras (Section \ref{sec:gauge-sparks}) will involve (rather subtle!) infinite-dimensional generalizations of the arguments from DW theory. Sparks on $\CN$ will come from framed Wilson lines, which generate algebraic functions on $G$. Sparks on $\CD$ will come from boundary $G_\pd$ symmetry defects, which generate the huge algebra of distributions on $G$, with a convolution product:
\be \begin{array}{rlll}
& \text{$G$ finite}:  & \CH_\CD=\C G\,, & \CH_\CN=\CO(G)\,; \\[.1cm] \leadsto &   \text{$G$ continuous}: & \CH_\CD = \text{Dist}(G)\,,& \CH_\CN = \CO(G)\,. \end{array} \label{gauge-spark-sum} \ee
While $\CH_\CN$ is a standard Hopf algebra, $\CH_\CD$ requires a topological completion in order for its coproduct to be defined (we'll explain carefully why). The module category for $\CH_\CD$ is just a (somewhat trivial) dg enhancement of the semisimple category $\text{Rep}(G)$; but the module category for $\CH_\CN$ is the highly nontrivial dg category $\QCoh(G)$, of quasi-coherent sheaves on the algebraic group $G$.

In the bulk (Section \ref{sec:gauge-bulk}, modules for the algebra $U \simeq \CH_\CD\ltimes \CH_\CN$ are equivalent to sheaves on $G$ equivariant for the adjoint action,
\be \CC_\CT \simeq U\text{-mod} \simeq \QCoh^{G_{ad}}(G)\,. \ee
As we'll explain, this is an expected description of bulk lines in $\CT$; \emph{e.g.} it's a simple case of the more general equivariant Rozansky-Witten theories used recently in \cite{OR-Chern,OR-TQFT}. It's moreover known that $\QCoh^{G_{ad}}(G)$ is the derived Drinfeld center \cite{Ben-Zvi:2008vtm} of $\text{Rep}(G)$ and is therefore braided --- the braiding's been described explicitly in \cite{bezrukavnikov2023equivariant}. It's much less clear that the standard braiding on $\QCoh^{G_{ad}}(G)$ obtained this way is the correct one for the physics of twisted 3d $\CN=4$ theory, especially when $G$ is nonabelian and the QFT is not free. Our spark analysis provides a direct construction of a generalized R-matrix and braiding element in $U$ that proves no quantum corrections enter to modify the expected answer: the braiding in B-twisted 3d $\CN=4$ theory \emph{does} coincide with the standard one on $\QCoh^{G_{ad}}(G)$.

Perturbatively, if we think of $\CT$ as a $\fg$ gauge theory rather than a $G_c$ gauge theory, the R-matrix has a simple form $\CR \sim \exp \sum_a \alpha^a\otimes\beta_a$, where $\{\alpha^a\}$ is a basis for $\fg$ and $\{\beta_a\}$ a dual basis for $\fg^*$. This connects our results to those of N. Aamand \cite{Aamand-BF} (and the related \cite{CWY-II}), who derived this R-matrix for perturbative BF theory from expectation values of crossed Wilson lines --- ultimately inspiring the current work.  We describe the full nonperturbative R-matrix in Section \ref{sec:gauge-R}.

In Section \ref{sec:gauge-cat}, we will discuss in some detail the categories of bulk and boundary line operators, fiber functors on them, and their relation to (putative) algebra objects $\kDk,\kNk,\CDCN$ and sparks. We analyze the parts of the formalism of ``Tannakian QFT'' that break, due ultimately to a (mild) failure of the dualizability conditions \AD$_k$ and \ADline. Note that \AD$_k$ \emph{must} fail in order to get infinite-dimensional spark algebras as in \eqref{gauge-spark-sum}. We also indicate how this failure can be circumvented in the current example.

We'll mention connections to some other perspectives on categories of line operators in B-twisted gauge theory in Section \ref{sec:gauge-persp}, including sheaves on loop spaces and modules for boundary VOA's. Boundary VOA's are not well understood when $G$ is nonabelian. A conjecture arising from 3d mirror symmetry is that $U\Mod$ should be equivalent to modules for Arakawa's ``chiral universal centralizer'' \cite{Arakawa-chiral} in this case.

\subsection{Fields and action} 
\label{sec:gauge-def}

One can recast the B twist of pure 3d $\CN=4$ gauge theory in twisted BV-BRST formalism, just as was done for pure matter. We review this briefly, for the sake of having self-contained computations; see \cite{Garner-formalism} for details.

Just like for the B twist of matter, the theory has a $\Z$-valued cohomological grading (physical $U(1)_H$ R-symmetry) and a separate $\Z_2$-valued fermion number that controls parity/signs.

The twisted fields are regrouped into two multiforms, valued in the \emph{complex} Lie algebra $\mathfrak g$ and its dual:
\be \mb A \in \mathfrak g\otimes  \Pi\Omega^\bullet(M)[1]\,,\qquad \mb B \in \mathfrak g^*\otimes \Pi\Omega^\bullet(M)[1]\,. \ee
In components, we denote
\be \mb A  = c + A_\mu dx^\mu + \ldots\qquad \mb B = b + J_\mu dx^\mu + \ldots\,. \ee
Here $A=A_\mu dx^\mu$ (bosonic, and degree zero) is a complexified $G$ gauge field, formed out of the physical $G_c$ gauge field and three scalars in the vectormultiplet that have been twisted into a one-form. The twisted theory has an enhanced complex $G$ symmetry. The scalar $c$ (fermionic, and degree 1) is its ghost. The zero-mode of $c$ is not a genuine local operator, but the derivatives of $c$ are, and are cohomologous to physical gauginos. The scalar $b$ (fermionic, degree 1) is also a gaugino. The complex one-form $J$ (bosonic, degree zero) is cohomologous to the Hodge-dual of the curvature $*F(A)$. In the interpretation of this theory as BF theory, $J$ is the ``$B$ field'' and might be thought of as a noncompact $\mathfrak g^*$ gauge connection. We denote it by $J$ because on the $\CD$ boundary condition it will become a boundary current.

The action and BRST transformations are
\be S   = \int_M \mb B F(\mb A) = \int_M \mb B d\mb A + \mb B \mb A^2
\qquad \begin{array}{l} Q(\mb A) = F(\mb A) \\ Q(\mb B) = d_{\mb A}\mb B = d\mb B+[\mb B,\mb A]\,, \end{array}
 \ee
with $Q = \{S,-\}_{\rm BV}$ more generally, where the BV bracket $\{\mb A(x),\mb B(y)\}_{\rm BV} = \delta^{(3)}(x-y)$ is induced by the kinetic term. Note that the equations of motion of the theory --- the RHS of the BRST variations --- set $\mb A$ to be flat and $\mb B$ (transforming in the coadjoint representation) to be covariantly constant. 

We could have included a coupling constant $\hbar$ in the action, like we did in the matter theory \eqref{matter-S}. (It can of course also be reabsorbed in the fields.) In the matter theory, this was useful in understanding scaling in commutation relations of the algebra $U$ and in the R-matrix. In nonperturbative gauge theory, there is a canonical normalization of sparks --- essentially due to integrality of $G$ representations, paired with integral `periodicity' of functions on $G$ --- so including a coupling constant turns out to be less meaningful.

\subsubsection{Descendants, local operators, and line operators}

Perturbatively, the structure of local operators and their descendants in pure gauge theory seems nearly identical to that for B-twisted matter. The $Q$-cohomology of local operators seems to be given by $\mathfrak g$-invariant polynomials in $c$ and $b$. One could then form (some) $Q$-closed line operators by integrating the first descendants of the local operators on curves $\gamma$ with $\pd\gamma=\oslash$. For abelian $\mathfrak g$, this includes operators like $\int_\gamma \mb A$ and $\int_\gamma \mb B$.

This structure is not quite correct nonperturbatively, if we take seriously the global form of the group $G$. Requiring $Q$-invariance in BV formalism only guarantees invariance under the Lie algebra $\mathfrak g$, and invariance under $G$ --- for example, under large/winding gauge transformations --- must further be imposed by hand. 

This has two important effects. First, as already mentioned, the ghost $c$ must be removed from local operators. (Derivatives of $c$ persist, but not in $Q$-cohomology.) Second, invariance under winding gauge transformations restricts line operators involving $\int\mb A$ to the usual physical Wilson lines
\be W_\rho(\gamma,\mb A)  = P\exp \int_\gamma \varphi_\rho(\mb A) \label{gauge-W} \ee
where $\rho$ is a holomorphic representation of the global group $G$ (analytically continued from a unitary representation of $G_c$),  and $\varphi_\rho:\mathfrak g\to \text{End}(\rho)$ the representation map. (For closed $\gamma$, one would further take a trace in $\rho$ to make \eqref{gauge-W} gauge-invariant.) For example, for $G=GL(1)$, only the operators
\be W_n(\gamma,\mb A) = e^{in\int_\gamma \mb A}\,,\quad n\in \Z \ee
are fully gauge invariant, as opposed to $n\notin \Z$ or the naive $\int_\gamma \mb A$.

In an abelian theory, lines of the form $\int_\gamma \mb B$, with any coefficient, still persist. In a nonabelian theory, they must be rendered gauge-invariant, \eg\ by using gauge-invariant polynomials in $\mb B$ or by dressing $\mb B$ with Wilson lines in the adjoint representation. A linked configuration of $\int_\gamma \mb B$ lines and Wilson lines can still be analyzed essentially the same way as in \eqref{def-PB}: linking is controlled by the propagator between $\mb A$ and $\mb B$, or by the higher Poisson bracket of perturbative local operators
\be \{\!\!\{ b_a,c^{a'}\}\!\!\} = \delta_a{}^{a'}\,. \ee
For example, in an abelian $G=GL(1)$ theory, if $\gamma,\gamma'$ form a Hopf link, then (in correlators)
\be \Big( \textstyle \int_\gamma \mb B\Big) \; W_n(\gamma',\mb A)  \sim \pm in W_n(\gamma',\mb A)\,. \ee
Another physical interpretation of this is that an insertion of $W_n(\gamma',\mb A)$ sources a monodromy singularity for the one-form $J$, which is measured by $\int_\gamma \mb B$.

\subsection{Boundaries, transversality, and completeness}
\label{sec:gauge-bdy}

The two basic boundary conditions we will use are Neumann and Dirichlet:
\begin{itemize}
\item[$\CN$:] $\mb B\big|_\pd =0$, $G$ gauge symmetry unbroken at boundary, $G$ bundles freely summed over
\item[$\CD$:] $\mb A\big|_\pd=0$, $G$-bundles trivialized at the boundary, $G$ broken to a boundary global symmetry $G_\pd$
\end{itemize}
These were defined in \cite{BDGH} as boundary conditions preserving 2d $\CN=(2,2)$ SUSY, and are compatible with both 3d $A$ and $B$ topological twists, becoming locally topologically invariant.

The gauge connection for the $G$ gauge symmetry on $\CN$ is clearly given by the 1-form component of $\mb A\big|_\pd$.  Dually, the current for the global $G_\pd$ symmetry on $\CD$ is given by the 1-form component of $\mb B\big|_\pd$ --- which is why we called it $J$.

The bulk theory has no framing anomaly, and we can check that the same is true for the boundary conditions, by computing boundary anomalies for the $SU(2)_C$ R-symmetry (broken at the boundary to $U(1)_C$) that's used to topologically twist. The analysis of \cite{dimofte2018dual} shows that the boundary anomaly polynomials are
\be \CN:\quad 2(\text{rank}\, \mathfrak g)\mb h\mb c\,,\qquad \CD:\quad - 2(\text{rank}\, \mathfrak g)\mb h\mb c\,, \ee
where $\mb h$, $\mb c$ are background curvatures for $U(1)_H$ and $U(1)_C$ R-symmetries, respectively. This is a mixed 't Hooft anomaly, and will not break $U(1)_C$.%
\footnote{The anomaly polynomials further show that if the boundary has nontrivial curvature, \emph{e.g.} if it is a sphere or a higher-genus Riemann surface, then $\mb c$ will necessarily be nonzero (as it's used to twist) and $U(1)_H$ will get broken,  thereby reducing the $\Z$ cohomological grading of the theory to $\Z/2$. This phenomenon is not relevant for the current analysis.}

It's also fairly straightforward to argue that the boundary conditions are 1-dualizable, with no essential difference between $\CN$ and $\ol\CN$, or between $\CD$ and $\ol \CD$; though doing this formally requires building up more technology than we will present here. 1-dualizability of such boundary conditions in gauge theory with matter was used extensively in \cite{OR-Chern,OR-TQFT}. The state spaces on discs bounded by $\CN$ and $\CD$ turn out to be remarkably small:
\be \raisebox{-.15in}{\includegraphics[width=1.2in]{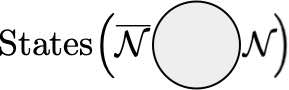}} = \C^G = \C \simeq \text{Ops}(\CN)\,,\quad \raisebox{-.15in}{\includegraphics[width=1.2in]{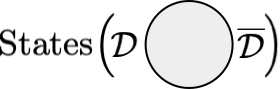}}  = \C[b] = \Lambda^\bullet \fg[-1]   = \text{Ops}(\CD) \label{gauge-disc} \ee
The theory on a disc bounded by $\CN$ can be reduced to B-twisted 1d $\CN=2$ pure gauge theory, whose state space is the $G$-invariant part of of the trivial representation $\C$ (namely $\C$ itself). By state-operator correspondence, this matches local operators on an $\CN$ boundary: there are none besides the identity. (Perturbatively, $c$ ghosts would survive, but nonperturbatively they are removed.) The theory on a disc bounded by $\CD$ can be reduced to B-twisted 1d $\CN=2$ fermi multiplets $(b,\bar b)$ valued in the Lie algebra $\mathfrak g^*$, and states look like polynomials in these fermions, $\C[b]$. This matches local operators on the $\CD$ boundary, where $b$ survives, and moreover can now form non-gauge-invariant combinations since gauge symmetry is broken on $\CD$.

\subsubsection{Sandwiches and transversality}
\label{sec:gauge-trans}

Let's consider some sandwiches between the $\CN$ and $\CD$ boundary conditions. One can analyze them almost exactly the same way that we did in Dijkgraaf-Witten theory, up to replacing the ``topological sigma models'' that appeared there with ``B models.''

Compactifying $\CT$ on an interval between $\ol\CN,\CN$ boundaries produces a 2d B-twisted pure gauge theory, \emph{a.k.a.} a 2d B-model with target $pt/G$.
\be \ol\CN\circ \CN \; \simeq\; \text{2d B-model to $pt/G$}\,. \label{gauge-Nsand} \ee
If the coordinate along the interval is `$s$' then the fields of the 2d B-model are $\mb A_{2d}=\mb A\big|_{ds=0}$ and $\bm\beta=\iota_{ds}\mb B$, with action $\int_{\R^2} \bm\beta F(\mb A_{2d})$.

To flip an $\CN$ boundary to a $\CD$ boundary, we couple it to a 2d B-model with target $G$ (as a complex manifold), using the gauge symmetry on $\CN$ to act one-sidedly on the additional $G$ target:
\be \ol \CD \; \simeq \; \CN \otimes_{\rm couple} (\text{2d B-model to $G$}) \label{gauge-flip} \ee
Since the one-sided action of $G$ on $G$ is free, with quotient $G\backslash G=pt$, the effect of coupling \eqref{gauge-flip} is precisely to break the gauge symmetry at the boundary. Moreover, we gain a new global boundary symmetry $G_\pd$, from the action of $G$ on $G$ on the other side. (If we couple $G$ to $\CN$ using the left action, we gain a $G_\pd$ from the right action.)

In case it is helpful, we can illustrate this coupling at the level of Lagrangians. The action of the 3d theory on a half-space with $\CN$ b.c. is just $\int_{\R^2\times(-\infty,0]} \mb B F(\mb A)$ (the boundary condition $\mb B|_0=0$ is just right for boundary terms in the EOM/$Q$-variation to vanish.) The 2d B-model to $G$ may be written $\int_{\R^2} \bm \Gamma \mb g^{-1} d\mb g$, where $\mb g \in \text{Maps}(\Omega^\bullet(\R^2),G)$, $\bm\Gamma\in \mathfrak g^*\otimes \Pi\Omega^\bullet(\R^2)[1]$. The total coupled action becomes 
\be \int_{\R^2\times(-\infty,0]} \mb B F(\mb A) + \int_{\R^2\times \{0\}} \bm \Gamma \mb g^{-1} d_{\mb A}\mb g =  \int_{\R^2\times(-\infty,0]} \mb B d\mb A + \int_{\R^2\times \{0\}} \bm \Gamma \mb g^{-1} \mb A\mb g + \ldots   \ee
The boundary EOM (or $Q$-variation) for $\bm \Gamma$ sets $\mb g^{-1}\mb A\big|_0\mb g=0 \; \Rightarrow\; \mb A\big|_0$, as we would like on a Dirichlet b.c. Moreover, boundary terms in the EOM for $\mb A$ set $\mb B\big|_0 = \frac{\pd}{\pd \mb A|_0} \bm  \Gamma \mb g^{-1} \mb A\big |_0\mb g = \mb g\bm \Gamma \mb g^{-1}$, relieving the constraint on $\mb B$ that $\CN$ had imposed. More precisely, it equates $J$ to the boundary B-model current for the right action of $G$ on $G$.

By using \eqref{gauge-flip} on both sides of a Neumann sandwich, we get a Dirichlet sandwich
\begin{align} \CD\circ\ol\CD\;&\simeq\; \text{(2d B-model to $G$)$\otimes_{\rm couple}$(2d B-model to $pt/G$)$\otimes_{\rm couple}$(2d B-model to $G$)} \notag \\
&=\; \text{2d B-model to $G/G\backslash G$} \notag \\
&\simeq\; \text{2d B-model to $G$} \label{gauge-Dsand} \end{align}
In the middle step, we identify a 2d B-model to the product $G\times G$ (from flipping on either side), with the right action on the first factor and the left action on the second factor gauged. In the second step, we simplify by performing the gauge quotient, which we can do since the gauge action is free. Another way to get the same answer, analogous to \eqref{DW-sandD} in DW theory, is to note that the degrees of freedom remaining after compactifying on a $(\CD,\ol\CD)$ sandwich come from the holonomy of the $G$ connection, from one boundary to the other --- measuring the relative trivializations. That's the factor of $G$ in \eqref{gauge-Dsand}.

Now, to analyze transversality, we flip an $(\ol\CN,\CN)$ sandwich on one side only:
\begin{align} \CD\circ\CN \;&\simeq\; \text{(2d B-model to $pt/G$)$\otimes_{\rm couple}$(2d B-model to $G$)} \notag \\
 &= \; \text{2d B-model to $G\backslash G$} \label{gauge-DNsand} \end{align} 
This is a B-model to $G$, with the left action gauged. Since the action is free and the quotient is a point, we get a global statement of transversality,
\be \CD\circ\CN\;\simeq \; \text{2d B-model to $pt$} = \text{id}_\oslash\,. \ee
The interface $k$ that implements this equivalence is a boundary condition for the $G\backslash G$ B-model. We can identify it by looking for a boundary condition that has only trivial local operators, with no residual gauge symmetry acting. We claim that $k$ is simply a Neumann boundary condition for the $G$ matter fields (for $\mb g$) and Neumann for the 2d gauge symmetry. Then its local operators are $G$-invariant functions on $G$, which are trivial.

Translating this to algebraic terms: The category of boundary conditions for the 2d B-model to $G\backslash G$ is equivariant coherent sheaves $\text{Coh}^G(G)$. The transversality interface $k$ is the structure sheaf
\be  k =\CO_G\,. \label{k-Og} \ee
It obeys $\text{End}_{\text{Coh}^G(G)}(k) = \CO(G)^G = \CO(G\backslash G) = \CO(pt) = \C$ (where $\CO(X)$ denotes algebraic functions on a space $X$).

Finally, we can open up the transversality sandwich:
\be  \raisebox{-.3in}{\includegraphics[width=2.2in]{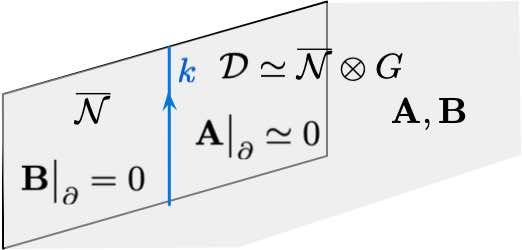}} \label{gauge-NkD}   \ee
Let us think of engineering the $\CD$ boundary by starting with $\ol \CN$ and coupling to a 2d B-model to $G$ on a half-boundary. In other words, the action, analogous to \eqref{SDkN}, looks like
\be S_{\CD_k\ol\CN} =  \hspace{-.6cm} \int_{\R_t\times\R_p\times\{s\geq 0\}}\hspace{-.5cm}\mb BF(\mb A) -  \hspace{-.5cm} \int_{\R_t\times \{p>0\}\times\{s=0\}} \hspace{-.5cm}  \bm \Gamma \mb g^{-1}d_{\mb A}\mb g\,,  \label{gauge-SDkN} \ee
with 2d Neumann boundary condition $\bm\Gamma\big|_{p=s=0} =0$  (matching \eqref{k-Og}) at the interface.

Beautifully, there are no further interface degrees of freedom to worry about, like there were for pure matter. Unfortunately, the strong dualizability condition \AD$_k$ for the $k$ interface between $\CD$ and $\ol \CN$ \emph{fails}. Neither capping off $\CD$ nor cupping ``into'' $\ol \CN$ are strictly defined. We'll explain this precisely in Section \ref{sec:gauge-gen}, after gaining more fluency with our categories.

\subsubsection{Completeness}
\label{sec:gauge-comp}

Completeness holds, just as in Dijkgraaf-Witten theory, because we can couple the global $G_\pd$ symmetry on a region $R$ of $\CD$ boundary to the gauge symmetry on an opposing region $\bar R$ of an $\CN$ boundary to glue the two together.

Perturbatively, this is accomplished by adding an interaction along the gluing region that couples the current on $\CD$ to the connection on $\CN$. Explicitly, the action compatible with the two boundary conditions on either side of the gluing (on $R=\R^2$, say) is
\be S_{\rm split} =  \int_{\R^2\times\{s\leq 0\}}^{(\CN)} \mb B F(\mb A) + \int_{\R^2\times\{s\geq 0\}}^{(\CD)} \big(\mb A' d\mb B' +\mb B' \mb A'{}^2 \big) \ee
Adding the gluing term
\be S_{\rm glue} = - \int_{\R^2} \mb A\big|_0 \mb B'\big|_0 = -\int_{\R^2} A\big|_0\wedge J'\big|_0 + \ldots  \label{gauge-Sglue} \ee
ensures, as in Section \ref{sec:matter-comp}, that the EOM/$Q$-variation of $\mb A$ in $S_{\rm split}+S_{\rm glue}$ has a boundary term that sets $\mb B\big|_0 = \mb B'\big|_0$; and the variation of $\mb B'$ in $S_{\rm split}+S_{\rm glue}$ has a boundary term  that sets $\mb A'\big|_0 = \mb A\big|_0$. These conditions ensure that the fields match across $s=0$, gluing the theory back together.

The operator $\mb A\mb B'$ is (really, the lowest component $cb'$, whose second descendant is what's integrated along the gluing surface) is the 2-Maurer-Cartan element that seems to control the gluing deformation.

Note that if we glue along a region $R$ that has boundaries, it must be well-bounded in the sense of Definition \ref{def-bounded} (Section \ref{sec:comp}). That's because, after setting $\mb A'\big|_0 = \mb A\big|_0$ to remove boundary terms at $s=0$,  the BRST variation
\be Q \int_R \mb A\big|_0 \mb B'\big|_0 = \int_R d\big(\mb A\big|_0 \mb B'\big|_0\big)  = \int_{\pd R} \mb A\big|_0 \mb B'\big|_0 \ee
may still have boundary terms on $\pd R$. Well-boundedness ensures that they vanish, without additional degrees of freedom being introduced.

One might further worry about nonperturbative corrections to the gluing, if the region $R$ has nontrivial topology. Gauging along a surface with nontrivial topology usually requires summing over all topological types of $G$-bundles there. However, the path integral in B-twisted gauge theory localizes to flat bundles (solutions of the EOM), so topological types that don't admit flat connections should not contribute at all. For flat bundles, the coupling \eqref{gauge-Sglue} is conceivably sufficient.

\subsection{Boundary spark algebras}
\label{sec:gauge-sparks}

We'll construct boundary spark algebras $\CH_\CN$ and $\CH_\CD$ in B-twisted gauge theory using a combination of the methods we employed in DW theory and in B-twisted matter.

To orient ourselves, we'll first analyze the vector spaces of states on rectangles --- this will tell us which vector spaces we're aiming for and also highlight main feature of B-twisted gauge theory: spark algebras are infinite dimensional, and $\CH_\CD$ and $\CH_\CN$ do not ``look the same.'' They are dual vector spaces in a topological sense.

We'll then produce the actual sparks and their Hopf-algebra structures. $\CH_\CN$ will come from framed Wilson lines, just like in DW theory; they can now also be thought of as exponentiated integrals of descendants. $\CH_\CD$ will be controlled by boundary $G_\pd$ symmetry, and will contain both infinitesimal symmetry transformations (which are integrated descendants) and finite ones (which are roughly their exponentials).

\subsubsection{Vector spaces}
\label{sec:gauge-vector}

Consider B-twisted gauge theory on a rectangle with two Neumann boundaries and two Dirichlet boundaries. Making the rectangle so that it's long and thin along the Dirichlet edges leads (via \eqref{gauge-Dsand}) to an effective 2d B-model with target $G$:
\be \raisebox{-.3in}{\includegraphics[width=4.2in]{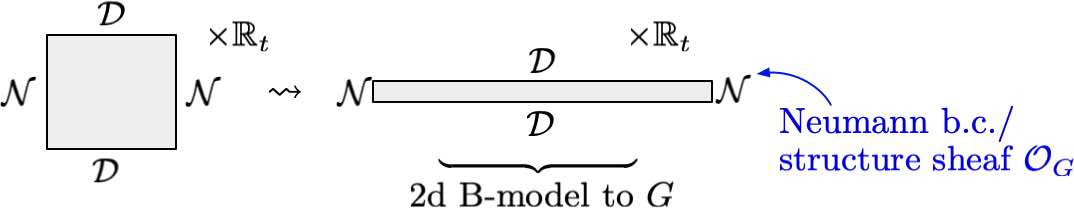}} \label{gauge-boxQMD} \ee
The short $\CN$ edges impose Neumann boundary conditions for this 2d B-model. Further compactifying on a strip with these two Neumann boundary conditions gives us a 1d B-twisted $\CN=2$ quantum mechanics with target the complex group $G$:
\be \text{$\CT$ on rectangle} \simeq \text{1d B-twisted QM to $G$}\,. \ee

The same answer could have been obtained by stretching the rectangle in the other direction:
\be \raisebox{-.3in}{\includegraphics[width=4.3in]{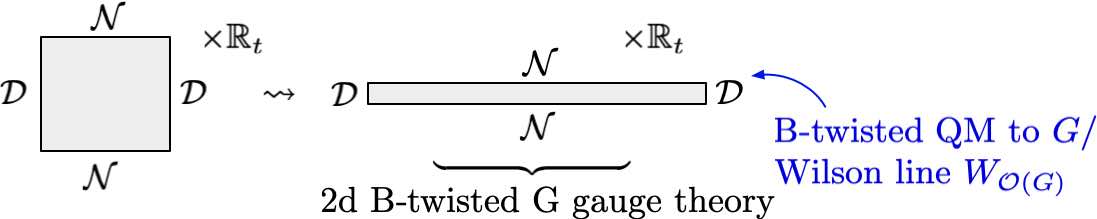}} \label{gauge-boxQMN} \ee
Via \eqref{gauge-Nsand}, the sandwich between Neumann b.c. produces a 2d B-twisted pure gauge theory. The short $\CD$ edges now effectively support 1d B-twisted quantum mechanics with target $G$, coupled to the 2d gauge theory. Further compactifying on the strip produces 1d B-twisted quantum mechanics with target $G\times G$ and gauge group $G$ acting diagonally (on the right of the first factor and the left of the second). Carrying out the quotient by this free gauge action reduces the theory to quantum mechanics with target~$G$.

B-twisted quantum mechanics with non-compact bosonic target space $G$ does not have a uniquely defined state space. In essence, there are multiple inequivalent choices of polarization. (This is the price of swapping the fermions of the pure matter theory from Section \ref{sec:matter-spaces} with bosons.) 

One standard choice of polarization identifies the $Q$-cohomology of the state space with Dolbeault cohomology $H_{\bar\pd}^{0,\bullet}(G)$, of forms with polynomial growth near infinity. As $G$ is affine, Dolbeault cohomology is supported in degree zero, and we just get the algebraic functions on $G$, $H_{\bar\pd}^{0,\bullet}(G) = \CO(G)$.

A different polarization produces the dual space --- Dolbeault homology of $G$, \emph{a.k.a.} the space of holomorphic distributions, which we'll denote $\text{Dist}(G)$. The space $\text{Dist}(G)$ is the continuous dual of $\CO(G)$. It contains a dense subspace
\be \C G = \C\langle\delta_g\rangle_{g\in G} \subset \text{Dist}(G)  \ee
consisting of Dirac delta-functions at every point of $G$, acting on functions by evaluation
\be \langle \delta_g , f \rangle = f(g) \,,\qquad f\in \CO(G)\,. \ee
It also contains all finite-order holomorphic derivatives of delta-functions. In particular, the derivatives at the identity $\delta_e$ may be identified (through their action on functions) with a copy of the enveloping algebra
\be U(\mathfrak g)\simeq \C\langle \pd_{\alpha_1}...\pd_{\alpha_k}\delta_e\rangle_{\alpha_i\in \mathfrak g} \subset \text{Dist}(G)\,,\quad \langle \pd_{\alpha_1}...\pd_{\alpha_k}\delta_e,\,f\rangle  := (-1)^k \pd_{\alpha_k}...\pd_{\alpha_1}f(e)\,, \ee
thinking of $\alpha_i\in\mathfrak g$ as tangent vectors at the identity on the RHS.

Physically, if $\CO(G)$ are chosen as the incoming states on the rectangle, then $\text{Dist}(G)$ are the dual outgoing states. We will momentarily find that $\CO(G)$ matches the spark algebra $\CH_\CN$, while $\text{Dist}(G)$ matches $\CH_\CD$.

\subsubsection{$\CH_\CN$ from Wilson lines}
\label{sec:gauge-HN}

To construct the actual spark algebra $\CH_\CN$, we look for operators supported on a strip of $\CN$ boundary. Since $\mb A$ is unconstrained at the $\CN$ boundary, this includes at least Wilson lines \eqref{gauge-W}. Just as in DW theory (Section \ref{sec:DW-HN}), Wilson lines can be stretched from one $k$ interface to another, where gauge symmetry is broken, to produce a gauge-invariant (and thus $Q$-closed) operator.

Recall that the holomorphic representations $\rho$ of $G$, which can label Wilson lines, are just analytic continuations of representations of the group $G_c$. They decompose into direct sums of finite-dimensional irreducibles. For each finite-dimensional $\rho$, we get to choose a vector $v\in\rho$ and a covector $w\in\rho^*$ to ``frame'' the Wilson line at its endpoints, where gauge symmetry is broken:
\be   \raisebox{-.4in}{\includegraphics[width=2.8in]{SparkDefD-DW.jpg}} \label{gauge-Nsparks} \ee
Just as in DW theory, each framed segment $wW_\rho v$ measures a matrix element $w\varphi_\rho(g)v$ of the holonomy $g$ of the gauge bundle from one $k$ boundary to the other.

Linear combinations of matrix elements for all possible $\rho$ span the space of algebraic functions on $G$, leading us to identify
\be \CH_\CN =  \bigoplus_{\rho\,\in\,\text{irrep}(G)} \rho^*\otimes \rho \simeq \CO(G)\,. \label{Wrr-gauge} \ee 
(Conversely: just as in the finite case, the space of functions $f\in\CO(G)$ with the left action $(g\cdot f)(h) = f(g^{-1}h)$ decomposes as a sum of all irreducible representations $\rho$, each appearing with multiplicity $\rho^*$.)

One may wonder whether we have found all the sparks this way --- whether stretched Wilson lines are all there is. One verification is that the space \eqref{Wrr-gauge} matches a reasonable quantization of the space of states on a rectangle (Section \ref{sec:gauge-vector}). Another, which jumps ahead of ourselves, is to observe that the category of lines on $\CN$ is equivalent to boundary conditions for the $\ol\CN\circ\CN$ sandwich theory \eqref{gauge-Nsand}, which is $\CC_\CN=\text{Coh}^G(pt) = \text{Rep}(G)$. This confirms that the \emph{only} line operators that exist on $\CN$ and could form sparks are Wilson lines labelled by $\rho\in \text{Rep}(G)$. There are also no nontrivial local operators on $\CN$ \eqref{gauge-disc}, and no junctions among Wilson lines labelled by different irreducible representations, so \eqref{gauge-Nsparks} for irreducible $\rho$ captures all of $\CH_\CN$.

The product on $\CH_\CN$ is commutative. This is clearly true classically: measuring the holonomy $g$ of the gauge bundle from one $k$ boundary to the other with matrix elements at nearby points just multiplies these matrix elements.
\be \raisebox{-.4in}{\includegraphics[width=2.1in]{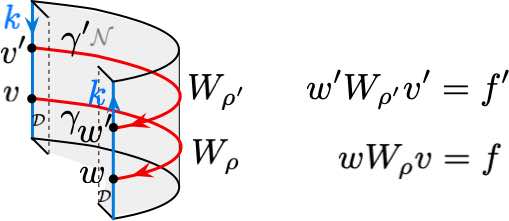}}\qquad   (f' \cdot f)(g) = f'(g)f(g) \label{gauge-HNprod} \ee 
Quantum mechanically, we recall that the theory localizes to flat bundles --- so the holonomy from one $k$ boundary to the other, whether the bundle is trivialized in a constant way, can never fluctuate; thus there can be no corrections to the classical product \eqref{gauge-HNprod}. One can also make this argument, perturbatively, by considering a correlation function of two boundary Wilson lines
\be \Big(P\exp\int_{\gamma'} \mb A\Big)\Big(P\exp\int_{\gamma} \mb A\Big) \label{W-correct} \ee
The propagator in our theory connects $\mb A$ to $\mb B$; and the only possible bulk vertex is of the form $\mb B\mb A\mb A$, which does not allow \emph{any} contractions, even with bulk vertex insertions, that would correct \eqref{W-correct}.

The remaining Hopf-algebra operations are straightforward.
The coproduct, as in DW theory, results from breaking a Wilson line $W_\rho$ at a $k$ boundary by inserting the identity $\id\in \text{End}(\rho) \simeq \rho\otimes \rho^*$, decomposed as $\id = \sum_i v^i \otimes w_i$ for a basis $\{v^i\}$ and dual basis $\{w_i\}$ of the finite-dimensional space $\rho$:
\be \Delta(w W_\rho v) = \raisebox{-.22in}{\includegraphics[width=3.5in]{Hopf-DW-Nv.jpg}}  = \sum_i  (w_i W_\rho v)\otimes (w W_\rho v^i)\,. \label{HN-D-gauge} \ee
More generally, for any function $f \in \CO(G)$ we have the convolution coproduct
\be \Delta:\CO(G)\to \CO(G)\otimes\CO(G)\,,\qquad  \Delta(f)(g_1,g_2) = f(g_2g_1)\,. \ee

The unit (the empty spark) is constant function $f=1$, \emph{a.k.a.} the unique matrix element of the trivial representation.

The counit is the partition function
\be  \raisebox{-.2in}{\includegraphics[width=2in]{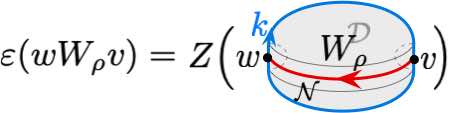}}   = w\varphi_\rho(e)v = w\cdot v\ee
The trivialization along $\CD$ on the back forces the holonomy of the bundle along the front to be trivial as well. Thus, this partition function measures the matrix element of the identity element in $G$. More generally, for any function
\be \varepsilon: \CO(G)\to \C\,,\qquad \varepsilon(f) = f(e)\,.  \label{gauge-eN} \ee

Finally, twisting sparks by $180^\circ$, reversing their orientation (and swapping $v$ and $w$), leads to a well-defined antipode. For any finite $\rho$, it sends a matrix element to the corresponding element of the inverse matrix. This implies on general functions that
\be (S f)(g) = f(g^{-1}) \label{gauge-SN} \ee

In summary, we have found the standard Hopf structure on functions on $G$:
\be \CH_\CN= \CO(G)\,:\quad
\begin{array}{c} (f\cdot f')(g) = f(g)f'(g) \\[.1cm] 1(g)\equiv 1 \end{array}\,,\quad \begin{array}{c} \Delta(f)(g_1,g_2) = f(g_2g_1) \\[.1cm] \varepsilon(f) =f(e) \end{array}\,,\quad S(f)(g)=f(g^{-1})\,.  \label{gauge-HN-Hopf} \ee

\subsubsection{$\CH_\CD$ from symmetry defects}
\label{sec:gauge-HD}

There are several sources of sparks on a strip of $\CD$ boundary.

One could consider local operators --- polynomials in the $b$'s --- but these can be brought to the $k$ boundaries, where they evaluate to zero.

In addition, there are integrals of $\mb B$ across the strip. For any $\alpha\in \mathfrak g$, we have a spark
\be \raisebox{-.4in}{\includegraphics[width=2in]{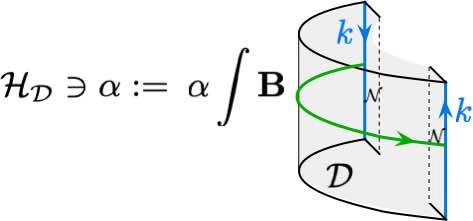}} \label{gauge-J} \ee
These sparks, which we'll just call `$\alpha$', are gauge invariant ($Q$-closed) on the $\CD$ boundary. They wouldn't generally be in the bulk, because $Q\mb B\sim d\mb B+[\mb B,\mb A]$, and the second term is problematic in a non-abelian theory; but on the $\CD$ boundary we've set $\mb A\big|_\pd = 0$.

Classically, the sparks $\alpha$ are commutative. However, the product does get a quantum correction, in non-abelian theories. An insertion of two $\int\mb B$ integrals on the boundary can be contracted with two $\mb A$'s in a bulk $\mb B\mb A\mb A$ vertex, using the structure constants of $\fg$, leading to a correction of the form $[\int\mb B,\int \mb B]\sim \int \mb B$. Fortunately, we do not need to calculate this directly, because symmetry comes to the rescue. The line integral $\int \mb B=\int J_\mu dx^\mu$ is nothing but the integral of the current for the boundary $G_\pd$ symmetry on $\CD$. Note that this symmetry is non-anomalous; it does not even have a boundary 't Hooft anomaly. Thus, the algebra of integrated currents must simply be the Lie algebra of $G_\pd$. For sparks, this means
\be \alpha\cdot \alpha' - \alpha'  \cdot \alpha =  [\alpha,\alpha']_{\mathfrak g}   \qquad\quad \raisebox{-.4in}{\includegraphics[width=2.2in]{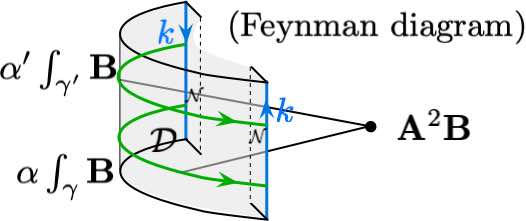}} \ee
All together, the sparks \eqref{gauge-J} generate a copy of the enveloping algebra
\be \C[\alpha]_{\alpha\in \mathfrak g} \simeq U(\mathfrak g) \subseteq \CH_\CD\,. \ee

However, this is not all. For every element $g\in G_\pd$ there is also a \emph{finite} symmetry defect $V_g$ we could insert on the $\CD$ strip, changing the trivialization of the bundle across it:
\be   \raisebox{-.4in}{\includegraphics[width=2.8in]{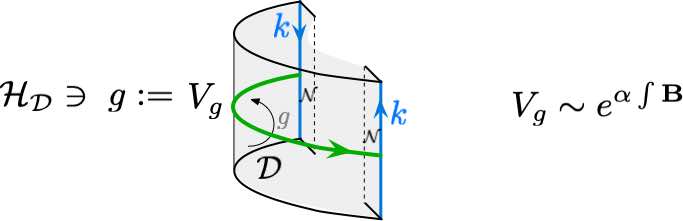}} \ee
We'll just call this spark `$g$'. When the finite group element can be expressed by exponentiating a Lie-algebra element, $g=e^\alpha$, we expect the corresponding defect to be realized by an exponentiated current $V_g = e^{\alpha \int_\gamma \mb B}$ (with point-splitting regularization given by shifting multiple copies of $\int\mb B$ in a direction normal to $\gamma$). The product of trivialization-changing defects is again completely controlled by symmetry, so we have
\be g\cdot g' = gg' \ee
Thus we find a copy of the discrete group algebra
\be \C[g]_{g\in G} \simeq \C G \subseteq \CH_\CD \ee

Altogether, integrated currents $\alpha$ and symmetry-changing defects $g$ generate the space we described in Section \ref{sec:gauge-vector} as holomorphic distributions on $G$. We propose that
\be \CH_\CD\simeq \text{Dist}(G)\qquad \text{with} \quad  g\leftrightarrow \delta_g\,,\quad  \alpha\leftrightarrow \pd_{-\alpha}\delta_e\,. \ee
We recall that the group algebra $\C G$ is a dense subalgebra in $\text{Dist}(G)$. The difference between $\C G$ and $\mathrm{Dist}(G)$ is precisely that $\mathrm{Dist}(G)$ remembers the geometry of $G$. While representations of $\C G$ are highly uncontrolled (they correspond to representations of $G$ as a discrete group), representations of $\mathrm{Dist}(G)$ are just algebraic representations of $G$. Jumping ahead slightly, this gives us some further confirmation that we've correctly identified all the sparks on $\CD$: as discussed in Section \ref{sec:gauge-HN}, we know that $\CC_\CN\simeq \text{Rep}(G)$, which precisely matches $\CH_\CD\text{-mod}$.

Each subalgebra $\C G$ and $U(\fg)$ is a Hopf subalgebra of $\CH_\CD=\text{Dist}(G)$. The Hopf operations can be computed in a straightforward way. For finite changes of trivialization, the arguments are identical to those in DW theory (Section \ref{sec:DW-HD}), and we get
\be \CH_\CD \supset \C G\,:\quad \begin{array}{c} g \cdot g' = gg' \\[.1cm]
1=e \end{array}\,,\qquad \begin{array}{c} \Delta(g) = g\otimes g \\[.1cm] \varepsilon(g) \equiv 1 \end{array}\,,\qquad S(g)=g^{-1}\,.  \label{gauge-CG-Hopf} \ee
For integrated currents, we just have the corresponding infinitesimal versions
\be \CH_\CD \supset U(\fg)\,:\quad \begin{array}{c} [\alpha,\alpha']= [\alpha,\alpha']_\fg \\[.1cm]
1=1 \end{array}\,,\qquad \begin{array}{c} \Delta(\alpha) = \alpha\otimes 1+1\otimes\alpha \\[.1cm] \varepsilon(\alpha) \equiv 0 \end{array}\,,\qquad S(\alpha)=-\alpha\,.  \label{gauge-UG-Hopf} \ee
For example, the coproduct just comes from splitting the integral in half:
\be \raisebox{-.4in}{\includegraphics[width=5in]{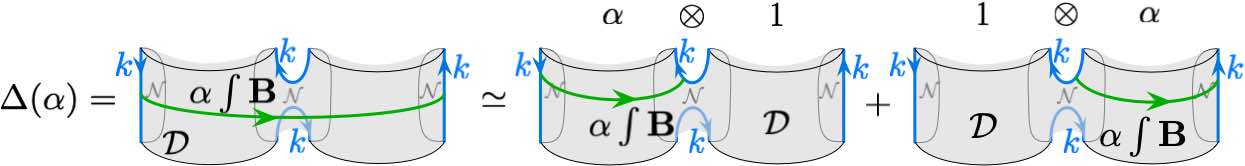}}  \ee

\subsubsection{Hopf pairing and topological completion}
\label{sec:gauge-Hopf}

The Hopf pairing now pairs functions and distributions, just as we expected from the dual quantizations of a rectangle in Section \ref{sec:gauge-vector}. For example, just like in DW theory, inserting a change-of-trivialization $V_g$ on the $\CD$ side of a solid ball will force the $G$-bundle to have holonomy $g$ across the $\CN$ side; so further inserting a framed Wilson line $wW_\rho v$ will measure the matrix element of $w\varphi_\rho(g)v$:
\be \raisebox{-.4in}{\includegraphics[width=2.5in]{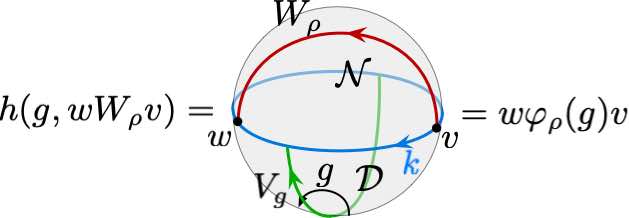}} \ee
or infinitesimally:
\be \raisebox{-.4in}{\includegraphics[width=5.5in]{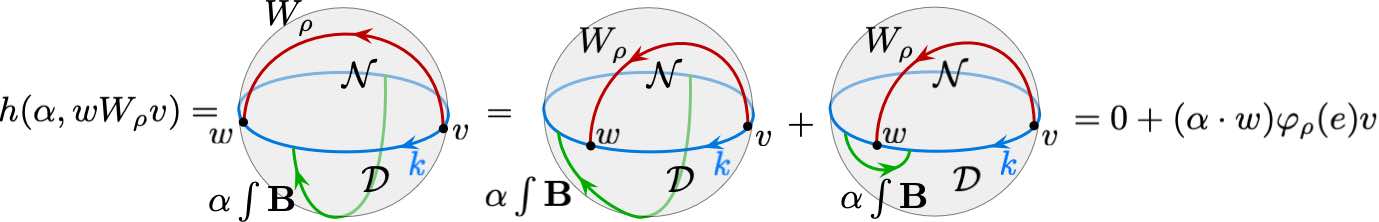}} \ee
More generally, $h(g,f) = f(g)$ and $h(\alpha,f) = \pd_\alpha f(e)$. The pairing is a Hopf pairing with respect to the operations \eqref{gauge-CG-Hopf}--\eqref{gauge-UG-Hopf} in $\CH_\CD$ and \eqref{gauge-HN-Hopf} in $\CH_\CN$.

An important final point to emphasize is that while $\CH_\CN=\CO(G)$, $\C G$, $U(\fg)$ are ordinary Hopf algebras (albeit infinite dimensional), the full $\CH_\CD=\text{Dist}(G)$ is not quite. Its coproduct requires a topological completion to define:
\be
\Delta: \mathrm{Dist}(G)\to \mathrm{Dist}(G)\wh\otimes \mathrm{Dist}(G)\,, \label{gauge-comp-cop}
\ee
essentially to make up for the difference between $\text{Dist}(G)$ and $\text{Dist}(G\times G)$. This topological completion is forced if one expresses $\mathrm{Dist}(G)$ as the continuous dual of $\CO(G)$, and then attempts to dualize the product in $\CO(G)$ to a coproduct on $\mathrm{Dist}(G)$. Explicitly, from the description of $\CO(G)$ in \eqref{Wrr-gauge}, we obtain the following description of $\text{Dist}(G)$:
\be\label{eq:Endrho}
\text{Dist}(G)=\prod_{\rho\,\in\,\text{irrep}(G)} \End (\rho)\,,
\ee
where $\End (\rho)$ is the algebra of endomorphisms of $\rho$. One can verify that this is an identification as topological algebras. The coproduct in this description comes from embeddings
\be
\Delta:\End (\rho)\to \prod_{\rho\subseteq \rho_1\otimes \rho_2} \End (\rho_1)\otimes \End (\rho_2)\,,  \label{rho-split}
\ee
defined as follows. Given any $E\in \End (\rho)$, and any $\rho_1, \rho_2$ such that $\rho\subseteq \rho_1\otimes \rho_2$, we decompose (since $\text{Rep}(G)$ is semisimple):
\be
\rho_1\otimes \rho_2=\rho^{\oplus n}\oplus \Big( \bigoplus_{\rho'\ne \rho}\rho'\Big)\,.
\ee
The matrix of $\Delta (E)$ will simply be $E$ acting on each $\rho$ and $0$ on the complement. One sees that the image of $\Delta (E)$ is not contained in $\mathrm{Dist}(G)\otimes \mathrm{Dist}(G)$ but rather in the completion, as in \eqref{gauge-comp-cop}.

The other Hopf operations are straightforward to describe under the identification \eqref{eq:Endrho}. The counit is zero on all $\End (\rho)$ except for $\rho=\C$, in which case it is the identity. The antipode maps
\be
S: \End (\rho)\to \End (\rho^*)
\ee
and is simply given by matrix transpose.

\subsection{The bulk spark algebra}
\label{sec:gauge-bulk}

We'll now derive the bulk spark algebra, from what we know about $\CH_\CD$ and $\CH_\CN$.

We know from the first part of \ref{Prop:decomp} (which does not depend on \AD$_k$ or finite-dimensionality) that the bulk spark algebra must satisfy
\be U \simeq \CH_\CD\otimes \CH_\CN  = \text{Dist}(G)\otimes \CO(G) \ee
as a vector space. Moreover, its Hopf-algebra operations are all derived from those in $\CH_\CD$ and $\CH_\CN$, as shown in Section \ref{sec:swap}, again without the use of \AD$_k$. In particular, the nontrivial formula \eqref{double-prod2} for the product in a Drinfeld double must hold. Let's spell this out.

To determine the product on $U$, it suffices to find the commutation relation of delta-distributions $g\in \C G$ and arbitrary functions $f\in \CO(G)$, since $\C G$ is a dense subalgebra of $\text{Dist}(G)$. We can do this algebraically with the aid of \eqref{double-prod2}. Since $\Delta^2 g = g\otimes g\otimes g$, we have
\be fg = \sum h(g, S f_{(1)})\, g f_{(2)}\, h(g, f_{(3)}) =  \sum f_{(1)}(g^{-1})\, g f_{(2)}\, f_{(3)}(g)\,. \ee
Evaluating both sides on an element $k\in G$ and using the coproduct for $f$, we get $f(k)g= g \sum  f_{(1)}(g^{-1}) f_{(2)}(k) f_{(3)}(g) = g f(gkg^{-1})$. Therefore:
\be (g f g^{-1})(k) = f(g^{-1}kg)  \label{gauge-adj} \ee
Consequently, the algebra $U$ is really a semi-direct product
\be
U=\mathrm{Dist}(G)\ltimes \CO (G)\,,
\ee
where the action of $\mathrm{Dist}(G)$ on $\CO (G)$ is given by the adjoint action of $G$ on $\CO (G)$. 

Physically, the relation \eqref{gauge-adj} comes just as it did in DW theory, from passing a symmetry defect $V_g$ across the framings at the endpoints of a Wilson line, where it will act on each one:
\be \raisebox{-.4in}{\includegraphics[width=5.7in]{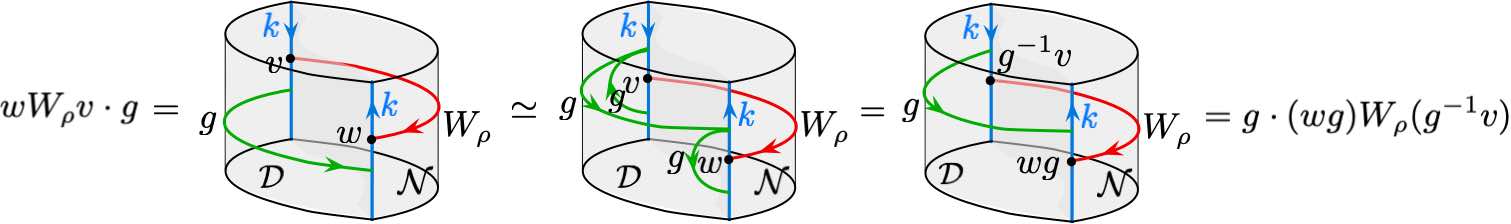}} \label{gauge-gW-act} \ee
This implies on matrix elements of $k\in G$ that $w\varphi_\rho(k) v\cdot g = g\cdot (wg)\varphi_\rho(k)(g^{-1}v) = g\cdot w\varphi_\rho(gkg^{-1})v$, and so on general functions $g^{-1}\cdot f(k)\cdot g = f(gkg^{-1})$ just as in \eqref{gauge-adj}.

In a similar fashion, one can show that the commutation relation of $U(\fg)$ with $\CO(G)$ is induced by the derivative of the conjugation action of $G$ on $\CO (G)$, namely
\be
[\alpha, f](h)=(\alpha f) (k)\,,\qquad \alpha\in \fg\,,\; f\in \CO (G)\,,\; k\in G\,,
\ee
where $\alpha f$ is the action of $\alpha$ on $f$ as a left-invariant vector field.

\subsubsection{The R-matrix}
\label{sec:gauge-R}

The R-matrix is significantly trickier to write down explicitly. Since \AD$_k$ fails and spark algebras are infinite-dimensional, we already expect that the R-matrix (at best) belongs to a completed tensor product $\CH_\CD \wh\otimes\CH_\CN = \text{Dist}(G)\wh\otimes \CO(G)$. Simply choosing a basis $\{f_i\}$ of $\CO(G)$ and a dual basis $\{f^i\}$ of $\text{Dist}(G)$ and writing
\be
\CR=\sum f^i\otimes f_i\in \text{Dist}(G) \wh\otimes \CO (G)
\ee
as predicted by Theorem \ref{Thm:double}, leads to an expression that is not entirely easy to compute. The $f^i$ here will be smeared, non-localized distributions.

Fortunately, we are approaching well-trodden territory mathematically, as the Drinfeld double of $\CO(G)$ (or dually, the Drinfeld centers of $\CO(G)\text{-mod}$ and $\text{Dist}(G)\text{-mod}=\text{Rep}(G)$) are well studied. We can also gain some intuition from our QFT construction.  Altogether, we have the following ways of understanding the R-matrix:

\begin{enumerate}[leftmargin=*]

\item Physically, the $\CR$-matrix should be obtained from taking the box-gluing deformation, and splitting apart its tensor factors. For our pure gauge theory, the gluing element is given by \eqref{gauge-Sglue}; and on a rectangle, a generalization of the Riemann bilinear identity analogous to \eqref{Riem-bil} gives us
\be S_{\rm glue} = -\int_{[0,1]\times[0,1]} \mb A\mb B' \sim - \int_{[0,1]} \mb A\int_{[0,1]}\mb B' + Q\text{-exact}\,.\ee
This suggests the R-matrix should take the form
\be \CR = e^{\int\mb B\otimes \int \mb A}\,, \label{gauge-Rphys} \ee
with the two integrals producing sparks on $\CD$ and $\CN$, respectively.
If $\int\mb A$ made sense by itself, rather exponentiated individually in a Wilson line, then \eqref{gauge-Rphys} would suffice. However, nonperturbatively, \eqref{gauge-Rphys} requires further interpretation.

\item Suppose that $\ell,\ell'$ are two bulk line operators --- two modules of $U$ --- and we want to act with $\CR$ on $\ell\otimes \ell'$ to find the braiding element. Then \eqref{gauge-Rphys} can be directly interpreted. Roughly, the symmetry defect $\int\mb B$ acts on $\ell$ in a particular representation, and the R-matrix then measures a matrix coefficient of the bundle around $\ell'$ in that representation.

More concretely, $\ell$ is a $\text{Dist}(G)$-module, hence an algebraic representation of $G$. Let $\varphi_\ell:G\to \text{End}(\ell)$ be the $G$ action, choose a basis $\{v^i\}$ for $\ell$, a dual basis $\{w_i\}$ for $\ell^*$, and observe that there's a map
\be \ell \to \ell\otimes \CO(G)\,,\qquad v^i\mapsto \sum_j v^j\otimes w_j  \varphi_\ell(-) v^i\,, \label{gauge-coaction}  \ee
where $w_j  \varphi_\ell(-) v^i$ is the function giving the $(ji)$ matrix coefficient. Then the R-matrix on $\ell\otimes \ell'$ is given by composing this with the action of $\CO(G)$ on $\ell'$,
\be \label{R-coaction}
\CR: \ell\otimes \ell'\to \big(\ell\otimes \CO (G)\big)\otimes \ell' \overset{\text{id}\otimes \text{act}}{\longrightarrow}  \ell\otimes \ell'\,.
\ee
This description of the $R$ matrix also follows from recognizing modules of $U$ as Yetter-Drinfeld modules of $\CO(G)$ (see \cite{schauenburg2002hopf}).

\item The formula \eqref{gauge-Rphys} also makes sense perturbatively. If we replaced the gauge group $G$ with the Lie algebra $\mathfrak g$ (allowing only infinitesimal gauge transformations), then $\CH_\CN$ would look like $\text{Sym}(\fg^*)$, generated by sparks $\beta\int\mb A$ (with $\beta\in\fg^*$) and $\CH_\CD$ would look like $U(\mathfrak g)$, generated by sparks $\alpha\int\mb B$ (with $\alpha\in\fg$). Given a basis $\{\alpha^a\}$ for $\fg$ and a dual basis $\{\beta_a\}$ for $\fg^*$, we now find
\be \CR = e^{\sum_a \alpha^a\otimes \beta_a}\,. \label{gauge-Rpert} \ee

The R-matrix \eqref{gauge-Rpert} is the one derived in \cite{Aamand-BF} from expectation values of crossed Wilson lines in perturbative BF theory, in the presence of transverse boundary conditions. Indeed, if one brings the Wilson lines of \cite{Aamand-BF} to the boundary, they simply coincide with our $e^{\beta\int \mb A}$ and $e^{\alpha\int \mb B}$\,.

\item If we consider the subalgebra $U_G$ generated by $\C G$ and $\CO (G)$, and embed $\CO (G)$ in the linear dual of $\C G$, we obtain another topological Hopf algebra $\wh U_G$. The linear dual of $\C G$ is the product $\prod_{g\in G} \C_g$ (here $\C_g$ is spanned by $1_g$, the delta function at $g$), and the algebra $\CO (G)$ embeds into this big product by mapping $f$ to $f(g)1_g$. A module of $\wh U_G$ becomes a module of $U$ if the action of $\C G$ integrates to an algebraic action of $G$. Such a module is supported at finitely many points on $G$, and the support must be invariant under conjugation by $G$. Given two such modules $\ell, \ell'$, the $R$ matrix acting on the tensor product is given by:
\be
\CR= \sum_{g\in G} g\otimes  1_g \quad \in\quad   \C G \,\wh\otimes\,  \prod_{g\in G} \C_g = \wh U_G \,. \label{gauge-RUG}
\ee

\item Finally, in the case when $G=GL(1)^r = (\C^*)^r$ is an abelian group, the $R$ matrix is even more explicit. Let's denote the abelian fields $\mb A_a,\mb B^a$ for $a=1,...,r$. We've got sparks
\be N^a := \int\mb B^a \in \fg\,,\qquad K_a^{\pm 1} := e^{\pm \int \mb A_a} \in \CO(G)\,, \label{gauge-NK} \ee
where $N^a$ is an infinitesimal symmetry defect (a generator for the $a$-th $\mathfrak{gl}(1)$) and $K_a^\pm$ are formed from Wilson lines of charge $\pm 1$ (since they are abelian Wilson lines, they have canonical framing vectors at the $k$ boundaries). \label{NK-gauge-ab}

A module for $U$ is a vector space $\ell$ such that
\begin{itemize}
\item the $N^a$ each act semisimply, with integer eigenvalues (so that $\ell$ is a $GL(1)^r$ representation);
\item there's an action of the Laurent polynomial ring $\CO(G)\simeq \C[K_a^{\pm}]$ on $\ell$\,.
\end{itemize}
The two structures/actions commute. Each $N^a$ weight space independently has an action of $\C[K_a^{\pm}]$. Within each weight space, $K_a$ is simply represented by an invertible matrix, of arbitrary Jordan form.

The braiding on a product $\ell\otimes \ell'$ is now given from \eqref{gauge-Rphys} by
\be
\CR=\prod_a (1\otimes K_a)^{N^a \otimes 1}: \ell\otimes \ell'\to \ell\otimes \ell'\,.  \label{gauge-Rab}
\ee
Here $N^a\otimes 1$ acts as an integer on each weight space of $\ell$, and we raise $(1\otimes K_a)$ to that integer power. 

\end{enumerate}

The algebra $U$ also admits a ribbon element, in a generalized/completed sense, just like the R-matrix. Each of the descriptions of the R-matrix above can be adapted to make sense of the expected relation
\be
v= m\circ ( S\otimes 1)(\CR_{21})=\sum S(f_i)f^i
\ee
(which doesn't quite make sense, due to the hugeness of $\text{Dist}(G)$). Physically, we expect
\be v \sim e^{-\int\mb B\cdot\int\mb A}\,. \ee
On a $U$-module $\ell$, this acts as 
\be v :\ell \to \ell\otimes \CO(G) \overset{\text{id}\otimes S}\longrightarrow  \ell\otimes \CO(G) \overset{\text{act}}\longrightarrow \ell\,, \ee
where the first map is \eqref{gauge-coaction}. In an abelian theory, we simply have
\be \CR = \prod_a K_a^{-N^a} \ee

It is not entirely trivial to rigourously verify the condition $Sv=v$, which ensures $v$ is actually a ribbon element. One way to do this is to use the topological Hopf algebra $\wh U_G = \C G\wh\otimes \Pi_g \C_g$ from (4) above. Both $U$ and $\wh U_G$ embed, as topological vector spaces, into a bigger space
\be U \hookrightarrow \mathrm{Dist}(G)\wh\otimes \lp \prod_{g\in G} \C_g \rp \hookleftarrow \wh U_G\,. \ee
Since the R-matrices of $U$ and that of $\wh U_G$ must agree inside $\mathrm{Dist}(G)\wh \otimes \lp \prod_g \C_g\rp$, we just need to verify $S(v)=v$ in $\wh U_G$. This is easy, since in $\wh U_G$ we've got  $v = \sum_{g\in G} 1_{g^{-1}} g = \sum_{g\in G}g  1_{g^{-1}}$ from  \eqref{gauge-RUG}, which obviously obeys $S(v)=v$.

\subsection{Categories and generators}
\label{sec:gauge-cat}

To conclude this section, we compare module categories for the spark algebras we've constructed with more standard geometric formulations for the categories of lines in B-twisted 3d $\CN=4$ gauge theory. We verify that modules for the spark algebras are indeed equivalent to (particular, and more precise) versions of the expected categories. We had expected this from completeness, even though our argument for Theorem \ref{Thm:faithful} doesn't strictly hold.

We then describe properties of the categories, the special objects $\kDk,\kNk,\CDCN$, and the fiber functors. Had the dualizability condition \AD$_k$ held, we all the $\kDk,\kNk,\CDCN$ would have been compact, dualizable generators, and fiber functors would be given by Hom-ing out of these objects. We explain what gets ``broken'' when \AD$_k$ is relaxed in this example --- and what still works. Our findings are summarized in the following table:
\be \label{gauge-cat-sum}
\hspace{-.5in}
\begin{array}{c|c|c|c}
& \CC_\CN = \text{Rep}(G) & \CC_\CD = \QCoh(G) & \CC_\CT = \QCoh(G/G_{ad}) \\\hline
\text{\ADline}\text{ holds?} & \surd & \times & \times \\
\text{generator} & \kDk = \CO(G) & \kNk = \CO_G & \CDCN = \CO_G  \otimes_{\CC_\CT} \CO(G)_e := \CV_\CN\otimes_{\CC_\CT} \CV_\CD \\
\text{gen. dualizable?} & \times & \times & \times \\
\text{gen. compact?} & \times & \surd & \times \\
\text{alg. ops.?} & \mb m\;\surd\quad \mb m^*\; \times & \mb m\;\times\quad \mb m^*\; \surd & \mb m\;\times\quad \mb m^*\; \times \\
\text{fiber functor} & \CF_\CN = \text{Hom}(\id,\kDk\otimes-) & \CF_\CD = \text{Hom}(\kNk,-) & \CF_\CT = \text{Hom}(\CV_\CN,\CV_\CD\otimes-) \\
& \multicolumn{3}{c}{\text{all $\CF$'s are faithful, continuous} } \\
\text{sparks} & \CH_\CD\simeq \text{End}(\kDk)^* &  \CH_\CN\simeq \text{End}(\kNk) & U \simeq \text{End}(\CV_\CD)^*\otimes \text{End}(\CV_\CN)
\end{array}
\ee

In the final part, Section \ref{sec:gauge-persp}, we connect the categories to a few other perspectives on 3d $\CN=4$ gauge theory including sheaves on derived loop spaces, and modules for boundary VOA's.

\subsubsection{Categories from spark algebras}
\label{sec:gauge-catspark}

First, we can identify the categories of all dg modules (no finiteness condition) for $\CH_\CD=\text{Dist}(G)$ and $\CH_\CN=\CO(G)$ as
\be
 \CH_\CD\text{-mod} \simeq \mathrm{Rep}(G) = \text{QCoh}(pt/G)\,,\qquad \CH_\CN\text{-mod}\simeq \QCoh (G)\,. \label{gauge-cat-id}
\ee
For $\CH_\CN$-mod, this is just the definition of quasi-coherent sheaves on the affine variety $G$: possibly infinite-rank modules for the algebra of functions $\CO(G)$. For $\CH_\CD$-mod, the identification with $\text{Rep}(G)$ follows rigorously from the construction of $\text{Dist}(G)$ as the dual of $\CO(G)$  (\cf\ \eqref{eq:Endrho}); then modules for $\text{Dist}(G)$ are comodules for $\CO(G)$, which are precisely the algebraic representations of $G$, whence $\text{Rep}(G)$.

Note that $\text{Rep}(G)$ is (the dg enhancement of a) semisimple category, albeit with infinitely many simple objects. $\QCoh (G)$ is gigantic, neither finite nor semisimple.

Physically, the $G$-representations in $\CH_\CD\Mod$ are Wilson lines on the $\CN$ boundary. In $\CH_\CN\Mod$, the skyscraper sheaves $\CO_g$ supported at points $g\in G$ are global symmetry defects $V_g$ on the $\CD$ boundary. General sheaves in $\QCoh(G)$ may be thought of as smeared, or averaged, symmetry defects.

The coproduct in $\CH_\CD$ induces the standard tensor product of $G$-representations. The coproduct in $\CH_\CN$ induces \emph{convolution} of sheaves on $G$, \emph{i.e.} the push-forward under the multiplication map
\be \ell\otimes \ell' := m_*(\ell'\boxtimes \ell)\quad\text{for}\; \ell,\ell'\in \text{QCoh}(G)\,,\qquad m:G\times G\to G\,. \label{gauge-convprod} \ee
In particular, skyscraper sheaves will obey $\CO_g \otimes \CO_{g'} = \CO_{gg'}$, which is exactly the way symmetry defects should behave.

As for the category $U\text{-mod}$, we note that an object there has an action of $\CO(G)$ --- hence forms a quasi-coherent sheaf over $G$ --- and possess a compatible action of $\mathrm{Dist}(G)$. Since $\C G$ acts by conjugation on $\CO (G)$, we find that
\be
U\text{-mod}\simeq \QCoh^{G_{ad}}(G) = \QCoh(G/G_{ad})\,, \label{gauge-catU}
\ee
where $G_{ad}$ denotes the adjoint action. The tensor product is again convolution. Moreover, it's known that $\QCoh^{G_{ad}}(G)$ is the derived Drinfeld center, in the sense of \cite{Ben-Zvi:2008vtm}, of either $\text{Rep}(G)$ or $\QCoh(G)$. Physically, sheaves supported at the identity $e\in G$, which must form $G_{ad}$ representations in the equivariant category, correspond to bulk Wilson lines. Sheaves supported on other conjugacy classes in $G$ correspond to bulk vortex lines (\emph{a.k.a.} monodromy defects, or `Gukov-Witten' defects analogous to those from \cite{GukovWitten}).

This all parallels the finite-group categories \eqref{DW-sandN}, \eqref{DW-sandD}, \eqref{DW-bulkcat}. We'd of course like to identify
\be \CC_\CN \simeq \CH_\CD\text{-mod}\,,\qquad \CC_\CT\simeq U\text{-mod}\,,\qquad \CC_\CD\simeq\CH_\CN\text{-mod}\,, \label{gauge-cat-match} \ee
as expected from completeness. To see that this is a reasonable match, we can compare with a \cite{KRS}-style construction of the categories of lines, as in \eqref{matter-book}. The 2d theory of an $(\ol \CN,\CN)$ sandwich is B-twisted $\CN=(2,2)$ $G$ gauge theory \eqref{gauge-Nsand}, and its category of boundary conditions (tentatively $\CC_\CN$) would usually be identified as $\text{Rep}(G)^{fd}$. The 2d theory of a $(\CD,\ol \CD)$ sandwich is a B-model with target $G$, and its category of boundary conditions (tentatively $\CC_\CD$) would usually be identified as $\text{Coh}(G)$. The 2d theory obtained from circle compactification around the link of a bulk line as in \eqref{DW-bulkcat} is a 2d B-model with target $G$ (the holonomy of the connection around $S^1$) and residual gauge group $G$ acting by conjugation. Its category of boundary conditions (tentatively $\CC_\CT$) would usually be identified as $\text{Coh}^{G_{ad}}(G)$.

We take this as sufficient confirmation of \eqref{gauge-cat-match}. Finiteness and dualizability play different roles in standard constructions of boundary conditions for 2d B-models than they do for lines in 3d --- in particular, the tensor product of lines in 3d is lost following the 2d sandwich/circle compactifications --- so we don't actually expect standard finiteness conditions for 2d B-models (\eg\ giving $\text{Rep}(G)^{fd}$ vs. $\text{Rep}(G)$, or $\text{Coh}(G)$ vs. $\text{QCoh}(G)$) to be particularly relevant.

Under the identification \eqref{gauge-cat-match}, we can identify compact objects, and dualizable objects as relevant for our 3d setting. Dualizable objects are just finite-dimensional modules for $\CH_\CD,\CH_\CN,U$. This means
\be
\begin{array}{c} \CC_\CN^{fd} \simeq \mathrm{Rep}(G)^{fd}\,,\qquad \CC_\CD^{fd} \simeq \QCoh (G)^{fd} = \text{Coh}(G)^{\text{fin supp}}\,, \\[.2cm]
\CC_\CT^{fd} = \QCoh^{G_{ad}}(G)^{fd} = \text{Coh}^{G_{ad}}(G)^{\text{fin supp}}\,. \end{array}
 \label{gauge-cat-fd}
\ee
Here $\text{Rep}(G)^{fd}$ just means finite-dimensional representations. $\QCoh (G)^{fd}$ is the category of sheaves with a finite-dimensional space of global sections: this is sheaves supported at finitely many points of $G$, with finite-dimensional stalk at each point. Roughly, $\QCoh (G)^{fd}$ is the category finitely generated by \emph{skyscraper sheaves}. Similarly, $U\text{-mod}^{fd}$ consists of $G_{ad}$-equivariant coherent sheaves on $G$ with finite support.

Finally, the subcategories of compact objects  are 
\be \begin{array}{c} \CC_\CN^{c} = \text{Rep}(G)^{fd}   =  \CC_\CN^{fd} \,,\qquad \CC_\CD^{fd} \simeq \text{Coh}(G) \supsetneq \CC_\CD^{fd}\,, \\[.2cm]
\CC_\CT^c \simeq \text{Coh}^{G_{ad}}(G) \supsetneq \CC_\CT^{fd}\,.
 \end{array} \label{gauge-cat-cpt}\ee
Notably, the axiom \ADline\ (Section \ref{sec:lines}) that we used to control the ``density'' of dualizable objects, and give a straightforward argument for the existence of antipodes (Section \ref{sec:Hopf-def}), \emph{fails} for $\CC_\CD$ and $\CC_\CT$. Of course, we \emph{did} find an antipode for all of $\CH_\CD,\CH_\CN,\CH_U$. The reason it worked, in retrospect, is that we were able to identify symmetries of the fiber functor $\CF_\CD:\CC_\CD\to \text{Vect}$ as the \emph{co-End} of the dual fiber functor $\CF_\CN:\CC_\CN\to \text{Vect}$, which was better behaved. (For some further discussion on the technicalities at play here, see Appendix \ref{app:weakenD}.)

\subsubsection{Generators}
\label{sec:gauge-gen}

We saw in Section \ref{sec:FF-Hom} that when strong dualizability \AD$_k$ of the $k$ interface holds, fiber functors can be represented as Hom's out of algebra objects $\kDk,\kNk,\CDCN$ in their respective categories. Also, spark algebras are finite dimensional. In B-twisted gauge theory, \AD$_k$ does not fully hold, and we'd like to explain what the consequences are.

It is easy enough to construct each of the special objects $\kDk,\kNk,\CDCN$ (whether or not they're related to fiber functors). We can identify $\kNk$ and $\kDk$ by taking the boundary conditions for the 2d sandwich theories $\CD\circ\ol \CD$ and $\ol\CN\circ\CN$ that were described in terms of quantum mechanics in \eqref{gauge-boxQMD} and \eqref{gauge-boxQMN}, respectively, and translating to geometry. The Neumann b.c. in \eqref{gauge-boxQMD} for a 2d B-model to $G$ becomes the structure sheaf
\be \kNk = \CO_G \;\; \in \text{QCoh}(G) = \CC_\CD\,. \ee
The quantum mechanics to $G$ in \eqref{gauge-boxQMN} defines a Wilson line in the regular representation $\CO(G)$ (with a left $G$ action), whence
\be \kDk = \CO(G)  \;\; \in \text{Rep}(G) = \CC_\CN\,. \ee
The bulk object is roughly the tensor product of these: 
\be \CDCN = \CO_G \otimes_{\CC_\CT} \CO(G)_e  \;\; \in \QCoh^{G_{ad}}(G) = \CC_\CT\,. \label{gauge-CDCN}  \ee
Here $\CO_G$ is the structure sheaf of $G$; $\CO(G)_e$ is a skyscraper sheaf at the identity whose stalk is a copy of the $G_{ad}$ module $\CO(G)$; and the tensor $\CO_G \otimes_{\CC_\CT} \CO(G)_e$ produces a sheaf supported on all of $G$ whose fiber at each point is a copy of $\CO(G)$.  As might have been expected, \emph{none} of $\kNk,\kDk,\CDCN$ are dualizable. The object $\kNk$ is compact (though the other two are not).  All of them are nonetheless generators.

We can also try to construct the multiplication $\mb m$ and comultiplication $\mb m^*$ that would make $\kDk,\kNk$ algebra objects. They are almost the same as for DW theory in Section \ref{sec:DW-mm}:
\begin{itemize}
\item For $\kDk=\CO(G) \in \text{Rep}(G)$, the putative multiplication $\mb m:\CO(G)\otimes \CO(G)\to \CO(G)$ and comultiplication $\mb m^*:\CO(G)\to \CO(G)\otimes \CO(G)$ should be induced from pull-back and push-forward along the diagonal map $\text{diag}:G\hookrightarrow G\times G$ as in \eqref{diag-push}. Pull-back of algebraic functions is always well defined, so $\mb m(f,f')(g):= f(g)f'(g)$ is fine. Push-forward is not strictly defined. One would want
\be (\mb m^* f)(g,g') := f(g)  \cdot \delta_{g=g'}\,, \ee
but the delta-distribution on the diagonal $\delta_{g=g'}$ is not an element of $\CO(G)\otimes \CO(G)$. For example, when $G=GL(1)=\C^*$, with $\CO(\C^*) = \C[x^{\pm 1}]$ (Laurent polynomials), the delta function could be written as $\delta_{x=y}=\sum_{n\in \Z} (x/y)^n$, but this requires a formal completion.

\item For $\kNk=\CO_G$, we have $\CO_G\otimes_{\CC_\CD}\CO_G \simeq \CO(G)\otimes_\C \CO_G$ (the structure sheaf of $G$, tensored with the vector space $\CO(G)$). Let's also write $\CO_G \simeq \C \otimes_\C \CO_G$.
 Then the putative maps $\mb m$ and  $\mb m^*$ come from maps of the vector spaces
\be \wt m: \CO(G)\to \C\,,\qquad \wt m^*: \C \to \CO(G)\,, \ee
tensored with the identity on $\CO_G$. The second map is just the inclusion of $\C$ as constant functions on $G$, which is fine. (So $(\wt m^*(1))(g):= 1$ for all $g$.) The first map requires an integration. Morally, it is given by restricting functions to the compact real locus $G_c\subset G$, and integrating against a normalized Haar measure
\be \wt m (f) := \oint_{G_c} f(g)\, dg\,,\qquad\text{normalization:}  \quad  \oint_{G_c}  1\,dg = 1\,. \ee
The subtle problem here is that, unless the group is abelian, the Haar measure is not algebraic. In order to intertwine the coproduct on sparks as in \eqref{mm*-D}, $\wt m$ also needs to induce a well-defined convolution of algebraic functions, of the form
\be * : \CO(G)\otimes\CO(G)\to \CO(G)\,,\qquad (f*f')(g) = \oint_{G_c} f(h^{-1}g)f'(h)\,dh\,,  \label{gauge-conv-int} \ee
and the RHS of \eqref{gauge-conv-int} does not land in $\CO(G)$ but a completion thereof (unless $G$ is abelian).

\end{itemize}
The lesson is that $\mb m$ is not strictly defined for $\kNk$ (unless $G$ is abelian) and $\mb m^*$ is not strictly defined for $\kDk$. Upon inspecting the thickened strips of $k$ interfaces that correspond to $\mb m$ and $\mb m^*$ (as in \eqref{mm*-strip}), this immediately implies that the following cap and cup formed from $k$ are not strictly defined:
\be \label{illcupcap}\hspace{-.5in} \text{ill-defined:}\qquad \raisebox{-.2in}{\includegraphics[width=2.3in]{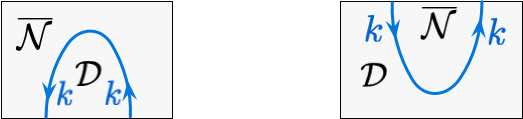}} \ee
This is a precise test of how the dualizability condition \AD$_k$ fails.

\subsubsection{Fiber functors}
\label{sec:gauge-FF}

We'll now identify our fiber functors in terms of the geometric categories $\CC_\CN=\text{Rep}(G)$, $\CC_\CD = \QCoh(G)$, $\CC_\CT = \QCoh(G/G_{ad})$, and relate them to the generating objects $\kDk,\kNk,\CDCN$.
Recall that in terms of modules for spark algebras, the fiber functors are always forgetful functors; we need to translate this to geometry.

The fiber functor $\CF_\CN$ is obviously given by the forgetful functor $\mathrm{Rep}(G)\to \mathrm{Vect}$. 
However, this does not correspond to Hom \emph{out} of $\kDk=\CO(G)$ as might have been expected, but rather by Hom \emph{into} the tensor of $\kDk$ and a given line:
\be
\CF_\CN (\ell )=\Hom_{\CC_\CN}(\id, \kDk\otimes \ell )\,. \label{gauge-FN}
\ee
This is the appropriate version of the state-operator correpondence for the current situation, replacing Section \ref{sec:FF-Hom}, when not all the cups/caps using $k$ make sense.

Since $\kDk$ is not dualizable, $\CF_\CN$ is \emph{not} equivalent to $\text{Hom}_{\CC_\CN}(\kDk,-)$. Relatedly, $\CF_\CN$ does not admit a left adjoint.%
\footnote{The existence of adjoints is how we will interpret \AD$_k$ in the mathematical setting of Appendix \ref{subsubsec:appendixfiber}.} %
This is the mathematical reason why $\End (\CF_\CN)=\mathrm{Dist}(G)$ is a topological Hopf algebra, with a completed coproduct, rather than an ordinary one. In \eqref{gauge-FN} the object $\kDk$ is being used as a \emph{co-generator} for $\CC_\CN$. The fiber functor $\CF_\CN$ is still faithful, as well as continuous. Curiously, its endomorphism algebra is still computed as $\text{End}_{\CC_\CN}(\kDk)$, since we have
\be \CH_\CD =\text{Dist}(G)  = \lp \CO (G)\otimes \CO(G)^*\rp^G = \text{End}_{\CC_\CN}(\kDk)=\text{End}(\CF_\CN) \,. \ee
This is true essentially because $\CC_\CN$ still satisfies \AD$_{\rm line}$ and therefore is self-dual (see Section \ref{app:weakenD}). 

On the other hand, $\CF_\CD$ is given by the global section functor $\QCoh (G)\to \mathrm{Vect}$, which \emph{is} Hom out of the structure sheaf $\kNk=\CO_G$,
\be
\CF_\CD (\ell )=\Hom_{\CC_\CD}(\kNk, \ell)\,.
\ee
$\kNk$ is compact, and a generator; thus $\CF_\CD$ is faithful and continuous and
\be \CH_\CN = \text{End}(\CF_\CD) = \text{End}_{\CC_\CD}(\kNk) = \CO(G) \ee
is a genuine Hopf algebra, rather than a topological one. However, $\CC_\CD$ breaks \AD$_{\rm line}$ due to \eqref{gauge-cat-cpt}. This is why $\End (\CF_\CD)$ is not finite-dimensional, and the fiber functor $\CF_\CD$ does not admit a right adjoint.

The situation becomes worse for $U\Mod=\QCoh (G/G_{ad})$. This category breaks \AD$_{\rm line}$ for the same reason $\CC_\CD$ does. The fiber functor admits neither left nor right adjoint, and is represented by neither an internal Hom in nor out of $\CDCN$. It is rather given by a combination
\be
\CF_\CT (\ell)=\Hom_{\CC_\CT} (\CO_G, \CO(G)_e\otimes \ell)\,,
\ee
splitting the two factors of $\CDCN$ from \eqref{gauge-CDCN}.
If we allow ourselves to take the dual of $\CO(G)_e$ and treat this as a limit of objects $\CO(G)^*=\lim_i (\rho_i)_e$ (skyscraper sheaves with fibers given by finite-dimensional reps $\rho_i$), then
\be
\CF_\CT (\ell)=\varinjlim_i \Hom_{\CC_\CT} (\CO_G \otimes (\rho_i)_e, \ell)\,.
\ee
The endomorphism of this limit of objects $\lim_i \CO_G\otimes (\rho_i)_e$ is precisely the topological Hopf algebra $U$. Although each individual object $\CO_G\otimes (\rho_i)_e$ is not a generator, the projective system is in a limit sense. This is intuitively why $\CF_\CT$ is also faithful.

\subsubsection{Connection to other perspectives}
\label{sec:gauge-persp}

We make some brief final remarks on the relation between our categories and other perspectives on lines in B-twisted gauge theory.

\begin{enumerate}[leftmargin=*]
    \item If we treat the B twist of pure gauge theory as RW theory to $T^*BG$, then a candidate for the bulk category is $\QCoh (\CL BG)$, where $\CL BG=\mathrm{Maps}(S^1, BG)$ is the loop space of the classifying space $BG$. This is a braided tensor category thanks to \cite{Ben-Zvi:2008vtm}, where it's shown to be the (derived) Drinfeld center of $\mathrm{Rep}(G)$. As algebraic varieties, $\CL BG$ is precisely $G/G_{ad}$, and we find equivalences
\be
U\Mod\simeq \QCoh (G/G_{ad})\simeq \QCoh (\CL BG)\,.
\ee
The braiding on $\QCoh (\CL BG)$ is not explicitly worked out in \cite{Ben-Zvi:2008vtm}.
A braiding for this category is explicitly described in \cite{bezrukavnikov2023equivariant}, which we expect to be equivalent to that given by our R-matrix. 

\item If we further simplify $T^*BG$ to $T^*B\fg$, and represent the latter as the space $\fg^*[-1]/\fg$, then just as in the case of affine spaces, we expect the infinitesimal braiding to be given by the Poisson bi-vector, where the Poisson structure is induced from the pairing between $\fg^*$ and $\fg$. We indeed find such an expression in the perturbative $R$ matrix of \eqref{gauge-Rpert}. 

\item If we ignore the geometry of $G$, and replace B-twisted gauge theory with DW theory for the infinite group $G$ (as an infinite discrete group), we get a huge category of bulk lines that looks like modules for the algebra $\wh U_G$ introduced in \eqref{gauge-RUG}. The R-matrix $\CR=\sum_{g\in G} g\otimes 1_g$ constructed there indeed takes the same form as in DW theory.

\item For $G=GL(1) = \C^*$, the bulk category $U\Mod$ is equivalent to modules for the commutative polynomial algebra $U'=\C[N,K^{\pm}]$, where $N$ acts semisimply with integer eigenvalues, as described below \eqref{gauge-NK}. This is a ribbon Hopf subalgebra of $U = \text{Dist}(\C^*)\otimes \CO(\C^*)$, with
\be \label{gauge-ab}
\begin{array}{c} \Delta(N) = N\otimes 1+1\otimes N\\ \Delta(K) = K\otimes K \end{array}\,,\quad 
\begin{array}{c} \varepsilon(N) = 0\\ \varepsilon(K)=1 \end{array}\,,\quad 
\begin{array}{c} S(N) = -N \\ S(K) = K^{-1} \end{array}\,,\quad 
\begin{array}{c} \CR = (1\otimes K)^{N\otimes 1} \\ v = K^{-N} \end{array}\,.
\ee

On the other hand, the boundary VOA for B-twisted abelian gauge theory was identified in \cite{BCDN}, as a $\Z$-lattice extension of the perturbative VOA constructed by \cite{CG}, which in turn is just a rank-two Heisenberg algebra. This boundary VOA is a well-studied algebra known as a half-lattice VOA ``$\Pi(0)$'' (see for instance \cite{Adamovic:2019tcl}). Its category of modules is studied in \cite{berman2002representations}, and the full subcategory generated by weighted modules is known to be equivalent to $U'\Mod$ as a ribbon category \cite{creutzig2020quasi}. (Typically the subcategory of dualizable objects, or even smaller categories where $K$ and $N$ both act semisimply, are the ones that appear in the VOA literature.)

\item The boundary VOA for nonabelian $G$ gauge theory is not known. Perturbatively, it is a $\fg\ltimes \fg^*$ current algebra \cite{CG}, but boundary monopole operators correct it in ways that are not yet well understood. By 3d mirror symmetry, we expect its module category to be equivalent to modules for the ``chiral universal centralizer'' at critical level $\mb I^{\kappa_c}_{{}^LG}$ defined by Arakawa \cite{Arakawa-chiral}. This comes from observing that 3d mirror symmetry relates B-twisted pure gauge theory with an A-twisted sigma-model to the Coulomb branch of pure gauge theory, \emph{a.k.a.} the universal centralizer for the Langlands-dual ${}^LG$ \cite{Teleman-ICM}; and the boundary VOA in the A-twist of a sigma-model should be a curved beta-gamma system on, \emph{a.k.a.} a chiralization of, the target \cite{CG}. Thus we have a conjecture:
\be \mb I^{\kappa_c}_{{}^LG}\Mod \;\overset{?}\simeq\; U\Mod\,. \ee
So far, the representation theory of $\mb I^{\kappa_c}_{{}^LG}$ has been difficult to study.

\end{enumerate}

\section{Gauge theory with matter}
\label{sec:GM}

In this final section, we consider the general case of B-twisted 3d $\CN=4$ gauge theory with gauge group $G_c$ and linear matter in a symplectic representation $T^*V$ of $G_c$. 
Just like our other 3d $\CN=4$ examples, this theory has multiple, equivalent descriptions, \eg
\begin{itemize}
    \item $G$-equivariant Rozansky-Witten theory to $T^*V$, \emph{a.k.a.} Rozansky-Witten theory whose target is the `symplectic stack' $T^*(V/G)$
    \item Derived BF theory with supergroup $G_c\ltimes \Pi V$ \cite{Kapustin:2009cd}
    \item Derived Chern-Simons theory with partially noncompact supergroup $T^*(G_c\ltimes \Pi V)\simeq  (G_c\ltimes \Pi V)\ltimes (\fg^* \times \Pi V^*)$, and a level given by the symplectic form on the cotangent bundle.
\end{itemize}
We'll mainly use the B-twisted 3d $\CN=4$ (or Rozansky-Witten) perspective, but will return to the supergroup perspective in Section \ref{sec:GM-supergroup}, as it unifies several constructions.

Gauge theory with matter compounds both the features and complications of the pure-matter and pure-gauge examples from  previous sections.  As in pure matter theory, some sparks come from integrated descendants, and require additional degrees of freedom (boundary fermions) along $k$ interfaces to be added. 
As in pure gauge theory, its spark algebras are infinite-dimensional (\AD$_k$ fails), so topology and completions come into play. The subtle dualizability discussions from pure gauge theory carry over directly to gauge theory with matter, and we won't repeat them here; instead, we focus on new features involving the coupling of gauge and matter.

One new feature is the presence of framing anomalies, which arise when $\det V$ is a nontrivial representation of $G$. We'll discuss in Section \ref{sec:GM-framing} how they can be compensated or avoided so as not to invalidate the analysis of the current paper. We note, though, that if one were to turn on FI parameters to resolve a Higgs branch (which is beyond the scope of this paper), and $\det V$ is nontrivial, framing anomalies will become impossible to avoid.

From the perspective of constructing spark algebras, the new features involve 1) needing to \emph{dress} matter sparks on $\CN$ with Wilson lines in order to make them gauge-invariant; and 2) nontrivial commutation relations between sparks on $\CD$ and sparks on $\CN$, due to these Wilson-line dressings, as well as to symmetry defects on $\CD$ acting nontrivially on matter. We describe this in Sections \ref{sec:GM-vect}--\ref{sec:GM-bulk}. The spark algebras we find take the form
\be \CH_\CD \simeq \text{Dist}(G)\ltimes \Lambda^\bullet V\,,\qquad \CH_\CN \simeq \CO(G) \times \Lambda^\bullet V^*\,, \ee
and, beautifully, the R-matrix and ribbon elements in the double $U\simeq \CH_\CD\otimes \CH_\CN$ are just products of those from pure-matter and pure-gauge theory:
\be \CR \simeq \CR_G \CR_V\,,\qquad v \simeq v_G v_V\,. \ee

In Section \ref{sec:GM-cat}, we'll argue that $U\Mod$ faithfully reproduces one of the expected descriptions of bulk line operators, as a category of equivariant matrix factorizations
\be \CC_\CT \simeq \text{MF}^G\big(G\times T^*[2](V[1]),W\big)\,,\qquad W = Y(\varphi_V(g)-1)X\,. \ee
Now, however, $U$ makes the braided ribbon structure on this matrix-factorization category fully explicit. We'll connect the braided structure coming from $U$ to braidings that are known to be induced from Drinfeld centers \cite{arkhipov2017equivariant} and from derived-loop-space \cite{Ben-Zvi:2008vtm} descriptions.  For abelian $G$, we also relate $U$ to the generalized quantum group found in \cite{creutzig2024kazhdan} using boundary vertex algebras.

\subsection{Action and boundary conditions}
\label{sec:GM-bc}

Twisted BV formalism for gauge theory with matter is summarized in \cite{Garner:2022rwe}.
The fields in the B twist of 3d $\CN=4$ $G_c$ gauge theory with $T^*V$ matter are the same as from Sections \ref{sec:matter}--\ref{sec:gauge},
\be \begin{array}{c}
\mb A \in \Pi\mathfrak g\otimes \Omega^\bullet(M)[1]\,,\qquad \mb B \in \Pi\mathfrak g^*\otimes \Omega^\bullet(M)[1]\,, \\[.1cm]
 \mb X \in V \otimes  \Omega^\bullet(M)[1]\,,\qquad \mb Y \in V^* \otimes  \Omega^\bullet(M)[1]\,. \end{array} \ee
with action%
\footnote{A coupling constant $\hbar$ --- parameterizing the symplectic form on $T^*V$ --- could be included, as in \eqref{matter-S}, by rescaling $\mb Y\to \hbar^{-1}\mb Y$.}
\be S = \int_M \mb B d F(\mb A) + \mb Y d_{\mb A}\mb  X =  \int_M \mb B d \mb A  +  \mb Y d\mb  X + \mb B\mb A^2 + \mb Y \varphi_V(\mb A)\mb X \label{GM-S} \ee
and BRST transformations
\be \begin{array}{c} Q\,\mb A = F(\mb A)\,,\quad Q\,\mb B= d\mb B+[\mb B,\mb A] + \varphi_V^*(\mb X\mb Y)\,, \\[.1cm]  Q\,\mb X = d\mb X+\varphi_V(\mb A)\mb X\,,\quad Q\,\mb Y = d\mb Y -\mb Y\varphi_V(\mb A)\,, \end{array} \label{GM-BRST} \ee
where $\varphi_V: \fg\to\text{End}(V)$ is the representation map, and $\varphi_V^*$ its adjoint (\ie\ transpose).

Our boundary conditions will be a product of the previous gauge and matter cases,
\begin{itemize}
\item[$\CN:$] Neumann for $G$, $\mb B\big|_\pd=\mb Y\big|_\pd = 0$\,,\qquad ($\mb A\big|_\pd,\mb X\big|_\pd$, gauge symmetry $G$ survive)
\item[$\CD:$] Dirichlet for $G$, $\mb A\big|_\pd=\mb X\big|_\pd = 0$\,,\qquad ($\mb B\big|_\pd,\mb Y\big|_\pd$, global symmetry $G_\pd$ survive)
\end{itemize}
These are again B-twists of half-BPS $\CN=(2,2)$ boundary conditions from \cite{BDGH}. 

Transversality now works the same way that it did in pure-gauge and pure-matter theories. The $k$ interface between $\CD$ and $\ol \CN$ (say) can be constructed explicitly by 1) ``resolving'' $\CD$ as  $\ol \CN$ coupled to an extra 2d boundary B-model with target $G$, and an extra boundary $\mb X\mb Y$ superpotential;  2) putting a Neumann b.c. for the boundary $G$ matter fields and the boundary $G$ gauge fields at $k$; 3) adding boundary fermions $\mb a,\bar {\mb a}\in T^*V$ at $k$, with $Q\mb a = \mb X$ and $Q\mb {\bar a}=\mb Y$. The only distinction between this and previous constructions of $k$ is that the fermions $a,\bar a$ are now charged under $G$.  (Their 1d QM action involves a coupling to the connection $\mb A\big|_k$.)
\be  \raisebox{-.3in}{\includegraphics[width=2.3in]{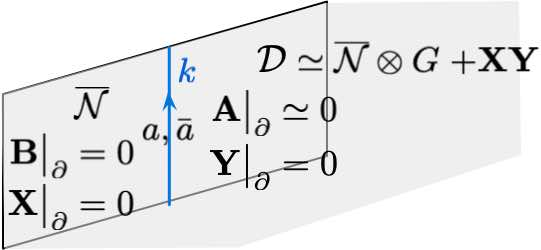}} \label{GM-NkD}   \ee

Just as in gauge theory, we expect that condition \AD$_k$ fails: the interface $k$ between $\CD$ and $\ol \CN$, or $\CN$ and $\ol\CD$, won't be dualizable, and spark algebras should be infinite dimensional.

Completeness also works the same way as before. To glue two a left half-space with $\CN$ b.c. (and fields $\mb A,\mb B,\mb X,\mb Y$) to a right half-space with $\CD$ b.c. (and fields $\mb A',\mb B',\mb X',\mb Y'$), we add a boundary action
\be S_{\rm glue} = - \int_{\R^2} \Big(\mb X\big|_0 \cdot \mb Y'\big|_0 + \mb A\big|_0\cdot \mb B'\big|_0\Big)\,. \label{GM-Sglue} \ee
This continues to make sense  for any gluing region $R$ that is well bounded in the sense of Def. \eqref{def-bounded}, as long as we introduce additional fermions along the edges of $\pd R$ as in \eqref{matter-SglueR}.

\subsubsection{Framing anomaly}
\label{sec:GM-framing}

There is a potential boundary framing anomaly in gauge theory with matter. If we view $\CD,\ol\CN$ as left boundary conditions and $\CN,\ol \CD$ as right boundary conditions, then the boundary anomaly polynomials computed as in \cite{dimofte2018dual} take the form
\be \begin{array}{l@{\quad\;\;}l}
 \CD:\;\;\mb t\cdot \mb f -  \text{Tr}(\varphi_V(\mb f))\mb c - 2(\text{rank}\,\fg)\mb h\mb c &  
 \CN:\;\; -  \mb t\cdot \mb f +\text{Tr}(\varphi_V(\mb f))\mb c + 2(\text{rank}\,\fg)\mb h\mb c \\[.1cm]
 \ol\CN:\;\; \mb t\cdot \mb f +\text{Tr}(\varphi_V(\mb f))\mb c + 2(\text{rank}\,\fg)\mb h\mb c &  
 \ol\CD:\;\;  -\mb t\cdot \mb f -\text{Tr}(\varphi_V(\mb f))\mb c - 2(\text{rank}\,\fg)\mb h\mb c\,,
\end{array}
 \label{GM-anom} \ee
where
\begin{itemize}
\item  $\mb c,\mb h$ are field strengths (\emph{a.k.a.} curvatures) for $U(1)_C$, $U(1)_H$ boundary R-symmetries (recall that a $U(1)_C$ background is turned on for the B-twist in curved space; while $U(1)_H$ charge is cohomological degree)
\item  $\mb f$ denotes the dynamical $G$ gauge field strength on $\CN,\ol\CN$, and denotes the $G_\pd$ global symmetry field strength on $\CD,\ol\CD$
\item $\mb t$ is the field strength for the global, abelian topological symmetry $G_{\rm top}$ (recall that $G_{\rm top}$ is dual to the continuous part of the center of $G$, in that $\fg_{\rm top}\simeq \text{Hom}_G(\fg,\C)$). 
\end{itemize}
We have not included flavor symmetry acting on hypermultiplets in this analysis, as it does not play an essential role.

We see two essential features from the above anomaly polynomials. If we want to use both $\CN$ and $\ol\CN$ boundaries, then whenever $\text{Tr}(\varphi_V(\mb f))\neq 0$ the symmetry $U(1)_C$ will be broken and there will be a framing anomaly. Since $\mb f$ is dynamical, the condition $\text{Tr}(\varphi_V(\mb f))\neq 0$ is equivalent to $\text{det}(V)$ being a nontrivial representation of $G$.

Dually, if we want to use both $\CD$ and $\ol \CD$ boundaries \emph{and} introduce boundary curvature so that $\mb c\neq 0$, then the boundary $G_\pd$ symmetry will be broken to a quotient that acts trivially on $\text{det}(V)$.

If one wants to avoid framing anomalies altogether, then one should choose a $V$ such that $\text{det}(V)$ is a trivial representation of $G$.

Alternatively, if we are careful, we can use the topological symmetry $G_{\rm top}$ to compensate locally for the anomaly.
Namely, in the presence of only $(\CN,\CD)$ boundary conditions, we observe from the terms $\pm(\mb t\cdot \mb f -  \text{Tr}(\varphi_V(\mb f))\mb c)$ in their anomaly polynomials that there is a diagonal combination $U(1)_C' \subset G_{\rm top}\times U(1)_C$ that does not couple to $\mb f$. If we use $U(1)_C'$ rather than $U(1)_C$ to twist near the boundaries, then we avoid both a framing anomaly on $\CN$ and a breaking of $G_\pd$ on $\CD$.

We could even introduce $(\ol\CN,\ol\CD)$ as well, if quickly we swap from twisting by $U(1)_C'$ near $(\CN,\CD)$ to twisting by the opposite combination $U(1)_C'' \subset G_{\rm top}\times U(1)_C$ that doesn't couple to $\mb f$ in $(\mb t\cdot \mb f +  \text{Tr}(\varphi_V(\mb f))\mb c)$ near $(\ol\CN,\ol\CD)$. This works --- in that it avoids framing anomalies and breaking of $G_\pd$ --- as long the $(\ol\CN,\ol\CD)$ and $(\CN,\CD)$ boundaries are separated. It will fail, \eg, if we form infinitesimal $\ol\CN\circ\CN$ or $\CD\circ \ol\CD$ sandwiches: the 2d sandwich theories have unavoidable anomalies. (An example is the 2d B-model in \eqref{GM-CN-exp} below.)

Twisting by $U(1)_C'$, or alternatively by $U(1)_C''$, near boundaries allows us to proceed with our standard constructions of spark algebras. The cost for this modified twisting is to introduce a nontrivial $G_{\rm top}$ background near boundaries, proportional to boundary curvature. This has two effects, both of which seem innocuous to sparks:
\begin{enumerate}
\item The $G_{\rm top}$ background could act on local operators (or operators of finite extent, such as sparks) that are charged under $G_{\rm top}$. However, there are no such operators in the B twist. This is because the current for $G_{\rm top}$, which is the abelian part(s) of $F(\mb A)$, is $Q$-exact due to  \eqref{GM-BRST}.
\item Curvature defects for $G_{\rm top}$ may be introduced, in the presence of sufficiently strong boundary curvature. The only nontrivial curvature defects for $G_{\rm top}$ are line-like, and are equivalent to abelian Wilson lines for $G$ (see \cite[Sec. 6]{BCDN}). Here we may find Wilson lines in powers of the representation $\text{det}(V)$.

These Wilson lines might modify the antipode in the spark algebra on $\CN$, but we'll see later that their corrections cancel out, essentially because they're abelian.
\end{enumerate}

\subsection{Expectations for categories, generators, and sparks}
\label{sec:GM-vect}

In the spirit of \cite{KRS}, we can do some quick reductions on intervals and rectangles to establish some expectations about boundary line operators (as categories), and boundary sparks (as vector spaces).

Reducing $\CT$ on a $(\CD,\ol\CD)$ sandwich produces a 2d B-model whose target is $G\times V^*[1]$. As in pure gauge theory, the factor of $G$ comes from the holonomy of the gauge connection from one $\CD$ boundary to the other; and as in pure matter theory, the factor of $V^*[1]$ comes from the $\mb Y$ fields that survive on the sandwich. With the shift `$[1]$' we keep track of the fact that functions on this bosonic $V^*$ have cohomological degree $1$.
Using a standard description of boundary conditions for a 2d B-model, we then expect
\be \CC_\CD \sim \text{Coh}(G\times V^*[1]) \label{GM-CD-exp} \ee
In particular, the boundary condition for the sandwich that's constructed from a thin strip of $\CN$, as on the RHS of \eqref{gauge-boxQM}, is Neumann for the $G$ holonomy and Dirichlet for $V^*$ fields (setting $Y=0$). In coherent sheaves, that identifies is as the structure sheaf of $G\times\{0\} \subset G\times V^*[1]$, whence we expect
\be \kNk \simeq \CO_{G\times \{0\}} \label{GM-kNk-exp} \ee

If we further compactify all the way down to 1d on a rectangle:
\be \raisebox{-.5in}{\includegraphics[width=5in]{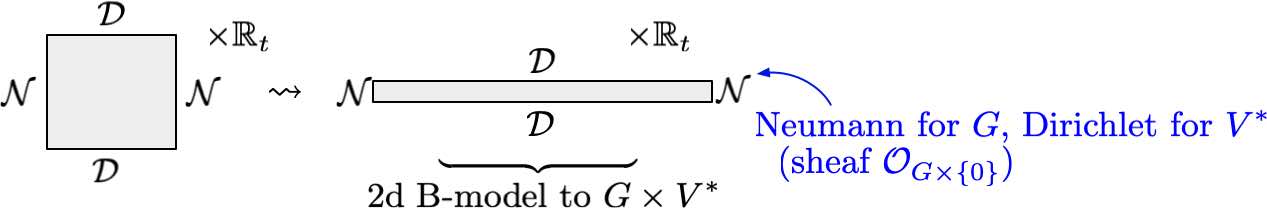}} \label{gauge-boxQM} \ee
we'll end up with a 1d B-twisted $\CN=2$ quantum mechanics with $G$-valued chiral multiplets and $T^*V$-valued fermi multiplets $(\psi_+,\psi^-)$. The fermi multiplets are just like in the pure-matter theory from Section \ref{sec:matter-spaces}. The state space on the rectangle has multiple quantizations, corresponding to inequivalent polarizations. The two that will be important for us are
\be \label{GM-states-exp} \begin{array}{l}  \text{States(rectangle)} \simeq \CO(G)\otimes \Lambda^\bullet V^*\qquad \text{(matching $\CH_\CN$)} \\[.1cm]
 \text{States(rectangle)} \simeq \text{Dist}(G) \otimes \Lambda^\bullet V \qquad \text{(matching $\CH_\CD$)} \end{array} \ee
These two polarizations are naturally dual. The quantizations of $\Lambda^\bullet V^* = \C[\psi^-_i]_{i=1}^{\text{dim}V}$ and $\Lambda^\bullet V = \C[\psi_+^i]_{i=1}^{\text{dim}V}$ and their inner product
\be h(\psi_+^i,\psi^-_j) = \delta^i{}_j\,,\qquad \text{induced (\emph{e.g.}) from}\; \psi_+^i = \frac\pd{\pd \psi^-_i} \;\text{on $\C[\psi^-_i]$} \ee
is identical to Section \ref{sec:matter-spaces}. The quantizations of algebraic functions $\CO(G)$ and distributions $\text{Dist}(G)$ and the inner product on them is identical to that in Section \ref{sec:gauge-vector}. Altogether, we have a pairing on dual states
\be h(g \otimes \psi_+^{i_1}...\psi_+^{i_k}, f\otimes \psi^-_{j_1}...\psi^-_{j_k}) =f(g)\, \frac{\pd}{\pd \psi^-_{i_1}}\frac{\pd}{\pd \psi^-_{i_2}}...\frac{\pd}{\pd \psi^-_{i_k}}(\psi^-_{j_1}\psi^-_{j_2}...\psi^-_{j_{k'}})\Big|_{\psi^-\equiv 0}\,.   \label{GM-pair-vect} \ee 

We can alternatively compactify $\CT$ on an $(\ol\CN,\CN)$ sandwich, getting a 2d B-model with target $V$ and gauge group $G$:
\be \raisebox{-.5in}{\includegraphics[width=5.3in]{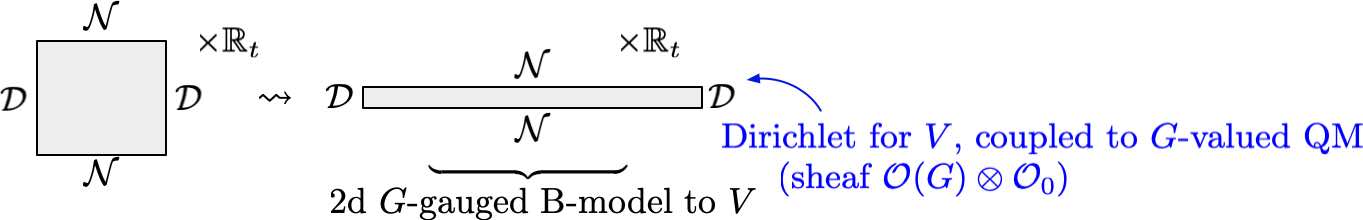}} \label{GM-boxQMN} \ee
The standard description of boundary conditions for this gauged B-model suggests that line operators on $\CN$ are $G$-equivariant coherent sheaves on $V[1]$,%
\footnote{As anticipated in Section \ref{sec:GM-framing}, this 2d B-model has a potential framing anomaly. It's expressed mathematically by the existence of a nontrivial Serre functor on the category $\text{Coh}^G(V)$, given by tensoring by the potentially nontrivial equivariant canonical bundle of $V$, which is the determinant bundle of $TV\to V$. This anomaly/Serre functor is not relevant for analyzing dualizability of objects in $\CC_\CN$, as lines on an $\CN$ boundary of a 3d theory. The anomaly is measuring what happens upon rotation of boundaries of the 2d sandwich theory after compactification (which does have monodromy), as opposed to rotation of lines in the plane of the boundary of the 3d theory (which we're actually interested in.)}
\be \CC_\CN \sim \text{Coh}^G(V[1])\,. \label{GM-CN-exp} \ee
The special boundary condition constructed from a strip of $\CD$ sets the $V$ matter fields to zero (hence Dirichlet), and also adds additional 1d $G$-valued matter. This lets us identify
\be \kDk \simeq \CO(G) \otimes \CO_0\,, \label{GM-kDk} \ee
a skyscraper sheaf at $0\in V[1]$, whose stalk is the regular representation of the group $G$.

Should we further compactify the $(\ol\CN,\CN)$ sandwich of \eqref{GM-boxQMN} to 1d, we find a second description of the effective theory on a rectangle. Now it appears as a 1d B-twisted $G$ gauge theory, with chiral matter $G\times G$, and a $T^*V$ fermi multiplet. Just as in pure gauge theory, the $G$ action on $G\times G$ (right on the first factor, left on the second) is free, and the quotient can be performed, leaving behind 1d B-twisted QM with $G$-valued chiral matter and a $T^*V$ fermi multiplet. That leads to the same description of rectangle state spaces as in \eqref{GM-states-exp}.

\subsection{$\CH_\CD$ from symmetry and matter}
\label{sec:GM-HD}

As in pure gauge theory, the spark algebra $\CH_\CD$ is controlled by the boundary $G_\pd$ symmetry.

Looking at anomaly polynomials \eqref{GM-anom}, we see that in the presence of nonzero $U(1)_C$ or $G_{\rm top}$ curvature near the boundary, the $G_\pd$ symmetry could be broken. However, if we perform the same alignment of $U(1)_C$ and $G_{\rm top}$ curvatures that we used to cancel a framing anomaly on $\CN$, we will also end up preserving the $G_\pd$ symmetry on $\CD$. We will assume such an alignment has been done.

Sparks on $\CD$ now contain different elements, all in cohomological degree zero:
\begin{enumerate}
\item Infinitesimal $\fg_\pd$ transformations, represented by integrated currents $\alpha\int\mb B$ for $\alpha\in \fg$
\item Finite $G_\pd$ symmetry defects, labelled by $g\in G$. If $g = e^\alpha$, then these are represented as exponentiated integrals $\exp \alpha\int\mb B$, with a point-splitting regularization in the vertical direction.
\item Integrated descendants of $Y$, corrected with fermions at the $k$ boundaries. Given a basis $e_i$ for $V^*$ and writing $\mb Y = \sum_i \mb Y^i e_i$, these sparks are represented as $\psi_+^i = \int Y^{i(1)}-\bar a^i+\bar b^i$, and are fermionic elements of the dual space $V$.
\end{enumerate}
\be \raisebox{-.5in}{\includegraphics[width=5in]{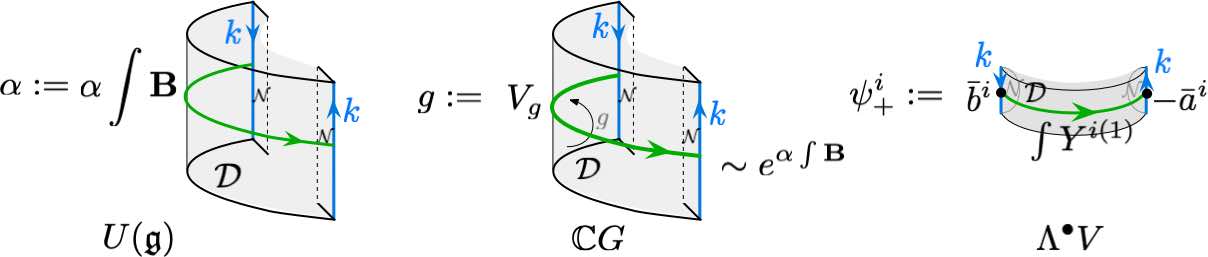}} \ee
The $G_\pd$ symmetry (or perturbatively: a 1-loop quantum correction) ensures that the associative algebra structure among $\alpha$'s and $g$'s is the same as described in pure gauge theory (Section \ref{sec:gauge-HD}). Namely, the product of $g$'s is the product in $G$ and the commutator of $\alpha$'s is the Lie bracket in $\fg$. Together, they generate $\text{Dist}(G)$, with its convolution product. 

The associative product of the fermionic sparks $\psi_+^i\in \Lambda^\bullet V$ is graded-commutative, just like it was for pure matter. There are no new corrections to this by coupling to gauge theory. In particular, the bulk interaction vertex $\mb Y\varphi_V(\mb A)\mb X$ in \eqref{GM-S} cannot be used to construct any Feynman diagrams that would correct the product of two $\psi_+$'s perturbatively; and the equations of motion in the B-twist don't allow nonperturbative instanton corrections.

The final piece of the associative structure involves commuting elements of $\text{Dist}(G)$ past $\psi_+$'s. Now the interaction vertex $\mb Y\varphi_V(\mb A)\mb X$ does contribute. Equivalently, the commutator is controlled by how boundary $G_\pd$ symmetry acts on $\psi_+$. Therefore,
\be g \psi_+^i g^{-1} = \varphi_V(g)^i{}_j \psi_+^j\,,\qquad [\alpha,\psi_+^i] = \varphi_V(\alpha)^i{}_j  \psi_+^j\,,
\qquad \raisebox{-.4in}{\includegraphics[width=2.4in]{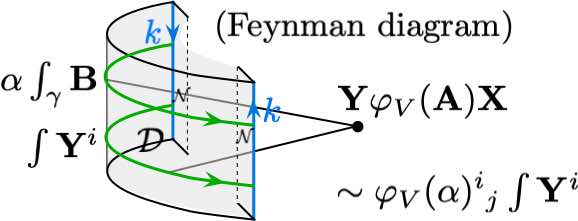}}
 \label{GM-HD-prod} \ee
where on the LHS $\varphi_V:G\to\text{End}(V)$ is the map defining the $G$-action on $V$, and on the RHS $\varphi_V:\fg\to \text{End}(V)$ is the induced map of Lie algebras. Altogether, this makes $\CH_\CD$ as an associative algebra be a semi-direct product
\be \CH_\CD \simeq \text{Dist}(G) \ltimes \Lambda^\bullet V\,. \label{GM-HD-semi} \ee

The remaining Hopf operations in $\CH_\CD$ are \emph{exactly} the same as in the pure-matter (Sec. \ref{sec:matter-sparks}) and pure-gauge (Sec. \ref{sec:gauge-HD}) discussions, by the same reasoning. We find:
\be \label{GM-HD} \begin{array}{c}
  \Delta(g) = g\otimes g \\
  \Delta(\alpha) = \alpha \otimes 1+1\otimes\alpha \\
  \Delta(\psi_+^i) = \psi_+^i\otimes 1 + 1 \otimes \psi_+^i \end{array},
 \qquad \begin{array}{c}
 \varepsilon(g) \equiv 1\\
 \varepsilon(\alpha)  \equiv 0 \\
 \varepsilon(\psi_+^i) \equiv 0 \end{array},
 \qquad
 \begin{array}{c}
   S(g) = g^{-1} \\
   S(\alpha) = -\alpha \\
   S(\psi_+^i) = -\psi_+^i \end{array}.
\ee
Just as in pure gauge theory, the coproduct requires a topological completion to be strictly defined, $\Delta:\CH_\CD\to \CH_\CD\wh\otimes\CH_\CD$, due to the appearance of $\text{Dist}(G)$.

\subsection{$\CH_\CN$ from dragging Wilson lines}
\label{sec:GM-HN}

Just as in pure gauge theory (Section \ref{sec:gauge-HN}), sparks on the $\CN$ boundary contain framed Wilson lines `$wW_\rho v$' for all holomorphic representations $\rho$ of $G$ and vectors $v\in \rho$, $w\in \rho^*$:
\be   \raisebox{-.4in}{\includegraphics[width=2.8in]{SparkDefD-DW.jpg}} \label{GM-Nsparks} \ee
These sparks may be identified with matrix elements of representations $\rho$. They continue to commute with each other: perturbatively, the new bulk interaction vertex $\mb Y\varphi_V(\mb A)\mb X$ does not modify this; and no nonperturbative instanton corrections are possible. More generally, it follows from the co-commutative coproduct \eqref{GM-HD} on $\text{Dist}(G)$ (which is an elementary consequence of linearity of spark integrals) and the expected pairing \eqref{GM-pair-vect} that the product on these sparks must be commutative. Thus, framed Wilson lines generate a commutative subalgebra $\CO(G)\subseteq \CH_\CN$.

We would expect the remainder of $\CH_\CN$ to come from integrated descendants of $\mb X$ at the boundary. Specifically, if we choose a basis $\{e^i\}$ for $V$, and decompose $\mb X = \sum_i \mb X_i e^i$, we expect sparks
\be \psi^-_i \;\overset?= \; \int \mb X_i + a_i - b_i\,, \ee
suitably decorated by fermions on the $k$ interfaces, as in \eqref{matter-X-ferm}.
There is a small but important problem with this, with a small but important solution. Namely, gauge symmetry is unbroken on the Neumann boundary, and $\mb X_i$ are not gauge-invariant operators. (They are not $Q$-closed under the new BRST variations \eqref{GM-BRST}.)

This can be fixed by attaching the $\mb X_i$ to Wilson lines. Explicitly, for any point $p$ on an $\CN$ strip, let
\be W_{V^*}(p) = P\text{exp}\int_k^p \varphi_{V^*}(\mb A)\qquad  \raisebox{-.2in}{\includegraphics[width=.9in]{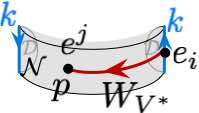}} \hspace{-.5in} \ee
denote the Wilson line in representation $V^*$ (which is the rep that the components/operators $\{\mb X_i\}$ transform in), stretched from the right $k$ boundary of the strip (where $G$ symmetry is broken) to $p$. Let $\{e_i\}$ be a dual-basis of $V^*$, which we'll use to frame the Wilson line at $k$. At the point $p$, we can frame it with a basis vector $e^j\in V$, getting an operator $e^j W_{V^*}(p) e_i$. Neither  $e^j W_{V^*}(p) e_i$ nor the insertion $X_i(p)$ is gauge-invariant. But the combination
\be \wt X_i(p) := \sum_j X_j(p) (e^j W_{V^*}(p) e_i)  \label{GM-X-W}   \ee
is invariant (thus $Q$-closed), and allows the definition of a modified spark
\be \psi^-_i := \int \wt X_i^{(1)} + a_i - \tilde b_i \;\; =\quad  \raisebox{-.23in}{\includegraphics[width=1.1in]{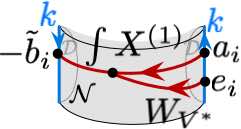}}   \;\;. \label{GM-psi-W} \ee
(Note that the boundary fermion $\tilde b_i = b W_{V^*} e_i$ gets dressed as well.)

The fermionic sparks \eqref{GM-psi-W} still anti-commute with each other, and commute with the framed Wilson lines \eqref{GM-Nsparks}. One can check that no Feynman diagrams can contribute. An additional confirmation is that the coproduct on the \emph{entire} dual algebra $\CH_\CD$ in \eqref{GM-HD} is co-commutative. Thus, as an associative algebra, we find
\be \CH_\CN \simeq \CO(G)\otimes \Lambda^\bullet V^*\qquad \text{(graded commutative)}\,. \ee

The Hopf operations on the $\CO(G)$ subalgebra are exactly the same as in pure gauge theory (Section \ref{sec:gauge-HN}). However, the operations on $\Lambda^\bullet V^*$ are twisted in interesting ways by the Wilson-line dressing factors.

In what follows, it will be useful to introduce the matrix of sparks
\be K^i{}_j := e_jW_V e^i  \; \in \CO(G)\,,\qquad  K^i{}_j=\raisebox{-.17in}{\includegraphics[width=1in]{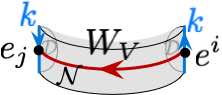}} \ee 
the matrix elements in representation $V$. If $V$ is a faithful representation, then polynomials in the $K^i{}_j$'s span all of $\CO(G)$ (they may have relations among them). The matrix elements of Wilson lines $W_{V^*}$ in the dual representation are related to $K$ by
\be  \raisebox{-.17in}{\includegraphics[width=1in]{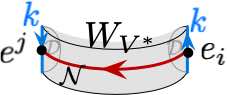}} = (K^{-1,T})_i{}^j = (K^{-1})^j{}_i\,.\ee
Moreover, from pure gauge theory, we know that
\be \Delta(K^i{}_j) =   \sum_k K^k{}_j\otimes K^i{}_k\,,\qquad \varepsilon( K^i{}_j) = \delta^i{}_j\,,\qquad   S(  K^i{}_j) = (K^{-1})^i{}_j\,. \label{cop-K} \ee

\subsubsection{Coproduct of matter sparks}

The coproduct of the matter spark $\psi^-_i$ gains an additional contribution.
Consider integrating along the waist of a pair of pants the dressed descendant of $X_i$, as in \eqref{GM-X-W}:
\be \raisebox{-.4in}{\includegraphics[width=5.7in]{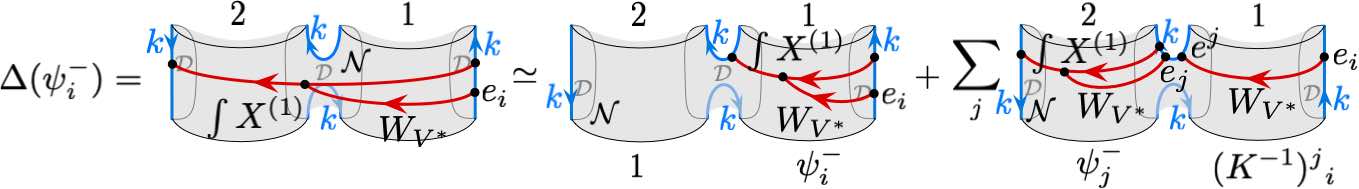}} \ee
The integral of the descendant does not fully decompose into a sum of contributions from individual legs, as in pure matter. Rather, when integrating the descendant along the second (left-most) leg, the Wilson-line dressing must stretch all the way across the first leg as well. We can split the Wilson line at the $k$ boundary by inserting a full set of states $\sum_j e_j\otimes e^j  = \text{id} \in \text{End}(V^*)$. After reordering factors according to our conventions for the coproduct, we find
\be \Delta(\psi_i^-) = \psi^-_i\otimes 1+\sum_j (K^{-1})^j{}_i \otimes \psi^-_j\,. \label{eq:coprodV*} \ee

\subsubsection{Antipode of matter sparks}

The antipode works in a similar way. Rotating a dressed matter spark by 180$^\circ$ rotates the basepoint of the Wilson line as well, from the right boundary to the left boundary:
\be \label{GM-S-rot} \raisebox{-.4in}{\includegraphics[width=5.2in]{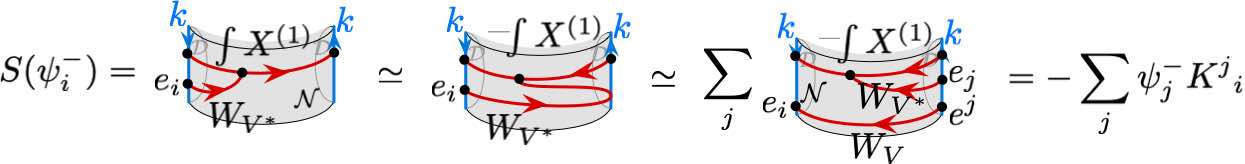}} \ee
To move the basepoint back to its standard position on the right, we can stretch the Wilson line across, then split it, into a product of a $W_V$ line depicted across the bottom of the strip, and the standard dressing factor $W_{V^*}$ coming in from the right. There is also an additional sign due to reversing the orientation of the $X^{(1)}$ integral.

There is one further potential correction to the antipode, arising from the framing anomaly discussed in Section \ref{sec:GM-framing}. The S-twist used to define the antipode topologically does introduce boundary curvature. If it's un-twisted, it not only rotates a spark around as in \eqref{GM-S-rot}, but also sandwiches it between two additional defects for the topological symmetry $G_{\rm top}$ used in cancelling a framing anomaly --- which are Wilson lines in the representation $\det(V)$ and its dual:
\be \raisebox{-.5in}{\includegraphics[width=3.3in]{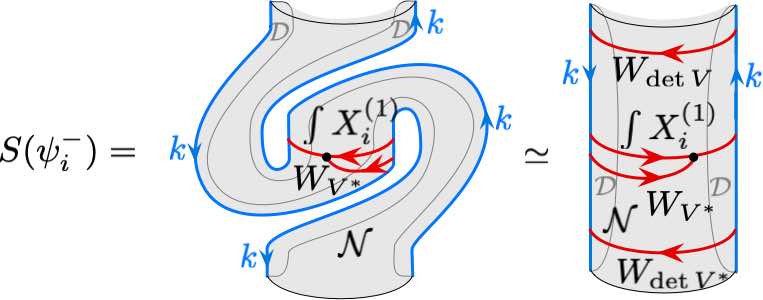}} \ee
Therefore, a more accurate formula for the antipode would be $S(\psi^-_i) = -(\det K) \sum_j^i K^j{}_i (\det K)^{-1}$. However, since $\CH_\CN$ is abelian, the factors of $\det K$ simply cancel, and we recover~\eqref{GM-S-rot}.

\subsubsection{Counit of matter sparks}

The counit is more straightforward. The dressing by Wilson lines does not seriously modify the basic argument we used in Section \ref{sec:matter-antipode} that a matter spark inserted in an otherwise empty ball could be deformed to a small closed loop, and then shrunk to zero size:
\be \raisebox{-.4in}{\includegraphics[width=4in]{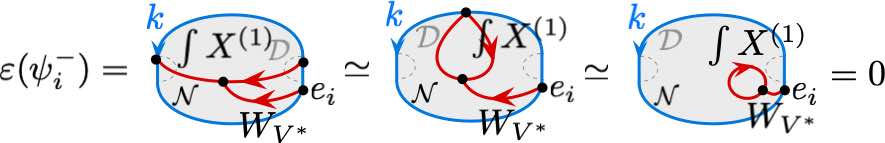}} \ee
Therefore, on all generators $\psi_i^-$, the counit evaluates to zero.

\subsubsection{Summary and Hopf pairing}

Collecting and summarizing our results for the boundary spark algebra $\CH_\CN$, we have found a graded-commutative algebra $\CH_\CN = \CO(G)\otimes \Lambda^\bullet V^*$, with
\be \label{GM-HN} \begin{array}{c}
\Delta(f)(g_1,g_2) = f(g_2 g_1) \\ 
\Delta(K^i{}_j) =  \sum_k K^k{}_j \otimes K^i{}_k \\ 
\Delta(\psi^-_i) = \psi^-_i\otimes 1 + \sum_j (K^{-1})^j{}_i \otimes \psi^-_j
\end{array}\,,
\begin{array}{c}
\varepsilon(f)(g) =f(e) \\
\varepsilon(K^i{}_j) = \delta^i{}_j \\
\varepsilon(\psi^-_i) = 0
\end{array}\,,
\qquad \begin{array}{c}
S(f)(g) = f(g^{-1}) \\ 
S(K^i{}_j) = (K^{-1})^i{}_j \\
S(\psi^-_i) = - \sum_j \psi^-_j K^j{}_i
\end{array}\,.
\ee
Here $f\in \CO(G)$ is a general function, and $K^i{}_j \in \CO(G)$ are the matrix elements of $V$.

One may now verify that the expected pairing \eqref{GM-pair-vect}, relating dual vector spaces, is indeed a Hopf pairing between $\CH_\CD$ from \eqref{GM-HD} (which is co-commutative but not commutative) and $\CH_\CN$ (which is commutative but not co-commutative). A subtle feature of the pairing is that, since $\CH_\CD$ is not commutative, it matters in which order elements are written. Physically, it is clear that we have
\be h(g,f) = f(g)\,,\qquad h(\psi_+^i,\psi^-_j) = \delta^i{}_j \label{GM-h-basic}  \ee
for all $g\in \C G$, $f\in \CO(G)$, $\psi_+^i\in \Lambda^\bullet V$, $\psi^-_i\in \Lambda^\bullet V^*$; and this determines $h$ on all other elements. It follows from \eqref{GM-h-basic} and the identifies of Hopf pairings \eqref{h-pair-def} that we established topologically in Section \ref{sec:Hopf-pair} that
\be  h(g \psi_+^{i_1}...\psi_+^{i_k}, f \psi^-_{j_1}...\psi^-_{j_{k'}}) = h(g,f) h(\psi_+^{i_1}...\psi_+^{i_k},\psi^-_{j_1}...\psi^-_{j_{k'}})\,,  \label{GM-h-order} \ee
in agreement with the vector-space pairing \eqref{GM-pair-vect}; but we warn readers that in the opposite order $h( \psi_+^{i_1}...\psi_+^{i_k} g, \psi^-_{j_1}...\psi^-_{j_{k'}}f) \neq h(\psi_+^{i_1}...\psi_+^{i_k},\psi^-_{j_1}...\psi^-_{j_{k'}}) h(g,f)$ doesn't hold.

\subsection{Bulk sparks}
\label{sec:GM-bulk}

As a vector space, the bulk spark algebra $U$ must be isomorphic to $\CH_\CD\otimes \CH_\CN$. The associative product on $U$ may be computed systematically using formula \eqref{double-prod2}.
There are several parts to this, each of which have physical meaning:
\begin{enumerate}
\item The matter sparks $\psi_+^i\in  \CH_\CD$ commute with the matrix elements $\CO(G)\subset \CH_\CN$ formed from Wilson lines. Physically, this is for the same reason that the $\psi^-_i\in \CH_\CN$ commute with $\CO(G)$: there are no quantum corrections to the product of either $\mb X$ or $\mb Y$ matter operators on the boundary with the connection $\mb A$.
\item The matter sparks $\psi_+^i\in \CH_\CD$ no longer commute with $\psi^-_j\in \CH_\CN$, like they did in \eqref{matter-pm-comm} for pure matter theory. This is due to dressing by Wilson lines on the $\CN$ boundary. Specifically, the fermion $\tilde a_j = (K^{-1})^k{}_j a_k$ at the end of a $\psi^-_j$ spark gets dressed, but the fermion $b_j$ at its beginning does not. This  means that when the two sparks move past each other, the $\delta^i_j$ factors commuting fermions at the two ends no longer cancel like they did in \eqref{double-cancel}; rather they give  
\be \hspace{-.5in} \raisebox{-.5in}{\includegraphics[width=5.5in]{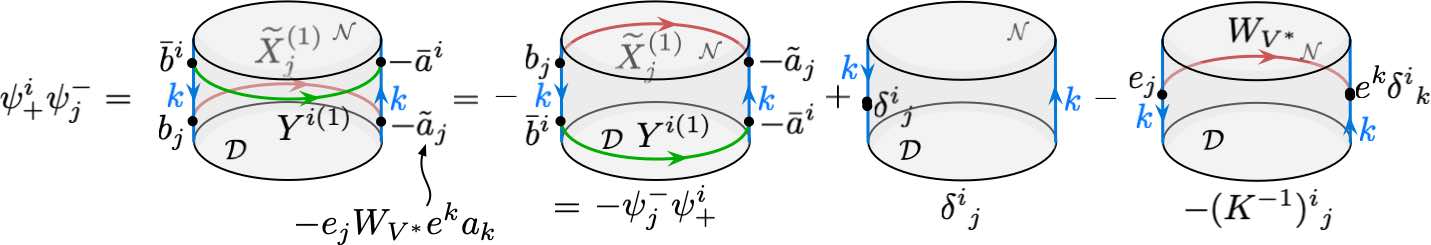}} \label{GM-Ucomm} \ee
In other words, the anti-commutator is $[\psi_+^i,\psi^-_j] = \delta^i{}_j - (K^{-1})^i{}_j$.

\item Elements of $\text{Dist}(G) \subset \CH_\CD$, which are defects for the global $G_\pd$ symmetry on $\CD$, commute with all sparks in $\CH_\CN$ according to the charges of their endpoints on the $k$ boundaries between $\CD$ and $\CN$. In pure gauge theory, we saw in \eqref{gauge-gW-act} that this induced the standard conjugation action on $f\in \CO(G)$,
\be g f(k)g^{-1} = f(g^{-1}kg)\,. \ee
Similarly, the dressed $\psi^-_i$ sparks have at their \emph{beginnings} three contributions: fermions $b_i$; basis vectors $e_i$ for the Wilson lines $W_{V^*}$ that dress the $X_i^{(1)}$ descendants; and basis vectors $e_i$ that dress the fermions $\tilde a_i = (e_iW_{V^*})\cdot a$ at the far ends. All three of these transform simultaneously in representation $V^*$. There is nothing at the far ends that transforms nontrivially under $G_\pd$. Graphically:
\be \hspace{-.4in} \raisebox{-.5in}{\includegraphics[width=6in]{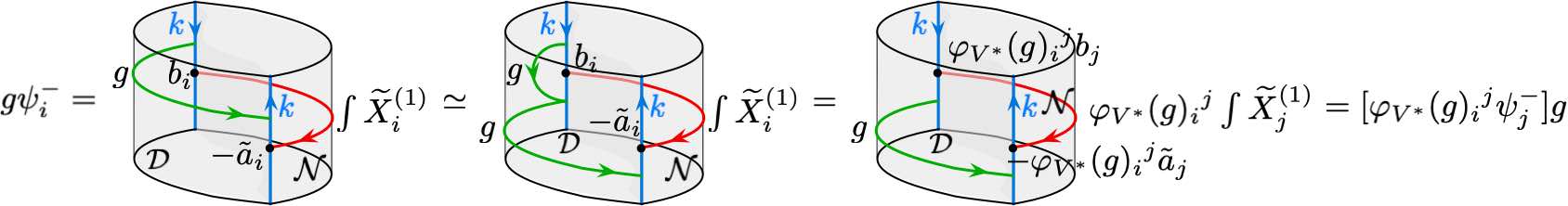}} \ee
We find (as might be expected) that the sparks $\psi^-_i$ simply transform in representation $V^*$,
\be g \psi^- g^{-1} = \varphi_{V^*}(g)\cdot \psi^-\,. \ee
\end{enumerate} 
We invite readers to verify all the formulas here algebraically as well, using \eqref{double-prod2}.

Altogether, we may summarize the associative algebra structure on $U\simeq \CH_\CD\otimes \CH_\CN = \text{Dist}(G)\otimes \Lambda^\bullet (T^*V) \otimes \CO(G)$ as follows.  The subalgebra  $ \Lambda^\bullet(T^*V) \otimes \CO(G)$ is a central extension
\be \CO(G) \hookrightarrow \Lambda^\bullet(T^*V) \otimes \CO(G) \to \!\!\!\!\!\to  \Lambda^\bullet(T^*V) \label{GM-U-central} \ee
in which the anti-commutators $[\psi_+^i,\psi^-_j]=0$ of pure matter are modified to \eqref{GM-Ucomm},
\be [\psi_+^i,\psi^-_j] = \delta^i{}_j - (K^{-1})^i{}_j\,. \ee
The entire algebra $U$ is a further semi-direct product
\be U = \text{Dist}(G) \ltimes \big[ \Lambda^\bullet(T^*V) \otimes \CO(G)\big]\,, \label{eq:Usemidirect}
 \ee
controlled by the $G$ action on $T^*V$ and its adjoint action on $\CO(G)$.

The Hopf operations on $U$ are straightforward. They are directly inherited from $\CH_\CD$ and $\CH_\CN$, as in \eqref{double-Hopf} (or \eqref{double-cop}--\eqref{double-S}). The coproduct requires a topological completion to be defined, due to the appearance of $\text{Dist}(G)$.

\subsubsection{The R-matrix and ribbon twist}
\label{sec:GM-R}

As in pure gauge theory, the R-matrix is valued in a completion $\CH_\CD \,\wh\otimes\, \CH_\CN$. Physically, we expect it to be given by integrating the gluing deformation  \eqref{GM-Sglue} over a rectangle,
\be \CR \sim \exp{\int_\square \big(\mb Y\otimes \mb X+\mb B\otimes \mb A} \big)\label{GM-Rphys} \ee
We claim that this can be further decomposed into sparks dressed by Wilson lines
\be \CR \simeq   \bigg(\exp\int_\square \mb B\otimes \mb A\bigg) \bigg( \exp\int_\square \mb Y\otimes \wt {\mb X}_{right}  \bigg) \simeq  \bigg( \exp\int_\square \mb Y\otimes \wt {\mb X}_{left}  \bigg)   \bigg(\exp\int_\square \mb B\otimes \mb A\bigg)\,,  \label{GM-Rphys-dressed} \ee
where the difference between the two expressions is whether the integrated $\mb X$ operators are dressed with Wilson lines attached to the right or left boundary of the $\CN$ strip that they're placed on.

It should be possible to derive \eqref{GM-Rphys-dressed} from \eqref{GM-Rphys} by carefully de-exponentiating and computing quantum corrections. We will instead argue for \eqref{GM-Rphys-dressed} using the general algebraic formalism we've developed in Section \ref{sec:bulk}.

Let's choose a basis $\{f_a\}$ of $\CO(G)$ and a dual basis $\{f^a\}$ in $\text{Dist}(G)$, under the Hopf pairing $h$. (The $f^a$ will be smeared, non-localized distributions.) Let's also choose a basis $\{\omega_I\}$ of $\Lambda^\bullet V^*$, and a dual basis $\{\omega^I\}$ of $\Lambda^\bullet V$. (The $\omega_I$ are monomials in $\psi^-_i$ and the $\omega^I$ are monomials in $\psi_+^i$.) Then, under the carefully ordered Hopf pairing \eqref{GM-h-order}, the $\{f^a \omega^I\}$ and $\{f_a\omega_I\}$ will be dual bases of $\CH_\CD$ and $\CH_\CN$, and the R-matrix is
\be \CR = \sum_{a,I}(-1)^{|\omega^I||\omega_I|} f^a\omega^I\otimes f_a\omega_I \ee
Since $\omega^I\otimes 1$ commute with $1\otimes f_a$, we may further factor this as
\be \CR \simeq  \Big(\sum_a f^a\otimes f_a\Big) \Big(\sum_I (-1)^{|\omega^I|} \omega^I \otimes \omega_I\Big) = \CR_G \CR_V\,, \label{GM-R-fact1} \ee
\vspace{-.2in}
\be \CR_G = \sum_a f^a\otimes f_a\,,\qquad \CR_V = \sum_I (-1)^{|\omega^I|}\omega^I \otimes \omega_I = \exp \lp -\sum_i \psi_+^i\otimes \psi^-_i \rp \notag \ee
where $\CR_G$ is the R-matrix of pure gauge theory that we gave many explicit forms of in Section 
\ref{sec:gauge-R}; and $\CR_V$ is the R-matrix of pure matter theory, though made of $\psi^-_i$ sparks that have been dressed by Wilson lines. Thus, \eqref{GM-R-fact1} matches the LHS of \eqref{GM-Rphys-dressed}.

To get the RHS of \eqref{GM-Rphys-dressed}, we use the explicit description \eqref{R-coaction} of the R-matrix of pure gauge theory to compute
\be \CR_G (\psi_+^i\otimes \psi^-_j) \CR_G^{-1} =   \textstyle\sum_k \psi_+^k \otimes K^i{}_k \psi^-_j\,.   \ee
Therefore,
\be \CR = \CR_G  \exp \lp -\sum_i \psi_+^i\otimes \psi^-_i\rp  =  \exp\lp -\sum_i \psi_+^i\otimes ({\textstyle \sum_j}\psi^-_j K^j{}_i)\rp \CR_G\,, \ee
where $\sum_j \psi^-_j K^j{}_i$ are the sparks on $\CN$ dressed by Wilson lines attached to the left $k$ boundary rather than the right one.

We also expect a ribbon element of the form 
\be
v=m \circ (S\otimes 1)(\CR_{21})\,\in\, \wh U\,.
\ee
Using the same basis that we used to describe the R-matrix above, we find
\be v = \sum_{a,I} S(f_a\omega_I) f^a\omega^I = \sum_I S(\omega_I) \bigg( \sum_a S(f_a) f^a \bigg) \omega^I =  \sum_I S(\omega_I) v_G \omega^I\,, \ee
where $v_G$ is the ribbon element for pure gauge theory. Using
\be v_G( \psi_+^i) v_G^{-1} = (K^{-1})^i{}_j \psi_+^j\,,\ee
we can simplify the above expression to
\begin{align} v &= \exp \lp \sum_i S(\psi^-_i) v_G(\psi_+^i) v_G^{-1}\rp  v_G =  \exp \lp \sum_i (-\psi^-_k K^k{}_i) ((K^{-1})^i{}_j\psi_+^j)\rp v_G \notag \\
 &=  \exp\lp \sum_i -\psi^-_i\psi_+^i\rp v_G = v_V v_G = v_G v_V\,, \label{GM-twist} \end{align}
where $v_V$ is the ribbon twist for pure matter theory.

\subsubsection{Supergroup gauge theory}
\label{sec:GM-supergroup}

As noted in the introduction of this section, B-twisted gauge theory with matter can be thought of as a supergroup gauge theory \cite{Kapustin:2009cd}. An easy derivation of this in BV formalism can be found in \cite{Garner:2022rwe}. This perspective can significantly simplify understanding of the Hopf-algebra structures on $\CH_\CN, \CH_\CD$, as well as $U$. We'd like to indicate how.

Let us first look at $\CH_\CD$. We have seen that it is a semi-direct product $\mathrm{Dist}(G)\ltimes \Lambda^\bullet V$. Let us consider the supergroup $H=G\ltimes \Pi V$, where $\Pi V$ is the vector space $V$ shifted to be odd/fermionic. Then $\CH_\CD$ is precisely distributions on the supergroup $H$,
\be
\CH_\CD= \mathrm{Dist}(G)\ltimes \Lambda^\bullet V=\mathrm{Dist}(H)\,.
\ee
This is in fact an identification as topological Hopf algebras. Similarly, one can identify $\CH_\CN$ with the algebra of functions on $H$, namely
\be
\CH_\CN=\CO (G)\otimes\Lambda^\bullet V^*=\CO (H)\,.
\ee
This is in fact also an identification of Hopf algebras. To see this, let us consider taking two elements $(g_1, v_1), (g_2, v_2)\in H$. Then their multiplication is
\be
(g_1, v_1)\cdot (g_2, v_2)=(g_1g_2, g_2^{-1}\cdot v_1+v_2)\,.
\ee
The coproduct on $\psi^-_i$ in equation \eqref{eq:coprodV*} is exactly dual to this multiplication. 

Finally, the quasi-triangular Hopf algebra $U$ is simply
\be
U=\mathrm{Dist}(H)\ltimes \CO (H)\,.
\ee
Its R-matrix is identical to the one discussed in pure gauge theory, in Section \ref{sec:gauge-R}. We can recover the R-matrix in equation \eqref{GM-R-fact1} via the decomposition $H=G\ltimes \Pi V$.

\subsection{Categories and connection to other perspectives}
\label{sec:GM-cat}

We now describe the module categories for our spark algebras geometrically, connecting them to other expected descriptions of the categories of line operators.

\subsubsection{Boundary categories}

Since $\CH_\CD=\text{Dist}(G)\ltimes \Lambda^\bullet V$ is a semi-direct product, and (as we know from pure gauge theory) $\text{Dist}(G)$-modules are just $G$-representations, we find an equivalence
\be
\CH_\CD\Mod\simeq (\Lambda^\bullet  V)\Mod_G =: \CC_\CN\,,
\ee
where the RHS is the category of $G$-equivariant dg modules for $\Lambda^\bullet  V$. We can apply equivariant Koszul duality, just as in the case of pure matter (\cf\ \eqref{CohV-better}--\eqref{CohV-Koszul}) to further obtain an equivalence
\be
\CH_\CD\Mod^{fd} \simeq \Coh^G (V[1]) = \CC_\CN^{fd}\,.
\ee
This is just as expected from \eqref{GM-CN-exp} in Section \ref{sec:GM-vect}. The tensor product dictated from coproduct in $\CH_\CD$ is just ordinary tensor product of sheaves on $V[1]$. Just as in pure gauge theory, the fiber functor (which now also implements Koszul duality along $V[1]$) may be expressed as $\CF_\CN = \text{Hom}(\id,\kDk\otimes-)$, with $\kDk=\CO(G)\otimes \CO_0$ from \eqref{GM-kDk}. (We also recall that $\kDk$ is not strictly a dualizable object in $\CC_\CN^{fd}$, but rather a limit of dualizable objects.)

On the other hand, for the commutative algebra $\CH_\CN = \CO(G)\otimes \Lambda^\bullet V^*$, we have
\be
\CH_\CN\Mod\simeq \QCoh (G)\otimes \big( \Lambda^\bullet V^*\Mod\big) =: \CC_\CD\,.
\ee
Applying a Koszul duaity to the second factor, we obtain
\be
\CH_\CN\Mod^{fd} \simeq \Coh (G)^{fd}\otimes \Coh (V^*[1]) \simeq \Coh(G \times V^*[1])^{fd} = \CC_\CD^{fd} \,.
\ee
This agrees with \eqref{GM-CD-exp}, up to the subtle dualizability condition along $G$: just as in pure gauge theory, dualizable sheaves must have finite support along $G$. The coproduct in $\CH_\CN$ induces a tensor product in $\Coh(G \times V^*[1])^{fd}$ that combines tensor of sheaves along $V^*$ with convolution along $G$.

\subsubsection{Bulk lines}

The category $\CC_\CT\simeq U\Mod$ is the derived Drinfeld center of either $\CH_\CN\Mod$ or $\CH_\CD\Mod$. Physically, a model for $\CC_\CT$ comes from compactifying the 3d theory on a circle linking bulk lines:
\be \raisebox{-.25in}{\includegraphics[width=3.2in]{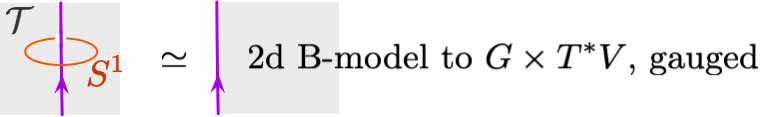}} \ee
Along the lines of \cite{KRS, OR-Chern, OR-TQFT}, we find a 2d B-model with target $G\times (T^*[2]V[1])$, where the factor of $G$ is a basepointed holonomy around the $S^1$, and the coordinates on $T^*V$ are values of $X,Y$ at the basepoint. There is a residual 2d gauge group $G$ acting acting as conjugation on the $G$ factor, and in the usual way on $X,Y\in T^*V$. There is also a superpotential $W = Y\cdot(\varphi_V(g)-1)X$ whose critical locus enforces twisted equations of motion: the fact that $X$ and $Y$ are covariantly constant sections around the $S^1$, and that the moment map $\varphi_V^*(XY)$ vanishes.

The category of boundary conditions for this 2d B-model is given by matrix factorizations, so we expect that
\be
\CC_\CT \sim \MF^G \big(\underset{g}{G}\times \underset{Y}{T^*[2]}(\underset{X}{V[1]}), W\big)\,, \qquad W= Y\cdot (\varphi_V(g)-1)X\,. 
\ee
This category is $\Z$-graded, since the functions $X_i,Y^i$ have cohomological degree 1 and $W$ has degree 2.

How is our category $U\Mod$ related to this category of matrix factorizations? We sketch an answer using a relative version of the so-called ``quadratic Koszul duality" \cite{Kapustin:2002bi}. Let $\mathrm{Cl}(W)$ be the Clifford algebra in $\Coh (G)$ generated by $V\oplus V^*$ (which are fermions) with commutator:
\be\label{eq:CliffW}
[v, v^*]=\frac{\pd^2 W}{\pd v\pd v^*}\in \CO_G\,.
\ee
Then the statement of this Koszul duality is that there is an equivalence
\be
\MF^G (G\times T^*[2](V[1]), W)\simeq \mathrm{Cl}(W)\Mod_G^{fg}\,,
\ee
where the RHS is the category of finitely-generated $G$-equivariant modules of $\mathrm{Cl}(W)$. This statement is proven in the dg setting by \cite{preygel2011thom}. 

If we examine the commutation relation of \eqref{GM-Ucomm}, we find that $\mathrm{Cl}(W)$ in this case is exactly the subalgebra of $U$ generated by $\psi_+=v$, $\psi^-=v^*$, and $\CO (G)$, up to inverting $g\mapsto g^{-1}$. Indeed, the double derivative
\be
\frac{\pd^2 W}{\pd v\pd v^*}=\frac{\pd^2}{\pd v\pd v^*}(v^*, (\varphi_V(g)-1)v)
\ee
is precisely the matrix element of $g-1$ acting on $V$. Moreover, the algebra $U$ is simply the semi-direct product of $\mathrm{Dist}(G)$ with $\mathrm{Cl}(W)$, signifying the fact that a module of $U$ is simply a $G$-equivariant module of $\mathrm{Cl}(W)$. Consequently, we find an equivalence
\be
\MF^G (G\times T^*[2](V[1]), W)\simeq U\Mod^{fg}\,.
\ee

The category $\CC_\CT$ has a monoidal structure, which from the point of view of $U\Mod$ is given by the co-product of $U$. In terms of matrix factorizations, this can be given geometrically via a convolution diagram as follows.  Let us denote $\CS=G\times T^*V$ and $\CP=G\times G\times T^*V$, and consider the following correspondence diagram:
\be
\btik
& \CP\arrow[dr, "q"]\arrow[dl, swap, "p"] &\\
\CS\times \CS & & \CS
\etik
\ee
where the map $p$ is given by $p (g_1, g_2, v, v^*)\mapsto (g_1, g_2v, v^*)\times (g_2, v, v^*)$ and the map $q$ given by $q(g_1, g_2, v, v^*)\mapsto (g_1g_2, v, v^*)$. Note that:
\be
p^*(W\otimes 1+1\otimes W)=(v^*, (g_1-1)g_2v)+(v^*, (g_2-1)v)=(v^*, (g_1g_2-1)v)=q^*(W)\,,
\ee
and therefore we have a well-defined functor\footnote{Technically this functor is only defined after we take an ind-completion of this category, however we will ignore this ind-completion in our discussions.}:
\be
q_*p^*: \MF_G(\CS, W)\boxtimes \MF_G(\CS, W)\to \MF_G(\CS, W)\,,
\ee
defining a monoidal structure on $\MF_G(\CS, W)$. We expect that this monoidal structure is equivalent to the one defined from $U$. 

$\CC_\CT$ should also have a braiding. From the perspective $U$ this is provided by the $R$ matrix. Unfortunately, it is rather unclear what the braiding is from the perspective of matrix factorizations. In principle it should follow from the fact that this category is a Drinfeld center.  

In fact, one can show that $\MF_G(\CS, W)$ is a Drinfeld center using yet another Koszul duality, the so-called  ``linear Koszul duality" of \cite{arkhipov2017equivariant}. Let $\CZ=G\times V$, we can view $\CS$ as a vector bundle over $\CZ$, such that $W$ is linear along the fiber (which is $V^*$). Let $\CA_\CZ$ be the dg algebra over $\CZ$ generated by $\Pi V$ (which are anti-commutative) together with differential $d\psi_+=\frac{\pd W}{\pd v}$. The statement of linear Koszul duality is the following equivalence
\be
\MF_G (X, W)\simeq \CA_\CZ\Mod_G^{fg}
\ee
where the RHS is the category of finitely-generated $G$-equivariant modules of the dg algebra $\CA_\CZ$. 

What are modules of $\CA_\CZ$? If one carefully analyze the differential, one finds that $\CA_\CZ$ is precisely functions on the following fiber product: 
\be
\btik
\mathrm{Spec}(\CA_\CZ)\rar\dar & V\dar{\Delta}\\
\CZ\rar{(g,v)\mapsto (v, gv)} & V\times V
\etik
\ee
Therefore, we find that objects in $\CA_\CZ\Mod_G$ are identified with sheaves on the stack $[\mathrm{Spec}(\CA_\CZ)/G]$, which is nothing but the loop space $\CL ([V/G])$ \cite{Ben-Zvi:2008vtm}. This category is shown to be the derived Drinfeld center for the category $\QCoh ([V/G])$, and therefore possess a braiding. Keeping track of gradings, we have an equivalence
\be
\MF_G (\CS, W)\simeq \Coh (\CL (V[1]/G))\simeq \CZ \lp \Coh (V[1]/G)\rp.
\ee
This shows that $\MF_G (\CS, W)$ is indeed a Drinfeld center. 

In summary, we have a triality of categories, related via Koszul dualities (and are equivalences of appropriate subcategories)
\be
\btik
& \MF_G (\CS, W)\arrow[dr]\arrow[dl] & \\
U\Mod & & \QCoh (\CL (V[1]/G))
\etik
\ee
We expect that these are all equivalences of braided tensor categories. Our $U$ gives an explicitly calculable braiding on these categories.

\subsubsection{Abelian gauge theory and VOA's}

In the situation where $G=\mathrm{GL}(1)^r=(\C^*)^r$ and $V=\C^n$, with $G$-module structure defined by an integral $n\times r$ charge matrix $\tau = \{\tau^a{}_i\}_{i=1,...,n}^{a=1,...,r}$, the Hopf algebra $U$ can be made very explicit.

We'll describe the subalgebra of $U$ where $\text{Dist}(G)$ is replaced by $U(\fg)$, recalling that modules for $\text{Dist}(G)$ are equivalent to modules for $U(\fg)$ with the restriction that the abelian generators $N^a\in \fg$ act semisimply with integer eigenvalues (see p. \pageref{NK-gauge-ab}).

The subalgebra of $U$ in this case is generated by
\be
N^a\in \fg\,, \qquad \psi_+^i\in \Pi V\,,\quad  \psi^-_i\in \Pi V^*\,, \qquad K_a^\pm \in \CO(G)\,, 
\ee
for $a=1,...,r$ and $i=1,...,n$. The non-zero commutation relations are
\be
[N^a, \psi_+^i]=\tau^a{}_i\psi_+^i, \qquad [N^a, \psi^-_i]=-\tau^a{}_i \psi^-_i, \qquad \{\psi_+^i,\psi^-_j\}=\delta^i{}_j\Big(1-\prod_a K_a^{-\tau^a{}_i}\Big)\,.
\ee
The Hopf structure is given by
\be \begin{array}{l}
\Delta(N^a) = N^a\otimes 1+1\otimes N^a \\
\Delta(K_a^\pm) = K_a^\pm \otimes K_a^\pm \\
\Delta(\psi^+_i) = \psi^+_i \otimes 1+1\otimes \psi^+_i \\
\Delta(\psi^-_i) = \psi^-_i \otimes 1 +\prod_a K_a^{-\tau^a_i}\otimes \psi^-_i \end{array}\;\;
\begin{array}{l} S(N^a)=-N^a \\ S(K_a^\pm) = K_a^\mp \\ S(\psi^+_i) = -\psi^+_i \\ S(\psi^-_i) = -\psi^-_i\prod_a K_a^{\tau^a{}_i} \end{array}\;\;
\begin{array}{l}
 \varepsilon(N^a) = 0 \\ \varepsilon(K_a^\pm) = 1 \\
 \varepsilon(\psi_+^i) = 0 \\ \varepsilon(\psi^-_i) = 0 \end{array}
\ee
We also find the following expression of the $R$ matrix
\be
\CR= \prod_a (1\otimes K_a)^{N_a\otimes 1} \prod_i (1-\psi_+^i\otimes \psi^-_i)\,.
\ee
This is identical to the quasi-triangular Hopf algebra defined in \cite{creutzig2024kazhdan}, upon mapping $\psi^-_i\mapsto \psi^-_i \prod_a K_a^{\tau^a{}_i}$. 
It was shown in \cite{creutzig2024kazhdan} that there is an equivalence of braided tensor categories between finite-dimensional modules of $U$ and modules of the boundary vertex algebra of \cite{CG, BCDN}. We've now used spark algebras to explain the origin of these quasi-triangular Hopf algebras.

\newpage

\appendix

\section{Tannakian QFT from line operators: some mathematical considerations}
\label{sec:math}

In this appendix, we offer a slightly different perspective on the framework of Section \ref{sec:setup} and the results of Sections \ref{sec:linefunctor}, \ref{sec:Hopf-bdy} and \ref{sec:bulk}. Namely, our construction of fiber functors and their endomorphism algebras can all be phrased entirely in terms of monoidal categories of bulk and boundary line operators, avoiding an explicit mention of 2-categories of boundary conditions, or 3-categories of 3d TQFT's. 

There are several motivations behind doing this. First, we hope to give a rigorous mathematical explanation of the results contained in Sections \ref{sec:linefunctor}--\ref{sec:bulk}, and further clarify how various assumptions are being utilized. Second, in many mathematical treatments of extended TQFT, the TQFT is most directly accessed/defined via categories of lines -- \emph{e.g.} Reshetikhin-Turaev \cite{RT} and Turaev-Viro \cite{TuraevViro,BW-spherical} theories. Third, we hope that this discussion will facilitate the application of our strategy of constructing fiber functors to general braided tensor categories, whether or not one views them as coming from TQFT.

\subsection{Tannakian reconstruction}
\label{A1:reconstruction}

Let $\CC$ be a monoidal category (abelian or dg) and let $\CF: \CC\to \mathrm{Vect}$ be a monoidal fiber functor, with monoidal isomorphism
\be
J:  \CF (X)\otimes \CF (Y) \overset\sim\longrightarrow  \CF (X\otimes Y)\,,\qquad  \forall\; X,Y\in \CC\,,
\ee
and sending $\id_\CC$ to $\id_{\rm Vect}=\C\in \mathrm{Vect}$. Then we know that $\CF$ lifts to a functor
\be
\wt{\CF}: \CC\to  \End (\CF)\Mod\,,
\ee
where $\End (\CF)$ is the algebra of endomorphisms of the fiber functor. In the abelian case, basically by definition, this algebra can be computed as the subalgebra of a very large algebra
\be
\End (\CF)\subseteq \prod_{X\in \CC} \End_\C (\CF (X))\,.
\ee
This sub-algebra is determined by requiring that for any $f: X\to Y$ and any $g:=\{g_X\}_{X\in \CC}\in \End (\CF)$, the following is true (the naturality condition)
\be
g_Y \CF (f)=\CF (f)g_X\,. 
\ee
In the dg case, one needs to put assumptions on $\CC$ and $\CF$ to make this algebra $\End (\CF)$ more controllable, which we will discuss later.

The essence of the reconstruction theorem (Tannakian formalism), is that the monoidal structure on $\CC$ induces extra structures on $\End (\CF)$. For example, the monoidal structure induces a map of algebras
\be
\Delta: \End (\CF)\to \End (\CF\otimes \CF)\,, \label{Delta-app}
\ee
where we send the natural transformation $g=\{g_X\}_{X\in \CC}$ to the element $\Delta (g)=\{g_{X\otimes Y}\}_{X,Y\in \CC}$, as a natural transformation for $\CF\otimes \CF$. Explicitly, given two elements $X, Y\in \CC$, the natural transformation $\Delta (g)$ on $\CF(X)\otimes \CF(Y)$ is given by
\be\label{eq:deltaJJ}
\btik
\CF (X)\otimes \CF (Y)\rar{J} & \CF (X\otimes Y)\rar{g_{X\otimes Y}} & \CF (X\otimes Y)\rar{J^{-1}} & \CF (X)\otimes \CF (Y)\,.
\etik
\ee
If $\End (\CF\otimes \CF)=\End (\CF)\otimes \End (\CF)$, then \eqref{Delta-app} endows $\End (\CF)$ with the structure of a bi-algebra.

Similarly, one can define a map of algebras
\be
\epsilon: \End (\CF)\to \End_\C (\CF (\id_\CC))=\C\,,
\ee
by sending $g=\{g_X\}$ to $g_{\id}$, where $\id$ is the unit object. 

 If moreover all objects in $\CC$ are dualizable, then $\CF$ preserves duals. For example, the evaluation map on $\CF (X)\otimes \CF (X^*)$ is defined by
\be
\btik
\CF (X)\otimes \CF (X^*)\rar{J}  & \CF (X\otimes X^*)\rar{\CF (\mathrm{ev}_\CC)} & \CF (\id_{\CC})\rar & \id_{\rm Vect}
\etik
\ee
Notice this is similar to equation \eqref{defEv}. The coevaluation is defined by
\be
\btik
\id_{\rm Vect}\rar & \CF (\id_{\CC}) \rar{\CF (\mathrm{coev}_\CC)} &  \CF (X^*\otimes X)\rar{J^{-1}} & \CF (X^*)\otimes \CF (X)
\etik
\ee
 The zigzag condition (S-move) is guaranteed by that $\CF$ maps identity morphism to identity morphism.  From this, we can define an antipode on $\End (\CF)$
\be
S: \End (\CF)\to \End (\CF)^{op}
\ee
 essentially by the same formula as equation \eqref{S-alg}. This makes $\End (\CF)$ into a Hopf algebra.

The arguments we presented in Sections \ref{sec:FF} and \ref{sec:Hopf-def} of the main paper gave a topological interpretation of the above constructions. Note, moreover, that nothing here requires $\CF$ to be faithful.

However, even in the case when $\CC$ is abelian, $\End (\CF\otimes \CF)$ may not always be equal to $\End (\CF)\otimes \End (\CF)$, but rather a completion of the latter. In this case, one must consider $\End (\CF)$ not as an ordinary Hopf algebra, but a topological Hopf algebra, whose topology comes from the inverse-limit topology. This problem persists in the dg case. In the following sections, we will explain how our various assumptions ensure the equality $\End (\CF\otimes \CF)=\End (\CF)\otimes \End(\CF)$ hold, allowing us to implement the above reconstruction physically (as well as when it fails, which results in topological Hopf algebras). We will also comment later how our completeness condition guarantees that $\CF$ is faithful and conservative, and induces an equivalence between $\CC$ and $\End (\CF)\Mod$ beyond the abelian case. 

Another problem that we have addressed in this paper, and will further clarify in the subsequent sections, is where fiber functors come from. Our answer in the main paper is that transverse boundary conditions give rise to fiber functors, and result in Drinfeld's double constructions. We will reinterpret this purely in terms of line operators.

\subsection{Category of line operators}
\label{app:line-setup}

\subsubsection{Dualizability assumptions on categories}

Suppose that, instead of a 3d topological QFT $\CT$ with two topological boundary conditions $(\CD,\CN)$, all we had access to were the three categories $\CC_\CT$ and $\CC_\CD,\CC_\CN$ of bulk and boundary line operators. What structures do our assumptions from Sections \ref{sec:top}--\ref{sec:lines} impose on these categories?

First of all, in order to apply Tannakian formalism, the difference between dualizable objects and all objects must be small. In Section \ref{sec:lines}, we were led to require that 
\begin{itemize}
\item[{[\AD]}] $\CC_\CD$, $\CC_\CN$ are tensor categories (monoidal with exact tensor product) and that $\CC_\CT$ is a ribbon category (a braided tensor category with ribbon twist).
\item[{[\AD]}] The subcategories $\CC_\CT^{fd},\CC_\CD^{fd},\CC_\CN^{fd}$ of dualizable objects have a canonical pivotal structure.
\item [{[\AD$_{\rm line}$]}] Each category is compactly generated and compact objects are rigid, \ie\ $\CC\simeq \text{Ind}(\CC^{c})$ and $\CC^c\subseteq \CC^{fd}$.
\end{itemize}
Moreover, we require that tensor product is continuous, in that it respects colimits. Indeed, we will work under the assumptions that all functors we consider are continuous. A consequence of this is that $\CC^c$ is a monoidal subcategory, and a continuous functor is determined by its value on $\CC^c$, which will eventually allow us to reduce the calculation of the endomorphism algebra to compact (therefore rigid by assumption) objects.  

There is additional information coming from the fact that $\CD,\CN$ are boundary conditions for $\CT$. The bulk-boundary map of line operators (colliding bulk line operators with either boundary) gives rise to monoidal functors (usually called the bulk-boundary map)
\be \CC_\CD\longleftarrow \CC_\CT : \beta_\CD\,,\qquad  \beta_\CN: \CC_\CT\longrightarrow \CC_\CN\,. \label{CT-act} \ee
In particular, both $\CC_\CD$ and $\CC_\CN$ naturally become \emph{module categories} for $\CC_\CT$. We'll use the convention that $\CC_\CT$ acts on the left on $\CC_\CN$ and on the right on $\CC_\CD$ to distinguish the different orientations of $\CN$ and $\CD$. Moreover, as bulk lines can be moved around boundary lines, the maps \eqref{CT-act} lift or refine to maps to the Drinfeld centers%
\footnote{Recall that, in the abelian case, an object of the Drinfeld center of $\CC$ is an object of $\CC$ together with a choice of half-braiding around all other objects. In the dg case, we should use a derived Drinfeld center as in \cite{Ben-Zvi:2008vtm}.} 
\be  \CZ(\CC_\CD)\longleftarrow \CC_\CT:\beta_\CD\,,\qquad \beta_\CN: \CC_\CT\longrightarrow \CZ(\CC_\CN)\,. \label{CT-act-Z} \ee
(There would be a `$\otimes$op' on $\CC_\CD$ if we had used the same action of $\CC_\CT$ on both sides.)
The relation \eqref{CT-act-Z} is sometimes known as a central structure for $\CC_\CD,\CC_\CN$.

Moreover, the fact that $\CD,\CN$ are boundary conditions --- as opposed to interfaces between $\CT$ and another nontrivial 3d theory --- means that the Drinfeld centers taken \emph{relative} to bulk lines must be trivial: $\CZ_{\CC_\CT}(\CC_\CN) \simeq \CZ_{\CC_\CT}(\CC_\CD) \simeq \mathrm{Vect}$. We collect the various relations of bulk and boundary categories in an additional assumption.

\begin{itemize}
\item[{[\textbf B]}]  The categories $\CC_\CD,\CC_\CN$ are endowed with a central structure $\beta_\CD,\beta_\CN$ as in \eqref{CT-act-Z} (which in particular makes both $\CC_\CD$ and $\CC_\CN$ module categories for $\CC_\CT$), such that relative Drinfeld centers are trivial:
\be \CZ_{\CC_\CT}(\CC_\CN) \simeq \CZ_{\CC_\CT}(\CC_\CD) \simeq \mathrm{Vect}\,. \ee
\end{itemize}

Finally, we note that assumption \AD$_{\rm line}$, implying that our three categories are ``rigid monoidal categories'' in the sense of \cite[App. D]{gaitsgory2015sheaves}, also implies that each category itself is dualizable, and in fact self-dual up to a reversal of monoidal structure
\be  (\CC_\CD)^* \simeq \CC_\CD^{\otimes \text{op}}\,,\qquad (\CC_\CN)^* \simeq \CC_\CN^{\otimes \text{op}}\,,\qquad (\CC_\CT)^* \simeq \CC_\CT^{\otimes \text{op}}\,. \ee
The dual $\CC_\CD^{\otimes op}$ corresponds to line operators on the flipped boundary $\ol\CD$; and $\CC_\CN^{\otimes op}$ to lines operators on $\ol\CN$. Moreover, we can access the 2-category of interfaces that can end at an $(\CN,\ol \CD)$ junction by expressing it as a category of left-right bimodules for $(\CC_\CN,\CC_\CD^{\otimes op})$, or simply a module category for $\CC_\CN\otimes \CC_\CD$\,:  
\be  \raisebox{-.5in}{\includegraphics[width=1in]{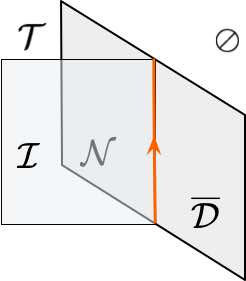}} \qquad \longleftrightarrow\qquad \CI \in \CC_\CN\otimes\CC_\CD\text{-mod}  \label{DN-int} \ee
Any such interface $\CI$ that is 1-dualizable (bendable around one axis) in the TQFT, \emph{a.k.a.} dualizable as a category, is dualizable as a module of $\CC_\CN\otimes \CC_\CD$ \cite[Prop. D.5.4]{gaitsgory2015sheaves}.

\subsubsection{Transversality and fiber functors}

Next, we rephrase the transversality axiom \AT\ in terms of lines.

Transversality was the statement that the composition $\CD\circ \CN$ is the trivial 2d theory. The category of line operators in the sandwich $\CD\circ\CN$ is the relative tensor product $\CC_\CD\otimes_{\CC_\CT}\CC_\CN$: namely it is obtained by taking lines from either boundary and identifying the action of lines in the bulk on the two sides, via the bulk-boundary maps. Thus, a statement of global transversality would be an equivalence
\be
\CC_\CD\otimes_{\CC_\CT}\CC_\CN\simeq \mathrm{Vect}\,. 
\ee
There are many levels to which one says two categories are equivalent (as plain categories, monoidal categories, etc). Based on Section \ref{sec:trans}, we are led to
\begin{itemize}
\item[{[\AT]}] There is an isomorphism $\CC_\CD\otimes_{\CC_\CT}\CC_\CN \simeq \text{Vect}$ as \emph{module categories} for $\CC_\CD\otimes \CC_\CN$.
Equivalently, $\Hom_{\CC_\CT\text{-mod}} (\CC_\CD^{\otimes\rm op}, \CC_\CN)\simeq \text{Vect}$ as module categories for $\CC_\CD\otimes \CC_\CN$.
\end{itemize}

In particular, there is an action of $\CC_\CD\otimes \CC_\CN$ on Vect. The interface $k$ as in \eqref{k-bdy} is represented by the one-dimensional vector space $\C\in\text{Vect}$. By acting on $k$ with either $\CC_\CN$ or $\CC_\CD$, we get monoidal fiber functors: 

\be \begin{array}{c}\CF_\CN:\CC_\CN\to \text{Vect} \\ \CF_\CN(\ell):= \ell\cdot k \in \text{Vect}\end{array}\,,\qquad \begin{array}{c} \CF_{\ol\CD} :\CC_{\CD}^{\otimes\rm op}\to \text{Vect} \\ \CF_{\ol\CD}(\ell) := k\cdot \ell\in \text{Vect} \end{array}\,,\label{eq:FDN-k} \ee
represented topologically by colliding lines on $\CD$ or $\ol\CN$ (say) with a $k$ interface:
\be \raisebox{-.4in}{\includegraphics[width=3.8in]{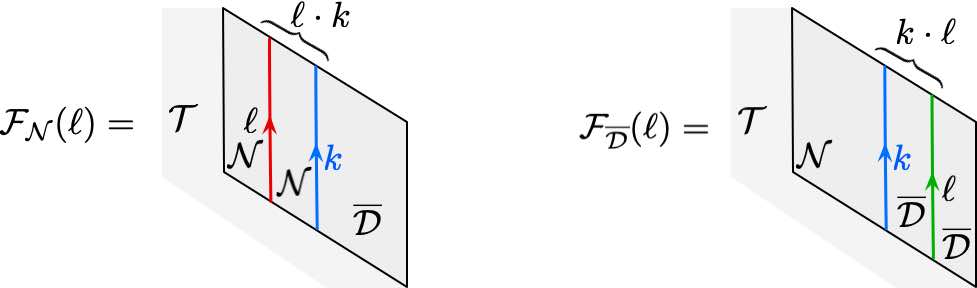}}  \label{DN-k-act} \ee
We've tried to be careful about orientations: in order to represent both functors are collisions with $k$ on a single (say, right) boundary of the bulk theory, we are forced to dualize one of the line-operator categories, say $\CC_\CD = \text{End}(\CD)$ to $\CC_\CD^{\otimes\rm op}=\text{End}(\ol \CD)$. The fiber functor on $\CC_\CD$ itself is a composition of $\CF_{\ol\CD}$ with the duality map, or (equivalently) a collision with $k$ on a left boundary: 
\be \CF_\CD = \big[ \CC_\CD \overset{\sim}\longrightarrow \CC_\CD^{\otimes \rm op} \overset{\CF_{\ol\CD}} \longrightarrow \text{Vect}\big]  \qquad \raisebox{-.55in}{\includegraphics[width=1.7in]{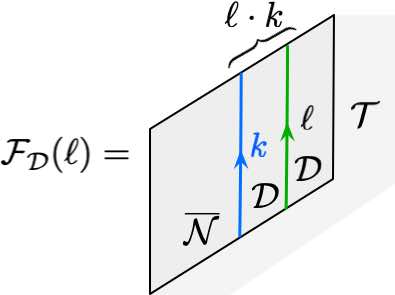}} \ee
We get a bulk fiber functor by composing these with either bulk-boundary map,
\be \CF_\CT(\ell) := \beta_\CN(\ell)\cdot k  = k\cdot \beta_{\ol\CD}(\ell)\,,\qquad i.e.\quad\CF_\CT=\CF_\CN\beta_\CN\cong \CF_{\ol\CD}\beta_{\ol\CD} = \CF_\CD\beta_\CD\,. \label{FT-bulk-iso} \ee
In TQFT, $\beta_{\ol\CD}$ is collision of a bulk line with the right b.c. $\ol\CD$ (as opposed to $\beta_\CD$, which is collision with the left b.c. $\CD$). Viewed this way, the isomorphism $\CF_\CN\beta_\CN\cong \CF_{\ol\CD}\beta_{\ol\CD}$ arises  from half-braiding a bulk line around the boundary interface $k$:
\be \raisebox{-.4in}{\includegraphics[width=1.5in]{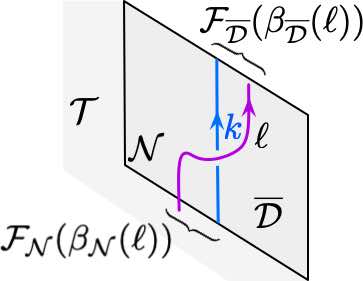}} \ee

There is one more piece of structure that transverse boundary conditions give, which will be essential to understanding the Hopf pairing. Note that the action of $\CC_\CN$ and $\CC_\CD^{\otimes \rm op}$ on $\mathrm{Vect}$ commute with each other --- essentially because they act on different sides of $k$. This commutation relation is an extra structure, realized by natural isomorphisms
\be\label{eq:naturalll}
\raisebox{-.5in}{\includegraphics[width=1.3in]{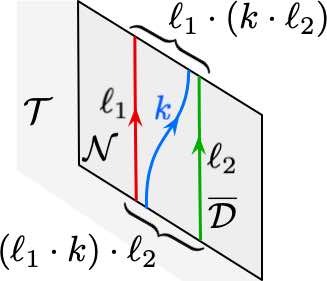}} \qquad
(\ell_1 \cdot k)\cdot \ell_2\cong \ell_1 \cdot (k\cdot \ell_2)\,,\qquad \ell_1\in \CC_\CN, ~\ell_2\in \CC_\CD^{\otimes\rm op} 
\ee
(In extended TQFT, this is an associator for 1-morphisms in the boundary 2-category.)
These natural isomorphisms induce functors between monoidal categories
\be\label{eq:commutator}
\CC_\CD^{\otimes \rm op}\to \End_{\CC_\CN} (\mathrm{Vect})\,, \qquad \CC_\CN\to \End_{\CC_\CD^{\otimes \rm op}} (\mathrm{Vect})\,.
\ee
Moreover, the natural transformations in equation \eqref{eq:naturalll} must be compatible with the central structure, in the sense that if $\ell_2=\beta_\CN (\ell_2')$ for some $\ell_2'\in \CC_\CT$ , then the above natural isomorphism must be coming from the braiding between $\CC_\CT$ and $\CC_\CD$ induced by $\beta_\CD$. Similar statement must be true for $\CC_\CN$.

We'll make one final assumption:
\begin{itemize}
    \item [{[\textbf{BF}]}] The functors in equation \eqref{eq:commutator} are surjective. 
\end{itemize}
This is reasonable from the following perspective.
When there is an underlying 2-category of right boundary conditions $\mb B$, containing objects $\CN,\ol \CD$, with $\CC_\CN=\text{End}_{\mb B}(\CN)$, $\CC_\CD^{\otimes \rm op}=\text{End}_{\mb B}(\ol\CD)$, $k\in \text{Hom}_{\mb B}(\ol\CD,\CN) = \text{Vect}$, the structures in \eqref{eq:FDN-k}, \eqref{eq:naturalll}, \eqref{eq:commutator} are all automatic from composition/associativity of 1-morphisms in $\mb B$. Consider the categorified Yondeda-like functors 
\be Y_\CN: \begin{array}{ccc}\mb B &\to& \CC_\CN\Mod\\
   \CB &\mapsto & \text{Hom}_{\mb B}(\CB,\CN)\end{array}
\,,\qquad Y_\CD: \begin{array}{ccc}\mb B &\to& \CC_\CD\Mod \\
\CB &\mapsto & \text{Hom}_{\mb B}(\ol\CD,\CB) \end{array}\,. \ee
One expects these to be surjective on objects and morphisms, as would be the case one category level lower. Then the two maps in \eqref{eq:commutator} are simply $Y_\CN$ applied to the 1-morphisms $\CC_\CD=\text{End}_{\mb B}(\ol\CD)^{\otimes \rm op}$ (on the LHS) and $Y_\CD$ applied to the 1-morphisms $\CC_\CN = \text{End}_{\mb B}(\ol\CN)$, whence both should be surjective.

Physically, \textbf{BF} is capturing the fact that $k$ is purely a boundary interface (with trivial extension into the bulk) --- as opposed to the more general possibility depicted in \eqref{DN-int} where a bulk interface $\CI$ ends on the boundary.

\subsubsection{Adjoints and algebra objects}\label{subsubsec:appendixfiber}

How do we compute the endomorphism algebras of the fiber functors? 
We sketched one answer to this, using only line-operator categories, in Section \ref{sec:FF-Hom}, which we'd now like to put on firmer mathematical footing.

Recall that in Section \ref{sec:linefunctor} we introduced three special ``thick'' line operators $\kNk\in \CC_\CD,\,\kDk\in \CC_\CN$, and $\CDCN\in \CC_\CT$ that are represented by strips on the boundaries and a carved-out cylinder in the bulk with both $\ol\CD,\ol\CN$ boundaries, respectively:
\be
\raisebox{-.4in}{\includegraphics[width=4.7in]{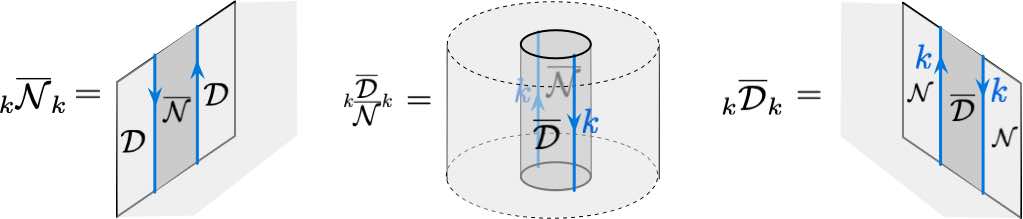}}  \label{alg-objects}
\ee
We argued that, if we assume \AD$_k$, then the fiber functors could be represented by taking Hom with these objects,
\be \CF_\CD = \text{Hom}_{\CC_\CD}\big(\kNk,-)\,,\quad \CF_\CN = \text{Hom}_{\CC_\CN}\big(\kDk,-)\,,\quad \CF_\CT = \text{Hom}_{\CC_\CT}\big(\CDCN,-)\,. \label{F-kDk-2} \ee
With this, we can compute the endomorphism algebra of each of the functor using the Yoneda lemma. 

Mathematically, working within line-operator categories, the ability to write $\CF$ as in \eqref{F-kDk-2} is guaranteed by the existence of adjoints. Specifically, the assumption \AD$_k$ (\cf\ illustrated in \eqref{k-superdual}) implies: 

\begin{itemize}

    \item [{[\AD$_k$]}] Each functor $\CF_\CN,\CF_\CD,\CF_\CT$  admits both left and right adjoints, and that the left and right adjoints agree. 
    
\end{itemize}

The existence of right adjoints is in fact already guaranteed by \AD$_{\rm line}$ (\cite[Prop. D.2.2]{gaitsgory2015sheaves}); and the existence of left adjoints implies that the right adjoints of each $\CF$ map compact objects to compact objects, and on these compact objects
\be\label{eq:LRadF}
\CF^L(V)=\CF^R(V^*)^*\,, \qquad \forall\; V\in \mathrm{Vect}^{fd}\,. 
\ee
The object representing the functor $\CF$ is simply $\CF^L (\id_{\rm Vect})$. This is true because we always have the following equalities, thanks to the adjunction between $\CF$ and $\CF^L$,
\be
\CF (X)=\Hom_{\mathrm{Vect}}(\id_{\rm Vect}\,, \CF (X))\cong \Hom_{\CC} (\CF^L (\id_{\rm Vect}), X)\,.
\ee
In particular, the ``thick'' line operators of \eqref{alg-objects} may all be defined as
\be \label{kDk-adj}
\kNk :=\CF_\CD^L (\id_{\rm Vect})\,,\qquad \kDk:=\CF_\CN^L (\id_{\rm Vect})\,,\qquad \CDCN :=\CF_\CT^L (\id_{\rm Vect})\,.
\ee
(Equivalently, we've also got $\kNkd = \CF_{\ol\CD}^L (\id_{\rm Vect})$, where $\kNkd$ is image of $\kNk\in \CC_\CD^{\otimes op} =\text{End}(\CD)$ in its dual category $\CC_\CD = \text{End}(\ol \CD)$). 

There are many consequences to \AD$_k$ beyond this. For example, since left and right adjoints are assumed to agree, $\CF^L (\id_{\rm Vect})=\CF^R (\id_{\rm Vect})$, and therefore by  equation \eqref{eq:LRadF}, the object $\CF^L (\id_{\rm Vect})$ is self-dual. Another consequence is that $\CF^L (\id_{\rm Vect})$ is a compact object. This is because
\be \label{FL-compact}
\begin{aligned}
    \Hom_{\CC} \lp \CF^L (\id_{\rm Vect}), \varinjlim \ell_i\rp& =\Hom_{\mathrm{Vect}} \lp\id_{\rm Vect}, \CF\varinjlim \ell_i\rp\\  (\text{continuity of } \CF \text{ and compactness of } \id_{\rm Vect})&=\varinjlim \Hom_{\mathrm{Vect}} \lp\id_{\rm Vect}, \CF\ell_i\rp\\ &= \varinjlim \Hom_{\mathrm{Vect}} \lp \CF^L (\id_{\rm Vect}), \ell_i\rp
\end{aligned}
\ee

We need one final assumption. Due to \AD$_{\rm line}$, the functors $\beta_\CN, \beta_\CD,$ $\CF_\CN,\CF_\CD$ (or the $\ol \CD$ versions) all admit continuous right adjoints, and fit into the following diagram
\be
\btik
 &\CC_\CD\otimes_{\CC_\CT}\CC_\CN & \\
\CC_\CD \arrow[ur, "\CF_\CD"]& &\CC_\CN\arrow[ul,swap,"\CF_\CN"]\\
 & \CC_\CT \arrow[ul, "\beta_\CD"]\arrow[ur, swap,"\beta_\CN"]& 
\etik 
\ee
Recall from \eqref{FT-bulk-iso} that there are isomorphisms of functors $\CF_\CN\beta_\CN \cong \CF_{\ol \CD}\beta_{\ol\CD}$ (using the $\ol \CD$ versions).
We can take right adjoints to obtain two natural transformations
\be\label{eq:basechange}
\beta_{\ol\CD}\beta_\CN^R\to \CF_{\ol\CD}^R\CF_\CN\,,\qquad \beta_\CN\beta_{\ol\CD}^R\to \CF_\CN^R\CF_{\ol\CD}\,.
\ee
For example, the first natural transformation is given by the following composition
\be \beta_{\ol\CD}\beta_\CN^R \longrightarrow  \CF_{\ol\CD}^R\CF_{\ol\CD}\beta_{\ol\CD}\beta_\CN^R \longrightarrow \CF_{\ol\CD}^R\CF_\CN \beta_\CN\beta_\CN^R \longrightarrow \CF_{\ol\CD}^R\CF_\CN\,, \ee
where the first map is the adjunction $\mathrm{Id}\to \CF_{\ol\CD}^R\CF_{\ol\CD}$, the second map is the isomorphism of functors $\CF_{\ol\CD}\beta_{\ol\CD}\cong \CF_\CN\beta_\CN$, and the last map is the adjunction $ \beta_\CN\beta_\CN^R\to \mathrm{Id}$. We require the following to be true.

\begin{itemize}
    \item [{[\textbf{BC}]}] The natural transformations of equation \eqref{eq:basechange} are isomorphisms. 
\end{itemize}
This condition is in some sense a base-change condition, and is satisfied when the categories involved are sheaves on algebraic varieties and functors are push-forward and pull-back functors.

In 3d TQFT, the base-change transformations \eqref{eq:basechange} have the following interpretation. The adjoints of the bulk-boundary maps take a boundary line $\ell$ (on either $\CN$ or $\ol\CD$) and place it on a drilled-out cylinder (wrapped by either $\CN$ and $\ol \CN$, or $\CD$ and $\ol\CD$) to create a bulk line:
\be \raisebox{-.35in}{\includegraphics[width=3.5in]{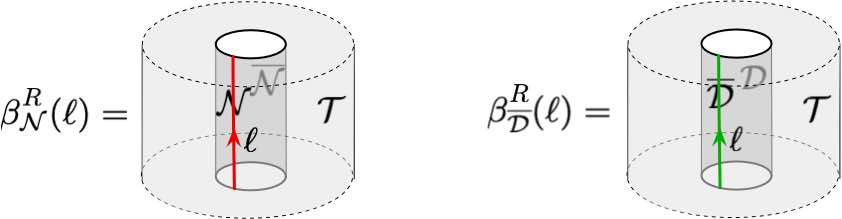}} \label{betaR} \ee 
The adjoints of the boundary fiber functors, represented by colliding with $k$ as in \eqref{DN-k-act}, send a vector space $V$ to its tensor product with either $\kDk$ or $\kNkd$, as appropriate:
\be  \raisebox{-.35in}{\includegraphics[width=3.5in]{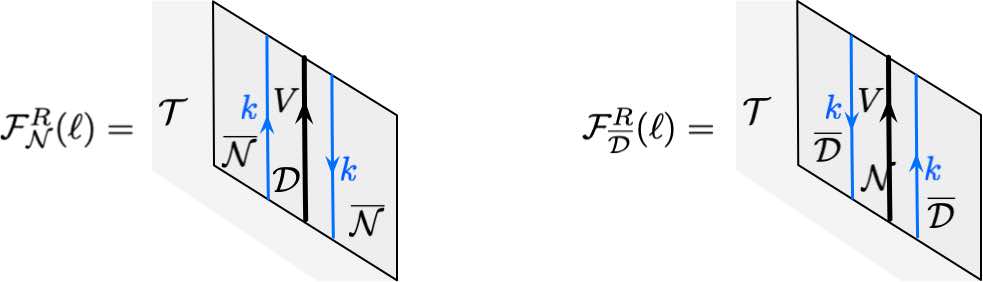}}  \ee
Then the base-change on (say) the LHS of \eqref{eq:basechange} relates 1) the collision of an $\CN$-cylinder bearing the line $\ell$ with a $\ol \CD$ boundary and 2) a $\kNkd$ strip on $\ol \CD$ bearing $\ell$. By transversality of $\ol\CN$ and $\ol\CD$, (1) and (2) should be isomorphic:
\be\raisebox{-.35in}{\includegraphics[width=4.8in]{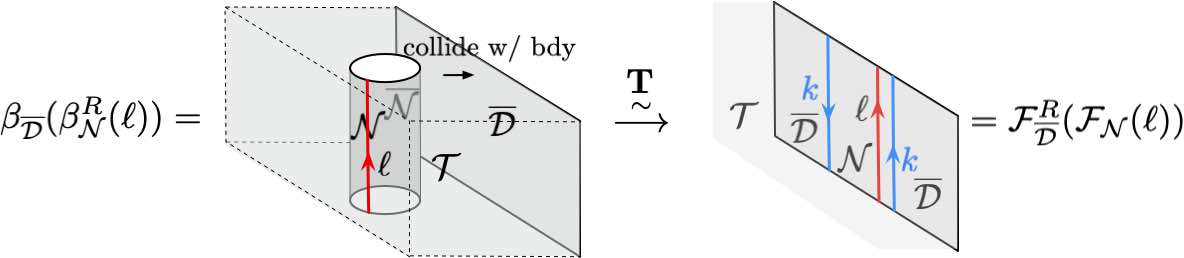}}  \ee

\subsubsection{Completeness and compact generators}\label{Appendix:complete}

It is unclear to us how to fully formulate completeness in the mathematical language of axiomatic TQFT. As we have mentioned, one can't hope for a morphism $\varphi: \CN\circ\CD\to \text{id}_\CT$; and even if such a morphism exists, one needs to require that it is ``surjective" on higher morphisms. 

However, for analyzing categories of line operators, we don't really need the full power of completeness: we just need to translate the strip gluing \AC$_{\rm strip}$. This also turns out to be the what is required to apply general Bar-Beck theory \cite[Sec. 3]{lurie2007derived}, allowing one to represent line-operator category using modules of an algebra. 

An application of strip gluing, discussed in great detail in Section \ref{sec:generators} shows that each of $\kNk$, $\kDk$, and $\CDCN$ must be \emph{generators} in their respective categories. For $X$ to be a generator means, in the abelian case, that every object lies in the image of $X$ tensored with a vector space; in the dg case it means that every object is a deformation of $X$ tensored with a dg vector space. 

Indeed, in Section \ref{sec:generators}, we argued that any object $\ell\in \CC_\CN$ can be identified with a deformation of $\kDk\otimes \CF_\CN (\ell)$ by the MC element $\mu_\CN$, identifying the action of $\CH_\CD$ on $\kDk$ and $\CF_\CN (\ell)$. The same holds true for $\CC_\CD$ and $\CC_\CT$. Moreover, we have argued that $\kNk$, $\kDk$, and $\CDCN$ are compact and rigid objects. Completeness then implies that

\begin{itemize}
\item[[\AC$_{\rm strip}$\!\!]] The objects  $\CF_\CD^L(\id_{\rm Vect})$, $\CF_\CN^L(\id_{\rm Vect})$, and $\CF_\CT^L(\id_{\rm Vect})$ are compact generators in their respective categories.
\end{itemize}

\subsection{Consequences of the assumptions}

\subsubsection{Hopf algebras from fiber functors and adjunctions}

In Section \ref{sec:spark-finite}, we argued that when \AD$_k$ was satisfied, spark algebras (the endomorphism algebras of our fiber functors) were finite dimensional. They gain involutive Hopf algebra structures from reconstruction, as discussed in Sections \ref{sec:Hopf-def}, \ref{sec:U-R} and Section \ref{A1:reconstruction}.
Adjunction provides a nice alternative perspective on defining the endomorphism algebras, as well as understanding their Hopf operations and their finite dimensionality. We describe this here.

Let us assume \AD$_k$. Then the fiber functors are equivalent to taking Hom's with the objects defined in \eqref{kDk-adj}, and their endomorphism algebras are identified with
\be \label{sparks-adj}
\CH_\CN= \End_{\CC_\CD} (\CF_\CD^L (\id_{\rm Vect}))^{\rm op}=\CF_\CD\CF^L_\CD (\id_{\rm Vect})\,,\quad \CH_\CD = \CF_\CN\CF_\CN^L(\id_{\rm Vect})\,,\quad U=\CF_\CT\CF_\CT^L(\id_{\rm Vect}) \,.
\ee
Recall from \eqref{FL-compact} each object $\CF^L(\id_{\rm Vect})$ is compact, and thus (by \AD$_{\rm line}$) rigid. Since each $\CF$ preserves rigid objects, we find that each algebra $\CH_\CN,\CH_\CD,U \in \text{Vect}$ is rigid, and thus necessarily finite dimensional.

The algebra structure these $\CH_\CN,\CH_\CD,U$ can be defined using adjunction. For example,
\be
\CH_\CN\otimes \CH_\CN \cong \CF_\CD\CF^L_\CD \CF_\CD\CF^L_\CD (\id_{\rm Vect})\longrightarrow \CF_\CD\CF^L_\CD(\id_{\rm Vect}) =\CH_\CN\,,
\ee
where the arrow above is the adjunction map (evaluation) $\CF^L_\CD \CF_\CD\to \mathrm{Id}$. The coalgebra structure is given by the fact that $\kNk=\CF^L_\CD (\id_{\rm Vect})=\CF^L_\CD \CF_\CD (\id_{\CC_\CD})$ has the structure of a coalgebra object in $\CC_\CD$. This coalgebra structure is again given by
\be
\kNk =\CF^L_\CD \CF_\CD (\id_{\CC_\CD})\longrightarrow \CF^L_\CD \CF_\CD\CF^L_\CD \CF_\CD (\id_{\CC_\CD})=\kNk \otimes \kNk.
\ee
Here the arrow above is the adjunction (coevaluation) $\mathrm{Id}\to \CF_\CD\CF_\CD^L$. Since $\CF$ is a monoidal functor, the image of $\kNk$, namely $\CH_\CN$, also has the structure of a coalgebra. These two structures must be compatible, thanks to the fact that $\CF_\CD$ is a monoidal functor.

This applies to $\CH_\CD$ and $U$ as well, and covers the situation of both abelian and dg categories. Also note that if we work exclusively with right b.c. and use the fiber functor $\CF_{\ol\CD}$ and its adjoint (as in \eqref{DN-k-act}), then, in the conventions of this paper, we will end up with Hopf algebra $\CH_\CN^{\rm coop}$ as its endomorphisms, where `coop' denotes reversed coproduct:
\be \CH_\CN^{\rm coop}  = \CF_{\ol \CD}\CF_{\ol \CD}^L(\id_{\rm Vect}) = \text{End}(\kNkd)\,. \ee
Finally, the algebra $U$ will be quasi-triangular, where the $R$ matrix comes from looking at the braiding action on
\be
U\otimes U= \CF_\CT \lp \CF_\CT^L(\id_{\rm Vect})\otimes \CF_\CT^L(\id_{\rm Vect})\rp. 
\ee
So far we have not used completeness. If we further assume \AC$_{\rm strip}$, then there are equivalences: 
\be\label{eq:completeequiv}
\CC_*\simeq \End (\CF_*)\Mod, \qquad *= \CN, \CD \text{ or } \CT.
\ee
These are now equivalences of monoidal categories (braided monoidal in the case of $\CT$).

\subsubsection{The pairing between $\CH_\CN$ and $\CH_\CD$}
\label{app:pair}

We have explained how we can arrive mathematically at spark algebras from dualizability assumptions, as well as Tannakian reconstruction. We explain now how the two spark algebras $\CH_\CN$ and $\CH_\CD$ are dual to each other.

The idea of this duality is as follows. Consider the two actions of $\CC_\CN$ and of $\CC_\CD$ on $\text{Vect}$  from \eqref{eq:FDN-k}, which commute with each other, in the sense of \eqref{eq:naturalll}. We have said in \ref{eq:commutator} that this implies the existence of a monoidal functor
\be
\CC_\CD^{\otimes \rm op} \longrightarrow \End_{\CC_\CN}(\mathrm{Vect}).
\ee
In the language of \cite{egno}, the category $\End_{\CC_\CN}(\mathrm{Vect})$ is called the dual category of~$\CC_\CN$. It was shown  (see \cite[Example 7.12.26]{egno}, or  \cite[Sec. 4.4]{DMNO-Witt}) that when $\CC_\CN$ is the category of modules of a Hopf algebra $\CH$, this dual category is precisely $\CH^{*, \rm coop}\Mod$.
Consequently, the above map induces a map of Hopf algebras $\CH_\CD^{*, \rm coop}\to \CH_\CN^{\rm coop}$, therefore $\CH_\CD^*\to \CH_\CN$. The argument in \textit{loc.cit.} is abelian in nature, but it can be adapted as follows to the dg context, with the help of Bar-Beck theory \cite[Sec. 3]{lurie2007derived}.

Recall that $\CF_\CN: \CC_\CN\to \mathrm{Vect}$ admits a right adjoint $\CF_\CN^R$, which is conservative. 
Let  $\kDk^*:=\CF_\CN^R(\id_{\rm Vect})=\CF_\CN^R(\CF_\CN (\id))$.%
\footnote{This object is the dual of $\kDk=\CF_\CD^L(\id_{\rm Vect})$. Using  \AD$_k$, left and right adjoints are isomorphic, and $\kDk^*\cong \kDk$. However, we'll consider weakening \AD$_k$ later, so we keep the dual notation here.} %
This object has the structure of an algebra in $\CC_\CN$, and the functor $\CF_\CN^R$ induces an equivalence\footnote{We would like to thank Theo Johnson-Freyd for pointing out to us that this equivalence requires $\CC_\CN$ to satisfy \AD$_{\rm line}$.}
\be
\wt \CF_\CN^R: \mathrm{Vect}\longrightarrow  \kDk^*\Mod (\CC_\CN)\,.
\ee
The action of $\kDk^*$ is given by
\be
\kDk^*\otimes \CF_\CN^R(V)=\CF_\CN^R(\CF_\CN ( \CF_\CN^R(V)))\to \CF_\CN^R(V)\,,\qquad \forall\; V\in \mathrm{Vect}\,.
\ee
Here the first equality follows from the condition of $\CF_\CN^R$ being a functor of module categories of $\CC_\CN$, and the second map is the adjunction $\CF_\CN\CF_\CN^R\to \mathrm{Id}$. 

Let us now compute $\End_{\CC_\CN}(\mathrm{Vect})$. Since we have identified $\mathrm{Vect}$ with $\kDk^*\Mod (\CC_\CN)$, the endomorphism becomes
\be
\End_{\CC_\CN} (\kDk^*\Mod (\CC_\CN))=\kDk^*\Mod\otimes_{\CC_\CN} (\kDk^*)^{\rm op}\Mod=\kDk^*\otimes (\kDk^*)^{\rm op}\Mod\,.
\ee
Since the fiber functor $\CF_\CN$ establishes an equivalence $\kDk^*\Mod\simeq \mathrm{Vect}$, we find that in the end, the above is equivalent to modules of the algebra $\CF_\CN\lp (\kDk^*)^{\rm op}\rp$. This algebra is precisely the co-End of $\CF$, and is identified with $\CH_\CN^{*}$ as an algebra.  

Of course, one could have started with $\kDk^*$, and obtain a map of Hopf algebras $\CH_\CD^*\to \CH_\CN$. The two maps are dual to each other.

\subsubsection{Non-degeneracy of the pairing and quasi-triangular Hopf algebra $U$}\label{subsubsec:ndpairU}

If we had two random Hopf algebras, there is no reason we could say anything about this map $\CH_\CD^*\to \CH_\CN$. For example, we could take the natural isomorphism in equation \eqref{eq:naturalll} to be identity, then the induced map $\CH_\CD^*\to \CH_\CN$ is simply the counit map. It will turn out that the requirement \textbf{BF} forces this map to be injective. 

To see this, consider the functor $\CC_\CD^{\otimes \rm op}\to \End_{\CC_\CN}(\mathrm{Vect})=\CH_\CD^{*, \rm coop}\Mod$. Of course this functor factors through
\be
\CC_\CD^{\otimes \rm op}\longrightarrow \CH_\CN^{\rm coop}\Mod \longrightarrow \CH_\CD^{*, \rm coop}\Mod.
\ee
Since the functor is surjective, there must be an object in $\CC_\CD^{\otimes \rm op}$ whose image under the above functor has a faithful action of $\CH_\CD^{*, \rm coop}$. Consequently $\CH_\CD^{*, \rm coop}\to \End (\CF_{\overline \CD})=\CH_\CN^{\rm coop}$ is an embedding. 

Similarly, the map $\CH_\CD^*\to \CH_\CN$ is an embedding. If the algebras in question are finite-dimensional (as we are assuming), then we come to the conclusion that these are isomorphisms of Hopf algebras. 

Finally, we consider the algebra $U=\End (\CF_\CT)$. The extended TQFT arguments presented in Section \ref{sec:bulk} imply that $U$ is a quasi-triangular Hopf algebra and is the double of $\CH_\CD$ and $\CH_\CN$. We now verify this in the current setting. 

First, we construct Hopf algebra embeddings $\CH_\CD, \CH_\CN\to U$. This is simply done by recognizing that $\CH_\CN=\End (\CF_\CD)\subseteq \End (\beta_\CD\circ \CF_\CD)=U$. Namely, an endomorphism of the functor $\CF_\CD$ automatically becomes an endomorphism of the composition $\beta_\CD\circ \CF_\CD$. We already see that they are dual to each other, so we are left to verify that $U$ is indeed the tensor product of the two. 

We use base-change \textbf{BC} to achieve this. We show that $U^*=\CH_\CN^*\otimes \CH_\CD^*$. 
\be
\begin{aligned}
    U^* & =\CF_\CT\CF_\CT^R (\id_{\rm Vect})\\ & =\CF_\CD\beta_\CD\beta_\CN^R\CF_\CN^R (\id_{\rm Vect}) \\ (\text{base-change \textbf{BC}}) &=\CF_\CD \CF_\CD^R \CF_\CN\CF_\CN^R(\id_{\rm Vect})\\ (\text{since } \CF^R (V)=V\otimes \CF^R (\id_{\rm Vect})) & = \CF_\CD \CF_\CD^R(\id_{\rm Vect})\otimes \CF_\CN\CF_\CN^R(\id_{\rm Vect}) \\
    &= \CH_\CN^*\otimes \CH_\CD^*
\end{aligned}
\ee
This verifies equation \eqref{DN-dualspaces}, and the statement that as soon as one of the three spark algebras is finite dimensional, then they all are (which we stated in Section \ref{sec:spark-finite}). 

To explain why the quasi-triangular Hopf structure (in particular the R-matrix) must come from the double construction, we note that the functor $\CC_\CT\to \CC_\CN$ upgrades to a functor into the center $\beta_\CN:\CC_\CT\to \CZ (\CC_\CN)$. Since $\wt \CF_\CN: \CC_\CN\to \CH_\CD\Mod$ is surjective, we obtain an induced braided monoidal functor
\be
\CC_\CT\overset{\beta_\CN}\longrightarrow \CZ (\CC_\CN)\overset{\wt\CF_\CN}\longrightarrow \CZ (\CH_\CD\Mod)=D (\CH_\CD)\Mod.
\ee
Since $U\cong D (\CH_\CD)$, the quasi-triangular Hopf algebra structure on $U$ must be from the double construction. 

\subsubsection{A summary}
\label{app:sum}

We end with a summary. To be able run our construction mathematically, purely in terms of categories of line operators, we have assumed
\begin{itemize}
    \item Dualizability conditions, in the form of \AD$_k$ and \AD$_{\rm line}$. 

    \item Transversality conditions, in the form of \AT, \textbf{BF}, and \textbf{BC}. 

    \item Completeness condition, in the form of \AC$_{\rm strip}$. 
    
\end{itemize}

The assumptions $\textbf{BF}$ and $\textbf{BC}$ shouldn't be thought of as additional assumptions on the TQFT, but rather conditions for $(\CC_\CT, \CC_\CD, \CC_\CN)$ to arise from a bulk-boundary TQFT. The other assumptions are genuine assumptions on the TQFT itself. Let us summarize how we have used these assumptions.

\begin{enumerate}

    \item We used dualizability \AD$_{\rm line}$ to be able to dualize categories and module-categories, as well as obtaining right adjoint functors.  

    \item We used dualizability condition \AD$_k$ to further guarantee the existence of left adjoint functors, which leads to finite-dimensionality of Hopf algebras. 

    \item We used \AT\, to obtain monoidal fiber functors. 

    \item We used \textbf{BF} to show that $\CH_\CN, \CH_\CD$ are dual to each other.

    \item We used \textbf{BC} to show that $U$ is the double of $\CH_\CN, \CH_\CD$. 

    \item Finally, we used \AC$_{\rm strip}$ to argue that $\CF$ upgrades to equivalences of tensor categories. 
    
\end{enumerate}

If one assumes that all $\CC_\CN,\CC_\CD,\CC_\CT$ are all finite semsimple fusion categories, then the only non-trivial conditions required are \AT\, and \AC$_{\rm strip}$!
In this case, the analysis essentially boils down to that of dual categories, from \cite[Sec 7.12]{egno}, \cite[Sec. 4.4]{DMNO-Witt}. Moreover, in this context, there are beautiful, specialized techniques for characterizing boundary conditions via Lagrangian subalgebras of the bulk fusion algebra --- and pairs of transverse boundaries via Lagrangian splittings, \cf\ \cite{CordovaZhang}. However, as we have just illustrated, constructions of Hopf algebras and their doubles apply far beyond the semisimple setting.

\subsection{Weakening dualizability conditions}
\label{app:weakenD}

When considering 3d $\CN=4$ gauge theories in Sections \ref{sec:gauge} and \ref{sec:GM}, we encountered examples that violated even the generous set of assumptions listed in \ref{app:sum}. In particular, as discussed in Section \ref{sec:gauge-cat}, dualizability \AD$_k$ of the $k$ interface was violated, and some of the boundary categories violated \AD$_{\rm line}$. Motivated by this, we conclude this appendix by considering possible weakenings of our dualizability assumptions. 

\subsubsection{Without dualizability of the interface}

What if we don't assume \AD$_k$? Then there is no guarantee that any of the endofunctors have left adjoints. However, assuming \AD$_{\rm line}$, one can still construct right adjoints of the fiber functors. The object $\CF^R(\id_{\rm Vect})$ is precisely the dual of $\CF^L(\id_{\rm Vect})$, if these are duaizable. 

In this case, the fiber functor naturally upgrades to a functor
\be
\wt \CF: \CC\longrightarrow \CF \CF^R(\id_{\rm Vect})\text{-coMod}\,,
\ee
where $\CF \CF^R(\id_{\rm Vect})$ is a co-algebra, usually called the ``co-End" of the functor $\CF$. The extra monoidal structure on $\CC$ gives $\CF \CF^R(\id_{\rm Vect})$ the structure of a bi-algebra, and in fact a Hopf algebra structure since $\CC$ is compactly generated by objects that are dualizable. Similar to the Hopf structure on $\CH$, one can describe this Hopf algebra structure using adjunctions. 

One can then simply take the linear dual of $\CF \CF^R(\id_{\rm Vect})$ to obtain the algebra $\CH=\End (\CF)$. The reason that this dual is precisely the endomorphism of $\CF$ follows from dualizability of $\CC_\CN$, which is guaranteed by \AD$_{\rm line}$. Indeed, in this case one can show that $\CC_\CN$ is equivalent to $\Hom (\CC_\CN, \mathrm{Vect})$, the category of continuous functors from $\CC_\CN$ to $\mathrm{Vect}$ \cite[D.1.2]{gaitsgory2015sheaves}. $\CF^R(\id_{\rm Vect})$ is precisely the object corresponding to $\CF$ under this equivalence, and we have
\be
\End (\CF)=\End_{\CC_\CN}(\CF^R(\id_{\rm Vect}))=\Hom_{\mathrm{Vect}}(\CF \CF^R(\id_{\rm Vect}), \id_{\rm Vect})=\CH\,.
\ee
Here we used adjunction in the second equality. 

However, in this case, $\wt\CF$ is not simply a functor from $\CC$ to modules of $\CH=\End (\CF)$. Rather, it maps to the category of smooth modules under a certain topology on $\CH$. In the abelian case, the topology is given by the inverse limit topology on the big product $\prod_{X\in \CC}\End (X)$. In the dg case, one simply notes that the Hopf algebra $\CF \CF^R(\id_{\rm Vect})$ is filtered by finite-dimensional vector spaces, since $\CF^R(\id_{\rm Vect})$ is filtered by compact objects. This gives the linear dual an inverse limit topology, with which one can define the category of smooth modules. Since $\CF$ maps $\CC^c$ to finite-dimensional comodules of $\CF \CF^R(\id_{\rm Vect})$, and therefore to finite-dimensional smooth modules of $\CH$, we obtain a functor
\be
\wt \CF: \CC\longrightarrow \CH\Mod^{\rm sm}\,,
\ee
to the category of smooth modules of $\CH$. 

Assuming only \AD$_{\rm line}$, the argument of duality of Hopf algebras from Section \ref{app:pair} still holds, assuming surjectivity of $\CC_\CN\to \End_{\CC_\CD^{\otimes \rm op}} (\mathrm{Vect})$ and $\CC_\CD^{\otimes \rm op}\to \End_{\CC_\CN} (\mathrm{Vect})$. In particular, there are still embeddings (where $\CH^*$ is the co-End here) 
\be
\CH_\CN^*\longrightarrow \CH_\CD\,, \qquad \CH_\CD^*\longrightarrow \CH_\CN\,. 
\ee
These must be continuous morphisms as the functors respect compact objects. The image of the embeddings must be dense otherwise they can't both be embeddings. On the other hand, since we assume that the functor $\CC_\CN\to \End_{\CC_\CD^{\otimes \rm op}} (\mathrm{Vect})$ is surjective, the algebra morphism $\CH_\CN^*\to \CH_\CD$ admits a splitting, since $\CH_\CN^*$ is a module of $\CH_\CD$. Density together with existence of splitting implies that these must be isomorphic. In particular, $\CH_\CN^*$ and $\CH_\CD$ must be finite-dimensional, since otherwise $\CH_\CD$ has a non-trivial topology. 

In essence, as long as we assume \AD$_{\rm line}$ of the two boundary conditions as well as \textbf{BF}, we are forced to have finite-dimensional Hopf algebras. 

\subsubsection{Without dualizability of boundary lines}

As has been discussed in Section \ref{sec:gauge-cat}, B twist of 3d $\CN=4$ gauge theories break \AD$_{\rm line}$ as well. How do we deal with this situation? What happened in this example is that one of the categories, say $\CC_\CN$, still satisfies \AD$_{\rm line}$, where-as the category $\CC_\CD$ does not. The fiber functor $\CF_\CN$ only admits a right-adjoint and $\CF_\CD$ only admits a left adjoint. The fiber functor $\CF_\CT$ admits neither the left or right adjoint. It turns out that this type of asymmetric set-up is what gives rise to topological Hopf algebras. 

Let us now work with the assumption that only $\CC_\CN$ satisfies \AD$_{\rm line}$ where as $\CC_\CD$ does not. In particular, $\CF_\CN$ has a right adjoint, and it upgrades to a functor
\be
\wt \CF_\CN: \CC_\CN \longrightarrow \CH_\CD\Mod^{\rm sm}\,,
\ee
where $\CH_\CD$ is the linear dual of $\CF_\CN\CF_\CN^R (\id_{\rm Vect})$, and is a topological Hopf algebra.

We now further assume that $\CF_\CD$ has a left adjoint, and therefore $\CF_\CD$ upgrades to a functor
\be
\wt \CF_\CD:\CC_\CD\longrightarrow \CH_\CN\Mod\,,
\ee
where $\CH_\CN=\CF_\CD\CF_\CD^L (\id_{\rm Vect})$ is a genuine bi-algebra. We do't know if this is a Hopf algebra anymore since $\CC_\CD$ is not generated by dualizable objects. We will still assume the reasonable assumption \text{BF}, which means there are surjective functors
\be
\CC_\CD^{\otimes \rm op}\longrightarrow \End_{\CC_\CN}(\mathrm{Vect})\,, \qquad \CC_\CN\longrightarrow \End_{\CC_\CD^{\otimes \rm op}}(\mathrm{Vect})\,.
\ee
Essentially the same argument of Section \ref{subsubsec:ndpairU} implies that the map $\CH_\CD^*\to \CH_\CN$ is an embedding of bi-algebras. Since $\mathrm{Vect}\simeq \CF_\CD^L (\id_{\rm Vect})\mathrm{-coMod}(\CC_\CD)$ in this situation, the same argument, applied to co-algebras rather than algebras, implies that
\be
\End_{\CC_\CD^{\otimes \rm op}}(\mathrm{Vect})\simeq \CH_\CN\mathrm{-coMod}\,. 
\ee
Surjectivity from $\CC_\CN$ to this category now imply that the map $\CH_\CD^*\to \CH_\CN$ must be surjective. Consequently, we have isomorphisms
\be
\CH_\CN^*\cong \CH_\CD, \qquad \CH_\CD^*\cong \CH_\CN\,,
\ee
in which the first is an isomorphism of topological bi-algebras and the second is an isomorphism of ordinary bi-algebras. In particular, $\CH_\CN$ is a Hopf algebra. 

Finally, the braided monoidal functor $\CC_\CT\to \CZ (\CC_\CN)$ leads to a braided monoidal functor to modules for the Drinfeld double
\be
\CC_\CT\longrightarrow D(\CH_\CD)\Mod\simeq D(\CH_\CN)\Mod\,,
\ee
as we were hoping.

In summary, instead of \AD$_{\rm line}$, we can relax our set-up to the following asymmetric condition. 
\begin{itemize}
    \item $\CC_\CN$ satisfies \AD$_{\rm line}$ whereas $\CF_\CD$ admits a left-adjoint. 
\end{itemize}
This condition is enough to guarantee topological (quasi-triangular) Hopf algebras from the fiber functors. Unfortunately, in this situation, we don't know how to give a natural generalization of completenesss assumption as in \AC$_{\rm strip}$. We believe that such a condition should exist and guarantees the equivalence of monoidal categories as in equation \eqref{eq:completeequiv}.

\section{Replacing boundaries with commutative algebra objects}
\label{sec:bdyE2}

In this section, we consider replacing boundary conditions with commutative algebras (generally $\E_2$ algebras) in the category $\CC_\CT$. This is particularly relevant in a situation where one is accessing a 3d TQFT via a boundary VOA, as such a commutative algebra object can be constructed from a VOA extension (see for instance \cite{Creutzig:2017anl}). We hope to further explore this connection in our future project \cite{sparkVOA}. 

Given a single topological boundary condition, say $\CN$, that's 1-dualizable, we can define a bulk line operator $\CV_\CN$ by carving out a cylinder and wrapping its boundary with $\CN$ and $\ol \CN$:
\be \raisebox{-.4in}{\includegraphics[width=3.1in]{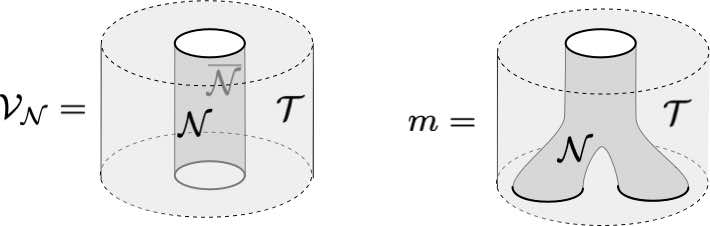}}  \label{alg-objects-V}
 \ee
This $\CV_\CN$ is an algebra object -- there's a product morphism $m: \CV_\CN\otimes\CV_\CN\to \CV_\CN$ defined by a pair of pants. The product is commutative, commuting with braiding on $\CV_\CN\otimes\CV_\CN$. In a dg/infinity setting, $\CV_\CN$ would be an $\E_2$ algebra object. If $\CV_\CN$ is dualizable and self-dual, the same sort of cobordisms one would draw in 2d TQFT \cite{MooreSegal} give it the structure of a Frobenius algebra object. Similarly, we may wrap an empty cylinder with $\CD,\ol\CD$ to define a commutative algebra object $\CV_\CD$.

In terms of the categories of line operators, the objects $\CV_\CN,\CV_\CD\in \CC_\CT$ are precisely realized as the images of the identity lines on the boundary, under the adjoints of the bulk-boundary map:
\be \CV_\CN = \beta_\CN^R(\id_{\CC_\CN})\,,\qquad \CV_\CD = \beta_\CD^R(\id_{\CC_\CD})\,. \ee
In general, $\beta_\CN^R(\ell)$ and $\beta_\CD^R(\ell)$ place a boundary line $\ell$ on the boundary of these cylinders in the bulk, as depicted in \eqref{betaR}. Adjunction is what turns $\CV_\CN,\CV_\CD$ into algebra objects, \emph{e.g.} with multiplication
\be m: \CV_\CN\otimes \CV_\CN \simeq \beta_\CN^R\beta_\CN \beta_\CN^R(\id_{\CC_\CN}) \longrightarrow \beta_\CN^R(\id_{\CC_\CN}) \ee
defined by the adjunction $\beta_\CN\beta_\CN^R\to\text{Id}$. Similarly, for any $\ell \in \CC_\CN$ (say) there is an action $\CV_\CN\otimes \beta_\CN^R(\ell) \to \beta_\CN^R(\ell)$, given algebraically by adjunction and depicted in TQFT by a pair of pants with $\ell$ along a leg:
\be
\CV_\CN\otimes \beta^R_\CN (\ell)\simeq\beta_\CN^R\beta_\CN\beta_\CN^R (\ell)\longrightarrow \beta_\CN^R(\ell) \qquad \raisebox{-.4in}{\includegraphics[width=1.35in]{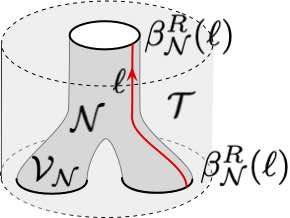}}
\ee
Altogether, this action defines monoidal functors from the category of lines on each boundary condition $(\CD,\CN)$ to modules for the respective algebra objects $(\CV_\CD,\CV_\CN)$ inside the braided tensor category $\CC_\CT$:
\be \rho_\CD: \CC_\CD\to \CV_\CD\text{-mod}(\CC_\CT)\,,\qquad \rho_\CN:\CC_\CN\to\CV_\CN\text{-mod}(\CC_\CT)\,. \label{wrapline-ND} \ee

Under suitable conditions, laid out in Bar-Beck theory \cite[Sec. 3]{lurie2007derived}, these are equivalences. It is a common practice in some mathematical treatments to simply substitute categories $\CC_\CD$ and $\CC_\CN$ with $\CV_\CD\Mod$ and $\CV_\CN\Mod$. Suppose we do this. These being boundary conditions require $\CZ_{\CC_\CT}(\CC_\CD)=\mathrm{Vect}$ and $\CZ_{\CC_\CT}(\CC_\CN)=\mathrm{Vect}$. Fortunately, this relative center is readily computable in either abelian or dg settings (see \cite{laugwitz2023constructing, chen2021extension}), and is given by
\be
\CZ_{\CC_\CT}(\CV\Mod)\simeq \CV\Mod^{\E_2},
\ee
where the RHS is the category of $\E_2$ (or local, in abelian case) modules. Therefore one requires that $\CV_*\Mod^{\E_2}\simeq \mathrm{Vect}$ for both $\CN, \CD$. 

Note that the tensor product
\be
\CV_\CD\otimes \CV_\CN\in \CC_\CT
\ee
is still an algebra object, though no longer $\E_2$. However, it is still true that $\CV_\CN\otimes \CV_\CD\text{-mod}(\CC_\CT)$ is a monoidal category, and there is an equivalence:
\be
\CV_\CD\text{-mod}(\CC_\CT)\otimes_{\CC_\CT}\CV_\CN\text{-mod}(\CC_\CT) \simeq \CV_\CD\otimes \CV_\CN\text{-mod}(\CC_\CT)\,.
\ee
Transversality then translates into an equivalence of monoidal categories:
\be
\AT:\qquad \CV_\CD\otimes \CV_\CN\text{-mod}(\CC_\CT)\simeq \mathrm{Vect},
\ee
and completeness \AC$_{\rm strip}$ implies that $\CV_\CD\otimes \CV_\CN$ is a generator in the category $\CC_\CT$. Indeed, an inspection of the picture representing $\CV_\CD\otimes \CV_\CN$ shows that, using transversality, it is equivalent to the thick line operator $\CDCN=\CF_\CT^L(\id)$ of \eqref{alg-objects}
\be \raisebox{-.4in}{\includegraphics[width=3.5in]{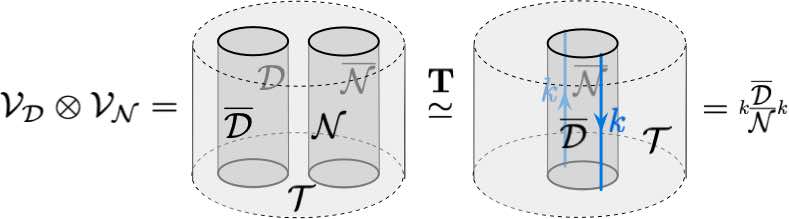}}  \label{alg-ND}  \ee
so the notions of completeness here and in Section \ref{Appendix:complete} coincide.

Under these assumptions, we construct a monoidal functor $\CF_\CT: \CC_\CT\to \mathrm{Vect}$ as
\be
\CF_\CT(\ell)=\CV_\CD\otimes \CV_\CN\otimes \ell \;\in\; \CV_\CD\otimes \CV_\CN\text{-mod}(\CC_\CT)\simeq \mathrm{Vect}\,.
\ee
This is precisely the induction functor from $\CC_\CT$ to the category of modules of the algebra $\CV_\CD\otimes \CV_\CN$. Since $\CV_\CD\otimes \CV_\CN\text{-mod}$ is equivalent to $\mathrm{Vect}$, we have:
\be
\CF_\CT(\ell)=\Hom_{\mathrm{Vect}}(\C, \CF_\CT(\ell))=\Hom_{\CV_\CD\otimes \CV_\CN\text{-mod}}(\CV_\CD\otimes \CV_\CN, \CF_\CT(\ell)).
\ee
Now by induction-restriction adjunction as well as the self-duality of $\CV_\CD\otimes \CV_\CN$, we find:
\be
\CF_\CT(\ell)=\Hom_{\CC_\CT}(\id, \CV_\CD\otimes \CV_\CN\otimes \ell)=\Hom_{\CC_\CT}(\CV_\CD\otimes \CV_\CN, \ell),
\ee
and therefore the functor $\CF_\CT$ can be represented by a hom functor, just as in \eqref{F-kDk-2}. By Yoneda lemma, we can then compute the endomorphism algebra of $\CF_\CT$ as $\End (\CF_\CT)=\End_{\CC_\CT}(\CV_\CD\otimes \CV_\CN)=:U$. The left and right adjoints of $\CF_\CT$ exist, and are canonically isomorphic, thanks to the fact that $\CV_\CD\otimes \CV_\CN$ is self-dual. Our assumption of completeness, namely that $\CV_\CD\otimes \CV_\CN$ is a generator, precisely guarantees that $\CF_\CT$ is conservative, and induces an equivalence $\CC_\CT\simeq U\Mod$.

We end this appendix by noting that these objects $\CV_\CN, \CV_\CD$ have appeared in all the examples we discussed. 

\begin{enumerate}
    \item In DW theory, where $\CC_\CT\simeq \text{Shv}^G(G)$, 
    the object $\CV_\CN=\CO_G$ is the regular representation supported at the identity $e\in G$, and $\CV_\CD=\CO (G)$ is the structure sheaf of~$G$.
    
    \item In B-twisted 3d $\CN=4$ $G$ gauge theory, with $G$ an algebraic group and $\CC_\CT\simeq \text{QCoh}^G(G)$, nearly the same description applies. $\CV_\CN$ is a skyscraper at the identity with stalk $\CO_G$; while $\CV_\CD$ is the dual of the structure sheaf $\CO(G)$, thought of as a projective system of objects. 

    \item In B-twisted pure matter theory, where $\CC_\CT^{fd} \simeq \text{Coh}((T^*V)[1])$, the objects 
$\CV_\CN$ and $\CV_\CD$ are structure sheaves of the two Lagrangian subspaces $V,V^*\subseteq T^*V$, respectively. In other words, they are simply sheaves supported on the Lagrangians that define the respective boundary conditions. This perspective is true intuitively for gauge theory with matter as well, if one thinks of gauge theory as a Rozansky-Witten theory to the stack $T^*(V/G)$.    
\end{enumerate}

\section{Koszul duality in Tannakian QFT}
\label{sec:Koszul}

We said in the introduction that that Koszul duality could be viewed as a perturbative manifestation of Tannaka duality. We take the opportunity here to expand on this.

We'll first review how Koszul duality arises in physics, following and combining perspectives from \cite{Costello-4dCS}, its recent generalizations \cite{CostelloPaquette, PaquetteWilliams}, as well as unpublished work of the first author presented in \cite{Dimofte-Koszul}.  Other recent works implementing Koszul duality in physics, with helpful discussions and references, include \cite{GMW, Aganagic-Koszul,BDGH}. Classic mathematics references include \cite{Priddy, beilinson1996koszul}.

When dealing with categories (with no additional structure, such as tensor products or braiding), Koszul duality is fundamentally a 2-dimensional phenomenon. Moreover, as traditionally formulated, it requires a perturbative description of local operators and boundary conditions, in a TQFT of cohomological type. In this paper, we have mainly been interested in categories of line operators in 3d TQFT's, which have additional structures. To connect them to Koszul duality, we'll need to compactify our 3d theory to 2d. We'll explain how we do this in Sections \ref{sec:K-2d} and \ref{sec:K-3d}.

One major goal of this Appendix is to show that, perturbatively, our spark algebras are the Koszul duals of algebras of local operators:
\be \CH_\CD \simeq A_\CN^!\,,\qquad \CH_\CN\simeq A_\CD^!\,,\qquad U \simeq A_\CT^! \ee
where $A_\CN,A_\CD$ are the $E_2$ algebras of local operators on $\CN$ and $\CD$ boundary conditions, and $A_\CT$ is the $E_3$ algebra of local operators in the bulk. Moreover, we'll explain that
\begin{itemize}
\item In this perturbative context, each spark algebra may be computed by integrating descendants of local operators. Mathematically, we get bar/Hochschild complexes: \vspace{-.3in}
\be \hspace{-.3in}\CH_\CD \simeq HH_\bullet(\CA_\CD,\C\otimes \C)\,,\quad \CH_\CN \simeq HH_\bullet(\CA_\CN,\C\otimes \C)\,,\quad U \simeq HH_\bullet(\CA_\CD\otimes\CA_\CN,\C\otimes \C)\,.\ee
\item With additional assumptions, the Hopf pairing between $\CH_\CD$ and $\CH_\CN$ (which determines the R-matrix on $U$) is computed by the $E_3$ bracket on $A_\CT$.
\end{itemize}

 This connects the current paper directly to the work of Costello and Costello-Paquette \cite{Costello-4dCS,CostelloPaquette}, who explained how line operators (in any dimension) should be represented as modules for the Koszul dual of local operators. It also connects the current paper to many mathematical results on Koszul duals of $E_n$ algebras. In particular, Tamarkin's quantization of Lie bi-algebras in \cite{Tamarkin-formality,tamarkin2007quantization} established (algebraically) an $E_2$-Hopf Koszul duality that seems to underlie the quantum double $U\simeq \CH_\CD\otimes\CH_\CN \simeq A_\CN^!\otimes A_\CD^!$ in our perturbative Tannakian TQFT.%
\footnote{We thank Kevin Costello for explaining this connection.} 
Other manifestations of $E_2$-Hopf duality include \cite{GerstenhaberVoronov,Fresse,Kadeishvili,Young-E2}.
$E_3$-quasitriangular Koszul duality follows formally from work of Lurie \cite{Lurie-DAGVI}, who established that the module category of an $E_3$ algebra is braided; it has also been considered in work of Ayala-Francis \cite{AyalaFrancis-Koszul}, and unpublished work of Costello-Francis-Gwilliam \cite{CostelloFrancisGwilliam}.

\subsection{Basics}
\label{sec:K-basics}

Koszul duality is fundamentally a phenomenon in 2-dimensional topological, cohomological, perturbative QFT. Suppose we have such a 2d QFT $\CT$ with a category $\CC_L$ of left boundary conditions and a category $\CC_R$ of right boundary conditions. Let $\mb B_L\in \CC_L$ and $\mb B_R\in \CC_R$ be two boundary conditions that are
\begin{enumerate}
\item Transverse, meaning that the state space on a strip bounded by $\mb B_L$ and $\mb B_R$ (some dg vector space) is quasi-isomorphic to the trivial space $\C$:
\be \text{States}(\mb B_L,\mb B_R) = \text{States}\bigg( \raisebox{-.17in}{\includegraphics[width=.65in]{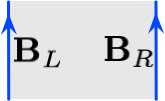}} \bigg) \simeq \C \label{bc-trans}  \ee

\item Both perturbative generators of their respective categories (a version of completeness), meaning that every object $b\in\CC_R$ can be obtained by tensoring $\mb B_R$ with a vector space $V_b$ (a Chan-Paton bundle) and turning on an interaction encoded in a Maurer-Cartan element $\mu_b$; and likewise for $\CC_R$. Schematically:
\be b \simeq (\wt V_b\otimes \mb B_L,\wt \mu_b)\quad \forall\, b\in \CC_L\,,\qquad  b \simeq (V_b\otimes \mb B_R,\mu_b)\quad \forall\, b\in \CC_R\,. \label{bc-mc} \ee
\end{enumerate}

Now denote the dg or  $A_\infty$ algebras of local operators on the two boundary conditions as $A_L := \text{End}(\mb B_L)$ and $A_R := \text{End}(\mb B_R)$, 
\be \raisebox{-.23in}{\includegraphics[width=4in]{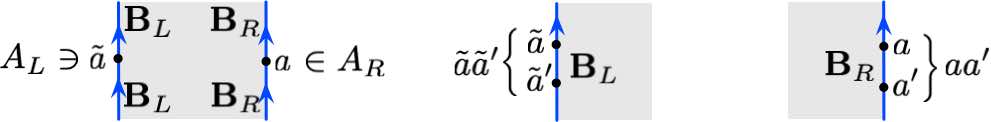}} \ee
Both $A_L$ and $A_R$ act on the state space on the $(\mb B_L,\mb B_R)$ strip, which is isomorphic to $\C$, and the two actions commute with each other. (Note that in order to make both actions explicit, simultaneously, one must use a resolution of $\C$ as a large dg vector space, whose cohomology happens to be trivial.) Thus, transversality ensures that
\be \text{End}_{A_L}(\C) \supseteq A_R\,,\qquad \text{End}_{A_R}(\C) \supseteq A_L\,. \ee
The second condition that both $\mb B_L,\mb B_R$ generate ensure that these inclusions are actually equalities:
\be \text{End}_{A_L}(\C) = A_R\,,\qquad \text{End}_{A_R}(\C) = A_L\,. \label{bc-KD} \ee
This is precisely the mathematical relation that identifies $A_L$ and $A_R$ as \emph{Koszul-dual} algebras.

More customarily, one might write $A_R \simeq A_L^!$, $A_L\simeq A_R^!$. The one-dimensional module $\text{States}(\mb B_L,\mb B_R)\simeq \C$, which exists due to transversality, provides what is known mathematically as an \emph{augmentation} of $A_L$, as well as $A_R$.

Assuming that $\mb B_L,\mb B_R$ are sufficiently nice (\emph{e.g.} compact) objects, this setup further implies an equivalence of categories
\be A_L\Mod\simeq \CC_R\,,\qquad \CC_L\simeq A_R\Mod\,. \label{bc-cats} \ee
This is just Tannaka duality! In a Tannakian context, we would say that (\eg) the first equivalence comes from having a fiber functor
\be \CF_R :\CC_R\to \text{Vect}\,,\qquad \CF_R(b) := \text{States}(\mb B_L,b) = \text{States}\bigg(  \raisebox{-.17in}{\includegraphics[width=.65in]{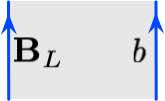}} \bigg)\,, \ee
and identifying the endomorphism algebra of the fiber functor as $\text{End}(\CF_R) = A_L$. The same is true for the second equivalence in \eqref{bc-cats}, using a fiber functor $\CF_L = \text{States}(-,\mb B_R)$.

A feature of the perturbative setting is that the representations \eqref{bc-mc} of objects in the two categories makes the fiber functors very concrete. For example, writing $b\simeq (V_b\otimes \mb B_R,\mu_b)$ (and assuming all operations involved are continuous) we get
\be \CF_R(b) = \big(V_b\otimes \text{States}(\mb B_L,\mb B_R),\mu_b\big) \simeq V_b\ee
In other words, $\CF_R$ \emph{measures} the vector spaces $V_b$ that are used to engineer each particular object $b$ from $\mb B_R$. The first equivalence of categories in \eqref{bc-cats} comes from a lift
\be \wt\CF_R:\CC_R \overset\sim\longrightarrow A_L\Mod\,, \ee
where the action of $A_L$ on $V_b$ is encoded, in a unique way, by the Maurer-Cartan element $\mu_b$. In other words, there is a 1-1 correspondence, up to isomorphism
\be \big\{\text{$A_L$ actions on $V_b$}\big\}\;\leftrightarrow\; \big\{\text{MC elements $\mu_b\in \text{End}_{\CC_R}(V_b\otimes \mb B_R)=\text{End}_\C(V_b) \otimes A_R$}\big\}\,. \ee
Similarly, in the second equivalence of \eqref{bc-cats}, objects $b\in \CC_L$ are identified with modules $(\wt V_b,\wt \mu_b)$ for $A_L$.

Finally, it's useful to note that with some dualizability of the 2d bulk theory, the two categories of left and right boundary conditions will be equivalent, up to reversing the order of Hom's: $\CC_R\simeq \CC_L^{\rm op}$ and vice versa. This leads to the final consequence of Koszul duality: Koszul-dual algebras are expected to have equivalent module categories
\be \CC_R\simeq\;\; A_L\Mod \simeq A_R\Mod^{\rm op} \;\; \simeq \CC_L^{\rm op}\,. \label{bc-ecats} \ee
The equivalence swaps representations $(V_b,\mu_b)\leftrightarrow(\wt V_b,\wt \mu_b)$ of the same boundary $b$.

\subsection{Boundary sparks and boundary local operators}
\label{sec:K-2d}

Now consider, say, the category of line operators $\CC_\CN$ on an $\CN$ boundary condition in 3d Tannakian QFT (satisfying the axioms of Section \ref{sec:setup}). To make this perturbative, suppose in addition that the theory is cohomological, and that all line operators $\ell\in \CC_\CN$ can be generated from the identity line $\id_\CN$ by tensoring with a vector space $V_\ell$ and deforming by a Maurer-Cartan element $\mu_\ell$,
\be \ell \simeq (V_\ell\otimes \id_\CN,\mu_\ell)\qquad\forall\,\ell\in \CC_\CN\,. \label{N1-gen} \ee

To relate to the 2d Koszul-duality setup above, let's compactify $\CT$ on an $\ol\CN\circ\CN$ sandwich. Recall that the category of boundary conditions for the sandwich $\ol\CN\circ\CN$ should be equivalent to $\CC_\CN$. Moreover, with the dualizability we've assumed in Section \ref{sec:setup}, the left and right categories of boundary conditions are equivalent.

There are two natural objects in $\CC_\CN$ that form transverse generators, as boundary conditions for $\ol\CN\circ\CN$:
\be \raisebox{-.25in}{\includegraphics[width=3in]{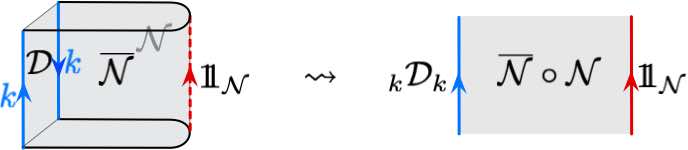}}    \label{K-2-trans} \ee
One of them is the image of the identity $\id_\CN$, viewed as a right boundary condition. The other is a thin strip of ${}_k\CD_k$, viewed as a left boundary condition.  Transversality of these effective boundary conditions for $\ol\CN\circ \CN$ follows immediately from transversality of $\CD$ and $\CN$ in Tannakian QFT. The fact that $\id_\CN$ generates was an \emph{assumption} in our perturbative setting. The fact that ${}_k\CD_k$ generates was established in Section \ref{sec:faithful}, as a consequence of completeness.%
\footnote{More precisely, in Section \ref{sec:faithful} we argued that $\kDk\in \CC_\CN$ is a generator. This is equivalent to ${}_k\CD_k \in \CC_\CN^{\otimes\rm op}$ generating the orientation-reversed category. Also note that due to the orientations we've chosen here we get $\CH_\CD=\text{End}({}_k\CD_k)$ as opposed to $\CH_\CD^{\rm op} = \text{End}(\kDk)$ as in Section \ref{sec:spark-Yo}.}

We've now got a setup for Koszul duality! The two algebras
\be \CH_\CD = \text{End}({}_k\CD_k)\,,\qquad  A_\CN := \text{End}(\id_\CN) \label{HDAN}  \ee
must be Koszul dual, as dg/$A_\infty$ algebras.
 But these now have intrinsic meaning. $\CH_\CD$ is of course the spark algebra on $\CD$, while $A_\CN$ is the algebra of local operators on $\CN$. Our lifted fiber functor
\be \wt \CF_\CN:\CC_\CN\to\CH_\CD\Mod\,,\qquad \wt\CF_\CN:\ell\mapsto (V_\ell,\mu_\ell) \ee
measures the vector space that's used to engineer $\ell$ from the identity line. We also expect an equivalence of categories
\be \CC_\CN\simeq \CH_\CD\Mod \simeq A_\CN\Mod^{\rm op}\,. \label{K-HD-AN} \ee

A major goal of this paper was to promote $\CH_\CD$ to a Hopf algebra, using manipulations in 3d space, so that the first equivalence in \eqref{K-HD-AN} captures not just categories, but monoidal categories with duals. In contrast, the algebra $A_\CN$ is \emph{not} a Hopf algebra. Since it's local operators on the $\CN$ boundary, it must be a commutative algebra --- or more accurately in a dg/infinity setting, a framed $E_2$ algebra. The framed $E_2$ structure includes, at the level of cohomology, a $(-1)$-shifted Poisson bracket, usually called the Gerstenhaber bracket; and a `BV' operator (\cf\ \cite{Getzler}).
 In $\CA_\CN\Mod^{\rm op}$, it's the framed $E_2$ structure on $A_\CN$ that controls the tensor product and duals on its module category. Under the Koszul duality of $\CH_\CD$ and $\CA_\CN$, the Hopf and framed $E_2$ structures are expected to be exchanged, \cf\ \cite{Tamarkin-formality,tamarkin2007quantization,GerstenhaberVoronov,Fresse,Kadeishvili,Young-E2}
 \be (\CH_\CD,\,\text{Hopf}) \;\leftrightarrow\; (A_\CN,\,\text{framed $E_2$})\,. \ee

The Koszul duality between $\CH_\CD$ and $A_\CN$, in particular the fact that they have mutually commuting actions on a (cohomologically) trivial vector space $\C$, also has a beautiful interpretation after using a state-operator correspondence. As we know from Section \ref{sec:spark-space}, $\CH_\CD$ is the space of states on a rectangle \eqref{Rectangles}. It acquires a (generally) non-commutative product from merging the `$\CD$' edges of the rectangle, as in \eqref{Hopf-cob}. Similarly, the algebra $A_\CN$ of boundary local operators is mapped by a state-operator correspondence to the space of states on a disc with $\CN$ boundary (or more accurately, $\CN$ and $\ol\CN$) all the way around. The product $A_\CN\otimes A_\CN\to A_\CN$ comes from merging discs:
\be \raisebox{-.3in}{\includegraphics[width=3.3in]{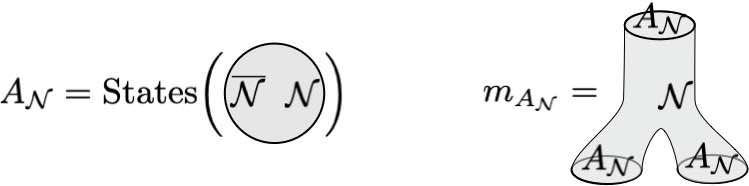}} \ee
and the fact that $A_\CN$ is a commutative (and more generally framed $E_2$) algebra comes from the fact that there's no preferred orientation in which to merge discs. The ``framed little disc'' that defined the framed $E_2$ operad is literally the disc (with a marking at the $\CN,\ol\CN$ junction) whose state space is $A_\CN$.

Now let $C$ denote the space of states on a bigon, \ie\ a disc with $\CD$ boundary on one side and $\CN$ boundary on the other:
\be C =    \text{States}\bigg( \raisebox{-.2in}{\includegraphics[width=.47in]{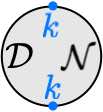}}\bigg) \simeq \C \label{bigon} \ee
By transversality, $C \simeq \C$. In this representation, however, it is clear that both $\CH_\CD$ and $A_\CN$ act on $C$. $\CH_\CD$ acts by merging with nearly all of the bigon except a small $\CN$ region; while $A_\CN$ acts by merging with the remaining small $\CN$ region:
\be \raisebox{-.4in}{\includegraphics[width=3.8in]{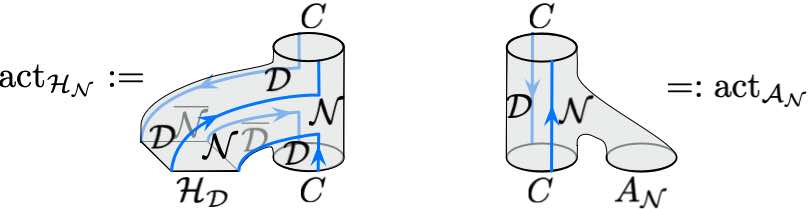}}   \label{Kos-act}  \ee
The actions commute because these regions don't overlap; and they should maximally commute because the regions are complements of each other.

\emph{Dually}, the object $C\in \text{Vect}$ (the category of line operators in the trivial 3d theory) defined by the bigon can itself be created by gluing together $\CH_\CD$ and $A_\CN$. Completeness ensures that there is a strip gluing \AC$_{\rm strip}$ that glues a bounded $\ol \CD$ strip on $\CH_\CD$ to an unbounded $\ol\CN$ strip on $A_\CN$:
\be \raisebox{-.4in}{\includegraphics[width=2.3in]{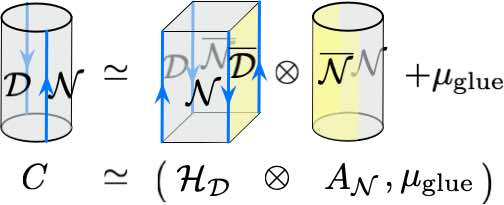}} \label{Kos-glue} \ee
In our perturbative context, the gluing must be implemented by deforming the tensor product $\CH_\CD\otimes A_\CN\in \text{Vect}$ by a Maurer-Cartan element $\mu_{\rm glue}$. (As a deformation of a dg vector space, this just means turning on a differential.) Since the vector space underlying $C$ is cohomologically trivial, we get $(\CH_\CD\otimes A_\CN,\mu_{\rm glue})\simeq \C$. Now the two actions from \eqref{Kos-act} are induced algebraically from $\CH_\CD$ and $A_\CN$ acting on themselves in the tensor product $\CH_\CD\otimes A_\CN$; this decomposition makes it clear that
\be A_\CN  = \text{End}_{\CH_\CD}(C)\,,\qquad \CH_\CD  = \text{End}_{A_\CN}(C)\,. \ee

\subsubsection{Example: enveloping algebras}

The pure matter theory of Section \ref{sec:matter} is entirely perturbative, and gives an easy example of the constructions here. One could also consider a perturbative version of B-twisted gauge theory, with gauge algebra $\fg$ rather than gauge group $G$.

In the case of pure matter $T^*V$, local operators on $\CN$ are given by $A_\CN = \text{Sym}^\bullet V^*[1] = \C[X_i]_{i=1}^n$ (polynomials in bosons of homological degree 1). The spark algebra on $\CD$ is $\CH_\CD = \Lambda^\bullet V = \C[\psi_+^i]_{i=1}^n$ (polynomials in fermions of degree zero). 
To get $(\CH_\CD\otimes A_\CN,\mu_{\rm glue}) \simeq \C$, we turn on the differential $\mu_{\rm glue} = \sum_i \psi_+^i X_i$.
As we've already noted in Section \ref{sec:matter}, this is classic Koszul duality from \cite{Priddy,beilinson1996koszul}.

For $\fg$ gauge theory with $T^*V$ matter, we are effectively dealing with a Lie superalgebra $\fh:= \fg\ltimes \Pi V$. Perturbative local operators on $\CN$ are now given by the Chevalley-Eilenberg complex of $\fh$, $A_\CN \simeq (\Lambda^\bullet (\fh^*[-1]),d_{CE})$, whereas local operators on $\CD$ are freely generated $\CA_\CD \simeq \Lambda^\bullet \fh[-1]$. Perturbative sparks on $\CD$ are given by the enveloping algebra $\CH_\CD \simeq U(\fh)$, whereas perturbative sparks on $\CN$ are given by functions on the Lie algebra $\CH_\CN\simeq \text{Sym}^\bullet \fh^*$.
 These are again classic examples of Koszul duals.

\subsubsection{Factorization homology and the bar construction}

In a perturbative setting, the ``factorization homology'' construction of sparks from Section \ref{sec:spark-HH} also gains extra significance. Once we assume that the identity line is a generator for boundary line operators, we can turn it sideways and use it to generate sparks --- choosing  `$L$' in \eqref{L-fact} to be the identity.

Let's explain what this means for (say) $\CH_\CD$. Suppose $\id_\CD$ is a generator of $\CC_\CD$, with endomorphism algebra $A_\CD$ (the local operators on $\CD$). Due to our fiber functor (or equivalently, by acting on the same bigon as in \eqref{bigon}), we also have $\C$ as both a left and right module for $A_\CD$. The construction of Section \ref{sec:spark-HH} then says that
\be \CH_\CD \simeq \C\otimes_{A_\CD}\C :=  HH_\bullet(A_\CD, \C\otimes \C)\,. \label{HD-bar} \ee
Namely, $\CH_\CD$ is Hochschild homology of $A_\CD$, with coefficients in the product of left and right modules $\C\otimes \C$.  Spelling it out, the RHS is a complex
\be \cdots \overset d\to \C\otimes A_\CD\otimes A_\CD\otimes A_\CD \otimes \C\overset d \to \C\otimes  A_\CD\otimes A_\CD \otimes \C \overset d\to \C\otimes  A_\CD \otimes \C \overset d\to \C\otimes \C\,, \label{bar-complex} \ee
with differentials given by alternating sums of product and action maps. For example, $d(1\otimes a\otimes b\otimes 1) = a(1)\otimes b\otimes 1 - 1\otimes ab\otimes 1 + 1\otimes a\otimes b(1)$. Physically, the cohomology classes in \eqref{HD-bar} all represent integrated descendants of local operators.

Let's recall schematically why \eqref{HD-bar} is a bialgebra. Hochschild homology of the form \eqref{HD-bar} for any $A_\infty=E_1$ algebra $A_\CD$ would be a coalgebra. Now $A_\CD$ is an $E_2$ algebra, meaning (by Dunn additivity) it has two compatible homotopy-associative products (and is thus commutative). One of the products --- corresponding to the horizontal direction in \eqref{L-fact} is used to construct the Hochschild complex. The other product remains and defines a product on Hochschild homology, making it a bialgebra.

Formula \eqref{HD-bar} gains extra significance when we can relate $A_\CD$ to local operators $A_\CN$ on the opposite boundary. They are related, of course, through interactions with the bulk. The $E_3$ algebra of bulk local operators $A_\CT$ has bulk-boundary maps to both $A_\CD$ and $A_\CN$. These maps can be derived in an ``$E_3$ way'' -- roughly meaning they get enhanced by integrating bulk descendants around configurations of boundary operators. We then expect that our transversality condition in ``Tannakian QFT'' implies that the derived tensor product of boundary algebras relative to the bulk is trivial:
\be  A_\CD \otimes_{A_\CT} A_\CN  \simeq \C\,.
\label{ADTN} \ee 
Similarly, completeness should imply that the $E_2$-derived-centers of either boundary algebra should recover the bulk:
\be  A_\CT \simeq \CZ^{E_2}(A_\CN) \simeq \CZ^{E_2}(A_\CD)\,. \label{ADTN-C} \ee

A somewhat trivial way to satisfy relations \eqref{ADTN}--\eqref{ADTN-C} is to just have
\be A_\CT \simeq A_\CD\otimes A_\CN\,,\qquad\text{with}\qquad A_\CD\simeq A_\CN^*\,, \label{ATDN-simple} \ee
\ie\ with $A_\CD$ and $A_\CN$ linear duals of each other. (The $E_3$ bracket in $A_\CT$ then pairs elements an element in $A_\CN$ with its dual in $A_\CD$.)
This is exactly what happened in our pure-matter example, where we had $A_\CN = \text{Sym}^\bullet V^*[-1]$ and $A_\CD = \text{Sym}^\bullet V[-1]$.
In this very special situation, when we can identify $A_\CD\simeq A_\CN^*$, the construction \eqref{HD-bar} coincides with the standard ``bar construction'' of the Koszul dual of $A_\CN$. It would be interesting to investigate the relation among \eqref{ADTN}, \eqref{ADTN-C}, and \eqref{HD-bar} more generally.

\subsection{Bulk sparks and bulk local operators}
\label{sec:K-3d}

Let's now treat line operators in the bulk perturbatively in a similar way, and reduce to two dimensions in order to connect with Koszul duality.

Suppose that the identity in the bulk, $\id_\CT$, generates bulk lines. So any $\ell\in \CC_\CT$ has a representation
\be \ell \simeq (V_\ell \otimes \id_\CT,\mu_\ell) \qquad\forall\;\ell\in \CC_\CT\,. \ee

We compactify to 2d by expressing the complement of a line in $\R^3$ as a circle fibration: $(\R^2\backslash \{(0,0)\})\times \R_t\simeq S^1\times \R_{>0}\times \R^t$, and reducing on the circle:
\be \raisebox{-.3in}{\includegraphics[width=3.8in]{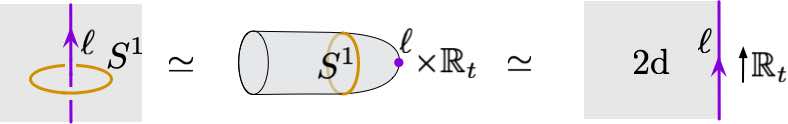}} \ee
The category of (say) right boundary conditions for the reduced theory may be identified with $\CC_\CT$. The generator $\id_\CT$ has a natural 3d origin, corresponding to smoothly capping off the compactification circle. In order to apply Koszul duality, we'd like to find a transverse generator to put on the other side, as a left boundary condition. In the work of \cite{Costello-4dCS,CostelloPaquette,PaquetteWilliams} the transverse boundary is defined asymptotically, by choosing an appropriate vacuum at infinity.

Alternatively, one may observe that after circle compactification, the reduced theory has local operators coming from 3d local operators \emph{and} from integrated descendants along $S^1$. Since the capped-off boundary labelled $\id_\CT$ kills all the descendants (while preserving all the 3d local operators), one may abstractly define a transverse boundary condition by the property that it kills/evaluates all the 3d local operators (while preserving all the integrated descendants). Geometrically, two such boundaries correspond to transverse Lagrangians with respect to a symplectic pairing between $H_0(S^1)$ and $H_1(S^1)$.%
\footnote{We thank S. Raghavendran for telling us about this perspective.}

If we have on hand a Tannakian QFT, however, we can define a boundary condition transverse to $\id_\CT$ directly in 3d. Indeed, we already did it, when constructing the fiber functor for bulk lines in \eqref{defFT}: the boundary condition $\CDCNd$ transverse to $\id_\CT$ is explicitly realized by placing $\CD$ along half of the compactification circle $S^1$, and $\CN$ along the other half:
\be \raisebox{-.4in}{\includegraphics[width=4.8in]{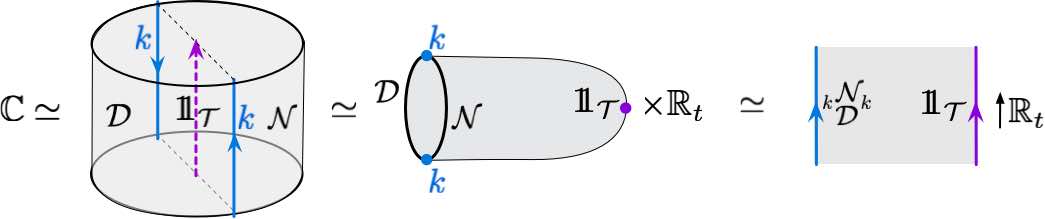}} \ee
Note that this hybrid $\CDCNd$ boundary has exactly the property we were after: all 3d bulk local operators are killed when brought to it, because they can be collided with one of the $k$ interfaces and evaluated to numbers; but all descendants of bulk operators integrated along $S^1$ will survive, either as sparks on the $\CD$ half or on the $\CN$ half.

We've now got our Koszul-duality setup. Setting $U = \text{End}(\CDCNd)$  (which of course is bulk sparks) and $A_\CT=\text{End}(\id_\CT)$ (which is bulk local operators), we must have
\be U \simeq A_\CT^!\,, \ee
with a fiber functor
\be \wt\CF_\CT:\CC_\CT\overset\sim\to U\Mod\,,\quad \wt\CF_\CT:\ell \mapsto (V_\ell,\mu_\ell)\,. \ee
We expect an equivalence of \emph{ribbon} categories
\be \CC_\CT\simeq U\Mod \simeq A_\CT\Mod\,, \ee
controlled by the ribbon Hopf algebra structure of $U$, and dually by the framed $E_3$ structure of $A_\CT$. It is expected that under Koszul duality these two algebraic structures precisely swap \cite{Lurie-DAGVI, Costello-4dCS}.

We can obtain a very concrete manifestation of this exchange of structure if we impose some additional assumptions. Suppose that bulk local operators take the simple form $A_\CT\simeq A_\CD\otimes A_\CN$ as a vector space, so that the bulk-boundary maps (colliding bulk local operators with the $\CD$ and $\CN$ boundaries)
\be   A_\CD \longleftarrow \CA_\CT \longrightarrow A_\CN  \label{ADN-surj} \ee
are 1) both surjective and 2) have kernels $A_\CN$ and $A_\CD$, respectively. (This is true for 3d $\CN=4$ pure matter theory, as well as perturbative 3d $\CN=4$ gauge+matter theory.) 
Let's also suppose that boundary spark algebras are generated by single integrated descendants of local operators, \emph{i.e.} by ``$HH_1$'' in the complex \eqref{bar-complex}. Let $\alpha\in \CH_\CD, \beta\in \CH_\CN$ be such generators, expressed as integrals
\be \alpha = \int_I a^{(1)}\,,\qquad \beta=\int_I b^{(1)} \qquad \text{for some}\qquad a\in A_\CD\,,\qquad b\in A_\CN\,.\ee
Then we can directly generalize the topological argument from Section \ref{sec:matter-Hopf} to analyze the Hopf pairing $h(\alpha,\beta)$. It is given by inserting the integrated descendants on the boundary of a ball:
\be  \raisebox{-.4in}{\includegraphics[width=5.2in]{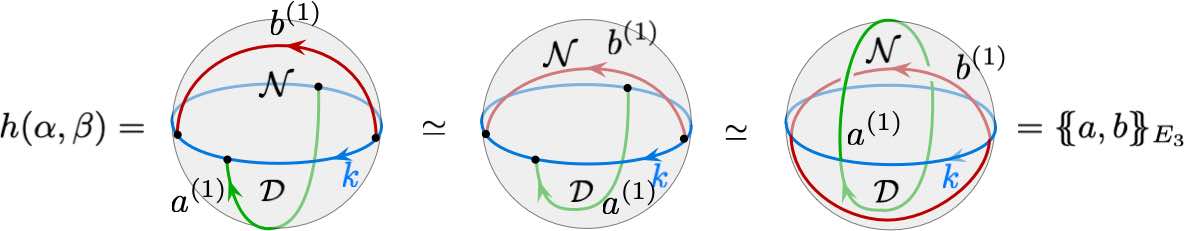}} \ee
By surjectivity of \eqref{ADN-surj}, the integrals can first be pulled into the bulk (as shown in the middle); and due to the kernels in \eqref{ADN-surj} the integrals can then be completed to closed loops (as shown on the RHS). The RHS is a Hopf link that defines the $E_3$ bracket of bulk local operators $a$ and $b$ \cite{descent}. Now by Theorem \ref{Thm:double}, the R-matrix in $U\simeq A_\CT^!$ inverts the bulk $E_3$ bracket in $A_\CT$.

If we combine our factorization $U\simeq \CH_\CD\otimes \CH_\CN$ (due to transversality) with Koszul duality for bulk and boundary local operators, we get a web of relations, such as
\begin{align} U &\simeq  A_\CT^! \simeq A_\CN^! \otimes A_\CD^! \overset{\eqref{HD-bar}}\simeq HH^\bullet(A_\CD,\C\otimes\C) \otimes HH^\bullet(A_\CN,\C\otimes\C)  \notag \\
&\simeq HH^\bullet(A_\CD\otimes A_\CN,\C\otimes \C)\,. \label{K-Uquant} \end{align}
These relations seem directly related to Tamarkin's approach \cite{tamarkin2007quantization} to producing formal quantizations of Lie bialgebras, via the $E_2$ operad.
In Tamarkin's work, $A_\CD\otimes A_\CN$ appears as the Chevalley-Eilenberg cochain complex associated to the classical Lie bialgebra that one would like to quantize. Then $U$, computed as Hochschild cohomology in \eqref{K-Uquant} (which is producing factorization cohomology of the $E_2$ operad) is the desired quantization.

\bibliographystyle{ytamsalpha}   
\bibliography{Spark}

\end{document}